\documentstyle[seceq,supplement,epsf]{ptptex}

\begin{document} 
\begin{center}
{\bf\Large Chapter 1}
\end{center}
\vskip 0.4cm
\begin{center}
{\bf\Large Black Hole Perturbation}
\end{center}
\bigskip

\centerline{\large
Yasushi Mino$^{1,2}$, Misao Sasaki$^1$, Masaru Shibata$^1$,
}

\medskip
\centerline{\large
Hideyuki Tagoshi$^3$ and Takahiro Tanaka$^1$
}
\bigskip

\begin{center}
{\em $^1$Department of Earth and Space Science,
 Osaka University, Toyonaka 560, Japan}\\
{\em $^2$Department of Physics, 
Kyoto University, Kyoto 606, Japan}\\
{\em $^3$National Astronomical Observatory,
Mitaka, Tokyo 181, Japan}\\
\end{center}
\bigskip
\centerline{\large\bf Abstract}
\medskip

In this chapter, we present analytic calculations of gravitational waves
from a particle orbiting a black hole. We first review the Teukolsky
formalism for dealing with the gravitational perturbation of a black
hole. Then we develop a systematic method to calculate higher order
post-Newtonian corrections to the gravitational waves emitted by an
orbiting particle. As applications of this method, we consider orbits
that are nearly circular, including exactly circular ones, slightly
eccentric ones and slightly inclined orbits off the equatorial plane of
a Kerr black hole and give the energy flux and angular momentum flux
formulas at infinity with higher order post-Newtonian corrections.
Using a different method that makes use of an analytic series
representation of the solution of the Teukolsky equation, we also give 
a post-Newtonian expanded formula for the energy flux absorbed by a Kerr
black hole for a circular orbit.

\section{Introduction}
\label{sec:intro}

In this chapter, we review recent progress in the analytic
calculations of gravitational waves from a particle orbiting a black
hole using a systematic post-Newtonian expansion method.
There has been substantial activities in this field recently and
there is a diversity of literature. Here we are mostly concerned with
the actual calculations of the gravitational waves from an orbiting
particle and we intend to make this chapter as self-contained as
possible. We do not, however, discuss much about implications of the
results to actual astrophysical situations.

In the black hole perturbation approach,
one considers gravitational waves from a particle of mass $\mu$ 
orbiting a black hole of mass $M$ assuming $\mu \ll M$. 
Although this method is restricted to the case when $\mu \ll M$, 
one can calculate very high order post-Newtonian corrections to 
gravitational waves using a relatively simple algorithm 
in contrast with the standard post-Newtonian analysis. 
This is because the fully relativistic effect of the spacetime 
curvature is naturally taken into account in the 
basic perturbation equation. 
We can also calculate numerically the gravitational waves without 
assuming the slow motion of its source. Then, we can easily 
investigate the convergence of the 
post-Newtonian expansion by comparing the result of the post-Newtonian 
approximation with the fully relativistic one. 
In this sense, the black hole perturbation method gives a very
important test of the post-Newtonian expansion. 
Further, since the effect of the spacetime curvature is naturally 
taken into account, 
we can easily investigate interesting relativistic effects such as 
tails of gravitational waves.

We consider the post-Newtonian wave forms and luminosity which 
are expanded by $v/c$, where $v$ is the order of the 
orbital velocity. 
The lowest order of the gravitational waves 
are given by the Newtonian quadrupole formula. 
We call the post-Newtonian formulas for the wave forms and 
luminosity which contain terms up to 
$O((v/c)^{n})$ beyond the Newtonian quadrapole formula 
as the $n/2$PN formulas.

Let us first briefly give a historical review.
The gravitational perturbation equation of a black hole using the 
Newman-Penrose formalism\cite{ref:NP} 
was derived by Bardeen and Press 
\cite{ref:BaPr} for the Schwarzschild black hole, and by 
Teukolsky \cite{ref:Teu} for the Kerr black hole. 
By using these equations, many numerical calculations of 
gravitational waves induced by the presence of a test particle have been
done. We do not list up all such works. Here, we only refer to three
articles;  Breuer\cite{ref:breuer}, Chandrasekhar\cite{ref:chandra},
and Nakamura, Oohara, and Kojima\cite{ref:NOK}. 

On the other hand,
analytic calculations of gravitational waves produced by the 
motion of a test particle have not been developed very much 
until recently. 
This direction of research was first done by 
Gal'tsov, Matiukhin and Petukhov\cite{ref:GMP} in which they 
considered a case when a particle moves a slightly eccentric orbit 
around a Schwarzschild black hole, and calculated the gravitational 
waves up to 1PN order.
Then, Poisson  \cite{ref:poisson} considered a case of circular orbit 
around a Schwarzschild black hole and calculated 
the wave forms and luminosity to 1.5PN order 
at which the tail effect appears. 
Cutler, Finn, Poisson and Sussman \cite{ref:CFPS} also worked on
the same problem numerically by using the least square fitting 
method, and obtained a formula for the luminosity to 2.5PN order. 
Subsequently, 
a highly accurate numerical calculation was carried out by 
Tagoshi and Nakamura \cite{ref:TN}. They obtained the 
formulas for the luminosity to 4PN order numerically by using 
the least square fitting method. They found the 
$\log v$ terms in the luminosity formula at 3PN and 4PN orders. 
They showed that, although the convergence 
of the post-Newtonian expansion is 
slow, the luminosity formula which is accurate to 
3.5PN order will be good enough to represent 
the orbital phase evolution of 
coalescing compact binaries accurately. 
After that, Sasaki \cite{ref:sasaki} found an analytic method and 
obtained the formulas which are needed to 
calculate the gravitational waves
to 4PN order. Then, 
Tagoshi and Sasaki\cite{ref:TS} obtained the 
gravitational wave forms and 
luminosity to 4PN order analytically, and confirmed the results of 
Tagoshi and Nakamura. 
These calculations were extended to 5.5PN order by Tanaka, Tagoshi, 
and Sasaki\cite{ref:TTS}.

In the case of orbits around a Kerr black hole, 
Poisson calculated the 1.5PN order corrections to the wave forms and
luminosity due to the rotation of the black hole and showed that the
result agrees with the standard post-Newtonian effect due to
spin-orbit coupling\cite{ref:poisson2}.
Then, Shibata, Sasaki, Tagoshi and Tanaka \cite{ref:SSTT} 
calculated the luminosity to 2.5PN order. They 
calculated the luminosity from a particle in circular orbit 
with small inclination from the equatorial plane. 
They used the Sasaki-Nakamura equation as well as the 
Teukolsky equation. 
This analysis was extended to 4PN order by 
Tagoshi, Shibata, Tanaka and Sasaki 
\cite{ref:TSTS} in which the orbit of 
the test particle was restricted to circular ones
on the equatorial plane. 
The analysis in the case of slightly eccentric orbit on the 
equatorial plane was also done by Tagoshi \cite{ref:tagoshi} 
to 2.5PN order. 

Tanaka, Mino, Sasaki, and Shibata \cite{ref:TMSS} 
considered the case when a spinning particle 
moves a circular orbits near the equatorial plane 
around a Kerr black hole, and derived the luminosity formula 
to 2.5PN order including the linear order effect of the particle's spin. 
They used the equations of motion of Papapetrou's \cite{ref:papa}
and the energy momentum tensor 
of the spinning particle given by Dixon \cite{ref:dixon}.

The absorption of gravitational waves into the black hole horizon, 
appearing at 4PN order in the Schwarzschild case, 
was calculated by Poisson and 
Sasaki in the case when a test particle 
is in a circular orbit\cite{ref:PoSa}. 
The black hole absorption in the 
case of rotating black hole appears at 2.5PN order \cite{ref:galtsov}. 
Recently a new analytic method to solve the homogeneous Teukolsky
equation was found by Mano, Suzuki, and Takasugi \cite{ref:MST}.
Using this method, the black hole absorption in the case of rotating
black hole was calculated by Tagoshi, Mano, and Takasugi \cite{ref:TMT}
to 6.5PN order beyond the quadrupole formula. 

If gravity is not described by the Einstein theory but by the
Brans-Dicke theory, there will appear scalar 
type gravitational waves as well as transverse-traceless 
gravitational waves. Such scalar type gravitational waves 
were calculated by Ohashi, Tagoshi and Sasaki\cite{ref:OTS}
in a case when a compact star is in a circular orbit on the equatorial
plane around a Kerr black hole.

The organization of this chapter is as follows.
We review the Teukolsky formalism for the black hole perturbation
in section \ref{sec:teukolsky}
and formulate a post-Newtonian expansion method of the Teukolsky
equation in section \ref{sec:pnexp}. 
Then we turn to the evaluation of gravitational waves by an orbiting
particle in the rest of sections.

First we consider circular orbits. 
In section \ref{sec:5.5PN}, we calculate the 
gravitational wave luminosity from a test particle in circular orbit
around a Schwarzschild black hole to 5.5PN order, based on Tanaka,
Tagoshi, and Sasaki\cite{ref:TTS}. This is the highest post-Newtonian
order achieved so far. Based on this result, we investigate the
convergence property of the post-Newtonian expansion in section
\ref{sec:convergence}. 
In section \ref{sec:kerrv8}, we consider circular orbits on the
equatorial plane around a Kerr black hole and calculate the luminosity
to 4PN order, based on Tagoshi, Shibata, Tanaka, 
and Sasaki\cite{ref:TSTS}. We find the luminosity contains the terms
which describe the effect of not only spin-orbit coupling but also the
effect of higher multipole moments of the Kerr black hole.

Next we consider slightly noncircular orbits. 
In section \ref{sec:schwaecc}, we calculate the $O(e^2)$ corrections to
the 4PN energy and angular momentum flux formulas in the case of a
slightly eccentric orbit around a Schwarzschild black hole, where $e$ is
the eccentricity. In section \ref{sec:kerrecc}, we consider a slightly
eccentric orbit on the equatorial plane around a Kerr black hole and
evaluate the $O(e^2)$ corrections to 2.5PN order, based on Tagoshi
\cite{ref:tagoshi}. 
Then in section \ref{sec:inclination}, we calculate the gravitational
waves induced by a test particle in circular orbit with small
inclination from the equatorial plane around a Kerr black hole and
evaluate the 2.5PN energy and angular momentum fluxes, based on Shibata,
Sasaki, Tagoshi, Tanaka\cite{ref:SSTT}.  
In section \ref{sec:Adbackreact}, we discuss the adiabatic orbital
evolution around a Kerr black hole under radiation reaction and show
that circular orbits will remain circular under adiabatic radiation
reaction but the stability of circular orbits can only be examined by an
explicit evaluation of the backreaction force.

In section \ref{sec:spin}, we consider the effect of the spin of a
particle. We first give a general formalism to treat the gravitational
radiation from a spinning particle orbiting a Kerr black hole. Then we
calculate the 2.5PN luminosity formula with the first order corrections
of the spin for circular orbits which are slightly inclined due to the
spin of the particle.

Finally, in section \ref{sec:bhabs}, 
we review a calculation of the black hole absorption based on 
Tagoshi, Mano, and Takasugi \cite{ref:TMT}. The black hole absorption
effect appears at $O(v^5)$ relative to the Newtonian quadrupole
luminosity for a Kerr black hole, while at $O(v^8)$ for a Schwarzschild
black hole. We show the energy absorption rate to $O(v^8)$ beyond the
lowest order for the Kerr case, i.e., $O(v^{13})$ or 6.5PN order beyond
the Newtonian quadrupole luminosity.

Since many of the calculations encountered in this black hole
perturbation approach are lengthy, various subsidiary equations and
formulas are deferred to appendices \ref{ap:spheroid} to
\ref{ap:absorpinx}.
In the rest of this chapter, we use the units of $c=G=1$. 
\section{Teukolsky formalism}
\label{sec:teukolsky}

In terms of the conventional Boyer-Lindquist coordinates, the metric of
a Kerr black hole is expressed as
\begin{eqnarray}
  \label{Kerr}
  ds^2&=&-{\Delta\over\Sigma}(dt-a\sin^2\theta\,d\varphi)^2
+{\sin^2\theta\over\Sigma}\left[(r^2+a^2)d\varphi-a\,dt\right]^2
\nonumber\\
&&+{\Sigma\over\Delta}dr^2+\Sigma d\theta^2\,,
\end{eqnarray}
where $\Sigma=r^2+a^2\cos^2\theta$ and $\Delta=r^2-2Mr+a^2$.
In the Teukolsky formalism\cite{ref:Teu}, 
the gravitational perturbations of a Kerr black 
hole are described by a Newman-Penrose quantity 
$\psi_4=-C_{\alpha\beta\gamma\delta}
n^{\alpha}\bar{m}^\beta n^\gamma \bar{m}^\delta$, where
$C_{\alpha\beta\gamma\delta}$
is the Weyl tensor, $n^\alpha=((r^2+a^2),-\Delta,0,a)/(2\Sigma)$
and
$m^\alpha=(ia\sin\theta,0,1,i/\sin\theta)/(\sqrt{2}(r+ia\cos\theta))$.

We decompose $\psi_4$ into Fourier-harmonic components according to
\begin{equation}
(r-i a \cos\theta)^4\psi_4=
\sum_{\ell m}\int d\omega 
e^{-i\omega t+i m \varphi} \ _{-2}S_{\ell m}(\theta)
R_{\ell m\omega}(r), \label{eq:psifour}
\end{equation}
The radial function $R_{\ell m\omega}$ and the angular function 
$_{s}S_{\ell m}(\theta)$ satisfy the Teukolsky 
equations with $s=-2$ as 
\begin{equation}
\Delta^2{d\over dr}\left({1\over \Delta}{dR_{\ell m\omega}\over dr}
\right)
-V(r) R_{\ell m\omega}=T_{\ell m\omega}, 
\label{eq:radTeukolsky}
\end{equation}
\begin{eqnarray}
& &\Bigl[{1 \over \sin\theta}{d \over d\theta}
 \Bigl\{\sin\theta {d \over d\theta} \Bigr\}
-a^2\omega^2\sin^2\theta
 -{(m-2\cos\theta)^2 \over \sin^2\theta}
\nonumber\\
& &~~~~~~~~~~~~~~~
+4a\omega\cos\theta-2+2ma\omega+\lambda\Bigr] 
{}_{-2}S_{\ell m}=0. \label{eq:spheroid}
\end{eqnarray}
The potential $V(r)$ is given by
\begin{equation}
V(r)=-{K^2+4i(r-M)K \over \Delta}+8i\omega r+\lambda,
\end{equation} 
where $K=(r^2+a^2)\omega-ma$ and $\lambda$ is the eigenvalue 
of $_{-2}S_{\ell m}^{a\omega}$. 
The angular function $_{s}S_{\ell m}(\theta)$ is 
the spin-weighted spheroidal harmonic which may be normalized as 
\begin{equation}
\int_0^{\pi} |_{-2}S_{\ell m}|^2 \sin\theta d\theta=1.
\label{eq:snorm}
\end{equation}
The source term $T_{\ell m\omega}$ is specified later. Here we
only mention that for orbits of our interest, which are bounded,
$T_{\ell m\omega}$ has support in a compact range of $r$.

We define two kinds of homogeneous solutions of the radial 
Teukolsky equation:
\begin{equation}
R^{\rm in}_{\ell m\omega} \to\cases{
B^{\rm trans}_{\ell m\omega}\Delta^2  e^{-ikr^*}\,&
for $r\to r_+$ \cr
r^3 B^{\rm  ref}_{\ell m\omega}e^{i\omega r^*}+
r^{-1}B^{\rm inc}_{\ell m\omega} e^{-i\omega r^*}\,&
for $r\to +\infty,$ \cr}
\label{eq:Rin} 
\end{equation} 
\begin{equation} 
R^{\rm up}_{\ell m\omega} \to\cases{ 
C^{\rm  up}_{\ell m\omega} e^{ik r^*}+ 
\Delta^2 C^{\rm ref}_{\ell m\omega} e^{-ik r^*}\,& 
for $r\to r_+,$ \cr
C^{\rm trans}_{\ell m\omega} r^3  e^{i\omega r^*}\,& 
for $r\to +\infty$\cr 
} 
\label{eq:Rup} 
\end{equation} 
where $k=\omega-ma/2Mr_+$ and 
$r^*$ is the tortoise coordinate defined by 
\begin{eqnarray}
r^{*}&=&\int {dr^*\over dr}dr \nonumber\\
&=&r+{2Mr_{+}\over {r_{+}-r_{-}}}\ln{{r-r_{+}}\over 2M}
-{2Mr_{-}\over {r_{+}-r_{-}}}\ln{{r-r_{-}}\over 2M},
\label{eq:rst}
\end{eqnarray}
where $r_{\pm}=M\pm \sqrt{M^2-a^2}$, and 
for definiteness, we have fixed the integration constant. 

We solve the radial Teukolsky equation by using the 
Green function method. A solution of the Teukolsky equation 
which has purely outgoing property at
infinity and has purely ingoing property at the horizon is given by 
\begin{equation}
R_{\ell m\omega}={1\over W_{\ell m\omega}}
 \left\{R^{\rm up}_{\ell m\omega}\int^r_{r_+}dr' 
 R^{\rm in}_{\ell m\omega}
T_{\ell m\omega}\Delta^{-2} 
+ R^{\rm in}_{\ell m\omega}\int^\infty_{r}dr' 
R^{\rm up}_{\ell m\omega} T_{\ell m\omega}\Delta^{-2}\right\}, 
\end{equation}
where the Wronskian $W_{\ell m\omega}$ is given by 
\begin{equation}
W_{\ell m\omega}=2 i \omega C^{\rm trans}_{\ell m\omega}
     B^{\rm inc}_{\ell m\omega}. 
\end{equation}
Then, the solution has an asymptotic property at the horizon as 
\begin{equation}
R_{\ell m\omega}(r\to r_+)\to 
{{B^{\rm trans}_{\ell m\omega} \Delta^2 e^{-i k r^*}} \over 
2i\omega C^{\rm trans}_{\ell m\omega}B^{\rm inc}_{\ell m\omega}}
\int^{\infty}_{r_+}dr' R^{\rm up}_{\ell m\omega} 
    T_{\ell m\omega}\Delta^{-2}
\equiv \tilde{Z}^{\rm H}_{\ell m\omega} \Delta^2 e^{-i k r^*}, 
\label{eq:zhole}
\end{equation}
The solution at infinity is also expressed as
\begin{equation}
R_{\ell m\omega}(r\to\infty)
 \to{r^3e^{i\omega r^*} \over 2i\omega 
   B^{\rm inc}_{\ell m\omega}}
\int^{\infty}_{r_+}dr'{T_{\ell m\omega}(r') 
R^{\rm in}_{\ell m\omega}(r') 
\over\Delta^{2}(r')}
\equiv \tilde Z_{\ell m\omega}^{\infty}r^3e^{i\omega r^*},
\label{eq:Rinfty}
\end{equation}
Here and in the following sections except for section \ref{sec:bhabs},
we focus on the gravitational waves emitted to infinity. Hence
$\tilde Z_{\ell m\omega}^\infty$ will be simply denoted as
$\tilde Z_{\ell m\omega}$. The gravitational waves absorbed into the
black hole horizon will be treated separately in section
\ref{sec:bhabs}. 

Now let us discuss the general form of the source term 
$T_{\ell m\omega}$. It is given by 
\begin{equation}
T_{\ell m\omega}
 =4\int d\Omega dt\rho^{-5}{\overline\rho}^{-1}(B_2'+B_2'^*)
e^{-im\varphi+i\omega t}{_{-2}S^{a\omega}_{\ell m} \over
\sqrt{2\pi}},
\label{eq:source}
\end{equation}
where
\begin{eqnarray}
B_2'&=&-{1 \over 2}\rho^8{\overline \rho}L_{-1}[\rho^{-4}L_0
(\rho^{-2}{\overline \rho}^{-1}T_{nn})]  \nonumber\\
&-&{1 \over 2\sqrt{2}}\rho^8{\overline \rho}\Delta^2 L_{-1}[\rho^{-4}
{\overline \rho}^2 J_+(\rho^{-2}{\overline \rho}^{-2}\Delta^{-1}
T_{{\overline m}n})],  \nonumber\\
B_2'^*&=&-{1 \over 4}\rho^8{\overline \rho}\Delta^2 J_+[\rho^{-4}J_+
(\rho^{-2}{\overline \rho}T_{{\overline m}{\overline m}})] \nonumber\\
&-&{1 \over 2\sqrt{2}}\rho^8{\overline \rho}\Delta^2 J_+[\rho^{-4}
{\overline \rho}^2 \Delta^{-1} L_{-1}(\rho^{-2}{\overline \rho}^{-2}
T_{{\overline m}n})], \label{eq:B2}\\
\nonumber
\end{eqnarray}
with
\begin{eqnarray}
\rho&=&(r-ia\cos\theta)^{-1}, \nonumber\\
L_s&=&\partial_{\theta}+{m \over \sin\theta}
-a\omega\sin\theta+s\cot\theta, \label{eq:rho}\\
J_+&=&\partial_r+{iK/\Delta}\,. \nonumber\\
\nonumber
\end{eqnarray}
In the above, $T_{nn}$, $T_{{\overline m}n}$ and 
$T_{{\overline m}{\overline m}}$  
are the tetrad components of the energy momentum tensor
($T_{nn}=T_{\mu\nu}n^\mu n^\nu$ etc.), and
the bar denotes the complex conjugation.

We consider $T_{\mu\nu}$ of a monopole particle of mass $\mu$.
The case of a spinning particle will be discussed in section
\ref{sec:spin} separately.
The energy momentum tensor takes the form,
\begin{equation}
T^{\mu\nu}
={\mu\over \Sigma \sin\theta dt/d\tau}{dz^{\mu}\over d\tau}
{dz^{\nu}\over d\tau}\delta(r-r(t))\delta(\theta-\theta(t))
  \delta(\varphi-\varphi(t)).
\end{equation}
where $z^\mu=\bigl(t,r(t),\theta(t),\varphi(t)\bigr)$ is a geodesic
trajectory and $\tau=\tau(t)$ is the proper time along the geodesic.
The geodesic equations in Kerr geometry are given by
\begin{eqnarray}
& &\Sigma{d\theta \over d\tau}
      =\pm\Bigl[C-\cos^2\theta \Bigl\{a^2(1-E^2)+
      {l_z^2 \over \sin^2\theta}\Bigr\}\Bigr]^{1/2} 
      \equiv \Theta(\theta),\nonumber\\
& &\Sigma{d\varphi \over d\tau}
      =-\Bigl(aE-{l_z \over \sin^2\theta}\Bigr)
      +{a \over \Delta}\Bigl(E(r^2+a^2)-al_z\Bigr) 
       \equiv \Phi, \nonumber\\
& &\Sigma{dt \over d\tau}=
      -\Bigl(aE-{l_z \over \sin^2\theta}\Bigr)a\sin^2\theta
      +{r^2+a^2 \over \Delta}\Bigl(E(r^2+a^2)-al_z\Bigr) 
      \equiv T, \nonumber\\
& &\Sigma{dr \over d\tau}=\pm\sqrt{R},  
\label{eq:geod}\\
\nonumber
\end{eqnarray}
where $E$, $l_z$ and $C$ are the energy, the $z$-component of the
angular momentum and the Carter constant of a test particle,
respectively.\footnote{These constants of 
motion are those measured in units of $\mu$. 
That is, if expressed in the standard units, $E$, $l_z$ and $C$ in 
Eq.~(\ref{eq:geod}) are to be replaced 
with $E/\mu$, $l_z/\mu$ and $C/\mu^2$,
respectively.}  $\Sigma=r^2+a^2\cos^2\theta$ and 
\begin{equation}
R=[E(r^2+a^2)-al_z]^2-\Delta[(Ea-l_z)^2+r^2+C].
\label{eq:Rgeo}
\end{equation}

Using Eq.~(\ref{eq:geod}), the tetrad components of the energy momentum 
tensor are expressed as
\begin{eqnarray}
T_{n\,n}&=&\mu{C_{n\,n} \over \sin\theta}
\delta(r-r(t)) \delta(\theta-\theta(t)) \delta(\varphi-\varphi(t)),
\nonumber\\
T_{{\overline m}\,n}&=&\mu{C_{{\overline m}\,n} \over \sin\theta}
\delta(r-r(t)) \delta(\theta-\theta(t)) \delta(\varphi-\varphi(t)),
\label{eq:tnn}\\
T_{{\overline m}\,{\overline m}}&=&\mu
{C_{{\overline m}\,{\overline m}} \over \sin\theta}
\delta(r-r(t)) \delta(\theta-\theta(t)) \delta(\varphi-
\varphi(t)),\nonumber\\
\nonumber
\end{eqnarray}
where
\begin{eqnarray}
C_{n\,n}&=&{1\over 4\Sigma^3 \dot t}\left[E(r^2+a^2)-al_z
+\Sigma{dr\over d\tau} \right]^2,\nonumber\\
C_{{\overline m}\,n}&=&
-{\rho \over 2\sqrt{2}\Sigma^2 \dot t}\left[E(r^2+a^2)-al_z
+\Sigma{dr\over d\tau} \right]
\left[i\sin\theta\Bigl(aE-{l_z \over \sin^2\theta}\Bigr)\right], 
\label{eq:cnn}\\
C_{{\overline m}\,{\overline m}}&=&
{\rho^2 \over 2\Sigma \dot t }
\left[i\sin\theta 
\Bigl(aE-{l_z \over \sin^2\theta}\Bigr)\right]^2,\nonumber\\
\nonumber
\end{eqnarray}
and $\dot t=dt/d\tau$.
Substituting Eq.~(\ref{eq:B2}) into Eq.~(\ref{eq:source}) and 
performing integration by part, we obtain
\begin{eqnarray}
T_{\ell m\omega}&=&{4\mu\over\sqrt{2\pi}}\int^{\infty}_{-\infty}
dt\int d\theta e^{i\omega t-im\varphi(t)} \nonumber\\
&\times&
\Bigl[-{1 \over 2}L_1^{\dag} \bigl\{ \rho^{-4}L_2^{\dag}(\rho^3 S)
\bigr\}
C_{n\,n}\rho^{-2}{\overline \rho}^{-1}\delta(r-r(t))
\delta(\theta-\theta(t)) \nonumber\\
&+&{\Delta^2 {\overline \rho}^2 \over \sqrt{2} \rho}
\bigl(L_2^{\dag} S+ia({\overline \rho}-\rho)\sin\theta S\bigr)
J_+ \bigl\{ C_{{\overline m}\,n}\rho^{-2}{\overline \rho}^{-
2}\Delta^{-1}
\delta(r-r(t))\delta(\theta-\theta(t)) \bigr\} \nonumber\\
&+&{1 \over 2\sqrt{2} }
L_2^{\dag}\bigl\{ \rho^3 S({\overline \rho}^2 \rho^{-4})_{,r} \bigr\}
C_{{\overline m}\,n}\Delta \rho^{-2}{\overline \rho}^{-2}
\delta(r-r(t))\delta(\theta-\theta(t)) \nonumber\\
&-&{1 \over 4}\rho^3 \Delta^2 S J_+\bigl\{\rho^{-4}
J_+\bigl({\overline \rho} \rho^{-2}C_{{\overline m}\,{\overline m}}
\delta(r-r(t))\delta(\theta-\theta(t))\bigr) \bigr\}
\Bigr],\label{eq:tass2}\\
\nonumber
\end{eqnarray}
where
\begin{equation}
L_s^{\dag}=\partial_{\theta}-{m \over \sin\theta}
+a\omega\sin\theta+s\cot\theta, 
\end{equation}
and $S$ denotes $_{-2}S_{\ell m}^{a\omega}(\theta)$ for simplicity.

We further rewrite Eq.~(\ref{eq:tass2}) as
\begin{eqnarray}
T_{\ell m\omega}&=&\mu\int^{\infty}_{-\infty}dt 
e^{i\omega t-i m \varphi(t)}
\Delta^2\Bigl[(A_{n\,n\,0}+A_{{\overline m}\,n\,0}+
A_{{\overline m}\,{\overline m}\,0})\delta(r-r(t)) \nonumber\\
&+&\left\{(A_{{\overline m}\,n\,1}+A_{{\overline m}\,{\overline m}\,1})
\delta(r-r(t))\right\}_{,r}
+\left\{A_{{\overline m}\,{\overline m}\,2}
\delta(r-r(t))\right\}_{,rr}\Bigr],
\label{eq:tone2} \\
\nonumber
\end{eqnarray}
where 
\begin{eqnarray}
A_{n\,n\,0}&=&{-2 \over \sqrt{2\pi}\Delta^2}
C_{n\,n}\rho^{-2}{\overline \rho}^{-1}
L_1^+\{\rho^{-4}L_2^+(\rho^3 S)\},\nonumber \\
A_{{\overline m}\,n\,0}&=&{2 \over \sqrt{\pi}\Delta} 
C_{{\overline m}\,n}\rho^{-3}
\Bigl[\left(L_2^+S\right)
\Bigl({iK \over \Delta}+\rho+{\overline \rho}\Bigr)
\nonumber \\
& &\qquad-a\sin\theta S {K \over \Delta}
   ({\overline \rho}-\rho)\Bigr],\nonumber \\
A_{{\overline m}\,{\overline m}\,0}
&=&-{1 \over \sqrt{2\pi}}\rho^{-3}{\overline \rho}
C_{{\overline m}\,{\overline m}}S\Bigl[
-i\Bigl({K \over \Delta}\Bigr)_{,r}-{K^2 \over \Delta^2}+
2i\rho {K \over \Delta}\Bigr],\nonumber \\
A_{{\overline m}\,n\,1}&=&{2\over \sqrt{\pi}\Delta }
\rho^{-3}C_{{\overline m}\,n}
[L_2^+S+ia\sin\theta({\overline \rho}-\rho)S],\nonumber \\
A_{{\overline m}\,{\overline m}\,1}
&=&-{2 \over \sqrt{2\pi}}
\rho^{-3}{\overline \rho}
C_{{\overline m}\,{\overline m}}S
\Bigl(i{K \over \Delta}+\rho\Bigr),\nonumber \\
A_{{\overline m}\,{\overline m}\,2}
&=&-{1\over \sqrt{2\pi}}\rho^{-3}{\overline \rho}
C_{{\overline m}\,{\overline m}}S. 
\label{eq:ann} 
\end{eqnarray}
Inserting Eq.~(\ref{eq:tone2}) into Eq.~(\ref{eq:Rinfty}), we obtain 
$\tilde Z_{\ell m\omega}$ as
\begin{equation}
\tilde Z_{\ell m\omega}=
{\mu\over2i\omega B^{\rm inc}_{\ell m\omega}}
\int^{\infty}_{-\infty}dt e^{i\omega t-i m \varphi(t)}
W_{\ell m\omega}\,,
\label{eq:zzq2}
\end{equation}
where 
\begin{eqnarray}
W_{\ell m\omega}&=&
\Bigl[R^{\rm in}_{\ell m\omega}\{A_{n\,n\,0}+A_{{\overline m}\,n\,0}
+A_{{\overline m}\,{\overline m}\,0}\}
\nonumber\\
& &-{dR^{\rm in}_{\ell m\omega} \over dr}\{ A_{{\overline m}\,n\,1}
+A_{{\overline m}\,{\overline m}\,1}\}
 +{d^2 R^{\rm in}_{\ell m\omega} \over dr^2}
 A_{{\overline m}\,{\overline m}\,2}\Bigr]_{r=r(t)}\,.
\label{eq:Wdef}
\end{eqnarray}

In this paper, we focus on orbits which are either circular (with or
without inclination) or eccentric but confined on the equatorial
plane. In either case, the frequency spectrum of $T_{\ell m\omega}$
becomes discrete. Accordingly $\tilde Z_{\ell m\omega}$ in
Eq.~(\ref{eq:zhole}) or (\ref{eq:Rinfty}) takes the form,
\begin{equation}
  \label{eq:tildeZ}
  \tilde Z_{\ell m\omega}=\sum_n \delta(\omega-\omega_n)
    Z_{\ell m\omega}\,.
\end{equation}
Then, in particular, $\psi_4$ at $r \to \infty$ 
is obtained from Eq.~(\ref{eq:psifour}) as
\begin{equation}
\psi_4={1\over r}\sum_{\ell m n}
Z_{\ell m\omega_n}{{}_{-2}S_{\ell m}^{a\omega_n} \over \sqrt{2\pi}}
e^{i\omega_n(r^*-t)+im\varphi}. 
\label{eq:psi4inf} 
\end{equation}
At infinity, $\psi_4$ is related to the two independent modes of
gravitational waves $h_+$ and $h_\times$ as
\begin{equation}
\psi_4={1\over2}(\ddot h_{+}-i\ddot h_{\times}). 
\label{eq:hplcr} 
\end{equation}
{}From Eqs.(\ref{eq:psi4inf}) and (\ref{eq:hplcr}), 
the luminosity averaged over $t \gg \Delta t$, where $\Delta t$ is the
characteristic time scale of the orbital motion (e.g., a period between
the two consecutive apastrons), is given by 
\begin{equation}
\left\langle{dE \over dt}\right\rangle=\sum_{\ell,m,n} 
{\Bigl|Z_{\ell m\omega_n}\Bigr|^2\over4\pi\omega_n^2} 
\equiv \sum_{\ell,m,n}\Bigl( {dE \over dt} \Bigr)_{\ell mn}.
\label{eq:Eflux} 
\end{equation}
In the same way, the time-averaged angular momentum flux is given by
\begin{equation}
\left\langle{dJ_z \over dt}\right\rangle=\sum_{\ell,m,n} 
{m\Bigl| Z_{\ell m\omega_n}\Bigr|^2\over4\pi\omega_n^3}
\equiv \sum_{\ell,m,n}\Bigl( {dJ_z \over dt} \Bigr)_{\ell mn}
=\sum_{\ell,m,n}{m\over\omega_n}\Bigl( {dE \over dt} \Bigr)_{\ell mn}\,.
\label{eq:Jflux} 
\end{equation}
\section{Post-Newtonian expansion of the ingoing wave solutions}
\label{sec:pnexp}
We consider the case when a test particle of mass $\mu$ is
in an orbit which is nearly circular around a Kerr black hole of mass
$M\gg\mu$ and describe a method to calculate the ingoing wave Teukolsky
functions $R^{\rm in}_{\ell m\omega}$ which 
are necessary to evaluate the 4PN formulas for the gravitational waves
energy and angular momentum fluxes emitted to infinity. In the
Schwarzschild case, we shall derive the
5.5PN luminosity formula in section \ref{sec:5.5PN}. A method to
calculate the ingoing wave solutions in 
this case is separately discussed in Appendix \ref{ap:RWthird} because
it is considerably more complicated than the method explained in this
section.

Using non-dimensional variables in the Teukolsky equation, 
we can see that the Teukolsky equation 
is expressed in terms of 
three basic variables, $z\equiv \omega r$, 
$\epsilon\equiv 2M\omega$ and
$a\omega\equiv q \epsilon/2$ where $q\equiv a/M$. 
In order to calculate the gravitational waves induced by 
a particle, we need to know the explicit form of the source 
terms $T_{\ell m\omega}(r)$. They will be given in the proceeding sections
for specified orbits. Here it is sufficient to note that they have
support only around $r=r_0$ where $r_0$ is the orbital radius 
for a circular orbit or the mean radius in the case of an eccentric
orbit (with small eccentricity). Hence from Eq.~(\ref{eq:Rinfty}), what
we need to know are the ingoing wave functions $R^{{\rm in}}(r)$ around
$r=r_0$, and their incident amplitudes $B^{{\rm inc}}$. 
Note that we do not need the transmission amplitudes 
$B^{{\rm trans}}$ to evaluate the gravitational waves at infinity. 
This fact considerably simplifies the calculations. 
Since we treat a test particle in a bound orbit which is nearly
circular, the contribution of $\omega$ to the Teukolsky functions 
comes from $\omega\sim m\Omega_\varphi$, where
$\Omega_\varphi\sim (M/r_0^3)^{1/2}$ is the orbital angular frequency.
We will evaluate $R^{\rm in}$ by setting 
three basic variables to be $z=\omega r_0\sim m (M/r_0)^{1/2}$ 
$\sim v$, $\epsilon\sim 2m (M/r_0)^{3/2}$ $\sim v^3$, and 
$a\omega$ $\sim qm (M/r_0)^{3/2}$ $\sim v^3$. 
Here, we have introduced a parameter $v\equiv (M/r_0)^{1/2}$ which 
represents the magnitude of the orbital velocity. 
We assume that $v$ is much smaller than the velocity of 
light; $v\ll 1$. Consequently, we also assume that 
$\epsilon\ll v\ll 1$. This relation is 
the basic assumption in obtaining the homogeneous solutions below. 

Now we calculate the ingoing wave solutions which are necessary to
calculate the luminosity to $O(v^8)$ beyond the lowest order for the
Kerr case. 
The method is mainly based on Shibata et al.\cite{ref:SSTT}
and Tagoshi et al.\cite{ref:TSTS}. 
An extension to $O(v^{11})$ calculations 
done by Tanaka, Tagoshi and Sasaki\cite{ref:TTS} for the Schwarzschild
case is given in Appendix \ref{ap:RWthird}.

First, we discuss the angular solutions. The angular solutions are the 
spin-weighted spheroidal harmonics. The angular equation 
(\ref{eq:spheroid}) contains only one small parameter $a\omega$. 
It is straightforward to 
calculate the spin-weighted spheroidal harmonic 
$_{-2}S_{\ell m}$ and its eigenvalue $\lambda$
by expanding the solution in power of $a\omega$. 
It can be done by the usual 
perturbation method\cite{ref:PT,ref:TSTS,ref:SSTT}. 
It is also possible to obtain them by using an expansion 
by means of Jacobi functions\cite{ref:fackerell}. 
The method and the results are given explicitly in Appendix
\ref{ap:spheroid}. 
Here we only show the eigenvalue $\lambda$ which is used to calculate 
the radial functions. 
The eigenvalue $\lambda$ is given by 
\begin{equation}
\lambda=\lambda_0+a\omega\lambda_1+a^2\omega^2 \lambda_2
+O((a\omega)^3)
\label{eq:pnlambda}
\end{equation}
where 
\begin{eqnarray}
&&\lambda_0=\ell(\ell+1)-2=(\ell-1)(\ell+2),
\nonumber\\
&&\lambda_1=-2m{\ell(\ell+1)+4\over\ell(\ell+1)}\,,
\nonumber\\
&&\lambda_2=-2(\ell+1)(c_{\ell m}^{\ell+1})^2
            +2\ell(c_{\ell m}^{\ell-1})^2
+{2\over 3}-{2 \over 3}{(\ell+4)(\ell-3)(\ell^2+\ell-3m^2) \over 
\ell(\ell+1)(2\ell+3)(2\ell-1)}\, ,
\end{eqnarray}
with
\begin{eqnarray}
& &c_{\ell m}^{\ell+1}={2 \over (\ell+1)^2}\Bigl\lbrack 
{(\ell+3)(\ell-1)(\ell+m+1)(\ell-m+1) \over (2\ell+1)(2\ell+3)}
 \Bigr\rbrack^{1/2},\nonumber\\
& &c_{\ell m}^{\ell-1}=-{2 \over \ell^2}\Bigl\lbrack 
{(\ell+2)(\ell-2)(\ell+m)(\ell-m) \over (2\ell+1)(2\ell-1)} 
\Bigr\rbrack^{1/2}.
\end{eqnarray}

Next we consider the homogeneous solution $R^{{\rm in}}$. 
We assume $\omega>0$ below. The solution for $\omega<0$ can be obtained
from the one for $\omega>0$ by using the symmetry of the homogeneous
Teukolsky equation which implies
$\overline{R_{\ell,m,\omega}}=R_{\ell,-m,-\omega}$. 
Here, we do not treat the Teukolsky equation directly. 
Instead, we transform the homogeneous Teukolsky equation to the
Sasaki-Nakamura equation \cite{ref:SN}, which is given by 
\begin{equation}
\Bigl[{d^2 \over dr^{*2}}-F(r){d \over dr^{*}}-U(r)\Bigr]
X_{\ell m\omega}=0. 
\label{eq:sneq}
\end{equation}
The function $F(r)$ is given by
\begin{equation}
F(r)={\eta_{,r} \over \eta}{\Delta \over r^2+a^2},
\label{eq:Fdef}
\end{equation}
where 
\begin{equation}
\eta=c_0+c_1/r+c_2/r^2+c_3/r^3+c_4/r^4,
\label{eq:etadef}
\end{equation}
with
\begin{eqnarray}
&c_0&=-12i\omega M+\lambda(\lambda+2)-12a \omega(a\omega-m),\nonumber\\
&c_1&=8ia[3a\omega-\lambda(a\omega-m)],\nonumber\\
&c_2&=-24ia M(a\omega-m)+12a^2[1-2(a\omega-m)^2 ],\nonumber\\
&c_3&=24ia^3(a\omega-m)-24Ma^2,\nonumber\\
&c_4&=12a^4.\label{eq:ceq}
\end{eqnarray}
The function $U(r)$ is given by
\begin{equation}
U(r)={\Delta U_1\over(r^2+a^2)^2}+G^2+{\Delta G_{,r}\over r^2+a^2}-FG,
\label{eq:Udef}
\end{equation}
where
\begin{eqnarray}
G&=&-{2(r-M) \over r^2+a^2}+{r\Delta \over (r^2+a^2)^2},\nonumber\\
U_1&=&V+{\Delta^2 \over \beta}
\Bigl[\Bigl(2\alpha+{\beta_{,r}  \over \Delta}\Bigr)_{,r}
-{\eta_{,r} \over \eta}
\Bigl(\alpha +{\beta_{,r} \over \Delta}\Bigr)\Bigr],\nonumber\\
\alpha&=&-i{K \beta \over \Delta^2}+3iK_{,r}
+\lambda+{6\Delta \over r^2},\nonumber\\
\beta&=&2\Delta\Bigl(-iK+r-M-{2\Delta \over r}\Bigr).\label{eq:ur}
\end{eqnarray}
When we set $a=0$, this transformation becomes the 
Chandrasekhar transformation \cite{ref:chandratrans} for the 
Schwarzschild black hole. 
The Sasaki-Nakamura equation was originally introduced, for the 
inhomogeneous case, to make the potential term short-ranged and 
to make the source term well-behaved at infinity. 
It is  not necessary to perform this transformation in this case,
since we are interested only in bound orbits. 
Nevertheless we choose to do this because
the lowest order solution becomes the spherical Bessel function
and we can apply the post-Newtonian expansion 
techniques developed for the Schwarzschild case by Poisson
\cite{ref:poisson} and Sasaki \cite{ref:sasaki}.

The relation between $R_{\ell m\omega}$
and $X_{\ell m\omega}$ is 
\begin{equation}
R_{\ell m\omega}={1 \over \eta}
\Bigl\{\Bigl(\alpha+{\beta_{,r} \over \Delta}\Bigr)\chi_{\ell m\omega}
-{\beta \over \Delta}\chi_{\ell m\omega,r} \Bigr\},\label{eq:xtor}
\end{equation}
where $\chi_{\ell m\omega}=X_{\ell m\omega} \Delta/(r^2+a^2)^{1/2}$.
Conversely, we can express $X_{\ell m\omega}$ in terms of 
$R_{\ell m\omega}$ as
\begin{equation}
X_{\ell m\omega}=(r^2+a^2)^{1/2}\,r^2\,J_{-}J_{-}
\left[{1\over r^2}R_{\ell m\omega}\right], \label{eq:rtox}
\end{equation}
where $J_{-}=(d/dr)-i(K/\Delta)$. 
Then the asymptotic behavior of 
the ingoing-wave solution $X^{{\rm in}}$ which corresponds
to Eq.~(\ref{eq:Rin}) is 
\begin{equation}
X^{\rm in}_{\ell m\omega}\to\cases{
A^{\rm ref}_{\ell m\omega} e^{i\omega r^*}+
A^{\rm inc}_{\ell m\omega} e^{-i\omega r^*}\,& 
for $r^* \to \infty$ \cr
A^{\rm trans}_{\ell m\omega} e^{-ik r^*}\, & for
$r^* \to -\infty$. }
\label{eq:xinasy}
\end{equation}
The coefficient 
$A^{\rm inc}$, $A^{\rm ref}$ and 
$A^{\rm trans}$ are respectively related to 
$B^{\rm inc}$, $B^{\rm ref}$ and 
$B^{\rm trans}$, defined in Eq.~(\ref{eq:Rin}), by 
\begin{eqnarray}
B^{\rm inc}_{\ell m\omega} &=&
   -{1\over 4\omega^2}A^{\rm inc}_{\ell m\omega}, 
\label{eq:ainbin}\\
B^{\rm ref}_{\ell m\omega}&=&-{4\omega^2\over c_0}
A^{\rm ref}_{\ell m\omega}, 
\label{eq:arefbref}\\
B^{\rm trans}_{\ell m\omega}&=&
 {1\over d_{\ell m\omega}}A^{\rm trans}_{\ell m\omega}, 
\label{eq:atransbtrans}\\
\end{eqnarray}
where $c_0$ is given in Eq.~(\ref{eq:ceq}) and
\begin{eqnarray*}
d_{\ell m\omega}&=&\sqrt{2Mr_+}[(8-24iM\omega-16M^2\omega^2)r_{+}^2 \\
& &+(12iam-16M+16amM\omega+24iM^2\omega)r_{+}
-4a^2m^2-12iamM+8M^2]. 
\end{eqnarray*}

Now we introduce the variable $z^*$ as 
\begin{eqnarray}
z^*&=&z+\epsilon\left[{z_{+}\over {z_{+}-z_{-}}}\ln(z-z_+)
-{z_{-}\over {z_{+}-z_{-}}}\ln(z-z_-)\right]\nonumber\\
&=&\omega r^*+\epsilon\ln\epsilon\,, \label{eq:zst}\\
\nonumber
\end{eqnarray}
where $z=\omega r$ and $z_\pm=\omega r_\pm$. 
To solve for $X^{{\rm in}}$, we set 
\begin{equation}
X^{\rm in}_{\ell m\omega}=\sqrt{z^2+a^2\omega^2}\xi_{\ell m}(z)
\exp\left(-i\phi(z)\right), 
\label{eq:xixi}
\end{equation}
where
\begin{eqnarray}
\phi(z) &=&\int dr\left({K\over \Delta}-\omega\right)\nonumber\\
&=&z^*-z-{\epsilon\over 2}mq{1\over {z_{+}-z_{-}}}
\ln{{z-z_{+}}\over {z-z_{-}}}\,.\label{eq:phase}\\
\nonumber
\end{eqnarray}
With this choice of the phase function, the ingoing wave boundary
condition at horizon reduces to that 
$\xi_{\ell m}$ is regular and finite at $z=z_+$.

Inserting Eq.~($\ref{eq:xixi}$) into Eq.~($\ref{eq:sneq}$) 
and expanding it in powers of $\epsilon=2M\omega$, we obtain
\begin{equation}
L^{(0)}[\xi_{\ell m}]=\epsilon L^{(1)}[\xi_{\ell m}]
+\epsilon Q^{(1)}[\xi_{\ell m}]+\epsilon^2 Q^{(2)}[\xi_{\ell m}]
+\epsilon^3 Q^{(3)}[\xi_{\ell m}]+\epsilon^4 Q^{(4)}[\xi_{\ell m}]
+O(\epsilon^5), \label{eq:eps}
\end{equation}
where $L^{(0)}$, $L^{(1)}$, $Q^{(1)}$ and $Q^{(2)}$ are differential 
operators given by
\begin{eqnarray}
L^{(0)}&=&{d^2\over dz^2}+{2\over z}{d\over dz}+\left(1-{\ell(\ell+1)
\over z^2}\right), \\
L^{(1)}&=&{1\over z}{d^2\over dz^2}
+\left( {{1}\over z^2}
+{2i\over z}\right){d\over dz}
-\left( {{4}\over {z^3}}-{i\over z^2}
+{1\over z}\right), \\
Q^{(1)}&=&{{iq\lambda_1}\over 2z^2}{d\over dz}
-{{4imq}\over {l(l+1)z^3}}, \\
\nonumber
\end{eqnarray} 
The formulas for $Q^{(2)}$, $Q^{(3)}$ and $Q^{(4)}$ are very 
complicated, and they are given explicitly in Appendix \ref{ap:Q}. 
Note that, when we set $a=0$, all $Q^{(n)}$ vanish. 

By expanding $\xi_{\ell m}$ in terms of $\epsilon$ as
\begin{equation}
\xi_{\ell m}=\sum_{n=0}^{\infty}\epsilon^n \xi^{(n)}_{\ell m}(z), 
\label{eq:expa}
\end{equation}
we obtain from Eq.~(\ref{eq:eps}) the iterative equations, 
\begin{equation}
  \label{eq:PNexpand}
L^{(0)}[\xi^{(n)}_{\ell m}]=z^{-2} W^{(n)}_{\ell m}\,,
\end{equation}
where
\begin{eqnarray}
& &W^{(0)}_{\ell m}=0, 
\label{eq:zeroji}\\
& &W^{(1)}_{\ell m}=z^2\left(L^{(1)}[\xi^{(0)}_{\ell m}]
+Q^{(1)}[\xi^{(0)}_{\ell m}]\right), 
\label{eq:first}\\
& &W^{(2)}_{\ell m}=z^2\left(L^{(1)}[\xi^{(1)}_{\ell m}]
+Q^{(1)}[\xi^{(1)}_{\ell m}]+Q^{(2)}[\xi^{(0)}_{\ell m}]\right),
 \label{eq:second}\\
& &W^{(3)}_{\ell m}=z^2\left(L^{(1)}[\xi^{(2)}_{\ell m}]
+Q^{(1)}[\xi^{(2)}_{\ell m}]+Q^{(2)}[\xi^{(1)}_{\ell m}]
+Q^{(3)}[\xi^{(0)}_{\ell m}]\right),
 \label{eq:third}\\
& &W^{(4)}_{\ell m}=z^2\left(L^{(1)}[\xi^{(3)}_{\ell m}]
+Q^{(1)}[\xi^{(3)}_{\ell m}]+Q^{(2)}[\xi^{(2)}_{\ell m}]
+Q^{(3)}[\xi^{(1)}_{\ell m}]+Q^{(4)}[\xi^{(0)}_{\ell m}]\right).
 \label{eq:forth}
\end{eqnarray}
As $\epsilon=2GM\omega$ if we recover $G$, the above expansion 
corresponds to the post-Minkowski expansion of the vacuum 
Einstein equations. 

The iterative equations (\ref{eq:PNexpand}) have been
obtained by expanding Eq.~(\ref{eq:sneq}) in powers of $\epsilon$ by
regarding $z=\omega r$ as the independent variable. Since the horizon is
at $z=z+=O(\epsilon)$, this procedure implicitly
assumes that $\epsilon\ll z$. Consequently, we cannot apply the above
expansion near the horizon where the ingoing wave boundary condition is
to be imposed. To implement the boundary condition correctly, we have to
consider a series solution of $\xi_{\ell m}$ which is valid near the
horizon as well as in the range $\epsilon\ll z$ and match it to the
series solution of the form (\ref{eq:expa}). Recently this matching
problem has been rigorously solved by Mano, Suzuki, and
Takasugi\cite{ref:MST} for the original Teukolsky equation. However, for
our present purpose, it is sufficient to resort to a simple
power-counting argument, by which it is possible to implement the
boundary condition of $\xi_{\ell m}$ at the horizon to the behaviors of
$\xi^{(n)}_{\ell m}$ at $z\gg\epsilon$ for $n\le2\ell$ (for
$n\le2\ell+1$ in the Schwarzschild case; see Appendix 
\ref{ap:RWthird}). 

Since the ingoing wave boundary condition is that $\xi_{\ell  m}$ 
is regular at horizon, if we introduce an independent variable
$x:=(z-z_+)/\epsilon$ we can expand $\xi_{\ell m}$ near the horizon as
\begin{equation}
  \label{Hexpand}
  \xi_{\ell m}=\sum_{n=0}^\infty\epsilon^n\xi^{\{n\}}_{\ell m}(x),
\end{equation}
This means we have
\begin{equation}
  \label{xiatH}
  \xi_{\ell m}(0)
=\xi^{\{0\}}_{\ell m}(0)\sum_{n=0}^\infty\epsilon^n C_n\,,
\end{equation}
where $C_n=\xi^{\{n\}}_{\ell m}(0)/\xi^{\{0\}}_{\ell m}(0)$. In other
words, $\xi_{\ell m}(0)$ should have a well-defined limit for
$\epsilon\to0$ except for the overall normalization factor
$\xi^{\{0\}}_{\ell m}(0)$.
Keeping this property in mind, let us consider the boundary
conditions for Eqs.~(\ref{eq:zeroji}) -- (\ref{eq:forth}).

The general solution to Eq.~(\ref{eq:zeroji}) is immediately obtained as 
\begin{equation}
\xi^{(0)}_{\ell m}
=\alpha_{\ell m}^{(0)} j_\ell+ \beta_{\ell m}^{(0)} n_\ell, 
\label{eq:xizero}
\end{equation} 
where $j_{\ell}$ and $n_{\ell}$ are the spherical Bessel functions.
The coefficients $\alpha_{\ell m}^{(0)}$ and $\beta_{\ell m}^{(0)}$ are 
to be determined by the boundary condition.
For convenience, we normalize the solution $X^{\rm in}$ such that 
the incident amplitude $A^{\rm inc}$ is of order unity.
Then both $\alpha_{\ell m}^{(0)}$ and $\beta_{\ell m}^{(0)}$ must be of
order unity. Since $j_{\ell}(z)\sim z^\ell$ and $n_{\ell}(z)\sim
z^{-\ell-1}$ for $z\ll 1$, the latter is $O(\epsilon^{2\ell+1})$ larger
than the former near the horizon where $z=O(\epsilon)$. Hence
Eq.~(\ref{xiatH}) implies we should set $\beta_{\ell m}^{(0)}=0$.
As for the value of $\alpha_{\ell m}^{(0)}$, since it only contributes
to the overall normalization of $X^{\rm in}$, we set 
$\alpha_{\ell m}^{(0)}=1$ for convenience.

Inspection of Eqs.~(\ref{eq:first}) -- (\ref{eq:forth}) reveals that
the solution $\xi_{\ell m}^{(n)}$ behaves as $z^{\ell-n}$
plus the homogeneous solution $\alpha_{\ell m}^{(n)} j_{\ell} 
+\beta_{\ell m}^{(n)} n_{\ell}$ for $z\to0$. As for 
$\alpha_{\ell m}^{(n)}$ ($n\ge1$), they simply contribute to 
renormalizations of $\alpha_{\ell m}^{(0)}$. Hence we put them to zero.
As for $\beta^{(n)}_{\ell m}$, from the same argument as given
above, we find they may become non-zero only for $n\ge2\ell+1$. Since
$\ell\ge2$ and $\epsilon=O(v^3)$, this implies that the near zone
contribution of $n_{\ell}\sim z^{-\ell-1}$, which is $O(v^{-(2\ell+1)})$
greater than the lowest order term $j_\ell$, to the gravitational waves
emitted to infinity may arise only at $O(v^{10})$ beyond the quadrupole
order. Since the post-Newtonian corrections we shall consider for the
Kerr case are those up to $O(v^8)$, we set
$\beta^{(n)}_{\ell m}=0$ and solve the iterative equations
(\ref{eq:PNexpand}) to $O(\epsilon^4)$ with the boundary conditions that
$\xi^{(n)}_{\ell m}\sim z^{\ell-n}$ at $z\to0$.
 We note that, in the Schwarzschild case which is discussed 
separately in Appendix \ref{ap:RWthird}, these boundary conditions
turn out to be appropriate for $n\le2\ell+1$; i.e., up to one power of
$\epsilon$ higher than the Kerr case.

To calculate $\xi^{(n)}_{\ell m}$ for $n\geq 1$, we rewrite
Eqs.~(\ref{eq:first}) -- (\ref{eq:forth}) in the indefinite
integral form by using the spherical Bessel functions as
\begin{equation}
\xi^{(n)}_{\ell m}=n_\ell\int^z dz j_\ell W^{(n)}_{\ell m}-
j_\ell\int^z dz n_\ell W^{(n)}_{\ell m}\,.
\label{eq:green}
\end{equation}
The calculation is straightforward but tedious.  
All the formulas which are needed to calculate the above integration 
to obtain $\xi^{(n)}_{\ell m}$ for $n\le2$ are shown in Appendix 
of Sasaki\cite{ref:sasaki}. 
They are recapitulated in an alternative way in Appendix 
\ref{ap:RWthird}.
Using those formulas, we have for $n=1$,
\begin{eqnarray}
\xi^{(1)}_{\ell m} 
  &=&{(\ell-1)(\ell+3)\over2(\ell+1)(2\ell+1)}j_{\ell+1}
    -\left({\ell^2-4\over2\ell(2\ell+1)}
        +{2\ell-1\over\ell(\ell-1)}\right)j_{\ell-1}
\cr
 &&~+R_{\ell,0} j_0
   +\sum_{m=1}^{\ell-2}\left({1\over m}+{1\over m+1}\right)
           R_{\ell,m} j_m
   -2D_{\ell}^{nj}+ij_\ell\ln z \nonumber\\
&+&{imq\over 2} \left({\ell^2+4}\over \ell^2(2\ell+1)\right)j_{\ell-1}
+{imq\over 2} \left({(\ell+1)^2+4}\over {(\ell+1)^2(2\ell+1)}\right)
j_{\ell+1},\label{eq:xsi}
\end{eqnarray}
where $D_{\ell}^{nj}$ is an extension of the spherical Bessel function 
defined in subsection \ref{tamadefinitions} of Appendix
\ref{ap:RWthird};
\begin{equation}
 D_{\ell}^{nj}={1\over 2}\left[j_{\ell} {\rm Si}(2z)-n_{\ell} 
\left({\rm Ci}(2z)-\gamma-\ln 2z\right)\right], 
\end{equation}
where $\gamma=0.5772\cdots$ is the Euler constant,
${\rm Ci}(x)=-\int^{\infty}_x dt\cos t/t$ and 
${{\rm Si}(x)}=\int^x_0 dt\sin t/t$,
and $R_{m,k}$ is a polynomial of the inverse power of $z$ defined 
by
\begin{eqnarray}
 R_{m,k}& = & z^2 (n_m j_k -j_m n_k) \cr
      & = & -\sum_{r=0}^{[(m -k -1)/2]} (-1)^r
    {(m-k-1-r)!~ \Gamma\left(m+{1\over 2}-r \right)
    \over r!~(m -k -1 -2r)!~ \Gamma\left(k +{3\over 2}+r\right)}
    \left({2\over z}\right)^{m-k-1-2r},    
\label{eq:Rmkdef}
\end{eqnarray}
for $m>k$ and it is defined by 
\begin{equation}
 R_{m,k}=-R_{k,m}.
\label{eq:Rmksym}
\end{equation}
for $m<k$. 

Next we consider $\xi^{(2)}_{\ell m}$. From Eqs.(\ref{eq:green}) and 
(\ref{eq:xsi}), we obtain $\xi^{(2)}_{\ell m}$ as 
\begin{equation}
\xi^{(2)}_{\ell m}=f^{(2)}_{\ell}+ig^{(2)}_{\ell}
+k^{(2)}_{\ell m}(q)\,,
\end{equation}
where $f^{(2)}_{\ell}$ and $g^{(2)}_{\ell}$ are the real 
and imaginary parts of $\xi^{(2)}_{\ell m}$ 
in the Schwarzschild limit, respectively, and 
$k^{(2)}_{\ell m}(q)$ is the correction term due to non-vanishing
$q=a/M$.
For $\ell=2$ and $\ell=3$, $f^{(2)}_{\ell}$ are given by
\begin{eqnarray}
  f^{(2)}_2
  &=&-{389\over70z^2}j_0-{113\over420z}j_1+{1\over7z}j_3
     +4 D_2^{nnj}\cr
  &&-{5\over 3z}D_2^{nj}+{10\over3}D_1^{nj}
    +{6\over z}D^{nj}_0+{107\over 105}D^{nj}_{-3}
\cr
  &&-{107\over210}j_2\ln z-{1\over2}j_2(\ln z)^2\,,
\cr
 f^{(2)}_3
  &=&{1\over4z}j_4+{323\over49z}j_2-{5065\over294z^2}j_1
    -\left({1031\over588z}+{445\over14z^3}\right)j_0
    +{65\over6z^2}n_0-{65\over6z}n_1
\cr
 &&-{3\over z}D^{nj}_3+{13\over 3}D^{nj}_2+{9\over z}D^{nj}_1
       +{30\over z^2}D^{nj}_0-{13\over21}D^{nj}_{-4}
\cr
 &&+4D_3^{nnj}
      -{13\over42}j_3\ln z-{1\over2}j_3(\ln z)^2\,,
\label{eq:f2f3}
\end{eqnarray}
where $D_{\ell}^{nnj}$ is defined in Appendix
\ref{ap:RWthird}, Eq.~(\ref{eq:tamadefD}). 
As explained in subsection \ref{tamaremarks} of Appendix
\ref{ap:RWthird}, the term $g^{(2)}_{\ell}$ is given for any
$\ell$ as 
\begin{equation}
g^{(2)}_{\ell}=-{1\over z} j_\ell+f^{(1)}_{\ell}\ln z. 
\label{eq:g2ell}
\end{equation}
The term $k^{(2)}_{\ell m}(q)$ is given for $\ell=2$ and $\ell=3$ by
\begin{eqnarray}
k^{(2)}_{2 m}&=&
{{191\,i}\over {180}}\,m\,q\,j_0 - {{{m^2}\,{q^2}\,j_0}\over {30}} - 
  {{m\,q\,j_1}\over {10}} \nonumber\\
& &- {{68\,i}\over {63}}\,m\,q j_2 - 
     {{{q^2}}\over {392}} j_2 - {{73\,{m^2}\,{q^2}}\over {1764}}\,j_2 
+ {{7\,m\,q\,j_3}\over {180}} - {i\over {72}}\,{q^2}\,j_3 + 
  {i\over {324}}\,{m^2}\,{q^2}\,j_3 \nonumber\\
& &+ {{11\,i}\over {420}}\,m\,q\,j_4 - {{{q^2}\,j_4}\over {392}} - 
  {{71\,{m^2}\,{q^2}\,j_4}\over {8820}} 
+  {{13\,i}\over 6}\,m\,q\,n_1 \nonumber\\
& &+ mq\left( -{{j_1 }\over 5} - 
     {{13\,j_3}\over {90}} \right) \,\ln z 
+   imq \left( {{-2}\over 5}\,D_1^{nj} - 
     {{13}\over {45}}\,D_3^{nj}\right), 
\end{eqnarray}
\begin{eqnarray}
k^{(2)}_{3 m}&=&
{{3527\,i}\over {840}}\,m\,q\,j_1 
   - {{2\,{m^2}\,{q^2}\,j_1}\over {315}} - 
  {{m\,q\,j_2}\over {36}} - {{5\,i}\over {504}}\,{q^2}\,j_2 + 
  {{5\,i}\over {2268}}\,{m^2}\,{q^2}\,j_2 \nonumber\\
& &- {{379\,i}\over {360}}\,m\,q\,j_3 - 
  {{{q^2}\,j_3}\over {360}} - {{7\,{m^2}\,{q^2}\,j_3}\over {720}} + 
  {{3\,m\,q\,j_4}\over {160}} - {i\over {140}}\,{q^2}\,j_4 + 
  {i\over {1120}}\,{m^2}\,{q^2}\,j_4 \nonumber\\ 
& &+ {{97\,i}\over {5040}}\,m\,q\,j_5 - {{{q^2}\,j_5}\over {360}} - 
  {{17\,{m^2}\,{q^2}\,j_5}\over {5040}} 
- {{103\,i}\over {48}}\,m\,q\,n_0 + 
  {{25\,i}\over 8}\,m\,q\,n_2 \nonumber\\ 
& &- \left( {{13\,m\,q\,j_2}\over {126}} + 
  {{5\,m\,q\,j_4}\over {56}} \right) \,\ln z
+ imq\left( {{-13}\over {63}}\,D_2^{nj} - 
     {{5}\over {28}}\,D_4^{nj} \right).
\end{eqnarray} 

As noted previously, 
the source term $T_{\ell m\omega}$ has support only around $r=r_0$,
hence around $z=\omega r_0=O(v)$. 
Therefore, to evaluate the source integral, we only need $X^{\rm in}$ at 
$z=O(v)\ll 1$, apart from the value of the incident amplitude 
$A^{{\rm inc}}$. 
Hence the post-Newtonian expansion 
of $X^{\rm in}$ corresponds to the expansion 
not only in terms of $\epsilon=O(v^3)$, but also of
$z$ by assuming $\epsilon\ll z\ll 1$. 
In order to evaluate the gravitational 
wave luminosity to $O(v^8)$ beyond the leading order, 
we must calculate the series expansion of $\xi^{(n)}_{\ell m}$ in powers
of $z$ for $n\le6-\ell$ for each $2\le\ell\le6$.
This follows from a simple power counting. The leading order
contribution of the $\ell$-th pole is $O(v^{2(\ell-2)})$ smaller
than that of the quadrupole, while the $n$-th post-Minkowski terms are
$O(\epsilon^n z^{-n})=O(v^{2n})$ relative to the lowest order terms in
the near-zone. Hence the leading term of $\xi^{(n)}_{\ell m}$
contributes at $O(v^{2(\ell-2)+2n})$ and $O(v^8)$ is attained for 
$\ell+n=6$ (see Appendix C of Ref.~\citen{ref:SSTT}). 

To evaluate $A^{{\rm inc}}$, we need to know the asymptotic 
behavior of $\xi^{(n)}_{\ell m}$ at infinity. Since the accuracy 
of $A^{{\rm inc}}$ we need is $O(\epsilon^2)$, we do not have to
calculate $\xi^{(3)}_{\ell m}$ and $\xi^{(4)}_{\ell m}$ 
in closed analytic form. We need only 
the series expansion formulas for $\xi^{(3)}_{\ell m}$ and
$\xi^{(4)}_{\ell m}$ around $z=0$, 
which are easily obtained from Eq. (\ref{eq:green}). 
This is also true for $\xi^{(2)}_{4 m}$ for $\ell=4$. 
Inserting $\xi^{(n)}_{\ell m}$ into Eq.~(\ref{eq:xixi}) and expanding it 
by $z$ and $\epsilon$ assuming $\epsilon\ll z\ll 1$, we obtain 
\begin{eqnarray}
\xi^{(3)}_{2 m}&=&
{{{-{q^2}}\over {30z}} - {i\over {30z}}\,m\,{q^3}}
+ {{-i}\over {30}} + {{7\,m\,q}\over {180}} - 
  {i\over {60}}\,{m^2}\,{q^2} + 
  {{m\,{q^3}}\over {36}} - 
  {{{m^3}\,{q^3}}\over {90}} \nonumber\\
& &
- {{m\,q\,\ln z}\over {30}} - 
  {i\over {30}}\,{m^2}\,{q^2}\,\ln z \nonumber\\
& &
+  z\,\left( {{319}\over {6300}} + 
     {{100637\,i}\over {441000}}\,m\,q - 
     {{{q^2}}\over {180}} 
+    {{17\,{m^2}\,{q^2}}\over {1134}} + 
     {{83\,i}\over {5880}}\,m\,{q^3} 
\right.\nonumber\\
& &\left.
-    {{61\,i}\over {13230}}\,{m^3}\,{q^3} + 
     {{\ln z}\over {15}} - 
     {{106\,i}\over {1575}}\,m\,q\,\ln z - 
     {i\over {30}}\,m\,q\,{{(\ln z)}^2} \right) 
\nonumber\\
& &
+O(z^2),
\label{eq:xi32}\\
\xi^{(4)}_{2 m}&=&{{{q^4}}\over {80\,{z^2}}}
+O(z^{-1}),
\label{eq:xi42}\\
\xi^{(3)}_{3 m}&=&
{{-i}\over {1260}}\,m\,q + 
  {{{m^2}\,{q^2}}\over {1890}} - 
  {i\over {1260}}\,m\,{q^3} - 
  {i\over {3780}}\,{m^3}\,{q^3}
 +O(z), 
\label{eq:xi33}\\
\xi^{(2)}_{4 m}&=&
\Bigl( {1\over {1764}} - {{11\,i}\over {15120}}\,m\,q + 
    {{{q^2}}\over {10584}} - {{19\,{m^2}\,{q^2}}\over {105840}} \Bigr) \,
  {z^2} +O(z^3). 
\label{eq:xi24}
\end{eqnarray} 
Inserting these $\xi^{(n)}_{\ell m}$ into Eq.~(\ref{eq:xixi}) and
expanding the result in terms of $\epsilon=2M\omega$, 
we obtain the post-Newtonian expansion of $X^{{\rm in}}$.  
The transformation from $X^{{\rm in}}$ to $R^{{\rm in}}$ 
is done by using Eq.~(\ref{eq:xtor}). 

Next, we consider $A^{{\rm inc}}$ to $O(\epsilon^2)$. 
Using the relation $j_{\ell+1}\sim$ $-j_{\ell-1}\sim$ 
$(-1)^{\ell+n}n_{2n-\ell}$ at $z\sim\infty$, etc., 
we obtain the asymptotic behaviors of 
$\xi^{(1)}_{\ell m}$ and $\xi^{(2)}_{\ell m}$
at $z\sim\infty$ as 
\begin{eqnarray}
\xi^{(1)}_{\ell m}&\sim& p^{(1)}_{\ell m} j_\ell+(q^{(1)}_{\ell m}
-\ln z)n_\ell+i j_\ell \ln z, \\
\xi^{(2)}_{\ell m}&\sim& 
\left(p^{(2)}_{\ell m}+q^{(1)}_{\ell m} \ln z-(\ln z)^2\right) j_\ell
+(q^{(2)}_{\ell m}-p^{(1)}_{\ell m} \ln z) n_\ell
\nonumber\\
& &
+i p^{(1)}_{\ell m}j_\ell \ln z+i(q^{(1)}_{\ell m}-\ln z) 
n_\ell \ln z
\end{eqnarray}
where
\begin{eqnarray}
p^{(1)}_{\ell m}&=&-{\pi\over 2}, \\
q^{(1)}_{\ell m}&=&{1\over 2}\left[\psi(\ell)+\psi(\ell+1)
+{{(\ell-1)(\ell+3)}\over \ell(\ell+1)}\right]-\ln 2
-{{2imq}\over {\ell^2(\ell+1)^2}}\,, \label{eq:qichi}\\
\psi(\ell)&=&\sum^{\ell-1}_{k=1}{1\over k}-\gamma, 
\end{eqnarray}
for any $\ell$ and 
\begin{eqnarray}
p^{(2)}_{2 m}&=&
{{457\,\gamma}\over {210}} - 
  {{{{\gamma}^2}}\over 2} + 
  {{5\,{{\pi }^2}}\over {24}} - 
  {i\over {18}}\,\gamma\,m\,q + 
  {{457\,\ln 2}\over {210}} \nonumber\\
& &
- \gamma\,\ln 2 - 
  {i\over {18}}\,m\,q\,\ln 2 - 
  {{{({\ln 2})^2}}\over 2}, \\
q^{(2)}_{2 m}&=&
     {{-457\,\pi }\over {420}} + 
     {{\gamma\,\pi }\over 2} + 
     {{5\,m\,q}\over {36}} + 
     {i\over {36}}\,m\,\pi \,q - 
     {i\over {72}}\,{q^2} + 
     {i\over {324}}\,{m^2}\,{q^2} + 
     {{\pi \,\ln 2}\over 2}, \\
p^{(2)}_{3 m}&=&
{{52\,\gamma}\over {21}} - {{{{\gamma}^2}}\over 2} + 
  {{5\,{{\pi }^2}}\over {24}} - {i\over {72}}\,\gamma\,m\,q + 
  {{52\,\ln 2}\over {21}} \nonumber\\
& &- \gamma\,\ln 2 - 
  {i\over {72}}\,m\,q\,\ln 2 - {{{({\ln 2})^2}}\over 2}, \\
q^{(2)}_{3 m}&=&
     {{-26\,\pi }\over {21}} + {{\gamma\,\pi }\over 2} + 
     {{67\,m\,q}\over {1440}} + {i\over {144}}\,m\,\pi \,q + 
     {i\over {360}}\,{q^2} - {{17\,i}\over {12960}}\,{m^2}\,{q^2} + 
     {{\pi \,\ln 2}\over 2}. 
\end{eqnarray}
Then noting that $\exp(-i\phi)\sim\exp(-i(z^*-z))$ at $z\sim\infty$, 
the asymptotic form of $X^{\rm in}$ is expressed as
\begin{eqnarray}
X^{\rm in}&=& \sqrt{z^2+a^2\omega^2}\exp(-i\phi)
\left\{f^{(0)}_{\ell m}+\epsilon\xi^{(1)}_{\ell m}
+\epsilon^2\xi^{(2)}_{\ell m}+...\right\}\nonumber\\
&\sim&e^{-i z^*}(z h_\ell^{(2)}e^{i z})\left[1
+\epsilon(p^{(1)}_{\ell m}+i q^{(1)}_{\ell m})
+\epsilon^2(p^{(2)}_{\ell m}+i q^{(2)}_{\ell m})\right]\nonumber\\
& &+e^{i z^*}(z h_\ell^{(1)}e^{-i z})\left[1
+\epsilon(p^{(1)}_{\ell m}-i q^{(1)}_{\ell m})
+\epsilon^2(p^{(2)}_{\ell m}-i q^{(2)}_{\ell m})\right],
\end{eqnarray}
where $h^{(1)}_{\ell}$ and $h^{(2)}_\ell$ are the spherical 
Hankel functions of the first and second kinds, 
respectively, which are given by
\begin{equation}
h^{(1)}_{\ell}=j_\ell+i n_\ell 
        \to (-1)^{\ell+1}{e^{i z}\over z},~ 
h^{(2)}_{\ell}=j_\ell-i n_\ell 
        \to (-1)^{\ell+1}{e^{-i z}\over z}.
\label{eq:hankel}
\end{equation} 
\ From these equations, noting $\omega r^*=z^*-\epsilon\ln\epsilon$, 
we obtain 
\begin{equation}
A^{{\rm inc}}={1\over 2}i^{\ell+1}e^{-i\epsilon\ln\epsilon}
\left[1+\epsilon(p^{(1)}_{\ell m}+i q^{(1)}_{\ell m})
+\epsilon^2(p^{(2)}_{\ell m}+i q^{(2)}_{\ell m})+...\right]. 
\end{equation}
The corresponding incident amplitude $B^{{\rm inc}}$ 
for the Teukolsky function is obtained from Eq.~(\ref{eq:ainbin}). 
\section{Gravitational waves to $O(v^{11})$ in Schwarzschild case}
\label{sec:5.5PN}

In this section we consider a circular orbit around a Schwarzschild black
hole and derive the 5.5PN formula for the energy flux emitted to infinity. 
In this case, we can take the orbit to lie on the equatorial plane 
($\theta=\pi/2$) without loss of generality. Then $E$ and $l_z$ are
given by setting $R(r_0)=\partial R/\partial r(r_0)=0$ where $R$
is given by Eq.~(\ref{eq:Rgeo}). This gives
\begin{equation}
E=(r_{0}-2M)/\sqrt{r_0(r_0-3M)},\quad
l_z=\sqrt{M r_{0}}/\sqrt{1-3M/r_0},
\label{eq:eq6}
\end{equation}
where $r_0$ is the orbital radius. The angular frequency is given by 
$\Omega_\varphi=(M/r_0^3)^{1/2}$. 

Defining $_s b_{\ell m}$ by 
\begin{eqnarray}
_0 b_{\ell m} & = & 
     {1\over2}\left[(\ell-1)\ell(\ell+1)(\ell+2)\right]^{1/2}
   {}_{0} Y_{\ell m}\left({\pi\over 2},0\right){E r_0\over r_0-2M}\,,
\cr
_{-1} b_{\ell m} & = & \left[(\ell-1)(\ell+2)\right]^{1/2}
    {}_{-1} Y_{\ell m}\left({\pi\over 2},0\right){l_z\over r_0}\,,
\cr 
_{-2} b_{\ell m} & = & _{-2} Y_{\ell m}\left({\pi\over 2},0\right)
            l_z\Omega_\varphi,
\label{eq:eq8}
\end{eqnarray}
where $_{s} Y_{\ell m}(\theta,\varphi)$ are the spin-weighted spherical 
harmonics\cite{ref:spin}, 
$\tilde Z_{\ell m\omega}$ is found to take the form 
\begin{equation}
 \tilde Z_{\ell m\omega}=Z_{\ell m}\delta(\omega-m\Omega_\varphi),
\label{eq:eq9}
\end{equation}
where
\begin{eqnarray}
Z_{\ell m}& = &{\pi\over i\omega r_0^2 B_{\ell\omega}^{in}}
  \Biggl\{\Biggl[
      -_0 b_{\ell m} -2i _{-1}b_{\ell m}
             \left(1+{i\over 2}\omega r_0^2/(r_0-2M)\right)
\cr &&
    + i _{-2} b_{\ell m}\omega r_0 (1-2M/r_0)^{-2}
     \left(1-M/r_0+{1\over 2}i\omega r_0\right)\Biggr] R_{\ell m}^{in}
\cr &&
    +\left[i_{-1}b_{\ell m}-_{-2}b_{\ell m}
         \left(1+i\omega r_0^2/(r_0-2M)\right)\right] 
           r_0 {R^{in}_{\ell\omega}}'(r_0)
\cr &&
      +{1\over 2} {}_{-2} b_{\ell m} r_0^2 {R_{\ell \omega}^{in}}''(r_0)
      \Biggr\}\quad, 
\label{eq:eq10}
\end{eqnarray}
where prime denotes the derivative with respect to the 
radial coordinate $r$. 
In terms of the amplitudes $Z_{\ell m}$, the gravitational wave 
luminosity at infinity is given by
\begin{equation}
 {dE\over dt}=\sum_{\ell=2}^{\infty} \sum_{m=-\ell}^{\ell}
   {\vert Z_{\ell m}\vert^2\over2\pi \omega^2},
\label{eq:lumiv11}
\end{equation}
where $\omega =m\Omega_\varphi$. 
Since the dominant frequency of the gravitational waves at
infinity is $2\Omega_\varphi$, an observationally relevant
post-Newtonian parameter is $x\equiv(M\Omega_\varphi)^{1/3}$.
We mention that our post-Newtonian expansion parameter is defined by
$v:=(M/r_0)^{1/2}$. In the case of a circular orbit around a
Schwarzschild black hole, however, we have $v=x$. Hence
the parameter $v$ is directly related to
the observable frequency in the present case.

Following the method given in section \ref{sec:pnexp},
instead of directly calculating $R^{\rm in}_{\ell\omega}$ from 
the homogeneous Teukolsky equation, 
we calculate the corresponding Regge-Wheeler
function $X^{\rm in}_{\ell\omega}$ first and then transform it
to $R^{\rm in}_{\ell\omega}$. 
The homogeneous Regge-Wheeler equation, which is given by setting $a=0$
in Eq.~(\ref{eq:sneq}), takes the form \cite{ref:RW},
\begin{equation}
\left[{{d^2\over dr^{*2}}+\omega^2-V(r)}\right]X_{\ell\omega}(r)
=0, \label{eq:rw}
\end{equation}
where 
\begin{equation}
V(r)=\left({1-{2M\over r}}\right)\left({{\ell(\ell+1)\over r^2}
-{6M\over r^3} }\right). 
\end{equation}
The transformation (\ref{eq:xtor}) reduces to 
\cite{ref:Chan}
\begin{equation}
R_{\ell m\omega}={\Delta\over c_0}
\left({{d \over dr^*}+i\omega}\right)
{r^2\over \Delta}\left({{d \over dr^*}+i\omega}\right)r 
X_{\ell\omega}, \label{eq:chantrans}
\end{equation}
where $c_0$, defined in Eq.~(\ref{eq:ceq}), reduces to
$c_0=(\ell-1)\ell(\ell+1)(\ell+2)-12iM\omega$. 
The inverse transformation (\ref{eq:rtox}) reduces to 
\begin{equation}
X_{\ell\omega}={r^5\over \Delta}
          \left({d \over dr^*}-i\omega\right){r^2\over\Delta}
          \left({d \over dr^*}-i\omega\right)
               {R_{\ell\omega}\over r^2}.
\label{eq:inconv}
\end{equation}
The asymptotic forms of $X^{\rm in}$ is the same as 
given in Eq.~(\ref{eq:xinasy}) except that now we have $k=\omega$.
The coefficients $A^{\rm inc}$, $A^{\rm ref}$
and $A^{\rm trans}$ are also respectively related to 
$B^{\rm inc}$, $B^{\rm ref}$ and 
$B^{\rm trans}$ as before. (See Eqs.~(\ref{eq:ainbin}), 
(\ref{eq:arefbref}) and (\ref{eq:atransbtrans}).)
Note that the coefficient that appears in Eq.~(\ref{eq:atransbtrans}) 
now reduces to 
\begin{equation}
d_{\ell m\omega}=16(1-2iM\omega)(1-4iM\omega).
\end{equation}

Corresponding to Eq.~(\ref{eq:zst}), we introduce the variable 
$z^*=z+\epsilon\ln(z-\epsilon)$. Then Eq.~(\ref{eq:xixi}) reduces to
\begin{equation}
X_{\ell}^{\rm in}=e^{-i(z^*-z)}z\xi_{\ell}(z)=
e^{-i\epsilon\ln(z-\epsilon)}z\xi_{\ell}(z), 
\label{eq:defxsi}
\end{equation}
and Eq.~(\ref{eq:eps}) becomes $L^{(0)}[\xi_{\ell}]
=\epsilon L^{(1)}[\xi_{\ell}]$. 
Thus Eq.~(\ref{eq:PNexpand}) simplifies considerably to become
\begin{equation}
L^{(0)}[\xi_{\ell}^{(n)}]=L^{(1)}[\xi_{\ell}^{(n-1)}].
\label{eq:rec}
\end{equation}
It may be worthwhile to note that the left hand side can be expressed
concisely as
\begin{equation}
  L^{(1)}[\xi_{\ell}^{(n-1)}]
=e^{-iz}{d\over dz}\left[{1\over z^3}{d\over dz}
\left(e^{iz}z^2\xi_{\ell}^{(n-1)}\right)\right].
\end{equation}

The calculations to $O(\epsilon^2)$ are already done 
in section \ref{sec:pnexp}. 
When we consider the gravitational wave luminosity to $O(v^{11})$, 
we need to calculate $A^{\rm inc}$ to $O(\epsilon^3)$ for $\ell=2$ 
and $3$ and to $O(\epsilon^2)$ and for $\ell=4$. 
Thus we need the closed analytic forms of $\xi^{(3)}_\ell$ 
for $\ell=2$ and 3 and $\xi^{(2)}_4$. The latter can be obtained in the
same way as in the previous section. The procedure to obtain
$\xi^{(3)}_\ell$ is explained in detail in Appendix \ref{ap:RWthird}. 

The real parts of $\xi^{(3)}_\ell$, $f^{(3)}_\ell$,
for $\ell=2$ and 3 are given as
\begin{eqnarray}
f_{2}^{(3)} = && 
 {{214\,{F_{2,0}[z(\ln z) j_{0}]}}\over {105}} 
- {{107\,D^{nj}_{-4}}\over {630}} - 
  {{457\,D^{nj}_{-2}}\over {70}} - {{2629\,D^{nj}_{0}}\over {630}} 
+ {16949\,D^{nj}_{2}\over {4410}} 
\cr &&
- {{107\,(\ln z)\,D^{nj}_{2}}\over {105}} 
  + {(\ln z)^2}\,D^{nj}_{2} - 
  {{2\,D^{nj}_{4}}\over {49}} - 12\,D^{nnj}_{-1} - 18\,D^{nnj}_{1} 
+ {{2\,D^{nnj}_{3}}\over 3} 
\cr &&
- 8\,D^{nnnj}_{2} 
  - {{197\,j_{-3}}\over {126}} + 
  {{2539\,j_{-1}}\over {3780}} + {{107\,(\ln z)\,j_{-1}}\over {70}} + 
  {{3\,{{(\ln z)}^2}\,j_{-1}}\over 2} + {{21\,j_{1}}\over {100}} 
\cr &&
+ {{349\,(\ln z)\,j_{1}}\over {140}} 
   + {{9\,{{(\ln z)}^2}\,j_{1}}\over 4} - 
  {{457\,j_{3}}\over {1050}} + {{29\,(\ln z)\,j_{3}}\over {252}} - 
  {{{{(\ln z)}^2}\,j_{3}}\over {12}} + {{j_{5}}\over {504}}\,,
\label{eq:thirdf2}
\end{eqnarray}
\begin{eqnarray}
f_{3}^{(3)} = && 
{{26\,F_{3,0}[z(\log z) j_{0}]}\over {21}} 
  + {{13\,D^{nj}_{-5}}\over {98}} + 
  {{14075\,D^{nj}_{-3}}\over {882}} 
\cr && - 
  {{1424\,D^{nj}_{-1}}\over {63}} - {{2511\,D^{nj}_{1}}\over {70}} + 
  {{269\,D^{nj}_{3}}\over {70}} - 
  {{13\,(\log z)\,D^{nj}_{3}}\over {21}} 
\cr && + 
  {{(\log z)}^2}\,D^{nj}_{3} - {{D^{nj}_{5}}\over {18}} + 
  60\,D^{nnj}_{-2} + 34\,D^{nnj}_{0} - 
  {{710\,D^{nnj}_{2}}\over {21}} + {{6\,D^{nnj}_{4}}\over 7} 
\cr && - 
  8\,D^{nnnj}_{3} + {{75\,j_{-4}}\over {28}} - 
  {{19\,j_{-2}}\over {56}} - {{65\,(\log z)\,j_{-2}}\over {14}} - 
  {{15\,{{(\log z)}^2}\,j_{-2}}\over 2} - {{2789\,j_{0}}\over {3780}} 
\cr && - 
  {{221\,(\log z)\,j_{0}}\over {84}} - 
  {{17\,{{(\log z)}^2}\,j_{0}}\over 4} - {{7495\,j_{2}}\over {5292}} + 
  {{4867\,(\log z)\,j_{2}}\over {1764}} 
\cr && + 
  {{355\,{{(\log z)}^2}\,j_{2}}\over {84}} - 
  {{10963\,j_{4}}\over {32340}} + {{15\,(\log z)\,j_{4}}\over {196}} - 
  {{3\,{{(\log z)}^2}\,j_{4}}\over {28}} + {{4\,j_{6}}\over {1485}}\,,
\label{eq:thirdf3}
\end{eqnarray}
where the definitions of the functions $D^{nnj}_\ell$ etc. are 
given in Eqs.~(\ref{eq:tamadefD}) and (\ref{eq:Fkelldef}) of Appendix
\ref{ap:RWthird}. The imaginary parts $g^{(3)}_\ell$ are expressed in
terms of $f^{(1)}_\ell$ and $f^{(2)}_\ell$ as given in
Eq.~(\ref{eq:tamaImxi}). As for the real part of $\xi^{(2)}_4$,
$f^{(2)}_4$, it is calculated to be
\begin{eqnarray}
f^{(2)}_4 &=
 &{56\over 165\,z}j_5+
   \left(-{{5036}\over {33\,{z^4}}}+{30334\over 1155\,z^2}\right)j_4
   +\left({35252\over 33\,z^5}-{30334\over 165\,z^3}
   +{14401\over 3465\,z}\right)j_3
\cr
 &&-\left({5036\over 11\,z^5} + {{45461}\over {693\,{z^3}}}
         +{{36287}\over {9240\,z}}\right)j_1
  +\left({{140}\over {{z^3}}} - {{5}\over {18\,z}}\right)n_0
  -{{49}\over {6\,z}}n_2
\cr
 &&-{21\over 5\,z}D^{nj}_4 +{149\over 30}D^{nj}_3
    -{10\over3\,z}D^{nj}_2
    +{105\over \,z^2}D^{nj}_1+{210\over z^3}D^{nj}_0
    -{20\over z}D^{nj}_0
    +{1571\over3465}D^{nj}_{-5}
\cr
 &&+4D^{nnj}_4 - {1571\over6930}\,j_4\,\ln z
  -{1\over2}\,j_4\,(\ln z)^2\,. 
\label{eq:f4}
\end{eqnarray}

Using the analysis given in subsection \ref{tamaBasymp} of Appendix
\ref{ap:RWthird}, the above results readily give us the asymptotic forms 
of $\xi^{(3)}_\ell$ ($\ell=2,3$) and $\xi^{(2)}_4$ at $z\to\infty$,
from which the amplitudes $A^{\rm inc}$ to the required order are
calculated. The results are
\begin{eqnarray}
 A^{\rm inc}_2
  &=&-{1\over2}ie^{-i\epsilon(\ln2\epsilon+\gamma)}
     \exp\left[i\left\{\epsilon{5\over3}-\epsilon^2{107\over420}\pi
 +\epsilon^3\left({29\over648}-{107\over1260}\pi^2+{\zeta(3)\over3}\right)
+\cdots \right\}\right]
\cr
  && \times \Biggl[1-\epsilon{\pi\over2}
  +\epsilon^2
  \left({25\over18}+{5\over24}\pi^2+{107\over210}(\gamma+\ln2)\right)
\cr
&&\quad 
+\epsilon^3
  \left(-{25\over 36}\pi -{107\over 420}(\gamma+\ln 2)\pi
        -{\pi^3\over 16}\right)+\cdots\Biggr], 
\cr
 A^{\rm inc}_3
  &=&{1\over2}e^{-i\epsilon(\ln2\epsilon+\gamma)}
    \exp\left[i\left\{\epsilon{13\over6}-\epsilon^2{13\over84}\pi
  +\epsilon^3\left(-{29\over810}-{13\over252}\pi^2+{\zeta(3)\over3}
  \right)  +\cdots\right\}\right]
\cr
  &&\times\Biggl[1-\epsilon{\pi\over2}
  +\epsilon^2
       \left({169\over72}+{5\over24}\pi^2+{13\over42}(\gamma+\ln2)
        \right)
\cr
&&\quad +\epsilon^3\left(
       -{169\over 144}\pi -{13\over 84}(\gamma+\ln 2)\pi
        -{\pi^3\over 16}\right)+\cdots\Biggr],
\cr
 A^{\rm inc}_4
  &=&{1\over2}ie^{-i\epsilon(\ln2\epsilon+\gamma)}
    \exp\left[i\left\{\epsilon{149\over60}
      -i\epsilon^2{1571\over13860}\pi+\cdots\right\}\right]
\cr
  &&\times\left[1-\epsilon{\pi\over2}
  +\epsilon^2
       \left({22201\over7200}+{5\over24}\pi^2
      +{1571\over6930}(\gamma+\ln2)
        \right)+\cdots\right].
\label{eq:Aincto3}
\end{eqnarray}
The corresponding amplitudes $B^{\rm inc}$ are readily obtained from
Eq.~(\ref{eq:ainbin}).  

As in the previous section, from
Eqs.~(\ref{eq:defxsi}) and (\ref{eq:chantrans}), it is also
straightforward to obtain the near-zone post-Newtonian expansion of
$X^{\rm in}$ and hence of $R^{\rm in}$, assuming $z\ll 1$.
As discussed there, we need the series expansion formulas for 
$R^{\rm in}$ for $2(n+\ell-2)\le11$, hence for $n\le7-\ell$ for each
$2\le\ell\le7$. The resulting $R^{\rm in}$ for $2\le\ell\le7$ which are
necessary to calculate the luminosity to $O(v^{11})$ are given in
Appendix \ref{ap:v11}.

Finally, from Eq.~(\ref{eq:lumiv11}), we obtain the luminosity to
$O(v^{11})$ as
\begin{eqnarray}
\left\langle{dE\over dt}\right\rangle=\left({dE\over dt}\right)_N
&&\Biggl[
1 - {1247\over 336}\,v^2 + 4\,\pi\,{v^3} -{44711\over 9072}\,{v^4} 
-{8191\,\pi \over 672}\,{v^5} 
\nonumber\\
&&+\left( {{6643739519}\over {69854400}} -{{1712\,\gamma }\over {105}}
 + {{16\,{{\pi }^2}}\over 3} -{{3424\,\ln 2}\over {105}} 
- {{1712\,\ln v}\over {105}} \right)\,{v^6}
\nonumber\\
&&-{16285\,\pi \over {504}}\,{v^7} 
\nonumber\\
&&+\left( -{{323105549467}\over {3178375200}} + 
   {{232597\,\gamma }\over {4410}}  -{{1369\,{{\pi }^2}}\over{126}}
\right.
\nonumber\\
&&\qquad
\left. + {{39931\,\ln 2}\over {294}} 
   - {{47385\,\ln 3}\over {1568}} + {{232597\,\ln v}\over {4410}}
          \right)\,{v^8}
\nonumber\\
&&+  \left( {{265978667519\,\pi }\over {745113600}}
  -{{6848\,{\gamma}\,\pi }\over {105}} 
  -{{13696\,\pi \,\ln 2}\over {105}} 
  -{{6848\,\pi\,{\ln v}}\over {105}} \right) v^9 
\nonumber\\
&& + \left( -{{2500861660823683}\over {2831932303200}} 
  + {{916628467\,{\gamma}}\over {7858620}} 
  - {{424223\,{{\pi }^2}}\over {6804}}\right. 
\nonumber\\
&&\qquad
 \left. - {{83217611\,\ln 2}\over {1122660}} 
  + {{47385\,\ln 3}\over {196}}
  + {{916628467\,{\ln v}}\over {7858620}}\right) v^{10}
\nonumber\\
&&+\left( {{8399309750401\,\pi }\over {101708006400}} 
  + {{177293\,{\gamma}\,\pi }\over {1176}}\right.
\nonumber\\
&&\qquad
 \left. + {{8521283\,\pi \,\ln 2}\over {17640}} 
  - {{142155\,\pi \,\ln 3}\over {784}}
  + {{177293\,\pi\,{\ln v}}\over {1176}}\right) v^{11} \Biggr],
\label{eq:dEdt11}
\end{eqnarray}
where $(dE/dt)_N$ is the Newtonian quadrupole luminosity given by
\begin{equation}
\left({dE\over dt}\right)_{\rm N}={32\mu^2M^3\over5r_0^5}
={32\over5}\left({\mu\over M}\right)^2v^{10}.
\label{eq:Enewton}
\end{equation}
To compare the above result with those obtained previously by the
standard post-Newtonian method, we note that
$v=x\equiv(M\Omega_\varphi)^{1/3}$ in the present case. Then we find our
result agrees with the standard post-Newtonian results up to
$O(x^5)$\cite{ref:WagonerW,ref:wisem,ref:BDIWW,ref:BDI,ref:blanchet2,ref:WW}
in the limit  $\mu/M$ $\ll 1$. The contributions to the luminosity from
individual $\ell$ modes are given in Appendix \ref{ap:v11}. 
\section{Convergence of the post-Newtonian expansion}
\label{sec:convergence}

Using the results obtained in the previous section, we compare the
formula for the gravitational wave flux with the corresponding numerical
results and investigate the accuracy of the post-Newtonian expansion.

A high precision numerical calculation of gravitational waves from a
particle in a circular orbit around a Schwarzschild black hole has been
performed by Tagoshi and Nakamura. \cite{ref:TN} Since no assumption was
made about the velocity of the test particle, their results are correct
relativistically in the limit $\mu\ll M$.  In that work, $dE/dt$ was
calculated only for $\ell=2\sim 6$.  Here, for the orbital radius
$r_0\le 100 M$, we calculate $dE/dt$ again for all modes of $\ell=2\sim
6$ and for $\ell=7$ with odd $m$.  The estimated accuracy of the
calculation is about $10^{-11}$, which turns out to be accurate enough
for the present purpose.  As for the radius $r_0>100 M$, we use the data
calculated by Tagoshi and Nakamura \cite{ref:TN} which contain modes
from $\ell=2$ to $6$. 

\begin{figure}[htb]
\epsfxsize=12cm
\epsfysize=12cm
\centerline{\epsfbox{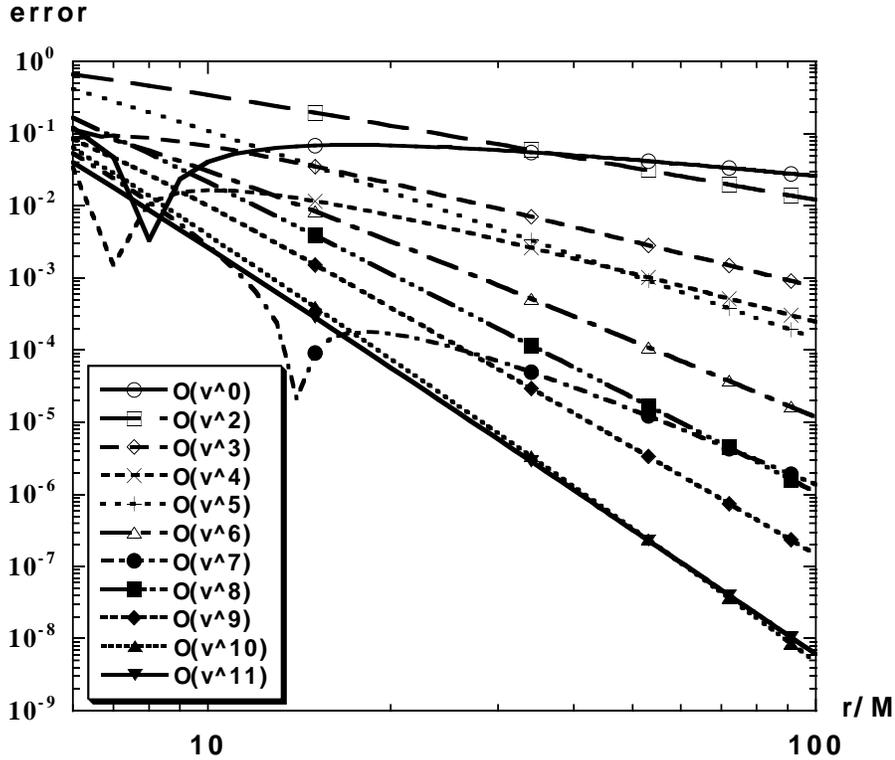}}
\caption{The error of the post-Newtonian formulas as
  functions of the Schwarzschild radial coordinate 
  $r$ for $6\leq r/M\leq 100$.  Contributions from $\ell=2$ to $7$ modes
  are included.}\label{fig:conv1}
\end{figure}

\begin{figure}
\epsfxsize=12cm
\epsfysize=12cm
\centerline{\epsfbox{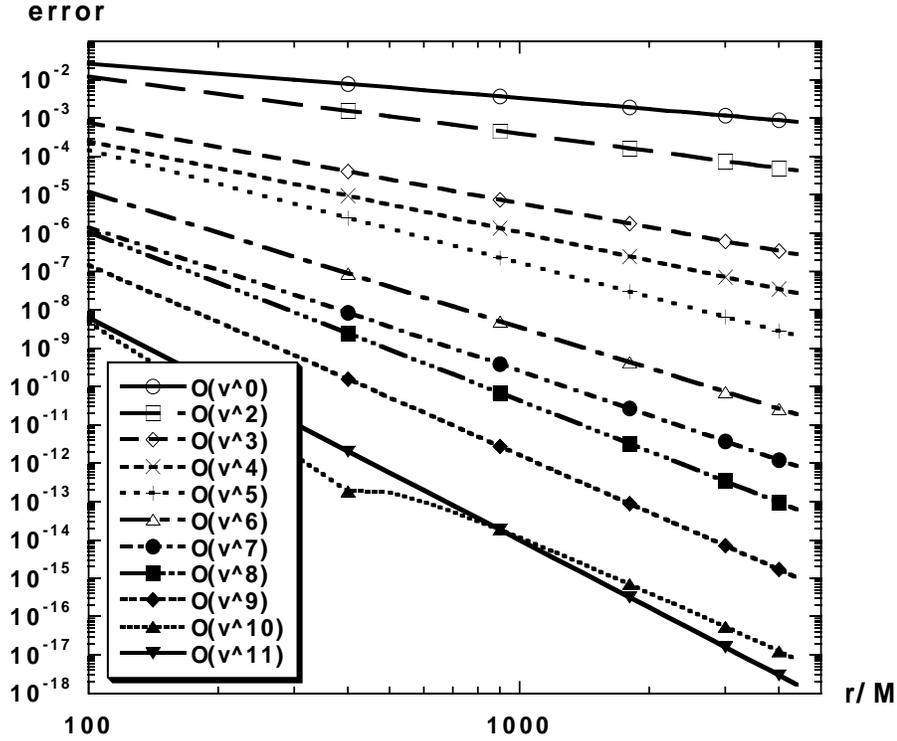}}
\caption{The same figure for $100\leq r/M$ $< 5000$.
  Contributions only from $\ell=2$ to $6$ are included in both the
  post-Newtonian formulas and in the numerical data. }\label{fig:conv2}
\end{figure}

In Figs. \ref{fig:conv1} and \ref{fig:conv2}, we show the error in the
post-Newtonian formulas as a function of the orbital radius $r$.
The error of the post-Newtonian formula is defined as
\begin{equation}
{\rm error}=\left|1-\left(dE\over dt\right)_{n}\bigg/
\left(dE\over dt\right)\right|, 
\end{equation}
where $(dE/dt)_{n}$ and $(dE/dt)$ denote the $(n/2)$PN formula and the
numerical result, respectively.  
In the plot of Fig.\ref{fig:conv2},
only the contributions from $\ell=2$ to $6$ are included in both the
post-Newtonian formulas and the numerical data.  We can see that, at
small radius less than $r\sim 10M$, the error of the 1PN and 2.5PN
formulas are larger than the other formulas.  On the other hand, the
Newtonian and the 2PN formulas are very accurate within this radius.
This is because those formulas coincide with the exact one accidentally
at a radius between $6M$ and $10M$. The error of each post-Newtonian
formula at the inner most stable circular orbit, $r=6 M$, becomes as
follows; $12\%$ (Newtonian), $66\%$ (1PN), $8.6\%$ (1.5PN), $3.4\%$
(2PN), $42\%$ (2.5PN), $11\%$ (3PN), $5.4\%$ (3.5PN), $17\%$ (4PN),
$8.4\%$ (4.5PN), $6.5\%$ (5PN), $4.1\%$ (5.5PN).  As is expected, the
errors of the post-Newtonian formulas decrease almost linearly up to
$r\sim 5000 M$ in a log-log plot.  This fact also suggests that the
numerical data have accuracy of at least $\sim 10^{-18}$ at $r\sim
5000M$.

In order to examine exactly to what order the post-Newtonian formulas
are needed to do accurate estimation of the parameters of a binary,
using data from laser interferometers, we must evaluate the systematic
error produced by incorrect templates.  However, here we simply
calculate the total cycle of gravitational waves from a coalescing
binary in a laser interferometer band and evaluate the error produced by
the post-Newtonian formulas.  It has been suggested that whether the
error in the total cycle is less than unity or not gives a useful
guideline to examine the accuracy of the post-Newtonian formulas as
templates \cite{ref:three} (see also Ref.~\citen{ref:poisson3}).

We ignore the finite mass effect in the post-Newtonian formulas and
interpret $M$ as the total mass and $\mu$ as the reduced mass of the
system.  The total cycle $N$ of gravitational waves from an inspiraling
binary is calculated by using the post-Newtonian energy loss
formula, $(dE/dt)_n$, and the orbital energy formula $(dE/dv)_n$ which
is truncated at $n/2$PN order as
\begin{equation} 
N^{(n)}=\int^{v_i}_{v_f}dv {\Omega_\varphi\over \pi} 
        {(dE/dv)_n\over |(dE/dt)_n|}, 
\end{equation} 
where $v_i=(M/r_i)^{1/2}$, $v_f=(M/r_f)^{1/2}$, and $r_i$ and $r_f$ are
the initial and final orbital separations of the binary. We define the
relative difference of cycle $\Delta N^{(n)}$ as $\Delta N^{(n)}\equiv$
$|N^{(n)}-N^{(n-1)}|$.  We adopt $r_f=6 M$ and $r_i$ is the one at which
the frequency of wave is 10Hz and which is given by $r_i/M\sim 347
(M_{\odot}/M)^{2/3}$.

The results for typical binary systems are given in Table
\ref{table:1}.
\begin{table}
\caption{The relative difference of cycle $\Delta N^{(n)}$ for 
typical coalescing compact binaries. The last line shows the cycle 
calculated by Newtonian quadrupole formula.}
\label{table:1}
\begin{center}
\begin{tabular}{ccccc} \hline\hline
$n$ & (1.4$M_\odot$,1.4$M_\odot$) & (10$M_\odot$,10$M_\odot$) &
(1.4$M_\odot$,10$M_\odot$) & (1.4$M_\odot$,70$M_\odot$) \\  \hline
2&356 &54  &216&212\\ 
3&228 &60  &208&296\\ 
4&11  &5   &15 &31 \\ 
5&12  &7   &20 &53 \\ 
6&11  &8   &22 &75 \\ 
7&1.2 &1.0 &2.6&10 \\ 
8&0.12&0.14&0.3&2.2\\ 
9&0.82&0.80&1.9&8.9\\ 
10&0.09&0.08&0.20&0.87\\ 
11&0.03&0.03&0.07&0.40\\ 
\hline
$N^{(0)}$&16000&600&3578&898\\
\hline
\end{tabular}
\end{center}
\end{table}
We only show the results for $q=0$.  The cases for
$q\neq 0$ are investigated in Shibata et al.\cite{ref:SSTT} and
Tagoshi et al.\cite{ref:TSTS}. This table suggests that we need the
3PN$\sim$4PN formula to obtain accurate wave forms for typical binaries
whose total mass are less than 20$M_\odot$. Although the required
post-Newtonian order is very high and it has not been achieved yet in
the standard post-Newtonian analysis, this results show that the
post-Newtonian approximation is applicable to the inspiral phase of
coalescing compact binaries.  In this sense, we can be optimistic. 

On the other hand, the convergence for the case of neutron star $-$
black hole binaries, whose mass is above several ten $M_\odot$, is very
slow. This is because $r_i/M$ become smaller for a larger mass black
hole, and the higher relativistic correction becomes more important.
{}From Table I, we might think that $N^{(n)}$ converges at $n=11$ even
for $(m_1,m_2)=$ (1.4$M_\odot$,70$M_\odot$).  However this is not
true. Note that Table I shows only the relative difference between the
post-Newtonian approximated cycles.  If we calculate the difference
between the post-Newtonian formula and the fully relativistic one, we
find that the 5.5PN formula is not accurate enough for the case
$(m_1,m_2)=$ (1.4$M_\odot$,70$M_\odot$), as pointed out by Tanaka,
Tagoshi and Sasaki\cite{ref:TTS}.

Finally we comment on the initial frequency.  The above results are
obtained by setting the initial frequency to 10Hz. However, it may be
difficult to observe gravitational wave at this frequency because of
the seismic noise. If we set the initial frequency higher than 10Hz, the
error $\Delta N$ becomes slightly smaller.  But since this dependence of
$\Delta N$ on the initial frequency is very weak, the above results are
insensitive to the variation of the initial frequency.

\section{Circular orbit on the equatorial plane 
            around a rotating black hole}
\label{sec:kerrv8}

In this section, we consider a circular orbit on the equatorial plane of
a Kerr black hole and calculate the 4PN luminosity formula.

We define the orbital radius as $r=r_0$. 
As in section \ref{sec:5.5PN}, we have $C=0$, and
$E$ and $l_z$ are determined by $R(r_0)=0$ and 
$\partial R/\partial r\mid_{r=r_0}=0$ as
\begin{equation}
E={{1-2v^2+q v^3}\over (1-3v^2+2qv^3)^{1/2}}\,,
\quad
l_z={{r_0v(1-2qv^3+q^2 v^4)}\over 
(1-3v^2+2qv^3)^{1/2}}\,,
\end{equation}
where $v=(M/r_0)^{1/2}$. From these, 
we can easily obtain $\varphi(t)$ as 
\begin{eqnarray}
\varphi(t)=\Omega_{\varphi} \, t\,;\quad
\Omega_{\varphi}={M^{1/2}\over r_0^{3/2}}
\left[1 - q v^3+q^2 v^6+O(v^9)\right]\,. 
\label{eq:Omphic}
\end{eqnarray} 
When $a=0$, this becomes $\Omega_{\varphi}=(M/r_0^3)^{1/2}$. 

The rest of the calculation is almost the same as 
in section \ref{sec:5.5PN}. 
The amplitude of the Teukolsky function 
$\tilde Z^{\infty}_{\ell m\omega}$ at infinity is expressed as
\begin{eqnarray}
\tilde Z^{\infty}_{\ell m\omega}&=&\mu
{{2 \pi \delta(\omega-m\Omega)}\over {2i\omega B^{{\rm inc}}}}
\Bigl[R^{{\rm in}}\{A_{n\,n\,0}+A_{{\bar m}\,n\,0}
+A_{{\bar m}\,{\bar m}\,0}\}
\nonumber\\
& &-{dR^{{\rm in}} \over dr}\{ A_{{\bar m}\,n\,1}
+A_{{\bar m}\,{\bar m}\,1}\}
 +{d^2 R^{{\rm in}} \over dr^2}
 A_{{\bar m}\,{\bar m}\,2}\Bigr]_{r=r_0,\theta=\pi/2},
\nonumber\\
&\equiv& 
\delta(\omega-m\Omega) Z^{\infty}_{\ell m\omega}\,,
\label{eq:zzq3}
\end{eqnarray} 
where $A_{nn0}$, etc. are given by Eq.~(\ref{eq:ann}). 

The total luminosity up to $O(v^8)$ is expressed as
\begin{eqnarray}
\left\langle{dE\over dt}\right\rangle
=\left({dE\over dt}\right)_{\rm N}
&&\left\{1 +\left(\hbox{$q$-independent terms}\right)
         - {73\,q\over 12}\,{v^3} + 
      + {{33\,{q^2}}\over {16}}\,{v^4} 
\right.
\nonumber\\
&+&{{3749\,q}\over {336}}\,{v^5} 
    -\left({{169\,\pi \,q}\over 6} + {{3419\,{q^2}}\over {168}}
        \right)\,{v^6}
\nonumber\\
&+& \left( {{83819\,q}\over {1296}} + {{65\,\pi \,{q^2}}\over 8} - 
         {{151\,{q^3}}\over {12}} \right) \,{v^7} 
\nonumber\\
&+& \left.
\left({{3389\,\pi \,q}\over {96}} - {{124091\,{q^2}}\over {9072}} 
     {{17\,{q^4}}\over {16}} \right)\,{v^8}\right\}\,,
\label{eq:dedtall8}
\end{eqnarray}
where $(dE/dt)_{\rm N}$ is the Newtonian quadrupole 
luminosity, Eq.~(\ref{eq:Enewton}), and the $q$-indepen\-dent terms are
identical to those in Eq.~(\ref{eq:dEdt11}). 

{}From an observational point of view, it is more convenient to express
the luminosity in terms of the variable
$x\equiv(M\Omega_\varphi)^{1/3}$. Using the relation between
$v\equiv(M/r_0)^{1/2}$ and $x$ given by Eq.~(\ref{eq:Omphic}), the
luminosity is expresses as
\begin{eqnarray}
\left\langle{dE \over dt}\right\rangle
 =\widetilde{\left({dE\over dt}\right)}_{\rm N}
&&\left( 1 + \left(\hbox{$q$-independent terms}\right)
 - {{11\,q}\over 4}\,{x^3} + {{33\,{q^2}}\over {16}} \,{x^4} 
\right.\nonumber\\
&-& {{59\,q}\over {16}}\,{x^5} 
+\left( - {{65\,\pi \,q}\over 6} + 
         {{611\,{q^2}}\over {504}}\right)\,{x^6} 
\nonumber\\
&+& \left( {{162035\,q}\over {3888}} +{{65\,\pi \,{q^2}}\over 8} 
          - {{71\,{q^3}}\over {24}} \right) \,{x^7} 
\nonumber\\
&+&\left.
 \left({{359\,\pi \,q}\over {14}} + {{22667\,{q^2}}\over {4536}} 
        + {{17\,{q^4}}\over {16}}\right)\,{x^8}\right),
\end{eqnarray}
where
\begin{equation}
  \label{eq:Enewtonx}
 \widetilde{\left({dE\over dt}\right)}_{\rm N}
={32\over 5}\left(\mu\over M \right)^2 x^{10}\,,
\end{equation}
and the $q$-independent terms are again identical to those in
Eq.~(\ref{eq:dEdt11}) with the replacement $v\to x$.
The partial luminosities for individual modes are given in Appendix
\ref{ap:v8}. The spin dependent term at $O(v^3)$ agrees with the
standard post-Newtonian result\cite{ref:KWW}.

Here it is interesting to investigate the origin of some of the 
spin-dependent terms. As an example, we consider the mode $\ell=|m|=2$. 
The luminosity from the $\ell=|m|=2$ modes is given by 
\begin{eqnarray}
\left({dE\over dt}\right)_{2,2}+\left({dE\over dt}\right)_{2,-2}
&&\cr
=\widetilde{\left({dE\over dt}\right)}_{\rm N}
&&\Biggl[1 +\left(\hbox{$q$-independent terms}\right)
     - {{8\,q}\over 3}\,{x^3} +2\,{q^2}\,{x^4} 
\nonumber\\
&+&{{52\,q}\over {27}}\,{x^5} 
 + \left( -{{32\,\pi \,q}\over 3} - {{1817\,{q^2}}\over {567}} 
    \right)\,{x^6}
\nonumber\\
&+&\left({{364856\,q}\over {11907}}+8\,\pi \,{q^2}-{8\,q^3\over 3}
          \right) \,{x^7} 
\nonumber\\
&+& \left({{208\,\pi \,q}\over {27}} 
    + {{105022\,{q^2}}\over {9261}} + {q^4}\right)\,{x^8} \Biggr]. 
\end{eqnarray}
We can derive some of the spin-dependent terms in the above formula from 
the quadru\-pole formula \cite{ref:Ryan}; 
$dE/dt=(32/5)\mu^2\hat{r}^4\Omega_\varphi^6$, 
where $\hat{r}$ is the orbital radius of a test 
particle in harmonic coordinates. 
If multipole moments of the black hole exist, 
the orbital radius changes 
due to the influence of those moments. 
The mass and mass current multipole moments of a 
Kerr black hole is given by 
$M_l+i S_l$ $=M(i a)^l$. We can express the orbital frequency of the 
test particle in harmonic coordinates. 
We find that the dominant effect of the multipole moments of a Kerr 
black hole to $dE/dt$ can be expressed as 
\begin{eqnarray} 
\left\langle{dE\over dt}\right\rangle
=\widetilde{\left({dE\over dt}\right)}_{\rm N}
\Bigl\{&&
1-{8\over 3}{S_1\over M^2} x^3-2 {M_2\over M^3} x^4
+\left(4 {S_3\over M^4}-{4\over 3}{M_2 S_1 \over M^5} \right)x^7 
\cr
&&
+\left(-{3\over 2}{M_2^2\over M^6} 
+{5\over 2}{M_4\over M^5}\right)x^8\Bigr\}\,.
\end{eqnarray} 
The terms in this expression agree with the corresponding terms of our
result such as $(-8/3) q x^3$, $2 q^2 x^4$, $(-8/3) q^3 x^7$ and
 $q^4x^8$. Thus, we may interpret the term $2 q^2 x^4$ as the effect of 
the quadrupole moment. The terms $(-8/3) q^3 x^7$ and $q^4 x^8$ 
are not due to a single multipole moment, but 
to combined effects of the multipole moments. 
\section{Slightly eccentric orbit 
             around a Schwarzschild black hole} 
\label{sec:schwaecc}

In this section, we present post-Newtonian
formulas of gravitational waves from a 
particle in slightly eccentric orbits around a Schwarzschild black hole. 
We derive the 4PN formulas of the energy and angular momentum fluxes 
to $O(e^2)$ where $e$ is the eccentricity of the orbit.

The solution of the geodesic equations for slightly 
eccentric orbits has been given by Apostolatos et al.\cite{ref:AKOP}. 
Here we briefly sketch the derivation of it. 
Since we may consider the orbit to lie on the equatorial plane without
loss of generality, we put $\theta=\pi/2$ and $C=0$ in the geodesic
equations (\ref{eq:geod}). 
Then we define a slightly eccentric orbit as follows: First of all, 
we assume that $E$ and $l_z$ are set to be such that $R(r)/r^4$, which
plays the role of an effective potential for the radial motion, has the 
minimum at $r=r_0$ and that the maximum value of 
the orbital radius is at $r=r_0(1+e)$, where $e \ll 1$. 
Thus, the following conditions hold;
\begin{equation}
{\partial(R/r^{4})\over \partial r}(r=r_0)=0 \quad{\rm and}\quad
R(r=r_0(1+e))=0.
\end{equation}
{}From these equations, $E$ and $l_z$ are expressed in terms of $r_0$
and $e$ as 
\begin{eqnarray}
& &E^2={(1-2v^2)^2 \over 1-3v^2}+{v^2 (1-6v^2) \over 1-3v^2}e^2 
-{2v^2 (1-7v^2) \over 1-3v^2}e^3+O(e^4),\\
& &l_z={M \over v\sqrt{1-3v^2}},
\end{eqnarray}
where $v \equiv \sqrt{M/r_0}$. For convenience, we also define
$\Omega\equiv v^3/M$. Then expanding the geodesic equations in powers of 
$e$, the solution is found to be\cite{ref:AKOP}
\begin{eqnarray}
& &r(t)=r_0[ 1+e \cos\Omega_rt \nonumber \\
&& \hskip 1cm +e^2
\{q_1(v)(1-\cos\Omega_rt)+q_2(v)(1-\cos 2 \Omega_rt)\}] + O(e^3),\\
& &\varphi(t)=\Omega_{\varphi}t-ep_1(v)\sin\Omega_rt \nonumber \\
&& \hskip 1cm+e^2\{p_2(v)\sin\Omega_r t+p_3(v)\sin 2\Omega_r t\}+O(e^3),
\label{eq:Seccsol}
\end{eqnarray}
where
\begin{eqnarray}
& &\Omega_r = \Omega(1-6v^2)^{1/2},\\
& &\Omega_{\varphi} = \Omega [1-f(v)e^2] ~
;\quad f(v)={3(1-3v^2)(1-8v^2) \over 2(1-2v^2)(1-6v^2)},
\label{eq:Omegaecc}\\
& &q_1(v)={1 -7v^2 \over 1-6v^2},~~~
q_2(v)={1 -11v^2+26v^4 \over 2(1-6v^2)(1-2v^2)},\\
& &p_1(v)={2(1 -3v^2) \over (1-2v^2)\sqrt{1-6v^2} },~~~
p_2(v)={2(1 -3v^2)(1-7v^2) \over (1-2v^2)(1-6v^2)^{3/2}}, \nonumber \\
& &p_3(v)
={5-64v^2+250v^4-300v^6 \over 4(1-2v^2)^2(1-6v^2)^{3/2}}.
\end{eqnarray}
As is well known, since $\Omega_r\neq\Omega_\varphi$, the orbit does not 
close.

Now we evaluate the source term of the Teukolsky equation.
In the present case, $A_{n\,n\,0}$ etc. given in Eqs.~(\ref{eq:ann})
reduce to
\begin{eqnarray}
& &A_{n\,n\,0}=-{E \over 2\sqrt{2}\Delta}S^{(0)}_{\ell m}
\biggl(1+{r^2 \over \Delta}{dr \over dt}\biggr)^2,\\
& &A_{\bar m\,n\,0}={ i l_z \over \sqrt{2\pi} }S^{(1)}_{\ell m}
\biggl({i\omega \over \Delta}+{2 \over r^3}\biggr)
\biggl(1+{r^2 \over \Delta}{dr \over dt}\biggr),\\
& &A_{\bar m\,\bar m\,0} ={l_z^2 \over 2\sqrt{2\pi}E } S^{(2)}_{\ell m}
\biggl({2i\omega (r-M) \over \Delta r^2}
  -{\omega^2 \over \Delta}\biggr),\\
& &A_{\bar m\,n\,1}={i l_z \over \sqrt{2\pi}r^2}S^{(1)}_{\ell m}
\biggl(1+{r^2 \over \Delta}{dr \over dt}\biggr),\\
& &A_{\bar m\,\bar m\,1}={l_z^2 \over \sqrt{2\pi}E}S^{(2)}_{\ell m}
\biggl({i\omega \over r^2}+{\Delta \over r^5}\biggr),\\
& &A_{\bar m\,\bar m\,2}
   ={l_z^2 \Delta\over 2\sqrt{2\pi}Er^4}S^{(2)}_{\ell m}\,,
\end{eqnarray}
where
\begin{eqnarray}
& &S^{(0)}_{\ell m}
    =L_1^{\dagger}L_2^{\dagger}S_{\ell m}({\theta=\pi/2}),\\
& &S^{(1)}_{\ell m}=L_2^{\dagger}S_{\ell m}({\theta=\pi/2}),\\
& &S^{(2)}_{\ell m}=S_{\ell m}({\theta=\pi/2}). 
\end{eqnarray}
Then noting that the orbits of our interest have the properties,
\begin{equation}
r(t+\Delta t_r)=r(t),\qquad 
\left.{d\varphi\over dt}\right|_{t=t+\Delta t_r}
=\left.{d\varphi\over dt}\right|_{t=t}, 
\end{equation}
where $\Delta t_r=2\pi/\Omega_r$, Eq.~(\ref{eq:zzq2}) 
can be rewritten as 
\begin{eqnarray}
\tilde Z_{\ell m\omega}&=&{\mu\over2i\omega B^{\rm inc}_{\ell m\omega}}
\int^{\infty}_{-\infty}dt e^{i\omega t-i m \varphi(t)}
W_{\ell m\omega} \nonumber\\
&=&{\mu\over2i\omega B^{\rm inc}_{\ell m\omega}} 
{2\pi\over {\Delta t_r}}\int^{\Delta t_r}_0 dt 
e^{i\omega t-i m \varphi(t)}
W_{\ell m\omega}\sum_n \delta(\omega-\omega_n),
\label{eq:Zecc}
\end{eqnarray}
where 
\begin{equation}
\omega_n=n\Omega_r+m\Omega_{\varphi} \quad (n=0,\pm 1, \pm 2, \ldots).
\label{eq:omegan}
\end{equation}
We see that $\tilde Z_{\ell m\omega}$ takes the form as given in
Eq.~(\ref{eq:tildeZ}) with $Z_{\ell m\omega}$ given by
\begin{equation}
\label{Zschecc}
Z_{\ell m\omega}={\mu\Omega_r\over2i\omega B^{\rm inc}_{\ell m\omega}} 
\int^{\Delta t_r}_0dt e^{i\omega t-i m \varphi(t)}W_{\ell m\omega}\,.
\end{equation}

When $Z_{\ell m \omega_n}$ are obtained, the energy and angular momentum
fluxes averaged over $t \gg \Delta t_r$ are calculated by using
Eqs.~(\ref{eq:Eflux}) and (\ref{eq:Jflux}), respectively.
Here, we show these fluxes accurate to $O(e^2)$ and to $O(v^8)$ beyond
Newtonian:
\begin{eqnarray}
\left\langle{dE \over dt}\right\rangle
         =\dot E^{(0)}+e^2\dot E^{(2)}+O(e^3),\\
\left\langle{dJ_z \over dt}\right\rangle
         =\dot J^{(0)}_z+e^2\dot J^{(2)}_z+O(e^3).
\end{eqnarray}
We note that $\dot J^{(0)}_z=\dot E^{(0)}/\Omega$ where $\Omega=v^3/M$.
We have already given the 5.5PN formula for $\dot E^{(0)}$ 
in section \ref{sec:5.5PN}, Eq.~(\ref{eq:dEdt11}).
Hence our task here is to evaluate the $O(e^2)$ corrections.
{}From the forms of $r(t)$ and $\varphi(t)$ given in
Eqs.~(\ref{eq:Seccsol}), we readily see that 
$Z_{\ell  m\omega_n}=O(e^{|n|})$ for $n=\pm1,\pm2$.
Thus, we only need the modes $n=0, \pm1$: As for the $n=0$ modes, 
we must retain terms up to $O(e^2)$, while for the $n=\pm 1$ modes, we 
only need terms up to $O(e)$. 
Then the 4PN formulas for $\dot E^{(2)}$ and $\dot J^{(2)}_z$ 
are found as
\begin{eqnarray}
e^2\dot E_{\rm PN}^{(2)}=&&e^2\left({dE\over dt}\right)_{\rm N} 
\nonumber\\
&&\times\Biggl\{ {{37}\over {24}} - 
     {{65\,{v^2}}\over {21}} + 
     {{1087\,\pi \,{v^3}}\over {48}} - 
     {{465337\,{v^4}}\over {9072}} - 
     {{118607\,\pi \,{v^5}}\over {1344}} - 
     {{17328779\,\pi \,{v^7}}\over {48384}} \nonumber \\
&&+\biggl( {{98546617999}\over {69854400}} - 
        {{65056\,{\gamma}}\over {315}} + 
        {{608\,{{\pi }^2}}\over 9} + 
        {{1712\,\ln 2}\over {315}} - 
        {{234009\,\ln 3}\over {560}} \nonumber \\
&&  - {{65056\,\ln v}\over {315}} \biggr)v^6  + 
\biggl( -{6653525574791 \over 2118916800 } + 
        {{118015\,{\gamma}}\over {98}} - 
        {{34093\,{{\pi }^2}}\over {126}} \nonumber \\
&&- {{1035547\,\ln 2}\over {4410}}  + 
        {{3986901\,\ln 3}\over {1120}} 
+ {{118015\,\ln v}\over {98}} \biggr)v^8  
                     \Biggr\},
\label{eq:dedtSecc}
\end{eqnarray}
and
\begin{eqnarray}
e^2\dot J^{(2)}_{z\,{\rm PN}}
=&&{e^2\over\Omega}\left({dE\over dt}\right)_{\rm N}
\nonumber\\
&&\times\Biggl\{ -{5\over 8} + 
     {{749\,{v^2}}\over {96}} + {{49\,\pi \,{v^3}}\over 8} - 
     {{232181\,{v^4}}\over {6048}} + 
     {{773\,\pi \,{v^5}}\over {336}} - 
     {{300637\,\pi \,{v^7}}\over {1008}} \nonumber \\
&&+\biggl( {{8017536229}\over {12700800}} - 
        {{19367\,{\gamma}}\over {210}} + 
        {{181\,{{\pi }^2}}\over 6} \nonumber \\ 
&&~+{{20009\,\ln 2}\over {210}} - 
        {{78003\,\ln 3}\over {280}} - 
        {{19367\,\ln v}\over {210}} \biggr){v^6} \nonumber \\
&&+\biggl( -{{12713730793}\over {61122600}} + 
        {{3463711\,{\gamma}}\over {8820}} - 
        {{14675\,{{\pi }^2}}\over {252}} 
\nonumber\\
&&~-{{2312441\,\ln 2}\over {980}}
+{{35449083\,\ln 3}\over {15680}}
+{{3463711\,\ln v}\over {8820}} \biggr)v^8\Biggr\}. 
\label{eq:djdtSecc}
\end{eqnarray}
In Appendix \ref{ap:v8ecc}, we show each $(\ell,m,n)$ 
component of the energy and angular momentum fluxes. Note that 
there was an error in the coefficients of the  
$e^2v^4$ terms in Ref.~\citen{ref:tagoshi}. This error is corrected in
Eqs.~(\ref{eq:dedtSecc}) and (\ref{eq:djdtSecc}) above.

To express the energy and angular momentum fluxes in terms of the
variable $x=(M\Omega_\varphi)^{1/3}$, we use Eq.~(\ref{eq:Omegaecc}).
To $O(e^2)$, it can be easily solved for $v$ as
\begin{equation}
  \label{eq:veqomega}
  v=x\left[1+{1\over3}f(x)e^2+O(e^3)\right].
\end{equation}
Then to $O(x^8)$ the energy and angular momentum fluxes are expressed as
\begin{eqnarray}
\left\langle{dE \over dt}\right\rangle
&=&\widetilde{\left({dE\over dt}\right)}_{\rm N}
\Biggl[1 + \left(\hbox{\rm $e$-independent terms}\right)
\nonumber\\
&&+ 
  {e^2}\,\left( {{157}\over {24}} - 
     {{6781\,{x^2}}\over {168}} + 
     {{2335\,\pi \,{x^3}}\over {48}} - 
     {{14929\,{x^4}}\over {189}} - 
     {{773\,\pi \,{x^5}}\over 3} \right.
\nonumber\\
&&+{{156066596771\,{x^6}}\over {69854400}} 
-{{106144\,{\gamma}\,{x^6}}\over {315}} 
+{{992\,{{\pi }^2}\,{x^6}}\over 9} 
\nonumber\\
&&- {{80464\,{x^6}\,\ln 2}\over {315}} 
-{{234009\,{x^6}\,\ln 3}\over {560}} 
-{{106144\,{x^6}\,\ln x}\over {315}} 
\nonumber\\
&&-{{32443727\,\pi \,{x^7}}\over {48384}} 
-{{3045355111074427\,{x^8}}\over {671272842240}} 
+{{507208\,{\gamma}\,{x^8}}\over {245}} 
\nonumber\\
&&
-{{31271\,{{\pi }^2}\,{x^8}}\over {63}} 
-{{151336\,{x^8}\,\ln 2}\over {441}} 
+{{12887991\,{x^8}\,\ln 3}\over {3920}} 
\nonumber\\
&&\left.
+{{507208\,{x^8}\,\ln x}\over {245}} 
\right) \Biggr],
\label{eq:esdedtx}
\end{eqnarray}
and
\begin{eqnarray}
\left\langle{dJ \over dt}\right\rangle
&=&\widetilde{\left({dJ\over dt}\right)}_{\rm N}
\Biggl[1+\left(\hbox{\rm $e$-independent terms}\right)
\nonumber\\
&&+{e^2}\,\left( {{23}\over 8} 
- {{3259\,{x^2}}\over {168}} 
+{{209\,\pi \,{x^3}}\over 8} 
-{{1041349\,{x^4}}\over {18144}} 
-{{785\,\pi \,{x^5}}\over 6} \right.
\nonumber\\
&&+{{91721955203\,{x^6}}\over {69854400}} 
-{{41623\,{\gamma}\,{x^6}}\over {210}} 
+{{389\,{{\pi }^2}\,{x^6}}\over 6} 
-{{24503\,{x^6}\,\ln 2}\over {210}} 
\nonumber\\
&&
-{{78003\,{x^6}\,\ln 3}\over {280}} 
-{{41623\,{x^6}\,\ln x}\over {210}} 
-{{91565\,\pi \,{x^7}}\over {168}} 
\nonumber\\
&&
-{{105114325363\,{x^8}}\over {72648576}} 
+{{696923\,{\gamma}\,{x^8}}\over {630}} 
-{{4387\,\pi^2\,{x^8}}\over {18}} 
\nonumber\\
&&\left.
- {{7051\,{x^8}\,\ln 2}\over {10}} 
+{{3986901\,{x^8}\,\ln 3}\over {1960}} 
+{{696923\,{x^8}\,\ln x}\over {630}} 
\right) \Biggr],
\end{eqnarray}
where $\widetilde{(dJ/dt)}_{\rm N}$ is the Newtonian angular momentum
flux expressed in terms of $x$, 
\begin{equation}
\widetilde{\left({dJ_z\over dt}\right)}_{\rm N}
={32\over 5}\left(\mu\over M\right)^2 M x^7,
\label{eq:Jnewtonx}
\end{equation}
and the $e$-independent terms in both $\langle dE/dt\rangle$ and 
$\langle dJ/dt\rangle$ are the same and are given by the terms in the
case of circular orbit, Eq.~(\ref{eq:dEdt11}), with the replacement
$v\to x$.

Finally, we consider the stability of circular orbits.
We note that the following relation holds:
\begin{equation}
\left\langle{dE \over dt}\right\rangle
-\Omega_{\varphi}\left\langle{dJ_z \over dt}\right\rangle 
=\sum_{\ell,m}{m\Omega_r \over 4\pi} 
\left( {|Z_{\ell m}^{(+)}|^2 \over \omega_+^3}
-{|Z_{\ell m}^{(-)}|^2 \over \omega_-^3} \right)e^2 +O(e^4)
\equiv H(v) e^2+O(e^4),
\end{equation}
where $\omega_{\pm}=m\Omega_\varphi\pm\Omega_r$ and
$Z_{\ell m}^{(\pm)}=Z_{\ell m\omega_{\pm}}$.
Here $H(v)$ is an important quantity which determines the stability 
of circular orbits under the radiation reaction.
Assuming the adiabatic evolution of the orbit,
the evolution equations for $r_0$ and $e$ due to the gravitational
radiation reaction are written as\cite{ref:AKOP} 
\begin{eqnarray}
\mu{d r_0 \over dt}& &
=-{2M(1-3v^2)^{3/2} \over v^4(1-6v^2)} \dot E^{(0)}+O(e^2),\\
\mu{d\ln e\over dt}& &=-{(1-2v^2)(1-3v^2)^{1/2} \over v^2(1-6v^2)}
\Bigl[g(v) \dot E^{(0)}+G(v)\Bigr]+O(e), 
\label{eq:dlnedt}
\end{eqnarray}
where
\begin{eqnarray}
g(v)&=&{2-27v^2+72v^4-36v^6 \over 2(1-2v^2)^2(1-6v^2)},\\
G(v)&=&\dot E^{(2)}-\Omega \dot J_z^{(2)}
          =H(v)-f(v)\dot E^{(0)}.
\label{eq:EH}
\end{eqnarray}
Using Eqs.(\ref{eq:dedtSecc}) and (\ref{eq:djdtSecc}), 
the 4PN formula of $G(v)$ is calculated as 
\begin{eqnarray}
G_{\rm PN}(v)=&&\left({dE\over dt}\right)_{\rm N}
\biggl[ {{13}\over 6} - {{2441\,{v^2}}\over {224}} + 
  {{793\,\pi \,{v^3}}\over {48}} - 
  {{234131\,{v^4}}\over {18144}} - 
  {{121699\,\pi \,{v^5}}\over {1344}} \nonumber \\
&& - 
  {{414029\,\pi \,{v^7}}\over {6912}} 
+   \biggl( {{36300112493}\over {46569600}} - 
     {{72011\,{\gamma}}\over {630}} + 
     {{673\,{{\pi }^2}}\over {18}} \nonumber \\
&&- 
     {{56603\,\ln 2}\over {630}} - 
     {{78003\,\ln 3}\over {560}} 
-    {{72011\,\ln v}\over {630}} \biggr)v^6  \nonumber \\
&&+ 
  \biggl( -{18638348721901 \over 6356750400} + 
     {{7157639\,{\gamma}}\over {8820}} - 
     {{17837\,{{\pi}^2}}\over {84}} \nonumber \\
&&+ 
     {{1085179\,\ln 2}\over {1764}} + 
     {{20367531\,\ln 3}\over {15680}} - 
     {{133120\,\ln 4}\over {441}} + 
     {{7157639\,\ln v}\over {8820}} \biggr)v^8\biggr].\nonumber \\
\end{eqnarray}
Note that $[g(v)\dot E^{(0)}+G(v)]/\dot E^{(0)}\to 19/6$ for $v\to0$;
i.e., in the Newtonian limit, the radiation reaction always reduces 
the eccentricity\cite{ref:peters}. By a numerical calculation,
Apostolatos et al.\cite{ref:AKOP} found that there exists a critical
radius $r_c$ below which the circular orbit becomes unstable;
$r_c\simeq6.6792M$. On the other hand, we find the use of the 
4PN formulas for $\dot E^{(0)}$ and $G(v)$ gives $r_c\sim7.38M$.
This indicates that a much higher order PN formula will be necessary to 
determine $r_c$ with good accuracy.
\section{Slightly eccentric orbit around a rotating black hole}
\label{sec:kerrecc}

In this section, we consider a slightly eccentric orbit on the
equatorial plane of a Kerr black hole and calculate the leading order
corrections of the eccentricity to the energy and angular momentum
fluxes up to $O(v^5)$ beyond Newtonian.  
The calculation is parallel to the one given in the previous section.

We consider the motion of a particle 
in the equatorial plane $\theta=\pi/2$, hence we have $C=0$. 
We define the radius $r=r_0$ as the one at which
the potential $R/r^4$ is minimum; 
$\partial(R/r^4)/\partial r\mid_{r=r_0}=0$.
We define the eccentricity $e$ such that $r=r_0(1+e)$ is 
a turning point of the radial motion at which $R(r=r_0(1+e))=0$. 
We assume $e \ll 1$. 
Using these definitions of $r_0$ and $e$, $E$ and $l_z$ 
are expressed as 
\begin{eqnarray*}
E &=& E^{(0)}+e E^{(1)}+e^2 E^{(2)}+O(e^3), \\
l_z &=& l_z^{(0)}+e l_z^{(1)}+e^2 l_z^{(2)}+O(e^3), \\ 
\end{eqnarray*} 
where $E^{(n)}$ and $l_z^{(n)}$ ($n=0,1,2$) are given by
\begin{eqnarray*}
E^{(0)}&=&{{1-2v^2+q v^3}\over (1-3v^2+2qv^3)^{(1/2)}}\,,\\ 
E^{(1)}&=&0\,,\\ 
E^{(2)}&=&{{v^2(1-3v^2+qv^3+q^2v^4)(-1+6v^2-8qv^3+3q^2v^4)} 
\over 2(1-3v^2+2qv^3)^{3/2}(-1+2v^2-q^2v^4)}\,,\\ 
l_z^{(0)}&=&{{r_0v(1-2qv^3+q^2 v^4)}\over 
(1-3v^2+2qv^3)^{(1/2)}}\,,\\ 
l_z^{(1)}&=&0\,,\\ 
l_z^{(2)}&=&{{q\, r_0\, v^5(q-3v+qv^2+q^2v^3) 
(-1+6 v^2-8 qv^3+3 q^2 v^4)} 
\over 2(1-3v^2+2qv^3)^{3/2}(-1+2v^2-q^2v^4)}\,,\\ 
\end{eqnarray*} 
where $v=(M/r_0)^{1/2}$.
The post-Newtonian expansions of 
$E^{(n)}$ and $l_z^{(n)}$ up to the required order are
\begin{eqnarray}
E&=&1 - {M\over 2r_0} 
+ {3M^2\over 8r_0^2} -
  {qM^{5/2}\over r_0^{5/2}}
+e^2\left( {M\over {2r_0}} 
- {5M^2\over 4r_0^2} +
   {3qM^{5/2}\over r_0^{5/2}}\right)+O(v^6), 
\label{eq:EE}\\
l_z&=&(Mr_0)^{1/2}\left[1+{3M\over 2r_0}-{3qM^{3/2}\over r_0^{3/2}}
+\left({27\over 8}+q^2\right){M^2\over r_0^2}
-{15q M^{5/2}\over 2r_0^{5/2}}\right.
\nonumber\\
& &\left. +e^2\left({q^2M^2\over 2 r_0^2}
          -{3 q M^{5/2}\over 2r_0^{5/2} }\right)
+O(v^6) \right].
\label{eq:lz}\\
\nonumber
\end{eqnarray}

Now we solve the geodesic equations for a slightly eccentric orbit. 
The radial equation is 
\begin{equation}
\left({dr\over dt}\right)^2={R\over T^2}.
\end{equation}
We expand $r(t)$ as 
\begin{equation}
r(t)=r_0\left[1+e r^{(1)}(t)+e^2 r^{(2)}(t)
+O(e^3)\right], \label{eq:rt}
\end{equation}
and $R/T^2$ in terms of $e$ and $v$ 
using Eq.~(\ref{eq:EE}) and (\ref{eq:lz}). 
Collecting terms of the equal order in $e$, we obtain 
\begin{equation}
\left({dr^{(1)}\over dt}\right)^2
=\Omega_r^2(1-(r^{(1)})^2), \label{eq:dr1}
\end{equation} 
and 
\begin{equation} 
{1\over \Omega_r^2}{dr^{(1)}\over dt}
  {dr^{(2)}\over dt}=-r^{(1)}r^{(2)}+q_0 
+q_1 r^{(1)}+q_2 (r^{(1)})^2, \label{eq:dr2}
\end{equation} 
where $\Omega_r$, $q_0$, $q_1$ and $q_2$ are given
in the post-Newtonian series forms as
\begin{eqnarray}
\Omega_r&=&{M^{1/2} \over r_0^{3/2}}
\left[1-{3M\over r_0}+{3 q M^{3/2}\over r_0^{3/2}}
-{(9+3 q^2)M^2\over {2 r_0^2}}
+{15 q M^{5/2}\over r_0^{5/2}}+O(v^6)\right], \\
q_0&=&
-1 + {M\over r_0} - {2qM^{3/2}\over r_0^{3/2}} + 
   {\left( 6 + q^2\right)M^2\over r_0^2} - 
   {20qM^{5/2}\over r_0^{5/2}} + O(v^6), 
\\
q_1&=&
 {2M\over r_0}\left[1 +{2M\over r_0} - 
   {3qM^{3/2}\over r_0^{3/2}} +{4M^2\over r_0^2} 
- {6qM^{5/2}\over r_0^{5/2}} + O(v^6)\right], 
\\
q_2&=&
1 - {3M\over r_0} +{2qM^{3/2}\over r_0^{3/2}} 
  -{(10 +q^2)M^2\over r_0^2} +{26qM^{5/2}\over r_0^{5/2}} 
+ O(v^6). 
\\
\nonumber
\end{eqnarray}
We obtain $r^{(1)}(t)$ from Eq.~(\ref{eq:dr1}) as
\begin{equation}
r^{(1)}(t)=\cos\Omega_r t, \label{eq:r1}
\end{equation}
where we set $r(t=0)=r_0(1+e)$. Substitution of Eq. (\ref{eq:r1})
into Eq. (\ref{eq:dr2}) and yields after integration
\begin{equation}
r^{(2)}(t)=q_3(1-\cos\Omega_r t)+q_4(1-\cos 2\Omega_r t), \label{eq:r2}
\end{equation}
where $q_3=-q_0$ and $q_4=q_2/2$. 

In the same way, we can solve the angular motion $\varphi(t)$. From
Eq.~(\ref{eq:geod}), we have $d\varphi/dt=\Phi/T$, which can be expanded
in terms of $e$ using Eqs.~(\ref{eq:EE}), 
(\ref{eq:lz}), (\ref{eq:rt}), (\ref{eq:r1}) and (\ref{eq:r2}). 
Integrating the resulting equation, we obtain 
\begin{equation}
\varphi(t)=\Omega_{\varphi} t 
+ e p_1 \sin \Omega_rt +e^2 p_2 \sin \Omega_rt
+e^2 p_3 \sin 2\Omega_r t+O(e^3), \label{eq:phi}
\end{equation}
where
\begin{eqnarray}
p_1&=& -2 - {4M\over r_0} + {6qM^{3/2}\over r_0^{3/2}} 
  -{(17+q^2)M^2\over r_0^2} +{48qM^{5/2}\over r_0^{5/2}}  
   + O(v^6), \nonumber\\
p_2&=& 2 + {2M\over r_0} -{2qM^{3/2}\over r_0^{3/2}} 
  +{( 1 - q^2)M^2\over r_0^2} +{6qM^{5/2}\over r_0^{5/2}} 
   + O(v^6), 
\nonumber\\
p_3&=& {5\over 4} + {M\over 4r_0} -{2qM^{3/2}\over r_0^{3/2}} 
   -{(9+7q^2)M^2\over 8r_0^2} +{59( -1+q^2)M^3\over 8r_0^3}
+ O(v^6), 
\label{eq:pp}\\
\nonumber
\end{eqnarray}
and 
\begin{eqnarray}
\Omega_{\varphi}&&=\left({M\over r_0}\right)^{1/2}
\left[1 - {{q\,{M^{3/2}}}\over {{{{r_0}}^{3/2}}}} \right.
\nonumber\\
& &\left. 
+e^2\left( -{{3}\over {2}} + 
   {9M\over 2r_0} - {9q M^{3/2}\over 2 r_0^{3/2}} + 
   {{3\left( 6 + {q^2} \right){M^{2}}}\over r_0^{2}} - 
   {60q M^{5/2}\over r_0^{5/2}}\right) 
    + O(v^6) \right]. 
\label{eq:omegaphi}\\
\nonumber
\end{eqnarray}
As in the case of the previous section, the fact that $\Omega_r\not=
\Omega_{\varphi}$ implies that these eccentric orbits are not closed. 

Using the above solution of the geodesic equations, 
we evaluate the source term of the Teukolsky equation. 
We set $\theta=\pi/2$ in the expressions of $A_{n\,n\,0}$ etc. in
Eqs.~(\ref{eq:ann}). 
Again, parallel to the discussion in section \ref{sec:schwaecc},
the orbits of our interest have the properties,
\begin{equation}
r(t+\Delta t_r)=r(t),\qquad 
\left.{d\varphi\over dt}\right|_{t=t+\Delta t_r}
=\left.{d\varphi\over dt}\right|_{t=t}, 
\end{equation}
where $\Delta t_r=2\pi/\Omega_r$. Hence Eq.~(\ref{eq:zzq2}) 
reduces to the form,
\begin{equation}
\tilde Z_{\ell m\omega}
=Z_{\ell m\omega}\delta(\omega-\omega_n),
\label{eq:ZeccKerr}
\end{equation}
where 
\begin{equation}
\omega_n=n\Omega_r+m\Omega_{\varphi}, \quad (n=0,\pm 1, \pm 2, \ldots),
\label{eq:omeganK}
\end{equation}
and
\begin{equation}
\label{eq:ZKecc}
  Z_{\ell m\omega}
={\mu\Omega_r\over2i\omega B^{\rm inc}_{\ell m\omega}} 
\int^{\Delta t_r}_0dt e^{i\omega t-i m \varphi(t)}W_{\ell m\omega}\,,
\end{equation}
with $W_{\ell m\omega}$ given by Eq.~(\ref{eq:Wdef}).

Using the solution of the geodesic equation for $r(t)$, 
we expand $W_{\ell m\omega}$ in terms of $e$.
The result takes the form,
\begin{eqnarray}
W_{\ell m\omega}&=&f_0+e\left(f_1 r^{(1)}
+f_2 {dr^{(1)}\over dt}\right)
+e^2\left(f_3 r^{(2)}+ f_4 {dr^{(2)}\over dt}+f_5 ({r^{(1)}})^2\right.
\nonumber\\
&+&\left.
 f_6 \left({dr^{(1)}\over dt}\right)^2
+f_7 r^{(1)} {dr^{(1)}\over dt}+f_8\right)
+O(e^3), \label{eq:wexpand}\\
\nonumber
\end{eqnarray}
where $f_0\sim f_8$ are time-independent coefficients.
Inserting this form to Eq.~(\ref{eq:ZKecc})
we obtain 
\begin{eqnarray} 
Z_{\ell m\omega_n}
&=& {\mu\pi\over i\omega_n B^{\rm inc}_{\ell m\omega_n}}
\left[\left\{ f_0+e^2\left({f_5\over 2}+f_8
-{{m^2 p_1^2}\over 4}f_0\right.\right.\right.
\nonumber\\
& &\left.\left. +(q_3+q_4)f_3
+{{im\Omega_r p_1}\over 2}f_2
+{\Omega_r^2\over 2}f_6 \right)\right]\delta_{n,0} 
\nonumber\\
& &+e \left({f_1\over 2}+{{m p_1}\over 2}f_0
-{{i \Omega_r}\over 2}f_2\right)\delta_{n,1}
\nonumber\\
& &+e \left({f_1\over 2}-{{m p_1}\over 2}f_0
+{{i \Omega_r}\over 2}f_2\right)\delta_{n,-1}
\nonumber\\
& &+e^2\left(
{{3\,{f_5}}\over 4} + {f_8} 
+ {{{f_1}\,m\,{p_1}}\over 4} - 
  {{{f_0}\,{m^2}\,{{{p_1}}^2}}\over 8} 
+ {{{f_0}\,m\,{p_3}}\over 2} + 
  {f_3}\,{q_3} 
\right.\nonumber\\
& &\left.+ {{{f_3}\,{q_4}}\over 2} 
- {i\over 4}\,{f_7}\,w + 
  {i\over 4}\,{f_2}\,m\,{p_1}\,w 
+ i\,{f_4}\,{q_4}\,w + 
  {{{f_6}\,{w^2}}\over 4}
\right)\delta_{n,2}
\nonumber\\
& &+e^2\left(
{{3\,{f_5}}\over 4} + {f_8} 
- {{{f_1}\,m\,{p_1}}\over 4} - 
  {{{f_0}\,{m^2}\,{{{p_1}}^2}}\over 8} 
- {{{f_0}\,m\,{p_3}}\over 2} + 
  {f_3}\,{q_3} 
\right.\nonumber\\
& &\left.\left.+ {{{f_3}\,{q_4}}\over 2} 
+ {i\over 4}\,{f_7}\,w + 
  {i\over 4}\,{f_2}\,m\,{p_1}\,w 
- i\,{f_4}\,{q_4}\,w + 
  {{{f_6}\,{w^2}}\over 4}
\right)\delta_{n,-2}\right],
\label{eq:Ilmw}
\end{eqnarray}
where $\delta_{n,n'}$ is the Kronecker delta.
We see from this equation that $Z_{\ell m\omega_n}=O(e^{|n|})$
just as in the Schwarzschild case.
Therefore we only need to retain the $n=0$, $\pm1$ modes
to evaluate the luminosity up to $O(e^2)$. 

We calculate the energy and angular momentum fluxes 
to $O(v^5)$ beyond the quadru\-pole formula and to $O(e^2)$ in the
eccentricity. The time-averaged energy and angular momentum fluxes
are given by Eqs.~(\ref{eq:Eflux}) and (\ref{eq:Jflux}), respectively.
In order to express the post-Newtonian corrections to the luminosity,
we define $\eta_{\ell mn}$ as
\begin{equation}
\left( {dE \over dt} \right)_{\ell mn}
\equiv{1\over2}\left({dE\over dt}\right)_{\rm N}\eta_{\ell,m,n}\,,
\label{eq:etalmw}
\end{equation}
where $(dE/dt)_{\rm N}$ is the Newtonian quadrupole 
luminosity given by Eq.~(\ref{eq:Enewton}). 
In the following, we show $\eta_{\ell mn}$
for $m\geq 0$ modes. $\eta_{\ell, m,n}$ for $m<0$ are
obtained from the symmetry $\eta_{\ell,m,n}=\eta_{\ell,-m,-n}$,
which follows from the property of $\omega_n$ in the present case,
given by Eq.~(\ref{eq:omegan}). 

For $\ell=2$, the 2.5PN formulas for $\eta_{\ell,m,n}$ 
are found to be\footnote{As mentioned in the previous section, we have
detected an error in the formula for $\eta_{2,2,0}$ in
Ref.~\citen{ref:tagoshi}. The term $-14270/147 e^2 v^4$ there is in
correct. Accordingly, formulas for $dE/dt$ and $dJ_z/dt$ below 
are also corrected in this paper.} 
\begin{eqnarray*}
\eta_{2,2,0}&=&
1 - {{107\,{v^2}}\over {21}} + 4\,\pi \,{v^3} - 
   6\,q\,{v^3} + {{4784\,{v^4}}\over {1323}} + 
   2\,{q^2}\,{v^4} \\
& & - {{428\,\pi \,{v^5}}\over {21}} + 
   {{4216\,q\,{v^5}}\over {189}} \\
& &+ {e^2}\,\left( -10 + {{932\,{v^2}}\over {21}} - 
      46\,\pi \,{v^3} + 84\,q\,{v^3} - 
      {{14270\,{v^4}}\over {147}} - 
      23\,{q^2}\,{v^4} \right.\\
& &\left.+ {{4748\,\pi \,{v^5}}\over {21}} + 
      {{2675\,q\,{v^5}}\over {189}} \right), \\ 
\eta_{2,2,1}&=&
{e^2}\,\left( {{729}\over {64}} - 
        {{3645\,{v^2}}\over {64}} + 
        {{2187\,\pi \,{v^3}}\over {32}} - 
        {{3645\,q\,{v^3}}\over {32}} \right.\\
& &\left.
   +    {{24057\,{v^4}}\over {256}} + 
        {{2187\,{q^2}\,{v^4}}\over {64}} - 
        {{6561\,\pi \,{v^5}}\over {16}} + 
        {{9477\,q\,{v^5}}\over {112}} \right), \\
\eta_{2,2,-1}&=&
{e^2}\,\left( {9\over {64}} + 
        {{1041\,{v^2}}\over {448}} + 
        {{9\,\pi \,{v^3}}\over {32}} - 
        {{153\,q\,{v^3}}\over {32}} \right.\\
& &\left.+ 
        {{2224681\,{v^4}}\over {112896}} + 
        {{99\,{q^2}\,{v^4}}\over {64}} + 
        {{615\,\pi \,{v^5}}\over {112}} - 
        {{27857\,q\,{v^5}}\over {336}} \right), \\
\eta_{2,1,0}&=&
{{{v^2}}\over {36}} - 
      {{q\,{v^3}}\over {12}} - 
      {{17\,{v^4}}\over {504}} + 
      {{{q^2}\,{v^4}}\over {16}} + 
      {{\pi \,{v^5}}\over {18}} - 
      {{793\,q\,{v^5}}\over {9072}} \\
& &+  {e^2}\,\left( {{-2\,{v^2}}\over 9} + 
         {{2\,q\,{v^3}}\over 3} + 
         {{93\,{v^4}}\over {112}} - 
         {{{q^2}\,{v^4}}\over 2} - 
         {{19\,\pi \,{v^5}}\over {36}} - 
         {{27113\,q\,{v^5}}\over {18144}} \right), \\ 
\eta_{2,1,1}&=&
{e^2}\,\left( {{4\,{v^2}}\over 9} - 
        {{4\,q\,{v^3}}\over 3} - 
        {{172\,{v^4}}\over {63}} + {q^2}\,{v^4} + 
        {{16\,\pi \,{v^5}}\over 9} + 
        {{2794\,q\,{v^5}}\over {567}} \right), \\ 
\eta_{2,0,\pm 1}&=&
{e^2}\,\left( {1\over {96}} - 
      {{145\,{v^2}}\over {672}} + 
      {{\pi \,{v^3}}\over {48}} + 
      {{3\,q\,{v^3}}\over {16}} + 
      {{282521\,{v^4}}\over {169344}} - 
      {{3\,{q^2}\,{v^4}}\over {32}} \right.\\
& &\left.- 
      {{83\,\pi \,{v^5}}\over {168}} - 
      {{1255\,q\,{v^5}}\over {504}} \right), \\
\end{eqnarray*}
and $\eta_{2,1,-1}$ becomes $O(v^6)$. 
Putting together the above results, 
we obtain $(dE/dt)_\ell \equiv \sum_{mn} (dE/dt)_{\ell mn}$ 
for $\ell=2$ as
\begin{eqnarray}
\left({dE\over dt}\right)_2&=&
\left({dE\over dt}\right)_{\rm N}\left\{
1 - {{1277\,{v^2}}\over {252}} + 
      4\,\pi \,{v^3} - {{73\,q\,{v^3}}\over {12}} + 
      {{37915\,{v^4}}\over {10584}} \right.\nonumber\\
& &
+  {{33\,{q^2}\,{v^4}}\over {16}} - 
      {{2561\,\pi \,{v^5}}\over {126}} + 
      {{201575\,q\,{v^5}}\over {9072}} \nonumber\\
& &+  {e^2}\,\left( {{37}\over {24}} - 
         {{2581\,{v^2}}\over {252}} + 
         {{1087\,\pi \,{v^3}}\over {48}} - 
         {{211\,q\,{v^3}}\over 6} + 
         {{325393\,{v^4}}\over {21168}} + 
         {{105\,{q^2}\,{v^4}}\over 8} \right.\nonumber\\
& &\left.\left.
-        {{29857\,\pi \,{v^5}}\over {168}} + 
         {{11293\,q\,{v^5}}\over {672}} \right) \right\}. 
\label{eq:dedt2}\\
\nonumber
\end{eqnarray}

For $\ell=3$, we obtain
\begin{eqnarray*}
\eta_{3,3,0}&=&
{{1215\,{v^2}}\over {896}} - 
      {{1215\,{v^4}}\over {112}} + 
      {{3645\,\pi \,{v^5}}\over {448}} - 
      {{1215\,q\,{v^5}}\over {112}} \\
& &+  {e^2}\,\left( {{-10935\,{v^2}}\over {448}} + 
         {{37665\,{v^4}}\over {256}} - 
         {{142155\,\pi \,{v^5}}\over {896}} + 
         {{134865\,q\,{v^5}}\over {448}} \right), \\
\eta_{3,3,1}&=&
{e^2}\,\left( {{640\,{v^2}}\over 
          {21}} - {{46720\,{v^4}}\over {189}} + 
        {{5120\,\pi \,{v^5}}\over {21}} - 
        {{1280\,q\,{v^5}}\over 3} \right), \\
\eta_{3,3,-1}&=&
{e^2}\,\left( {{15\,{v^2}}\over {14}} + 
        {{1055\,{v^4}}\over {126}} + 
        {{30\,\pi \,{v^5}}\over 7} - 
        {{435\,q\,{v^5}}\over {14}} \right), \\
\eta_{3,2,0}&=&
{{5\,{v^4}}\over {63}} - 
      {{40\,q\,{v^5}}\over {189}} + 
      {e^2}\,\left( {{-65\,{v^4}}\over {63}} + 
         {{520\,q\,{v^5}}\over {189}} \right), \\
\eta_{3,2,1}&=&
{e^2}\,\left( {{3645\,{v^4}}\over 
          {1792}} - {{1215\,q\,{v^5}}\over {224}} \right),\\
\eta_{3,2,-1}&=&
{e^2}\,\left( {{5\,{v^4}}\over {1792}} 
 - {{5\,q\,{v^5}}\over {672}} \right), \\
\eta_{3,1,0}&=&
{{{v^2}}\over {8064}} - 
      {{{v^4}}\over {1512}} + 
      {{\pi \,{v^5}}\over {4032}} - 
      {{17\,q\,{v^5}}\over {9072}} \\
& &+  {e^2}\,\left( {{-{v^2}}\over {4032}} + 
         {{65\,{v^4}}\over {16128}} - 
         {{\pi \,{v^5}}\over {1152}} + 
         {{199\,q\,{v^5}}\over {36288}} \right), \\
\eta_{3,1,1}&=&
{e^2}\,\left( {{{v^2}}\over {126}} - 
        {{23\,{v^4}}\over {126}} + 
        {{2\,\pi \,{v^5}}\over {63}} + 
        {{122\,q\,{v^5}}\over {567}} \right), \\
\eta_{3,0,\pm 1}&=&
{e^2}\,\left( {{{v^4}}\over {2688}} - 
        {{q\,{v^5}}\over {1008}} \right), \\
\end{eqnarray*}
and $\eta_{3,1,-1}$ becomes $O(v^6)$. Thus we obtain
\begin{eqnarray}
\left({dE\over dt}\right)_3&=&
\left({dE\over dt}\right)_{\rm N}\left\{
{{1367\,{v^2}}\over {1008}} - 
      {{32567\,{v^4}}\over {3024}} + 
      \left( {{16403\,\pi }\over {2016}} - 
         {{896\,q}\over {81}} \right) \,{v^5} \right.\nonumber\\
& &\left.
+ {e^2}\,\left( {{1801\,{v^2}}\over {252}} - 
         {{78509\,{v^4}}\over {864}} + 
         \left( {{40083\,\pi }\over {448}} - 
            {{8913\,q}\over {56}} \right) \,{v^5}
          \right)\right\}.
\label{eq:dedt3}\\
\nonumber
\end{eqnarray}

For $\ell=4$, we have
\begin{eqnarray*}
\eta_{4,4,0}&=&
{{1280\,{v^4}}\over {567}} - 
  {{37120\,{e^2}\,{v^4}}\over {567}}\,,\\
\eta_{4,4,1}&=&
{{48828125\,{e^2}\,{v^4}}\over {580608}}\,,\\
\eta_{4,4,-1}&=&
{{32805\,{e^2}\,{v^4}}\over {7168}}\,,\\
\eta_{4,2,0}&=&
{{5\,{v^4}}\over {3969}} - 
  {{25\,{e^2}\,{v^4}}\over {3969}}\,,\\
\eta_{4,2,-1}&=&
{{5\,{e^2}\,{v^4}}\over {254016}}\,,\\
\eta_{4,0,\pm 1}&=&
{{{e^2}\,{v^4}}\over {225792}}\,,\\
\end{eqnarray*}
and $\eta_{4,2,1}$ becomes $O(v^6)$.
Hence we have 
\begin{equation}
\left({dE\over dt}\right)_4=
\left({dE\over dt}\right)_{\rm N}\left\{
{{8965\,{v^4}}\over {3969}} + 
  {{2946739\,{e^2}\,{v^4}}\over {127008}}\right\}.
\label{eq:dedt4}
\end{equation}

Finally, gathering all the above results, we have the luminosity
up to $O(v^5)$ as 
\begin{eqnarray}
\left\langle{dE\over dt}\right\rangle&=&
\left({dE\over dt}\right)_{\rm N}\left\{
1 - {{1247\,{v^2}}\over {336}} + 
      4\,\pi \,{v^3} - 
      {{73\,q\,{v^3}}\over {12}} - 
      {{44711\,{v^4}}\over {9072}} + 
      {{33\,{q^2}\,{v^4}}\over {16}} \right.\nonumber\\
& &
-  {{8191\,\pi \,{v^5}}\over {672}} + 
      {{3749\,q\,{v^5}}\over {336}} + 
      {e^2}\,\left( {{37}\over {24}} - 
         {{65\,{v^2}}\over {21}} + 
         {{1087\,\pi \,{v^3}}\over {48}} - 
         {{211\,q\,{v^3}}\over 6} \right.\nonumber\\
& &\left.\left.
-        {{465337\,{v^4}}\over {9072}} + 
         {{105\,{q^2}\,{v^4}}\over 8} - 
         {{118607\,\pi \,{v^5}}\over {1344}} - 
         {{95663\,q\,{v^5}}\over {672}} \right)\right\}.
\label{eq:dedtalle}
\end{eqnarray}
If we set $q=0$, the $e^2$ correction terms in the above formula
completely agree with the corresponding terms in Eq.~(\ref{eq:dedtSecc}) 
in the previous section.

To compare our results with those derived in the 
standard post-Newtonian method, it is convenient to change the parameter
from $v$ to $x\equiv (M\Omega_{\varphi})^{1/3}$. 
The relation between $v$ and $x$ is given by
\begin{equation}
v=x\left[1+{q\over 3}x^3+e^2\left\{{1\over 2}-{3\over 2}x^2
+{8\over 3}qx^3
-6x^4-q^2x^4+{31\over 2}qx^5\right\}\right]\,.
\label{eq:vptov} 
\end{equation}
Then we obtain
\begin{eqnarray}
\left\langle{dE \over dt}\right\rangle
&=&\widetilde{\left({dE\over dt}\right)}_{\rm N}
\Bigl\{1-{\frac {1247}{336}}\,{x}^{2}
-{11\over 4}\,q{x}^{3}+4\,{x}^{3}\pi 
-{\frac {44711}{9072}}\,{x}^{4}+{\frac {33}{16}}\,{x}^{4}{q}^{2}
\nonumber\\
&&
-{\frac {59}{16}}\,q{x}^{5}-{\frac {8191}{672}}\,{x}^{5}\pi
+\bigl({\frac {157}{24}}
-{\frac {6781}{168}}\,{x}^{2}
-{\frac {2009}{72}}\,q{x}^{3}
+{\frac {2335}{48}}\,{x}^{3}\pi
\nonumber\\
&&
-{\frac {14929}{189}}\,{x}^{4}
+{\frac {281}{16}}\,{x}^{4}{q}^{2}
-{\frac {2399}{56}}\,q{x}^{5}-{\frac {773}{3}}\,{x}^{5}\pi\bigr){e}^{2}
\Bigr\}, 
\end{eqnarray}
where $\widetilde{(dE/dt)}_{\rm N}$ is the quadrupole flux expressed
in terms of $x$, Eq.~(\ref{eq:Enewtonx}).
We find that the terms which are proportional to
$e^2$ agree with the formulas derived by 
Peters and Mathews \cite{ref:PM} 
at leading order, Galt'sov et al.\cite{ref:GMP} and 
Blanchet and Sch\"{a}fer at $v^2$ order \cite{ref:BS}, 
Blanchet and Sch\"{a}fer at $v^3$ order for $q=0$\cite{ref:BS2} and
Shibata at $v^3$ order for $q\neq0$\cite{ref:shibata}, if we expand
their formulas by $e$ assuming $e\ll 1$ and $\mu/M\ll 1$.

{}From Eq.~(\ref{eq:Jflux}), the partial mode contributions to the
angular momentum fluxes for $\ell=2$, 3 and 4 are calculated to be
\begin{eqnarray*}
\left({dJ_z\over dt}\right)_2&=&
\left({dJ_z\over dt}\right)_{\rm N}\left\{
1 - {{1277\,{v^2}}\over {252}} + 
      4\,\pi \,{v^3} - {{61\,q\,{v^3}}\over {12}} + 
      {{37915\,{v^4}}\over {10584}} + 
      {{33\,{q^2}\,{v^4}}\over {16}} \right.\\
& &-  {{2561\,\pi \,{v^5}}\over {126}} + 
      {{22229\,q\,{v^5}}\over {1296}} + 
      {e^2}\,\left( -{5\over 8} + 
         {{137\,{v^2}}\over {24}} + 
         {{49\,\pi \,{v^3}}\over 8} - 
         {{57\,q\,{v^3}}\over 4} \right.\\
& &\left.\left.
        -{{235675\,{v^4}}\over {14112}} + 
         {{203\,{q^2}\,{v^4}}\over {32}} - 
         {{20437\,\pi \,{v^5}}\over {504}} - 
         {{164449\,q\,{v^5}}\over {4536}} \right)\right\}.\\
\left({dJ_z\over dt}\right)_3&=&
\left({dJ_z\over dt}\right)_{\rm N}\left\{
{{1367\,{v^2}}\over {1008}} - 
      {{32567\,{v^4}}\over {3024}} + 
      \left( {{16403\,\pi }\over {2016}} - 
         {{88049\,q}\over {9072}} \right) \,{v^5}\right.\\
& &\left.+ {e^2}\,\left( {{67\,{v^2}}\over 
           {32}} - {{66497\,{v^4}}\over 
           {2016}} + 
         \left( {{43193\,\pi }\over {1008}} - 
            {{1675571\,q}\over {18144}} \right) \
           {v^5} \right)\right\},\\
\left({dJ_z\over dt}\right)_4&=&
\left({dJ_z\over dt}\right)_{\rm N}\left\{
{{8965\,{v^4}}\over {3969}} + 
   {{478195\,{e^2}\,{v^4}}\over {42336}}\right\}, 
\end{eqnarray*}
where $(dJ_z/dt)_{\rm N}$ is defined by 
\begin{equation}
\left({dJ_z\over dt}\right)_{\rm N}
={32\mu^2M^{5/2}\over5r_0^{7/2}}
={32\over5}\left({\mu\over M}\right)^2Mv^7.
\label{eq:Jnewton}
\end{equation}
Total angular momentum luminosity is then given by
\begin{eqnarray}
\left\langle{dJ_z\over dt}\right\rangle&=&
\left({dJ_z\over dt}\right)_{\rm N}\left\{
1 - {{1247\,{v^2}}\over {336}} + 
      \left( 4\,\pi  - {{61\,q}\over {12}} \right
         ) \,{v^3} + 
      \left( -{{44711}\over {9072}} + 
         {{33\,{q^2}}\over {16}} \right) \,{v^4} \right.\nonumber\\
& &  + \left( {{-8191\,\pi }\over {672}} + 
         {{417\,q}\over {56}} \right) \,{v^5} + 
      {e^2}\,\left( -{5\over 8} + 
         {{749\,{v^2}}\over {96}} + 
         \left( {{49\,\pi }\over 8} - 
            {{57\,q}\over 4} \right) \,{v^3} \right.\nonumber\\
& &\left.\left.
+  \left( -{{232181}\over {6048}} + 
            {{203\,{q^2}}\over {32}} \right) \,
          {v^4} + \left( {{773\,\pi }\over 
              {336}} - {{28807\,q}\over {224}} 
              \right) \,{v^5} \right)\right\}.
\label{eq:djdtall}
\end{eqnarray}
The $e^2$ terms in the above also agree with the corresponding terms in
Eq.~(\ref{eq:djdtSecc}). 
The angular momentum flux expressed in terms of
 $x=(M\Omega_\varphi)^{1/3}$ is given by
\begin{eqnarray}
\left\langle{dJ_z \over dt}\right\rangle
&=&\widetilde{\left({dJ_z\over dt}\right)}_{\rm N}
\Bigl\{
1-{\frac {1247}{336}}\,{x}^{2}
-{11\over 4}\,q{x}^{3}+4\,{x}^{3}\pi 
-{\frac {44711}{9072}}\,{x}^{4}
+{\frac {33}{16}}\,{x}^{4}{q}^{2}
\nonumber\\
&&
-{\frac {59}{16}}\,q{x}^{5}-{\frac {8191}{672}}\,{x}^{5}\pi
+\bigl({\frac {23}{8}}
-{\frac {3259}{168}}\,{x}^{2}
-{\frac {371}{24}}\,q{x}^{3}+{\frac {209}{8}}\,{x}^{3}\pi
\nonumber\\
&&
-{\frac {1041349}{18144}}\,{x}^{4}+{\frac {171}{16}}\,{x}^{4}{q}^{2}
-{\frac {243}{8}}\,q{x}^{5}
-{\frac {785}{6}}\,{x}^{5}\pi \bigr){e}^{2}\Bigr\},
\end{eqnarray}
where $\widetilde{(dE/dt)}_{\rm N}$ is the Newtonian flux expressed in
terms of $x$, Eq.~(\ref{eq:Jnewtonx}).
\section{Circular orbit with small inclination 
               from the equatorial plane}
\label{sec:inclination}
In this section, we consider the case of a circular orbit at $r=r_0$
with small inclination from the equatorial plane. We evaluate
$\langle dE/dt\rangle$ and $\langle dJ_z/dt\rangle$ to $O(v^5)$ beyond
Newtonian. 

In this case, the orbital plane precesses around the symmetric axis.
The degree of precession is determined by the value of the Carter
constant $C$. If $r_0$ and $C$ are given, 
the energy $E$ and the $z$-component of the angular momentum $l_z$ 
are obtained by the two equations, $R=0$ and $\partial R/\partial r=0$,
where $R$ is a function defined by Eq.~(\ref{eq:Rgeo}). 
We introduce a dimensionless parameter $y$ defined by
\begin{equation}
y={C\over Q^2}\,;\quad Q^2=l_z^2+a^2(1-E^2). 
\end{equation}
We assume $y$ is a small number.
Since $Q^2\sim l_z^2$ and $C \sim l_x^2+l_y^2$, 
this is physically equivalent to assuming $l_x^2+l_y^2 \ll l_z^2$. 
Since we do not need the exact expressions for $E$ and $l_z$
in terms of $r_0$ and $y$, we show them 
to the first order in $y$ as well as to $O(v^5)$. 
They are given by
\begin{eqnarray}
E&=&1-{M \over 2r_0}+{3M^2 \over 8r_0^2}
-{M^{3/2}a\over r_0^{5/2}}(1-{y\over2})+O(v^6),\\
l_z&=&(Mr_0)^{1/2}\Biggl[\Bigl(1-{y \over 2}\Bigr)
+{3 M\over 2r_0}(1-{y\over2})-{3M^{1/2}a\over r_0^{3/2}}(1-y)
\nonumber\\
&&\quad+{27M^2\over8 r_0^2}(1-{y\over2})+{a^2\over r_0^2}(1-2y)
-{15M^{3/2}a\over 2r_0^{5/2}}(1-y)+O(v^6)\Biggr],
\end{eqnarray}
where note that $a=Mq$ ($|q|<1$).

To solve the geodesic equations under the assumption 
$y\ll1$, we first set $\theta=\pi/2+y^{1/2} \theta'$ 
and consider the geodesic equation for 
$\theta$. It then becomes
\begin{equation}
\Bigl({d\theta' \over d\tau}\Bigr)^2 
={1\over\Sigma^2}\left[Q^2 - {\sin^2(y^{1/2} \theta')\over y} 
\Bigl\{a^2(1-E^2)+{l_z^2 \over \cos^2(y^{1/2} \theta') } 
\Bigr\}\right]. \label{eq:jkl}
\end{equation}
Since the right hand side of Eq.~(\ref{eq:jkl}) contains only
even-functions of $y^{1/2}\theta'$, we can solve it iteratively by
expanding $\theta'$ as
\begin{equation}
\theta'=\theta_{(0)}+y\theta_{(1)}+y^2\theta_{(2)}+\cdots. 
\end{equation}
This method is similar to the one we have used in section
\ref{sec:schwaecc} or \ref{sec:kerrecc}.
However, here we only consider the lowest order solution 
$\theta_{(0)}$. This means we take into account 
the effect of inclination up
to $O(y)$, as seen from the structure of the geodesic equations. 
The equation for $\theta_{(0)}$ is
\begin{equation}
\Bigl({d\theta_{(0)}\over d\tau}\Bigr)^2=
  {Q^2\over\Sigma^2}(1-\theta_{(0)}^2),
\end{equation}
or dividing it by $(dt/d\tau)^2$,
\begin{equation}
\Bigl({d\theta_{(0)}\over dt}\Bigr)^2
={Q^2\over\sigma^2}(1-\theta_{(0)}^2),
\end{equation}
where 
\begin{equation}
\sigma\equiv -a(aE-l_z)+{a^2+r_0^2 \over \Delta(r_0)} 
\bigl\{ E(r_0^2+a^2)-al_z \bigr\}. 
\end{equation}
Then the solution is easily obtained as
\begin{equation}
\theta_{(0)}=\sin(\Omega_{\theta} t);\quad
\Omega_{\theta}={Q \over \sigma}\,,
\label{eq:exi}
\end{equation}
where we have chosen $\theta_{(0)}=0$ at $t=0$. 
Thus we have
\begin{equation}
\theta={\pi\over2}+y^{1/2}\sin(\Omega_{\theta}t).
\label{eq:thesol}
\end{equation}
Note that the solution
(\ref{eq:thesol}) implies that the inclination angle 
$\theta_i$ is indeed given by
$\theta_i=y^{1/2}$ in the present approximation.

Next, we consider the geodesic equation for $\varphi$. 
Taking account of the terms up to $O(y)$, it becomes
\begin{eqnarray}
{d\varphi \over dt}
&=&{\kappa \over \sigma}\left[1+\Bigl({l_z \over \kappa}
           -{a^2E\over \sigma}\Bigr)y\theta_{(0)}^2\right]
\cr
&=&\Omega_\varphi-y{\Omega_2\over2}\cos(2\Omega_\theta t), 
\label{eq:vap}
\end{eqnarray}
where 
\begin{equation}
\kappa \equiv -(aE-l_z)+{a \over \Delta(r_0)} 
\{ E(r^2_0+a^2)-al_z \},
\end{equation}
and
\begin{equation}
\Omega_{\varphi}={\kappa \over \sigma}+{1\over2}y\Omega_2\,,
\quad
\Omega_2={\kappa \over \sigma}
\Bigl({l_z \over \kappa}-{a^2E \over \sigma}\Bigr).
\end{equation}
The solution to Eq.~(\ref{eq:vap}) with $\varphi=0$ at $t=0$ is
\begin{equation}
\varphi=\Omega_{\varphi}t
 -y{\Omega_2 \over 4\Omega_{\theta}}\sin(2\Omega_{\theta} t).
\label{eq:solv}
\end{equation}
Note that $\Omega_{\varphi} \neq \Omega_{\theta}$.
This means the precession of a test particle orbit around the 
spin axis of the black hole. 
Specifically, to the order required for the present purpose, 
we have 
\begin{eqnarray}
\Omega_{\varphi}&=&{M^{1/2} \over r_0^{3/2}}
\left[1-{M^{1/2}a \over r_0^{3/2}}
+{3 \over 2}y\Bigl({M^{1/2}a \over r_0^{3/2}}-{a^2 \over r_0^2}\Bigr)
 + O(v^6)\right],\cr
\Omega_{\theta}&=&{M^{1/2} \over r_0^{3/2}}
\left[1-{3M^{1/2}a \over r_0^{3/2}}+{3a^2 \over 2r_0^2}
+O(v^6)+O(y)\right].
\label{eq:omeeq}
\end{eqnarray}
We see that $\Omega_{\varphi}-\Omega_{\theta} \to 2Ma/r_0^3$ 
for $r_0 \to \infty$ and $y \to 0$, 
which is just the Lense-Thirring precessional
frequency\cite{ref:membran}. 

Now we are ready to calculate the source integral for the amplitude
$\tilde Z_{\ell m\omega}$.
Analogous to the case of an eccentric orbit considered
in section \ref{sec:schwaecc} or \ref{sec:kerrecc},
Eq.~(\ref{eq:zzq2}) can be simplified further by noting that
the orbits of our interest have the properties,
\begin{equation}
\theta(t+\Delta t_\theta)=\theta(t),~~~~~~~~
\varphi(t+\Delta t_\theta)=\varphi(t)+\Delta \varphi,
\end{equation}
where $\Delta t_\theta$ is the orbital period of the motion in the
$\theta$-direction and $\Delta \varphi$ is the phase advancement during
$\Delta t_\theta$. In other words, we have
\begin{equation}
 \Omega_\theta={2\pi\over\Delta t_\theta}\,,\quad
 \Omega_\varphi={\Delta\varphi\over\Delta t_\theta}\,.
\label{eq:Omegadef}
\end{equation}
Then we obtain
\begin{equation}
\tilde Z_{\ell m\omega}=\sum_{n}\delta(\omega-\omega_n)
      Z_{\ell m\omega_n},
\end{equation}
where
\begin{equation}
  \label{omeganth}
  \omega_n=n\Omega_\theta+m\Omega_\varphi\quad(n=0,\pm1,\pm2,\cdots),
\end{equation}
and
\begin{equation}
Z_{\ell m\omega_n}=
{\mu\Omega_\theta\over2i\omega_n B^{\rm inc}_{\ell m\omega_n}}
\int^{\Delta t}_0 dt e^{i\omega t-im\varphi (t)}
W_{\ell m\omega_n}\,,
\label{eq:zzq}
\end{equation}
with $W_{\ell m\omega_n}$ being given by Eq.~(\ref{eq:Wdef}).

Let us discuss the final form of $Z_{\ell m\omega_n}$.
In the present case, up to $O(y)$ the integrand $W_{\ell m\omega_n}$
has the form,
\begin{eqnarray}
W_{\ell m\omega_n}&=& g_0+y^{1/2} g_1\theta_{(0)}
+y g_2\theta_{(0)}^2
\cr
&&+y^{1/2}g_3{d\theta_{(0)}\over dt}
+yg_4\theta_{(0)}{d\theta_{(0)}\over dt} 
+yg_5\Bigl( {d\theta_{(0)}\over dt} \Bigr)^2 
+O(y^{3/2}), \label{eq:iieq}
\end{eqnarray}
where $g_0\sim g_5$ are complicated functions of $r_0$. 
Using an approximation,
\begin{equation}
e^{i\omega_n t-im\varphi(t)} = e^{in\Omega_{\theta} t}
\Bigl(1+y{m \Omega_2 \over 8\Omega_{\theta}}
(e^{2i\Omega_{\theta} t}-e^{-2i\Omega_{\theta} t})+O(y^{3/2})\Bigr),
\label{eq:eimw}
\end{equation} 
we have 
\begin{eqnarray}
\int^{\Delta t}_0 &&dte^{i\omega t-im\varphi(t)}W_{\ell m\omega_n}
\cr
&&={2\pi\over\Omega_\theta}\biggl[\Bigl\{ \delta_{n0}
+y{m\Omega_2\over 8\Omega_{\theta}}
(\delta_{n,-2}-\delta_{n,2}) \Bigr\}g_0
+y^{1/2}{1\over 2i}(\delta_{n,-1}-\delta_{n,1})g_1
\cr
&&+y{1\over 4}(2\delta_{n,0}-\delta_{n,-2}-\delta_{n,2})g_2
+y^{1/2}{\Omega_\theta\over 2}(\delta_{n,-1}+\delta_{n,1})g_3
\cr
&&+y{\Omega_\theta\over 4i}(\delta_{n,-2}-\delta_{n,2})g_4
+y{\Omega_\theta^2\over 4}
(2\delta_{n,0}+\delta_{n,-2}+\delta_{n,2})g_5\biggr]
\cr
&&+O(y^{3/2}). 
\end{eqnarray}
Thus the amplitude $Z_{\ell m\omega_n}$ is found to have the form,
\begin{eqnarray}
Z_{\ell m\omega_n}
&=&\Bigl[\bigl(Z^{0,0}+yZ^{0,2}) \delta_{n,0}
  +y^{1/2}\bigl(Z^{1,1}\delta_{n,1}+Z^{1,-1}\delta_{n,-1}\bigr) 
\cr
&&+y\bigl(Z^{2,2}\delta_{n,2}
         +Z^{2,-2}\delta_{n,-2}\bigr)+O(y^{3/2})\Bigr], 
\label{eq:Zform}
\end{eqnarray}
where $Z^{i,j}$ are functions of $r_0$. 
Here, it is worth noting the symmetry of $Z_{\ell m\omega_n}$.
The spin weighted spheroidal harmonics have 
a property ${}_{-2}S_{\ell m}^{a\omega_n}(\theta) 
= (-1)^{\ell}{}_{-2}S_{\ell-m}^{a\omega_{-n}}(\pi-\theta)$. 
Then, from Eqs.~(\ref{eq:zzq}) and (\ref{eq:eimw}), we have 
$Z_{\ell-m\omega_{-n}}=(-1)^{n+\ell}{\overline Z}_{\ell m\omega_n}$. 

Now we evaluate the energy and angular momentum fluxes at infinity. 
The energy and angular momentum fluxes averaged over 
$t \gg \Delta t_\theta$ are give by Eqs.~(\ref{eq:Eflux}) and
(\ref{eq:Jflux}), respectively. 
Then we see from Eq.(\ref{eq:Zform}) that the $n=\pm2$ modes contribute
to the luminosity at $O(y^2)$. Thus, when we calculate the luminosity to 
$O(y)$, we need to include only the $n=0,\pm1$ modes. 
In order to express the post-Newtonian corrections to the luminosity,
we define $\eta_{\ell mn}$ as
\begin{equation}
\left( {dE \over dt} \right)_{\ell mn}
\equiv{1\over2}\left({dE\over dt}\right)_{\rm N}\eta_{\ell mn}\,,
\end{equation}
where $(dE/dt)_{\rm N}$ is the Newtonian quadrupole luminosity given by 
Eq.~(\ref{eq:Enewton}).

For $\ell=2$, the results are as follows.
If $|m+n|>2$ or $m+n=0$, 
$\eta_{\ell mn}$ becomes of $O(v^6)$ or higher.
The remaining $\eta_{\ell mn}$ which contribute to the 2.5PN
luminosity formula are given by
\def\pmz{\hphantom{\pm}0} 
\begin{eqnarray}
\eta_{2\pm 2\pmz}&=&
1-{107 \over 21}v^2+4\pi v^3-6q v^3+{4784 \over 1323}v^4
+2q^2v^4-{428\pi \over 21}v^5+{4216 \over 189}qv^5
\cr
&&+y \Bigl( -1+{170 \over 21}v^2-4\pi v^3+15qv^3-{4784 \over 1323}v^4
  -11q^2v^4
\cr
&&\qquad+{428\pi \over 21}v^5-{13186\over 189}qv^5 \Bigr), 
\cr
\eta_{2\pm2\mp1}&=&
y\Bigl( {1 \over 36}v^2 - {17 \over 504}v^4
+{\pi \over 18}v^5+{17 \over 1134}q v^5 \Bigr), \cr
\eta_{2\pm1\pmz}
&=&{1 \over 36}v^2-{1 \over 12}q v^3-{17 \over 504}v^4
+{1 \over 16}q^2v^4+{\pi \over 18}v^5-{793 \over 9072}q v^5
\cr
&&+y \Bigl( -{5 \over 72}v^2 +{1 \over 8}q v^3 +{85 \over 1008}v^4
-{1 \over 32}q^2v^4-{5\pi \over 36}v^5+{13931 \over 18144}qv^5 \Bigr),
\cr
\eta_{2\pm1\pm1}&=&
y\Bigl( 1- {170 \over 21}v^2 +4\pi v^3 -12q v^3
+{4784 \over 1323}v^4+{11 \over 2}q^2v^4 
\cr
&&\qquad-{428\pi \over 21}v^5+{11078 \over 189}q v^5 \Bigr),
\cr
\eta_{2\pmz\pm1}&=&
y\Bigl(  {1 \over 24}v^2  -{1 \over 12}q v^3
-{17 \over 336}v^4+{1 \over 24}q^2v^4
+{\pi \over 12}v^5-{745 \over 1008}qv^5 \Bigr).
\end{eqnarray}
Putting together the above results, 
we obtain $(dE/dt)_\ell \equiv \sum_{mn} (dE/dt)_{\ell mn}$ 
for $\ell=2$ as
\begin{eqnarray}
\Bigl( {dE \over dt} \Bigr)_2&=&
\left({dE\over dt}\right)_{\rm N}
\Bigl\{ 1 -{1277 \over 252}v^2 +4\pi v^3 
-{73 \over 12}qv^3\Bigl(1-{y \over 2}\Bigr)
+{37915 \over 10584}v^4 \cr
&&+{33 \over 16}q^2v^4 -{527 \over 96}q^2v^4y
-{2561\pi \over 126}v^5+{201575 \over 9072}qv^5\Bigl(1-{y\over2}\Bigr)
\Bigr\}. 
\end{eqnarray}

For $\ell=3$, the non-trivial $\eta_{\ell mn}$ are given by
\begin{eqnarray}
\eta_{3\pm3\pmz}&=&
{1215 \over 896}v^2 -{1215 \over 112}v^4
+{3645\pi \over 448}v^5 -{1215 \over 112}qv^5
\cr
&&+y \Bigl( -{3645 \over 1792}v^2 +{3645 \over 224}v^4
-{10935\pi \over 896}v^5 +{3645 \over 112}qv^5 \Bigr),
\cr
\eta_{3\pm3\mp1}&=&{5 \over 42}v^4 y,
\cr
\eta_{3\pm2\pmz}&=&{5 \over 63}v^4 -{40 \over 189}qv^5 
+y\Bigl( -{20 \over 63}v^4+{100 \over 189}qv^5\Bigr), 
\cr
\eta_{3\pm2\pm1}&=&
y\Bigl( {3645 \over 1792}v^2 -{3645 \over 224}v^4
+{10935\pi \over 896}v^5-{6075 \over 224}qv^5\Bigr), 
\cr
\eta_{3\pm2\mp1}&=&y\Bigl( {5 \over 16128}v^2 
-{5 \over 3024}v^4 
+{5\pi \over 8064}v^5+{25 \over 18144}qv^5\Bigr), 
\cr
\eta_{3\pm1\pmz}&=&{1 \over 8064}v^2 -{1 \over 1512}v^4
+{\pi \over 4032}v^5 -{17 \over 9072}qv^5 
\cr
&&+y \Bigl( -{11 \over 16128}v^2 +{11 \over 3024}v^4
-{11\pi \over 8064}v^5 +{95 \over 9072}qv^5 \Bigr),
\cr
\eta_{3\pm1\pm1}&=&
y\Bigl( {25 \over 126}v^4 -{80 \over 189}qv^5\Bigr), 
\cr
\eta_{3\pmz\pm1}&=&
y\Bigl( {1 \over 2688}v^2 -{1 \over 504}v^4
+{\pi \over 1344}v^5-{11 \over 1008}qv^5 \Bigr).
\end{eqnarray}
The other $\eta_{\ell mn}$ are of $O(v^6)$ or higher.
Then we obtain
\begin{equation}
\Bigl( {dE \over dt} \Bigr)_3=
\left({dE\over dt}\right)_{\rm N}
\Bigl\{  {1367 \over 1008}v^2 -{32567 \over 3024}v^4
+{16403\pi \over 2016}v^5-{896 \over 81}qv^5\Bigl(1-{y \over 2}\Bigr)
 \Bigr\}.
\end{equation}

For $\ell=4$, we have 
\begin{eqnarray}
&&\eta_{4\pm4\pmz}={1280 \over 567}v^4(1-2y),\cr
&&\eta_{4\pm3\pm1}={2560 \over 567}v^4 y, \cr
&&\eta_{4\pm3\mp1}={5 \over 1134}v^4 y, \cr
&&\eta_{4\pm2\pmz}={5 \over 3969}v^4(1-8y), \cr
&&\eta_{4\pm1\pm1}={5 \over 882}v^4 y, 
\end{eqnarray}
and the others are of $O(v^6)$ or higher.
Hence we obtain
\begin{equation}
\Bigl( {dE \over dt} \Bigr)_4=
\left({dE\over dt}\right)_{\rm N}
\times {8965 \over 3969}v^4. 
\end{equation}

Finally, gathering all the terms, the total energy flux 
up to $O(v^5)$ is found to be
\begin{eqnarray}
\left\langle {dE \over dt} \right\rangle &=&
\left({dE\over dt}\right)_{\rm N}
\biggl( 1 -{1247 \over 336}v^2+4\pi v^3 
 -{73 \over 12}qv^3\left(1-{y\over2}\right)
 -{44711 \over 9072}v^4 \cr
&&+{33 \over 16}q^2v^4-{527 \over 96}q^2v^4 y -{8191 \pi \over 672}v^5
+{3749 \over 336}qv^5\left(1-{y\over2}\right) \biggr). 
\end{eqnarray} 

Using the above results for $\eta_{\ell m n}$, the time-averaged angular
momentum flux is calculated from Eq.~(\ref{eq:Jflux}). 
The partial mode contributions of the $\ell=2$, 3 and 4 modes
are calculated to give 
\begin{eqnarray} 
\left({dJ_z\over dt}\right)_2
&=&\left({dJ_z\over dt}\right)_{\rm N}
 \biggl[1-{y\over 2}
-{1277\over 252}v^2\left(1-{y\over2}\right)
+4\pi v^3\left(1-{y\over 2}\right)
\cr
&&-qv^3\left({61\over 12}-{61\over 8}y\right)
+{37915\over 10584}v^4\left(1-{y\over2}\right)
+q^2v^4\left({33 \over 16}-{229 \over 32}y\right) 
\cr
&&-\pi v^5{2561 \over 126}\left(1-{y\over2}\right)
+qv^5\left({22229 \over 1296}-{27809\over864}y\right)
\biggr],
\cr
\left({dJ_z\over dt}\right)_3
&=&\left({dJ_z\over dt}\right)_{\rm N}
\biggl[{1367\over 1008}v^2\left(1-{y \over 2}\right)
-{32567 \over 3024}v^4\left(1-{y\over2}\right)
\cr
&&
+\pi v^5{16403\over 2016}\left(1-{y\over 2}\right)
-qv^5\left({88049 \over 9072}-{9817\over756}y\right)
\biggr], 
\cr
\left({dJ_z\over dt}\right)_4
&=&\left({dJ_z\over dt}\right)_{\rm N}
\biggl[{8965\over 3969}v^4\left(1-{y\over 2}\right)\biggr],
\end{eqnarray}
where $(dJ_z/dt)_{\rm N}$ is defined in Eq.~(\ref{eq:Jnewton}).
The total angular momentum flux is then given by
\begin{eqnarray}
\left\langle{dJ_z\over dt}\right\rangle
&=&\left({dJ_z\over dt}\right)_{\rm N}
\biggl[
\left(1-{y\over2}\right)-{1247\over336}v^2\left(1-{y\over2}\right)
+4\pi v^3\left(1-{y\over2}\right)
\cr
&&-{61\over12}qv^3\left(1-{y\over2}\right)
-{44711\over9072}v^4\left(1-{y\over2}\right)
+q^2v^4\left({33\over16}-{229\over32}y\right)
\cr
&&-{8191\over672}\pi v^5\left(1-{y\over2}\right)
+qv^5\left({417\over56}-{4301\over224}y\right)
\biggr].
\label{eq:Jtot}
\end{eqnarray}
We note that the result is proportional to $(1-y/2)$
in the limit $q\to0$. This is simply because
the orbital plane is slightly tilted from the equatorial plane
by an angle $\theta_i\sim y^{1/2}$, hence
$dJ_z/dt\sim (dJ_{tot}/dt)\cos\theta_i$.
\section{Adiabatic backreaction} 
\label{sec:Adbackreact}
In the preceding sections, we have evaluated the energy flux 
$\langle dE/dt\rangle$ and the $z$-component of the angular
momentum flux $\langle dJ_z/dt\rangle$ emitted to infinity by a particle
for various cases. By emitting gravitational waves, a particle orbit
will suffer from radiation reaction. In the limit of small $\mu/M$, the
reaction time scale will be much longer than the characteristic orbital
time scale; $t_{react}\sim M^2/\mu\gg\Delta t$. Hence the evolution of
the orbit will be well described by the adiabatic backreaction. 

In the case of orbits around a Schwarzschild black hole or orbits
confined on the equatorial plane around a Kerr black hole, it is
straightforward to calculate the evolutionary path under radiation
reaction because the orbits are completely specified by the energy $E$
and the $z$-component of the angular momentum $l_z$, hence their time
derivatives can be simply evaluated by equating them with $-\langle
dE/dt\rangle$ and $-\langle dJ_z/dt\rangle$, respectively. However, once
we consider motions off the equatorial plane of a Kerr black hole, the
orbits cannot be specified by $E$ and $l_z$ alone but the specification
of the Carter constant $C$ becomes necessary. Unlike $E$ or $l_z$, since
$C$ is not associated with the Killing vector of the spacetime, one
cannot calculate the radiation reaction to $C$ by simply calculating the 
gravitational waves at infinity. This implies that we have to derive a
local radiation reaction force term to the geodesic equation by
evaluating the metric perturbations around the particle, as is done in
the derivation of radiation reaction force in the standard
post-Newtonian method. For almost Newtonian orbits, applying a
post-Newtonian radiation reaction force, Ryan derived the evolution
equation for the Carter constant\cite{ref:Ryan2}. However, no
relativistic treatment has been done so far. This is a challenging
issue. An approach to this issue is discussed in Chapter 7.

In this section, instead of attacking this very difficult problem,
we discuss some general properties of the adiabatic radiation reaction
in a restricted class of orbits. Namely we consider orbits which are
circular or those having small eccentricity. We clarify
the conditions for circular orbits to remain circular under radiation
reaction. A detailed discussion on this matter has been given by
Kennefick and Ori\cite{ref:KenOri}. We give a less detailed but
more general discussion below.

We recall that the radial velocity $u^r=dr/d\tau$ is written in terms of
the first integrals of motion in the test particle limit as 
\begin{equation}
(\Sigma u^r)^2 = R(I^i,r) 
\label{eq:mtn}, 
\end{equation}
where $I^i=(E,l_z,C)$ and $R(I^i,r)$ is independent of $\theta$ and
$\phi$. First let us consider orbits which are circular in the
test particle limit. These orbits are determined by the conditions,
\begin{equation}
  \label{eq:circcond}
  R(I^i,r)=0,\quad {\partial R\over\partial r}(I^i,r)=0.
\end{equation}
Eliminating $r$ from these equations gives an implicit relation among
$I^i$'s. For example,
\begin{equation}
  \label{circspace}
  f(I^i)=R(I^i,r(I^i))=0,
\end{equation}
where $r(I^i)$ is obtained by solving the second of
Eq.~(\ref{eq:circcond}) for $r$. 
This equation determines a two-dimensional hypersurface $S$ in the
3-dimensional space $M$ of $I^i$'s. The adiabatic evolution of an orbit
is characterized by slow evolution of $I^i$, i.e., $\dot{I^i}=O(\mu)$,
where $\mu$ is the mass of the particle. 
Then a necessary condition for circular orbits to remain circular under
radiation reaction is that we have 
$\dot f(I^i)=f_{,i}\dot{I^i}=O(\mu^2)$. In other words, the vector
$\dot{I^i}$ on $S$ is tangent to $S$ to $O(\mu)$.
This condition can be shown to hold by the following theorem.

\medskip\noindent
{\it Theorem}: If the radiation reaction to the $r$-component of
the acceleration is of order $\mu$; $a^r:= du^r/d\tau=O(\mu)$,
i.e., the $r$-component of the radiation reaction force is well-defined
and finite, then for orbits which are circular in the test particle
limit; i.e., $u^r=O(\mu)$, the radiation reaction to $I^i$ is
constrained by the equation,
\begin{equation}
{\partial R\over \partial I^i}(I^i,r) \dot{ I^i }=O(\mu^2),
\label{eq:idot}
\end{equation} 
where the argument $r$ is to be replaced by $r(I^i)$ after
differentiation.

\medskip\noindent
{\it Proof}: It is almost trivial. Just taking the $ \tau $-derivatives of
both hand sides of Eq.~(\ref{eq:mtn}) gives Eq.~(\ref{eq:idot}).
Q.E.D.

\medskip\noindent
Thus, since 
\begin{equation}
  \dot f(I^i)=\left({\partial R\over\partial I^i}(I^i,r(I^i))
+{\partial R\over\partial r}(I^i,r(I^i))
{\partial r\over\partial I^i}\right)\dot{I^i},
\end{equation}
and the second term in the parentheses vanishes by definition,
we have $\dot f=O(\mu^2)$. 

This theorem alone, however, does not mean that circular orbits remain
circular, since we have constrained the first integrals to be those for
circular orbits from the beginning. Let us explain the reason.
Since we may regard $\dot{I^i}$ a vector field in $M$, what we need for
circular orbits to remain circular is the regularity of $\dot {I^i}$ in 
the vicinity of the hypersurface $S$. In other words, if the vector
field $\dot{I^i}$ is not differentiable on $S$, an orbit on $S$ may
spontaneously deviates away from $S$. A simple illustrative example is
the case $\dot I_{\perp}=\sqrt{I_\perp}$ at $I_\perp=0$ where $I_\perp$
is the component of $I^i$ perpendicular to $S$.

Thus, provided $\dot{I^i}$ is regular in an open neighborhood of $S$,
the above theorem implies that a circular orbit in the test particle
limit remains circular under adiabatic radiation reaction. In this case,
the radiation reaction to the Carter constant, $\dot{C}$, is determined
by the radiation reaction to the energy, $\dot{E}$, and the
$z$-component of the angular momentum, $\dot{l_z}$. Specifically we
have 
\begin{eqnarray}
 \dot{C} &=& \left( {2 \over \Delta} \left( E \left( r^2 + a^2 
 \right) -al_z \right) \left( r^2 + a^2 \right) - 2a \left( Ea -l_z
 \right) \right) \dot{E} \nonumber \\
 && + \left( -2 {a \over \Delta} \left( E \left( r^2 + a^2 \right) 
 -al_z \right) +2 \left( Ea -l_z\right) \right) \dot{l_z}.
\end{eqnarray}

Yet this is not the end of the story. What we have shown is that
$\dot I^i$ lies on $S$. But if $\dot I^i$ slightly off the hypersurface
$S$ is diverging away from $S$, circular orbits will be unstable.
Thus the condition for the stability of circular orbits is that $S$ is
an attractor plane of the vector field $\dot{I^i}$. 
However, the notion of divergence or convergence of a vector depends on
the metric of the space $M$, but we have no guiding principle to
determine the metric. This implies that the notion of the attractor or
the stability is ambiguous.

Nevertheless, extrapolating from the case of Newtonian orbits,
there seems to exist a natural choice of the metric. Namely,
as the distance of the orbit from the hypersurface $S$ of circular
orbits, we define the eccentricity of an orbit as given in
sections \ref{sec:schwaecc} and \ref{sec:kerrecc}.
With this choice of the metric, let us consider the adiabatic radiation
reaction problem in more specific terms.

Let us parametrize an orbit in terms of the mean radius $r_0$, the
eccentricity $e$ and the square root of the Carter constant
$y:=C^{1/2}$, instead of the energy $E$, the angular momentum $l_z$ and
the Carter constant $C$.
The mean radius $r_0$ is defined by the equation,
\begin{equation}
0=R'(I^i,r_0),
 \label{eq:rzero}
\end{equation}
where the prime denotes the partial derivative with respect 
to $r$. This definition says that $\dot r$ is maximum at $r=r_0$.
The eccentricity $e$ is defined by setting the maximum radius 
to $r=r_0(1+e)$, i.e.,
\begin{equation}
0=R(I^i,r_0(1+e)),
  \label{eq:ecc}
\end{equation}
This definition guarantees that $e=0$ corresponds to a circular orbit. 
Assuming $e\ll1$, the above equation can be expanded in powers of $e$ as
\begin{equation}
0=R(I^i,r_0)+{1 \over 2}R''(I^i,r_0)(r_0e)^2
+{1 \over 3!} R^{(3)}(I^i,r_0)(r_0e)^3 
+{1 \over 4!} R^{(4)}(I^i,r_0)(r_0e)^4 +\cdots,
  \label{eq:eccexpand}
\end{equation}
where $R^{(n)}$ is the $n$-th derivative of $R$ with respect to $r$.
The parameters $(r_0,e,y)$ are chosen because the geodesic trajectory
$x^\mu=z^\mu(\tau)$ allows perturbative expansion in powers of $e$ and $y$ 
at least for $e\ll1$ and $y\ll1$. Therefore the first integrals of
motion $I^i=(E,l_z,C)$ will be regular functions of the parameters
$(r_0,e,y)$. On the other hand, if we consider 
$(r_0,e,y)$ as functions of $I^i$, it should be noted that $e$ is not a
regular function of $I^i$ in a neighborhood of circular orbits because
of the absence of a term linear in $e$ in the right hand side of
Eq.~(\ref{eq:eccexpand}).

Now we consider the adiabatic evolution of $e$ under the radiation
reaction. Taking the $ \tau$-derivative of Eqs.~(\ref{eq:rzero}) and
(\ref{eq:eccexpand}), we get
\begin{eqnarray}
0 &=& R'_{0,i}\dot I^i +R''_0 \dot{r_0}\,,
\label{eq:rdot} 
\\
0 &=& R_{0,i}I^i +R''_0 
\left(r_0 \dot r_0 e^2 +r_0^2 e\dot e\right) 
+ {1\over 2}\left(R''_{0,i}\dot I^i+R^{(3)}_0 \dot r_0 \right) 
r_0^2e^2 
\nonumber \\ 
&&+{1 \over 2}R^{(3)}_0 r_0^3e^2\dot e 
+{1 \over 3!} R^{(4)}_0 r_0^4e^3\dot e
+O \left(e^3,\dot e e^4 \right), 
\label{eq:rdott}
\end{eqnarray}
where $R'_{0,i}=\partial^2 R/\partial I^i\partial r$ etc. 
Equation (\ref{eq:rdot}) determines $\dot{r_0}$ as
\begin{equation}
\dot{r_0}=-{R'_{0,i}\over R''_0}\dot I^i. 
\end{equation} 
Substituting this into Eq.~(\ref{eq:rdott}), 
we obtain the expression for $\dot{e}$ as 
\begin{eqnarray}
\dot{e} =&& {1\over e r_0^2 R''_0}
 \Biggl[-R_{0,i}+{e\over 2}{r_0 R^{(3)}_0\over R''_0}R_{0,i}
\nonumber \\
&& + e^2\left\{-{1\over 2}{r_0}^2 R''_{0,i}+ 
 \left(r_0+{1\over 2}{r_0^2 R^{(3)}_0\over R''_0}\right)
 R'_{0,i}+r_0^2\left({1\over 6}{R^{(4)}_0\over R''_0} 
 -{1\over 4}\left({R^{(3)}_0\over R''_0}\right)^2\right)
  R_{0,i}\right\}
\nonumber\\
&&+O(e^3)\Biggr]\dot I^i. 
\end{eqnarray}

Since the trajectory $z^\mu(\tau)$ is assumed to be analytic in
$(r_0,e,y)$, it is reasonable to further assume that $\dot I^i$ are 
regular functions of $(r_0,e,y)$. Then we can expand $\dot I^i$ with
respect to $e$ as 
\begin{equation}
\dot{I}^i(r_0,e,y) = \dot{I}^{i(0)}(r_0,y) 
+e \dot{I}^{i(1)}(r_0,y) 
+e^2 \dot{I}^{i(2)}(r_0,y) + \cdots.
\end{equation}
Then we obtain
\begin{eqnarray}
\dot{e} &=&{1 \over r_0^2 R''_0}
 \Biggl[-{1\over e}R_{0,i}\dot I^{i(0)}
 +\left(-R_{0,i}\dot I^{i(1)}
 +{1\over 2}{r_0 R'''_0\over R''_0}R_{0,i}\dot I^{i(0)}\right)
\nonumber \\
&& +e\Biggl(-R_{0,i}\dot I^{i(2)}
+{1 \over 2}{r_0R'''_0\over R''_0}R_{0,i}\dot I^{i(1)} 
\nonumber \\ 
&& 
\quad +\left\{-{1\over 2}{r_0}^2 R''_{0,i}+
 \left(r_0+{1\over 2}{{r_0}^2 R'''_0\over R''_0} \right)
 R'_{0,i}+r_0^2\left({1\over 6}{R^{(4)}_0\over R''_0} 
 -{1 \over 4}\left({R'''_0 \over R''_0}\right)^2 \right) 
 R_{0,i}\right\}\dot I^{i(0)}
 \Biggr)
\nonumber\\
&&+O(e^2)\Biggr].
\end{eqnarray}
As shown by the theorem above, Eq.~(\ref{eq:idot}), the leading term
of order $e^{-1}$ vanishes provided the radiation reaction force is
finite: 
\begin{equation}
R_{0,i} \dot{I}^{i(0)}=0.
 \label{eq:zero}
\end{equation}
The next order term determines 
whether the circular orbit remains circular or not. If it does not
vanish, the eccentricity will spontaneously develop as
\begin{equation}
\dot{e}= -R_{0,i} \dot{I}^{i(1)}. 
\end{equation}
Here the regularity of $\dot I^i$ comes into play. As noted above,
$e$ is singular on the hypersurface $S$. Hence if $\dot I^i$ is regular
on $S$, $\dot I^{(1)}(r_0,y)$ should vanish. By a detailed analysis, it
is shown in Ref.~\citen{ref:KenOri} that this is indeed the case. The
physical reason is rather simple: If one considers a slightly eccentric
orbit, there appears a frequency of wobbling motion due to the
eccentricity, say $\Omega_e$. In general the ratio of $\Omega_e$ to the
frequency of the motion in the $\theta$ or $\varphi$ direction is an
irrational number. Hence the part of the metric perturbation 
which is proportional to $e$ will have frequencies that are integer
multiples of $\Omega_e$, and the same property is shared by the
corresponding term of the backreaction force linear in $e$. Since any
sinusoidal oscillation has zero mean when averaged over time longer than 
its period, this implies there will be no term linear in $e$ in the
adiabatic expression of $\dot I^i$.

Thus we have
\begin{equation}
\dot{e} = \left[-R_{0,i}
 \dot{I}^{i(2)} + 
\left\{-{1\over 2}{r_0}^2 R''_{0,i}+
 \left(r_0+{1\over 2}{r_0^2 R'''_0\over R''_0} \right)
 R'_{0,i} \right\}\dot I^{i(0)}\right]e+O(e^2),
\end{equation}
and circular orbits will remain circular under radiation reaction.
As for the stability of circular orbits, whether the eccentricity
decreases or increases is determined by the sign of the coefficient of
$e$ in the right hand side. Thus it is necessary to calculate the
radiation reaction to the Carter constant to determine the stability.
As mentioned in the beginning of this section, this is a challenging
issue. Finally, we should again note that the meaning of stability does
depend on the definition of the eccentricity, i.e., how we define the
distance from the hypersurface of circular orbits $S$.

\section{Spinning particle}
\label{sec:spin}

So far we have considered only a monopole particle orbiting a black
hole. However, in a realistic binary system of compact bodies such as
a neutron star-neutron star, black hole-neutron star or black hole-black 
hole binary, both bodies may have non-negligible spin angular
momenta. Hence it is desirable to take into account not only the spin of
a black hole but also the spin of a particle in the calculations of
gravitational waves from a particle orbiting a black hole.

To incorporate the spin of a particle, one must know (1) the equations of 
motion and (2) the energy momentum tensor of a spinning particle. 
Fortunately, we know that (1) have been derived by
Papapetrou\cite{ref:papa}, Dixon\cite{ref:dixon} and Wald\cite{ref:wald}
and (2) has also been derived by Dixon\cite{ref:dixon}. Hence,
by using the expression for the energy momentum tensor of a spinning
particle as the source term in the Teukolsky formalism\cite{ref:Teu}, 
we can calculate the gravitational waves emitted by a spinning small
mass particle orbiting a rotating black hole. 
One may regard this particle as a model of a small Kerr black hole, but
it may be appropriate here to give a word of caution.
A Kerr black hole of mass $\mu$ and the spin parameter $S$, 
where $S$ is defined so that $\mu S$ gives the spin angular momentum,
has quadrupole $(\ell=2)$ and higher multipole moments $(\ell>2)$
proportional to $\mu S^{\ell}$ as well.
Since we neglect the contributions of these higher multipole moments 
here, our treatment will be valid only up to $O(S)$ if we regard the
particle as a Kerr black hole. 
To incorporate the contributions of all higher multipole moments 
to represent the Kerr black hole is a future problem to 
be investigated. 

Here we review the results obtained by Tanaka et al.\cite{ref:TMSS}.
We concentrate on the leading effect due to the spin of 
the small mass particle. We consider a class of circular orbits
which stay near the equatorial plane with the inclination solely due to
the spin of the particle, i.e., those orbits which would be confined in
the equatorial plane if the spin were zero. Then we calculate the
gravitational wave luminosity to $O(v^5)$ with linear corrections due to
the spin. 

\subsection{Equation of Motion and Source 
               Term of a spinning particle}

To give the source term of the Teukolsky equation, 
we need to solve the equations of motion of a spinning particle 
and also to give an expression for the energy momentum tensor. 
In this section we give the necessary expressions, following
Refs.~\citen{ref:dixon,ref:wald,ref:MSTspin}.

Neglecting the effect of the higher multipole moments, 
the equations of motion of a spinning particle are given by
\begin{eqnarray}
{D\over d\tau}p^\mu(\tau)  & = & 
-{1\over 2} R^\mu_{~\nu\rho\sigma}(z(\tau))
 v^\nu(\tau) S^{\rho\sigma}(\tau), 
\nonumber \\
{D\over d\tau}S^{\mu\nu}(\tau) & = & 
2 p^{[\mu}(\tau)v^{\nu]}(\tau), 
\label{sp:eqofmot}
\end{eqnarray}
where $v^{\mu}(\tau)=dz^{\mu}(\tau)/d\tau$, $\tau$ is a parameter
which is not necessarily the proper time of the particle,
and, as we will see later, the vector $p^{\mu}(\tau)$ and the 
antisymmetric tensor $S^{\mu\nu}(\tau)$ represent 
the linear and spin angular momenta of the particle, 
respectively. Here $D/d\tau$ denotes the covariant derivative along
the particle trajectory.

We do not have the evolution equation for $v^{\mu}(\tau)$ yet.
In order to determine $v^\mu(\tau)$,
we need to impose a supplementary condition which determines the center 
of mass of the particle\cite{ref:dixon},
\begin{equation}
S^{\mu\nu}(\tau)p_{\nu}(\tau)=0. 
\label{sp:centmass}
\end{equation}
Then one can show that $p_\mu p^\mu={\rm const}.$ and
$S_{\mu\nu}S^{\mu\nu}={\rm const}.$ along the particle 
trajectory\cite{ref:wald}.
Therefore we may set
\begin{eqnarray}
p^\mu&=&\mu u^\mu\,,\quad u_\mu u^\mu=-1,
\nonumber\\
S^{\mu\nu}&=&\epsilon^{\mu\nu}{}_{\rho\sigma}p^\rho S^\sigma\,,
\quad p_\mu S^\mu=0,
\nonumber\\
S^2&=&S_\mu S^\mu={1\over2\mu^2}S_{\mu\nu}S^{\mu\nu},
\end{eqnarray}
where $\mu$ is the mass of the particle,
 $u^\mu$ is the specific linear momentum,
and $S^\mu$ is the specific spin vector
with $S$ its magnitude. Note that if we use $S^\mu$ instead of
$S^{\mu\nu}$ in the equations of motion, the center of mass condition
(\ref{sp:centmass}) will be replaced by the condition
\begin{equation}
  \label{sp:pscond}
 p_\mu S^\mu=0. 
\end{equation}

Since the above equations of motion are invariant under 
reparametrization of the orbital parameter $\tau$,
 we can fix $\tau$ to satisfy  
\begin{equation}
 u^\mu(\tau)v_\mu(\tau)=-1. 
\label{sp:normvel}
\end{equation}
Then, from Eqs.~(\ref{sp:eqofmot}), (\ref{sp:centmass}) 
and (\ref{sp:normvel}), 
$v^\mu(\tau)$ is determined as\cite{ref:dixon}
\begin{equation}
v^\mu(\tau)-u^\mu(\tau)
={1\over 2}
\left(\mu^2+{1\over 4}R_{\chi\xi\zeta\eta}(z(\tau))S^{\chi\xi}(\tau)
S^{\zeta\eta}(\tau)\right)^{-1} S^{\mu\nu}(\tau)
R_{\nu\rho\sigma\kappa}(\tau)u^\rho(\tau)S^{\sigma\kappa}(\tau). 
\label{sp:vprelation}
\end{equation}
With this equation, the equations of motion (\ref{sp:eqofmot}) 
completely determine the evolution of the orbit and the spin.
Note that $v^\mu=u^\mu+O(S^2)$, hence $v^\mu$ and $u^\mu$ are identical
to each other to $O(S)$.

As for the energy momentum tensor, Dixon\cite{ref:dixon} gives it in 
terms of the Dirac delta-function on the tangent space at 
$x^\mu=z^\mu(\tau)$. For later convenience, in this paper
 we use an equivalent but
alternative form of the energy momentum tensor, given in
terms of the Dirac delta-function on the coordinate space:
\begin{eqnarray}
T^{\alpha\beta}(x) = &&\int d\tau\Biggl\{ 
p^{(\alpha}(x,\tau)v^{\beta)}(x,\tau)
{\delta^{(4)}\left(x-z(\tau)\right)\over\sqrt{-g}}
\nonumber \\
&&\quad -\nabla_{\gamma}\left(S^{\gamma(\alpha}(x,\tau)v^{\beta)}(x,\tau)
{\delta^{(4)}
\left(x-z(\tau)\right)\over\sqrt{-g}}\right)\Biggr\}, 
\label{sp:dixontensor}
\end{eqnarray}
where $v^{\alpha}(x,\tau)$, $p^{\alpha}(x,\tau)$ 
and $S^{\alpha\beta}(x,\tau)$ are bi-tensors 
which are spacetime extensions of $v^{\mu}(\tau)$, $p^{\mu}(\tau)$
and $S^{\mu\nu}(\tau)$ which are defined only along the world line, 
$x^\mu=z^\mu(\tau)$.\footnote{In the rest of this section, we use 
$\mu,\nu,\sigma,\cdots$ as the tensor indices associated 
with the world line $z(\tau)$ and $\alpha,\beta,\gamma,\cdots$ as 
those with a field point $x$, and suppress the coordinate indices
of $z(\tau)$ and $x$ for notational simplicity.}
To define $v^{\alpha}(x,z(\tau))$, $p^{\alpha}(x,z(\tau))$
and $S^{\alpha\beta}(x,z(\tau))$ 
we introduce a bi-tensor $\bar g^{\alpha}_{~\mu}(x,z)$ which satisfies 
\begin{eqnarray}
&&\lim_{x\to z}\bar g^{\alpha}_{~\mu}(x,z(\tau))
=\delta^{\alpha}_{~\mu}\,, 
\nonumber \\
&&\lim_{x\to z}\nabla_{\beta}
    \bar g^{\alpha}_{~\mu}(x,z(\tau))=0.  
\end{eqnarray}
For the present purpose, further specification of 
$\bar g^{\alpha}_{~\mu}(x,z)$ is not necessary. 
Using this bi-tensor $\bar g^{\alpha}_{~\mu}(x,z)$, we define 
$p^\alpha(x,\tau)$, $v^\alpha(x,\tau)$ and $S^{\alpha\beta}(x,\tau)$ as 
\begin{eqnarray}
p^\alpha(x,\tau)&=&\bar g^{\alpha}_{~\mu}
      \bigl(x,z(\tau)\bigr)p^\mu (\tau), 
\nonumber \\
v^\alpha(x,\tau)&=&\bar g^{\alpha}_{~\mu}
      \bigl(x,z(\tau)\bigr)v^\mu (\tau), 
\nonumber \\
S^{\alpha\beta}(x,\tau)&=&\bar g^{\alpha}_{~\mu}\bigl(x,z(\tau)\bigr)
\bar g^{\beta}_{~\nu}
      \bigl(x,z(\tau)\bigr)S^{\mu\nu}(\tau). 
\end{eqnarray}

It is easy to see that the divergence free condition of 
this energy momentum tensor gives the equations of motion
(\ref{sp:eqofmot}). 
Noting the relations,
\begin{eqnarray}
\nabla_\beta\bar g^{\alpha}_{~\mu}(x,z(\tau)) 
\delta^{(4)}(x,z(\tau)) & = & 0, 
\nonumber \\
v^{\alpha}(x)\nabla_{\alpha}\left({\delta^{(4)}\bigl(x,z(\tau)\bigr)\over
\sqrt{-g}}\right) & = & 
  -{d\over d\tau}\left({\delta^{(4)}\bigl(x,z(\tau)\bigr)
\over\sqrt{-g}}\right), 
\end{eqnarray}
the divergence of Eq.~(\ref{sp:dixontensor}) becomes
\begin{eqnarray}
\nabla_{\beta}T^{\alpha\beta}(x)
=&&\int d\tau \bar g^{\alpha}_{~\mu}\bigl(x,z(\tau)\bigr) 
{ \delta^{(4)} \bigl(x-z(\tau)\bigr) \over \sqrt{-g} }
\left( {d\over d\tau}p^{\mu}(\tau)
+{1\over 2}R^{\mu}_{~\nu\sigma\kappa}\bigl(z(\tau)\bigr)
v^{\nu}(\tau) S^{\sigma\kappa}(\tau) \right)
\nonumber \\ 
&&+{1\over 2} \int d\tau \nabla_\beta\left(
\bar g^{\alpha}_{~\mu}\bigl(x,z(\tau)\bigr) 
\bar g^{\beta}_{~\nu}\bigl(x,z(\tau)\bigr) 
{ \delta^{(4)} \bigl(x-z(\tau)\bigr) \over \sqrt{-g} } \right)
\nonumber\\
&&\quad\times\left( {d\over d\tau}S^{\mu\nu}(\tau)
-2p^{[\mu}(\tau)v^{\nu]}(\tau) \right). 
\end{eqnarray}
Since the first and second terms in the right-hand side must 
vanish separately, 
we obtain the equations of motion~(\ref{sp:eqofmot}).

In order to clarify the meaning of $p^{\mu}$ and $S^{\mu\nu}$, 
we consider the volume integral of this energy momentum tensor such as 
$\displaystyle \int_{\Sigma(\tau_0)} \bar g^{~\mu}_{\alpha}
         T^{\alpha\beta} d\Sigma_{\beta}$, 
where we take the surface ${\Sigma(\tau_0)}$ to be
perpendicular to $u^{\alpha}(\tau_0)$. 
It is convenient to introduce a scalar function $\tau(x)$ 
which determines the surface ${\Sigma(\tau_0)}$ by the equation
$\tau(x)=\tau_0$, and 
${\partial\tau/\partial x^{\beta}} =-u_{\beta}$ at $x=z(\tau_0)$. 
Then we have
\begin{eqnarray}
\int_{\Sigma(\tau_0)} \bar g^{~\mu}_{\alpha} 
         T^{\alpha\beta} d\Sigma_{\beta}
& = &
\int d^4 x \sqrt{-g} {\partial\tau\over \partial x^{\beta}}
 \delta(\tau(x)-\tau_0) \bar g^{~\mu}_{\alpha}T^{\alpha\beta}(x)
\nonumber \\
&=&\int d\tau' \Biggl\{\delta(\tau'-\tau_0) 
 \left[p^{\mu}+p^{[\mu} v^{\nu]}u_{\nu}
 -{1\over 2} {D u_{\nu}\over d\tau} S^{\nu\mu}\right] \Biggr\}
\nonumber \\
& = &p^{\mu}(\tau_0).
\end{eqnarray}
where we used the center of mass condition and the equation of motion
for $S^{\mu\nu}$. We clearly see $p^\mu$ indeed represents the linear
momentum of the particle.

In order to clarify the meaning of $S^{\mu\nu}$, 
following Dixon \cite{ref:dixon}, 
we introduce the relative position vector
\begin{equation}
 X^{\mu}:= -g^{\mu\nu}\partial_\nu\sigma(x,z), 
\end{equation}
where $\sigma(x,z)$ is the squared geodetic interval between $z$ and $x$ 
defined by using the parametric form of a 
geodesic $y(u)$ joining $z=y(0)$ and $x=y(1)$ as 
\begin{equation}
\sigma(x,z):={1\over 2}\int_0^1 g_{\alpha\beta}
       {dy^{\alpha}\over du}{dy^{\beta}\over du} du.
\end{equation}
Then noting the relations 
\begin{equation}
\lim_{x\to z}X^{\mu}=0, \quad
\lim_{x\to z}X^{\mu}_{~,\beta}=\delta^{\mu}_{\beta},
\end{equation}
it is easy to see that 
\begin{equation}
S^{\mu\nu}=2\int_{\Sigma_{\tau_0}}
      X^{[\mu}\bar g^{~\nu]}_{\alpha} T^{\alpha\beta} d\Sigma_{\beta}.
\end{equation}
Now that the meaning of $S^{\mu\nu}$ is manifest. 
{}From the above equation, it is also easy to see
that the center of mass condition (\ref{sp:centmass}) is the 
generalization of the Newtonian counter part,
\begin{equation}
 \int d^3 x \rho(x) x^i =0,
\end{equation}
where $\rho$ is the matter density. 

Before closing this subsection, we mention several conserved quantities 
of the present system. We have already noted that $p_\mu p^\mu=-\mu^2$
and $S_\mu S^\mu=S^2$ are 
constant along the particle trajectory on an arbitrary spacetime.
There will be an additional conserved
quantity if the spacetime admits a Killing vector field $\xi_{\mu}$, 
\begin{equation}
 \xi_{(\mu;\nu)}=0.
\end{equation}
Namely, the quantity
\begin{equation}
 Q_\xi:=p^{\mu}\xi_{\mu}-{1\over 2}S^{\mu\nu}\xi_{\mu;\nu}, 
\label{sp:cons}
\end{equation}
is conserved along the particle trajectory\cite{ref:dixon}. 
It is easy to verify that $Q_\xi$ is conserved by directly
using the equations of motion.

\subsection{Circular orbits near the equatorial plane}

Let us consider circular orbits around a Kerr black hole
with a fixed Boyer-Lindquist radial coordinate, $r=r_0$.
We consider a class of orbits that would stay on the equatorial plane
if the particle were spinless. Hence we assume that 
$\tilde \theta:=\theta -\pi/2 \sim O(S/M)\ll 1$.
Under this assumption, we write down the equations of motion
and solve them up to the linear order in $S$.

In order to find a solution representing a circular orbit, 
it is convenient to introduce the tetrad frame defined by 
\begin{eqnarray}
e^{0}_{~\mu}   &= &\Bigl(\sqrt{{\Delta \over \Sigma}}, ~0, ~0, 
-a\sin^2\theta\sqrt{{\Delta \over \Sigma}}\Bigr),
\nonumber \\
e^{1}_{~\mu}   &= &\Bigl(0, ~\sqrt{{\Sigma \over \Delta}}, ~0, ~0 \Bigr),
\nonumber \\
e^{2}_{~\mu}&= & \Bigl(0,~0, ~\sqrt{\Sigma}, ~0 \Bigr),
\nonumber \\
e^{3}_{~\mu}&= &\Bigl( -{a \over \sqrt{\Sigma} }\sin\theta 
,~0 ,~0 ,~{ r^2+a^2 \over \sqrt{\Sigma}} \sin\theta \Bigr),
\end{eqnarray}
where 
$\Sigma=r^2+a^2\cos^2\theta$, and
$e^a_{~\mu}=(e^a_{~t},e^a_{~r},e^a_{~\theta},e^a_{~\varphi})$ 
for $a=0\sim 3$. Hereafter,
we use the Latin letters to denote the tetrad indices. 

For convenience, we introduce $\omega_1 \sim \omega_6$ to represent 
the tetrad components of the spin coefficients, 
$\omega_{ab}^{~~c}=e^{~\mu}_a e^{~\nu}_b e^c{}_{\nu;\mu}$,
near the equatorial plane:
\begin{eqnarray}
 \omega_{01}^{~~0} & = &\omega_{00}^{~~1}=
 \omega_1+O(\tilde \theta^2),\quad 
 \omega_1=
 {a^2 -Mr\over r^2 \Delta^{1/2}},
\nonumber \\
 \omega_{31}^{~~0} & = &\omega_{30}^{~~1}=
 \omega_{13}^{~~0} = \omega_{10}^{~~3}=
 \omega_{03}^{~~1} = -\omega_{01}^{~~3}=
 \omega_2+O(\tilde \theta^2),\quad
 \omega_2
 :={a\over r^{2}}
\nonumber \\
 \omega_{22}^{~~1} & = &-\omega_{21}^{~~2}=
 \omega_{33}^{~~1} = -\omega_{31}^{~~3}=
 \omega_3+O(\tilde \theta^2),\quad
 \omega_3
 :={\Delta^{1/2}\over r^2},
\nonumber \\
 \omega_{02}^{~~0} & = &\omega_{00}^{~~2}=
 \omega_{12}^{~~1} = -\omega_{11}^{~~2}=\tilde \theta~ \omega_4
 +O(\tilde \theta^2),\quad
 \omega_4
 :=-{a^2\over r^3},
\nonumber \\
 \omega_{32}^{~~0} & = &\omega_{30}^{~~2}=
 -\omega_{23}^{~~0} = -\omega_{20}^{~~3}=
 \omega_{03}^{~~2} = -\omega_{02}^{~~3}=
 \tilde \theta~ \omega_5+O(\tilde \theta^2),\quad  
 \omega_5
 :=-{a\Delta^{1/2}\over r^{3}}, 
\nonumber \\
 \omega_{33}^{~~2} & = &-\omega_{32}^{~~3}=
 \tilde \theta~ \omega_6+O(\tilde \theta^2),\quad
 \omega_6 
 := -{(r^2+a^2)\over r^{3}}\,.
\end{eqnarray}
Since the following relation holds for an arbitrary vector $f^{\mu}$, 
$$
 e^a_{~\mu} {D \over d\tau}f^{\mu}=
 {d \over d\tau}f^{a}-\omega_{bc}^{~~a}v^b f^c, 
$$
the tetrad components of $Df^\mu/d\tau$ along a circular orbit
are given explicitly as 
\begin{eqnarray}
 e^0_{~\mu} 
{D\over d\tau}f^{\mu}&=&\dot f^{0} 
 -\left( A f^1 +\tilde\theta C f^2\right)
+O(\tilde \theta^2),
\nonumber \\
e^1_{~\mu} 
 {D\over d\tau}f^{\mu}&=&\dot f^{1} -\left( A f^0 +B f^3 +E f^2\right)
+O(\tilde \theta^2),
\nonumber \\
e^2_{~\mu} 
 {D\over d\tau}f^{\mu}&=&\dot f^{2} -\left( 
         \tilde\theta C f^0 
+\tilde\theta D f^3-Ef^1\right)+O(\tilde \theta^2),
\nonumber \\
e^3_{~\mu} 
 {D\over d\tau}f^{\mu}&=&\dot f^{3} 
  -\left( -B f^1 -\tilde\theta D f^2\right)
+O(\tilde \theta^2), 
\end{eqnarray}
where $A$, $B$, $C$, $D$ and $E$ are defined by\footnote{The 
symbols $A\sim E$ used here to define the auxiliary variables
are applicable only in this subsection, and not to be
confused with quantities defined with the same symbols such as
$E$ for energy, in the other sections.}
\begin{eqnarray}
  A&:=&\omega_1 v^0 +\omega_2 v^3,
\nonumber \\
  B&:=&\omega_2 v^0 +\omega_3 v^3, 
\nonumber \\
  C&:=&\omega_4 v^0 +\omega_5 v^3,
\nonumber \\
  D&:=&\omega_5 v^0 +\omega_6 v^3,
\nonumber \\
  E&:=&\omega_3 v^2, 
\end{eqnarray}
and we have assumed that $v^1=0$ and $v^2=O(\tilde \theta)$. 

For convenience, we rewrite the equations of motion by changing the spin 
variable. Instead of the spin tensor, we introduce a unit vector
parallel to the spin, $\zeta^a$, defined by
\begin{equation}
  \zeta^a:={S^a\over S}=-{1\over 2\mu S}\epsilon^a_{~bcd} u^b S^{cd},
\end{equation}
or equivalently by
\begin{equation}
 S^{ab}= \mu S \epsilon^{ab}_{~~cd} u^c \zeta^{d},
\end{equation}
where $\epsilon_{abcd}$ is the completely antisymmetric symbol with
the sign convention $\epsilon_{0123}=1$.
As noted in the previous subsection, if we use the spin vector as an 
independent variable, the center of mass condition (\ref{sp:centmass})
is replaced by Eq.~(\ref{sp:pscond}), that is 
\begin{equation}
\zeta^a u_a=0.
\end{equation} 
Then the equations of motion reduce to 
\begin{eqnarray}
 {d u^a\over d\tau} & = & \omega_{bc}^{~~a}v^b u^c -SR^a,
\nonumber \\
 {d \zeta^a\over d\tau} & = & \omega_{bc}^{~~a}v^b \zeta^c 
                       -Su^a \zeta^b R_{b},
\label{sp:tetradeq}
\end{eqnarray}
where 
\begin{equation}
  R^a:=R^{*a}{}_{bcd}v^b u^c \zeta^d
   ={1\over 2\mu S}R^a{}_{bcd} v^b S^{cd},
\end{equation}
and $ R^{*}_{abcd}={1\over 2}R_{abef}\epsilon^{ef}{}_{cd}$ 
is the right dual of the Riemann tensor.
It will be convenient to write explicitly the tetrad components of 
$ R^{*}_{~abcd}$.  
Since $\tilde\theta=O(S)$, we only need $R^{*}_{abcd}$ at 
$O(\tilde\theta^0)$. Then the non-vanishing components of
$R^{*}_{~abcd}$ are given by 
\begin{equation}
 -{1\over 2}R^{*}_{0123} =- R^{*}_{0213}=
 R^{*}_{0312}= R^{*}_{1203}=
 -R^{*}_{1302}= -{1\over 2}R^{*}_{2301}=
 -{M\over r^3}+O(\tilde \theta^2).
\end{equation}
Although we do not need them, we note that the following components are 
not identically zero but are of $O(\tilde \theta)$. 
$$
R^{*}_{1212},~ R^{*}_{1313},~ R^{*}_{1010},~ R^{*}_{2323},~
R^{*}_{2020}, ~{\rm and~} R^{*}_{3030}\,. 
$$
Further, we may set $u^\mu=v^\mu$ in the equations of motion
(\ref{sp:tetradeq}). 

\subsubsection{Lowest Order in $S$}

We first solve the equations of motion for a circular orbit at $r=r_0$
at the lowest order in $S$. 
For notational simplicity, we omit the suffix $0$ of $r_0$
in the following.
For the class of orbits we have assumed, we have $v^1=0$ and
$v^2=O(\tilde \theta)$.Then the non-trivial equations are 
\begin{eqnarray}
 {d\over d\tau}v^1&=& A v^0 + B v^3 =0,
\label{sp:v1} \\
\nonumber\\
 {d\over d\tau} \zeta^2 &= &0,
\quad
 {d\over d\tau} \left(
  \begin{array}{c}
   \zeta^0 \\ \zeta^1 \\ \zeta^3
  \end{array}\right)
 = \left(
  \begin{array}{ccc}
    0 & A & 0 \\
    A & 0 & B \\
    0 & -B & 0 \\
  \end{array}\right)
   \left(\begin{array}{c}
   \zeta^0 \\ \zeta^1 \\ \zeta^3
  \end{array}\right).
\label{sp:spineq0}
\end{eqnarray}
The equation (\ref{sp:v1}) determines the rotation velocity of the 
orbital motion. By setting $\xi:=v^3/v^0$, we obtain the equation
\begin{equation}
 \omega_1 +2\omega_2 \xi +\omega_3 \xi^2=0,  
\end{equation}
which is solved to give
\begin{equation}
 \xi={\pm\sqrt{Mr}-a\over \sqrt{\Delta}}. 
\end{equation}
The upper (lower) sign corresponds to the case that 
$v^3$ is positive (negative).
Then, with the aid of the normalization condition of the four momentum, 
$v^{\mu} v_{\mu}=-1+O(S^2)$, we find
\begin{equation}
 v^0={1\over\sqrt{1-\xi^2}},\quad  v^3={\xi\over\sqrt{1-\xi^2}}.
\label{sp:xtov}
\end{equation}
Note that, in this case, the orbital angular frequency $\Omega_\varphi$ 
is given by a well known formula,
\begin{equation}
\Omega_\varphi={\pm \sqrt{M} \over r^{3/2} \pm \sqrt{M}a}\,. 
\end{equation}
On the other hand, the equations of spin (\ref{sp:spineq0}) are
solved to give
\begin{equation}
 \zeta^2=-\zeta_{\perp}\,,
\qquad
\left(\begin{array}{c}
   \zeta^0 \\ \zeta^1 \\ \zeta^3
  \end{array}\right)
 =\zeta_\parallel\left(
  \begin{array}{c}
   \alpha\sin(\phi+c_1)+\beta c_2\\
    \cos(\phi+c_1)  \\
   -\beta\sin(\phi+c_1)-\alpha c_2
\end{array}\right).
\end{equation}
where $\zeta_{\perp}$, $\zeta_{\parallel}$, $c_1$ and $c_2$ are
constants, and
\begin{eqnarray}
\alpha&=&{A\over\sqrt{B^2-A^2}}=\mp v^3,
\quad \beta={B\over\sqrt{B^2-A^2}}=\pm v^0, 
\nonumber \\
 \phi & = & \Omega_p \tau,\quad \Omega_p=\sqrt{B^2-A^2} =\sqrt{{M \over
r^3}}\,.
\end{eqnarray}
The supplementary condition $v^a \zeta_a=0$ requires that $c_2=0$. The
condition $\zeta_a\zeta^a=1$ implies
$\zeta_\perp^2+\zeta_\parallel^2=1$.  Further since the origin of the
time $\tau$ can be chosen arbitrarily, we set $c_1=0$. Thus, we
obtain
\begin{equation}
\zeta^2=-\zeta_\perp\,,\qquad\left(
  \begin{array}{c}
 \zeta^0 \\ 
 \zeta^1 \\ 
 \zeta^3 \end{array}
\right)
=\zeta_{\parallel}
\left( \begin{array}{r} 
   \alpha\sin\phi \\ 
        \cos\phi \\
 -\beta\sin\phi
\end{array}\right).
\end{equation}
Here, we should note that $\Omega_p \not= \Omega_\varphi$ in general if 
$a\not= 0$ or $S \not=0$ (see below).

\subsubsection{First order in $S$}

Having obtained the leading order solution with respect to $S$, 
we now turn to the equations of motion up to the linear order in $S$. 
We assume that the spin vector components are expressed in 
the same form as were in the leading order but consider 
corrections of $O(S)$ to the coefficients $\alpha, \beta$ and
$\Omega_p$.
As we have noted, Eq.~(\ref{sp:vprelation}) tells us that $v^a$ can be
identified with $u^a$ to $O(S)$.
In order to write down the equations of motion up to the linear 
order in $S$, we need the explicit form of $R^a$, which can be evaluated 
by using the knowledge of the lowest order solution.
They are given as
\begin{eqnarray}
 R^0 & = & R^3 =O(\tilde \theta), 
 \nonumber \\
 R^1 & = & 3{M\over r^3} v^0 v^3 \zeta^2+O(\tilde \theta), 
\nonumber \\
 R^2 & = & 3{M\over r^3} v^0 v^3 \zeta^1+O(\tilde \theta).
\end{eqnarray}

First we consider the orbital equations of motion.
With the assumption that $v^1=0$ and $v^2=O(\tilde \theta)$, 
the non-trivial equations of the orbital motion are
\begin{eqnarray}
 \dot v^1 & = & A v^0+B v^3- SR^1=0,
 \label{sp:v1sec} \\
 \dot v^2 & = & (C v^0+D v^3)\tilde \theta - SR^2. 
 \label{sp:v2sec} 
\end{eqnarray}
The first equation gives the rotation velocity as before,
while the second equation determines the motion in the 
$\theta$-direction.

Again using the variable $\xi=v^3/v^0$, Eq.~(\ref{sp:v1sec}) 
is rewritten as 
\begin{equation}
\omega_1+2 \omega_2 \xi+\omega_3 \xi^2+3 {S_{\perp}M\over r^3}\xi=0, 
\end{equation} 
where $S_{\perp}:=S \zeta_{\perp}$. 
The solution of this equation is 
\begin{equation}
\xi=\left({\pm\sqrt{Mr}-a\over\sqrt{\Delta}}\right)
         \left(1\mp {3S_{\perp}\sqrt{M}\over 2 r^{3/2}}\right)+O(S^2).
\end{equation}
Using the relations (\ref{sp:xtov}), 
it immediately gives $v^0$ and $v^3$.
{}From the definition of the tetrad, we have the following relations,  
\begin{eqnarray}
 v^0 = \sqrt{\Delta\over \Sigma}\left[{dt\over d\tau}
    -a\sin^2\theta{d\varphi\over d\tau}\right],
 \nonumber \\
 v^3 = {\sin\theta\over \sqrt{\Sigma}}\left[-a{dt\over d\tau}
    +(r^2+a^2){d\varphi\over d\tau}\right]. 
\label{sp:v0v3}
\end{eqnarray}
Thus, the orbital angular velocity observed at infinity is calculated 
to be
\begin{eqnarray}
 \Omega_\varphi:=&&{d\varphi\over dt}
={a+\xi\sqrt{\Delta} \over r^2+a^2+a\xi\sqrt{\Delta}}+O(\tilde \theta^2)
\nonumber\\
    =&&\pm{\sqrt{M}\over r^{3/2}\pm a\sqrt{M}}
  \left[1-{3 S_{\perp}\over 2}{\pm\sqrt{Mr}
          -a\over r^2\pm a\sqrt{Mr}}\right]+O(\tilde \theta^2).
  \label{sp:Omegaori}
\end{eqnarray}

\begin{figure}[htb]
  \epsfxsize=12cm
  \centerline{\epsfbox{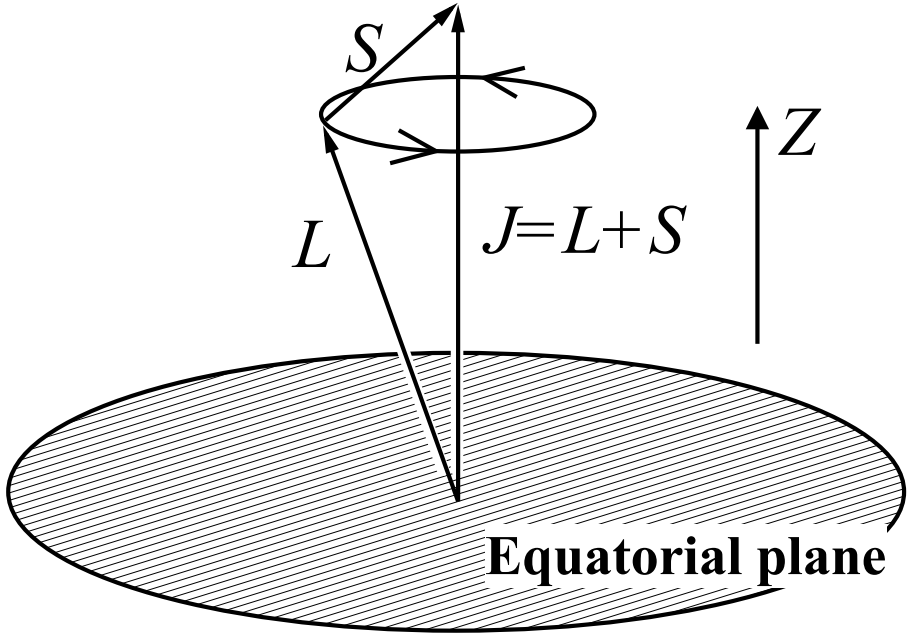}}
\caption{A schematic picture of the precession of orbit and spin
vector, to the leading order in $S$. The vector {\protect{\mib J}} 
represents the total angular momentum of the particle.
The vector {\protect{\mib L}} is orthogonal to the orbital plane and
reduces to the orbital angular momentum in the Newtonian limit.
In the relativistic case, however, these vectors should not be regarded
as well-defined.}\label{fig:precession}
\end{figure}

In order to solve the second equation (\ref{sp:v2sec}),
we note that $v^2=\sqrt{\Sigma}\dot\theta\simeq r 
\dot {\tilde\theta}$ and
\begin{equation}
 Cv^0+Dv^3=-{M\over r^2}{1+2\xi^2\over 1-\xi^2} +O(S).
\end{equation}
Then we find that Eq.~(\ref{sp:v2sec}) reduces to 
\begin{equation}
 r\ddot{\tilde\theta}
=-{M\over r^2}{1+2\xi^2\over 1-\xi^2}{\tilde\theta}
     -3 {S_{\parallel}M\over r^3}{\xi\over 1-\xi^2}\cos\phi\,, 
\label{sp:tildetheta}
\end{equation}
where $S_{\parallel}=S \zeta_{\parallel}$.
This equation can be solved easily by setting $\tilde\theta
=\theta_0 \cos\phi$. 
Recalling that $\Omega_p^2=M/r^3+O(S)$, we obtain
\begin{equation}
 \theta_0= -{S_{\parallel}\over r\xi}.
\end{equation}
Thus we see that the orbit will remain in the equatorial plane if
$S_\parallel=0$, but deviates from it if $S_\parallel\neq0$.
We note that
there exists a degree of freedom to add a homogeneous solution of 
Eq.~(\ref{sp:tildetheta}),
whose frequency, $\Omega_{\theta}=\displaystyle\sqrt{{M\over r^3}{
1+2\xi^2\over 1-\xi^2}}$, is different from $\Omega_p$ 
and which corresponds to giving a small inclination angle to the orbit,
indifferent to the spin.
Here, we only consider the case when 
this homogeneous solution to $\tilde\theta$ is zero, i.e., those orbits
which would be on the equatorial plane if the spin were zero.
Schematically speaking, the orbits under consideration
are those with the total angular momentum \hbox{\mib J} being
parallel to the $z$-direction, which is sum of the orbital and spin 
angular momentum $\hbox{\mib J}=\hbox{\mib L}+\hbox{\mib S}$ 
(see Fig.~\ref{fig:precession})).

Next we consider the evolution of the spin vector. 
To the linear order in $S$, the equations to be solved are 
\begin{eqnarray}
 \dot \zeta^0 & = & A\zeta^1+C\zeta^2 \tilde\theta-Sv^0 \zeta^a R_a,
\nonumber \\
 \dot \zeta^1 & = & A\zeta^0+B\zeta^3+ E\zeta^2,
\nonumber \\
 \dot \zeta^2 & = & \left(C\zeta^0+D\zeta^3\right)
                    \tilde\theta - E\zeta^1,
\nonumber \\
 \dot \zeta^3 & = & -B\zeta^1-D\zeta^2\tilde\theta-Sv^3 \zeta^a R_a. 
\end{eqnarray}
The third equation is written down explicitly as 
\begin{equation}
\dot \zeta^2=-\bar\theta\zeta_{\parallel}\kappa \sin\phi\cos\phi,
\end{equation}
where
\begin{equation}
   \kappa:=\alpha D -\beta C-\Omega_p \omega_3 r.
\end{equation}
Thus we find
\begin{equation}
\zeta^2=-\zeta_{\perp}
+{\theta_0\zeta_{\parallel}\kappa \over 4\Omega_p}\cos 2\phi.
\end{equation}
Since the spin vector $S^a$ is itself of $O(S)$ already, 
the effect of the second term is always unimportant as long as 
we neglect corrections of $O(S^2)$ to the orbit.

The remaining three equations
determine $\alpha$, $\beta$ and $\Omega_p$. 
Corrections of $O(S)$ to $\alpha$ and $\beta$ are less interesting  
because they remain to be small however long the time passes. 
On the other hand, the correction to $\Omega_p$ will cause a big effect 
after a sufficiently long lapse of time
 because it appears in the combination of
$\Omega_p \tau$. The small phase correction will be accumulated to
become large. 
Hence, we solve $\Omega_p$ to the next leading order. 
Eliminating $\zeta^0$ and $\zeta^3$ from these three equations, 
we obtain
\begin{equation}
\left[(B^2-A^2)-\Omega_p^2\right]
   ={S_{\perp}\over \xi}
    \left({AC-BD\over r}-\Omega_p^2\omega_3 \right). 
\end{equation}
Then after a straightforward calculation, we find
\begin{equation}
\Omega_p^2={M\over r^3}\left\{
          1-{3 S_{\perp}\over r^{3/2}}
         {\pm\sqrt{M}\left(2r^2-3Mr+a^2\right)
         +ar^{1/2}(M-r)\over r^2-3Mr\pm2a\sqrt{Mr}}\right\}.
\end{equation}
As noted above, $\Omega_p\neq\Omega_\varphi$ for
 $S_{\perp} \not=0$.
The difference $\Omega_p-\Omega_\varphi$ gives the angular velocity of
 the precession of the spin vector, as depicted in
 Fig.~\ref{fig:precession}.

\subsection{Gravitational waves and energy loss rate}

We now proceed to the calculation of the source terms in the 
Teukolsky equation and evaluate the gravitational wave flux. 
For this purpose, we must write down the expression of the 
energy momentum tensor of the spinning particle explicitly. 
We rewrite the tetrad components of the energy momentum tensor 
in the following way:
\begin{eqnarray}
 T^{ab} & = & \mu\int d\tau \left\{
   u^{(a} v^{b)} {\delta^{(4)} (x-z(\tau))\over \sqrt{-g}}
   -e^{(a}_{~\,\nu} e^{b)}_{~\rho} \nabla_{\mu} S^{\mu\nu} v^{\rho} 
   {\delta^{(4)} (x-z(\tau))\over \sqrt{-g}}\right\}
\nonumber \\
 & = & \mu\int d\tau\Biggl\{\left[u^{(a} v^{b)}
         +\omega_{dc}^{~\,(a} v^{b)} S^{dc} -\omega_{dc}^{~\,(a}
    S^{b)d} v^c \right] {\delta^{(4)} (x-z(\tau))\over \sqrt{-g}}
\nonumber\\ 
&&\qquad- {1\over \sqrt{-g}}\partial_\mu\left(
    S^{\mu (a} v^{b)} \delta^{(4)} (x-z(\tau))\right)\Biggr\}
\nonumber \\
 & =: & \mu\int d\tau \left\{A^{ab} 
        {\delta^{(4)} (x-z(\tau))\over \sqrt{-g}} 
    + {1\over \sqrt{-g}}\partial_\mu\left(
      B^{\mu ab}\delta^4 (x-z(\tau))\right)\right\}.
\end{eqnarray}
The last line gives the definition of $A^{ab}$ and $B^{\mu ab}$. 
Then the source term of the Teukolsky equation 
is given by Eq.~(\ref{eq:source}) with Eqs.~(\ref{eq:B2}). 

As we will see shortly, the terms proportional to 
$S_{\parallel}$ in the energy momentum tensor do not contribute 
to the energy or angular momentum fluxes at linear order 
in $S$. In other words, the energy and angular momentum fluxes 
are the same for all orbits having the same $S_{\perp}$. 
Thus, we ignore these terms in the following discussion. 
Further we recall that the particle can stay 
in the equatorial plane if $S_{\parallel}=0$. 
Hence we fix $\theta=\pi/2$ in the following 
calculations. 

Using the formula (\ref{eq:Rinfty}), 
we obtain the amplitude of gravitational waves at infinity as 
\begin{equation}
\tilde Z^{\infty}_{\ell m\omega} =
  \tilde Z^{nn}_{\ell m\omega}
 +\tilde Z^{\bar m n}_{\ell m\omega}
 +\tilde Z^{\bar m \bar m}_{\ell m\omega},
\end{equation}
where
\begin{eqnarray}
 \tilde Z^{nn}_{\ell m\omega} && =
 {i\sqrt{2\pi}\over \omega B^{in}_{\ell m\omega}}
   \delta\left(\omega-m\Omega\right)
   \left({dt\over d\tau}\right)^{-1} \left[
   A_{nn}-i\omega B^t_{nn}
   +imB^{\varphi}_{nn}-B^r_{nn}{\partial\over \partial r}\right]
\nonumber \\
   \times&&\left[L^{\dag}_1 \rho^{-4}\left(L^{\dag}_2 \rho^3 
    {}_{-2}S^{a\omega}_{\ell m}\right)\right]_{\theta=\pi/2}
    {1\over r\Delta} R^{in}_{\ell m\omega}
    \Biggr\vert_{r=r_0},
\nonumber \\     
 \tilde Z^{\bar m n}_{\ell m\omega} & = & 
 {i\sqrt{\pi}\over \omega B^{in}_{\ell m\omega}}
   \delta \left(\omega-m\Omega\right)
   \left({dt\over d\tau}\right)^{-1} \left[
   A_{\bar m n}-i\omega B^t_{\bar m n}
   +imB^{\varphi}_{\bar m n}-B^r_{\bar m n}
   {\partial\over \partial r}\right]
\nonumber \\
 \times&&\left(L^{\dag}_2 {}_{-2}S^{a\omega}_{\ell m}
   \right)_{\theta=\pi/2}{1\over \sqrt{\Delta}}
   \left[2{\partial\over \partial r}-{2iK\over\Delta}-{4\over r}\right]
     R^{in}_{\ell m\omega}\Biggr\vert_{r=r_0},
\nonumber \\     
 \tilde Z^{\bar m \bar m}_{\ell m\omega} & = & 
 {i\sqrt{\pi}\over \omega B^{in}_{\ell m\omega}}
   \delta\left(\omega-m\Omega\right)
   \left({dt\over d\tau}\right)^{-1} \left[
   A_{\bar m \bar m}-i\omega B^t_{\bar m \bar m}
   +imB^{\varphi}_{\bar m \bar m}-B^r_{\bar m \bar m}
   {\partial\over \partial r}\right]
\nonumber \\
 \times&&\left({}_{-2}S^{a\omega}_{\ell m}\right)_{\theta=\pi/2}
   \left[{\partial^2\over \partial r^2}
   -2\left({1\over r}+{iK\over\Delta}\right){\partial\over \partial r}
    -\left({iK\over\Delta}\right)_{,r}+{2iK\over\Delta r}
    -{K^2\over\Delta^2}\right]R^{in}_{\ell m\omega}\Biggr\vert_{r=r_0},
\nonumber\\
\label{sp:Ztilde}
\end{eqnarray}
and
\begin{eqnarray}
 A_{nn} & = & {1 \over 4}{1\over 1-\xi^2}\left\{
     1 - S_{\perp} \left( (2\omega_1 +\omega_3)\xi +\omega_2
     \right)\right\},
\nonumber \\
 B^{\mu}_{nn} & = & {1\over 4r}S_{\perp}{1\over 1-\xi^2}
   \left( {r^2 + a^2\over\sqrt{\Delta}}\xi+a,
       ~-\sqrt{\Delta}\xi,~ 0, ~{a\over\sqrt{\Delta}}\xi+ 1\right),
\nonumber \\
 A_{\bar m n} & = & {i \over 4\sqrt{2}}
    {1\over 1-\xi^2}\left\{
     2\xi -S_{\perp} \left(\omega_1\xi^2 -4\omega_2\xi -\omega_3
     \right)\right\},
\nonumber \\
 B^{\mu}_{\bar m n} & = & {i \over 4\sqrt{2}r}S_{\perp}
   {1\over 1-\xi^2} \left( {r^2 +a^2\over\sqrt{\Delta}}\xi^2+a\xi,
 ~-\sqrt{\Delta}(1+\xi^2), ~0, ~{a\over\sqrt{\Delta}}\xi^2+ \xi\right),
\nonumber \\
 A_{\bar m \bar m} & = & -{1\over 2}
    {1\over 1-\xi^2}\left\{
     \xi^2 +S_{\perp} \left(\omega_2 (1+2\xi^2) +\omega_3\xi
     \right)\right\},
\nonumber \\
 B^{\mu}_{\bar m \bar m} & = & {1\over 2r}S_{\perp} {1\over 1-\xi^2}
   \left(0, ~\sqrt{\Delta}\xi, ~0, ~0\right).
\end{eqnarray}
The Lorentz factor $dt/ d\tau$ which appears in Eqs.~(\ref{sp:Ztilde}) 
is calculated from Eqs.~(\ref{sp:v0v3}) as
\begin{equation}
 {dt\over d\tau} = {1\ \over r\sqrt{1-\xi^2}}\left(
           a\xi+ {r^2+a^2\over\sqrt\Delta}\right).
\end{equation}

In general, as we have seen in the preceding sections, when the orbit
is quasi-periodic the Fourier components of gravitational waves will
have a discrete spectrum;
\begin{equation}
\tilde Z_{\ell m\omega}
    =\sum_{n}\delta(\omega-\omega_n)Z_{\ell m\omega_n}\,.
\end{equation}
Then the time-averaged energy flux and the $z$-component of the angular
momentum flux are given by the formulas (\ref{eq:Eflux})
and (\ref{eq:Jflux}), respectively. 
In the present case, since we may regard the orbits to be on
the equatorial plane, 
the index $n$ degenerates to the angular index $m$ and $\omega_n$ 
is simply given by $m\Omega_\varphi$ $(n=m)$. Hence we eliminate the
index $n$ in the following discussion. 
Here we mention the effect of nonzero $S_{\parallel}$. 
If we recall that all the terms which are proportional to
$S_{\parallel}$ have the time dependence of $e^{\pm i\Omega_p \tau}$, we
find that they give the contribution to the side bands. 
That is to say, their contributions in $\tilde Z_{\ell m\omega}$ are 
all proportional to $\delta(\omega-m\Omega\pm\Omega_p)$. 
Then, since the energy and angular momentum fluxes are quadratic in 
$Z_{\ell m\omega_n}$, they are not affected by the presence of 
$S_{\parallel}$ as long as we are working only up to linear order 
in $S$. 

As before, in order to express the post-Newtonian corrections to the 
energy flux, we define $\eta_{\ell m\omega}$ as
\begin{equation}
 \left({dE\over dt}\right)_{\ell m}
   ={1\over 2}\left({dE\over dt}\right)_{\rm N}\eta_{\ell m},
\end{equation}
where $(dE/ dt)_{\rm N}$ is the Newtonian quadrupole formula
defined by Eq.~(\ref{eq:Enewton}).

We calculate $\eta_{\ell m}$ up to 2.5PN order. 
Keeping the $S$-dependent terms, the results are
\begin{eqnarray}
 \eta^{(s)}_{2\pm2} & = & \left(-{19 \over 3} v^3
           9q v^4 +{2134\over 63}v^5 \right) \hat s,
\nonumber \\
 \eta^{(s)}_{2\pm 1} & = &\left(
 -{1 \over 12} v^3 -{1\over 8}q v^4
              -{535\over 1008}v^5 \right)\hat s,
\nonumber \\
 \eta^{(s)}_{3\pm 3}& = & 
               -{10935\over 896} v^5 \hat s,
\nonumber \\
 \eta^{(s)}_{3\pm 2}& = & 
         {20\over 63} v^5 \hat s,
\nonumber \\
 \eta^{(s)}_{3\pm 1}& = & 
           -{1\over 8064}v^5 \hat s,
\end{eqnarray}
where $q=a/M$ and $\hat s:=S_{\perp}/M$. 
The rest of $\eta^{(s)}_{\ell m}$ are all of higher order. 
We should mention that if we regard the spinning particle 
as a model of a black hole or neutron star, 
$S$ is of order $\mu$. Therefore the correction due to 
$S$ is small compared with the $S$-independent terms in 
the test particle limit $\mu/M\ll 1$. 

Putting all together, we obtain 
\begin{eqnarray}
 \left\langle dE\over dt\right\rangle
   = \left(dE\over dt\right)_N
     \Biggl[&& 1-{1247\over 336} v^2 
             +\left(4\pi -{73\over 12}q -{25 \over 4}\hat s \right) v^3
\nonumber \\
             && +\left(-{44711\over 9072} +{33\over 16} q^2 
                  +{71\over 8} q\hat s\right) v^4
\nonumber\\
           && +\left(-{8191\over 672}\pi +{3749\over 336} q
                   +{2403\over 112}\hat s\right) v^5\Biggr].
\label{sp:dEdtv}
\end{eqnarray}
Since $v$ is defined in terms of the coordinate radius of the orbit,
the expansion with respect to $v$ does not have a clear gauge-invariant
meaning. In particular, for the purpose of the comparison with the
standard post-Newtonian calculations it is better to write the result 
by means of the angular velocity observed at infinity. 
Using the post-Newtonian expansion of Eq.~(\ref{sp:Omegaori})
\begin{equation}
 M\Omega_\varphi = v^{3}
   \left(1-\left({3\over 2}\hat s+q \right)v^{3}
    +{3\over 2}q\hat s v^{4}
    +O\left(v^6\right)\right), 
\end{equation}
Eq.~(\ref{sp:dEdtv}) can be rewritten as 
\begin{eqnarray}
 \left\langle dE\over dt\right\rangle
= \widetilde{\left({dE\over dt}\right)}_{\rm N}
   \Biggl[&& 1-{1247\over 336} x^2
        +\left(4\pi -{11\over 4}q -{5 \over 4}\hat s \right) x^3
\nonumber \\
         && +\left(-{44711\over 9072} +{33\over 16} q^2 
                  +{31\over 8} q\hat s\right)x^4
\nonumber\\
         &&+\left(-{8191\over 672}\pi +{59\over 16} q
                   -{13\over 16} \hat s\right) x^5\Biggr],
\label{sp:dedt}
\end{eqnarray}
where $x=(M\Omega_\varphi)^{1/3}$ and 
$\widetilde{(dE/dt)}_{\rm N}$ is the Newtonian quadrupole formula
expressed in terms of $x$, Eq.~(\ref{eq:Enewtonx}).
Since there is no sideband contribution in the 
present case, the angular momentum flux is simply given by 
$\langle dJ_z/dt\rangle=\Omega_\varphi^{-1}
\langle dE/dt\rangle_{\rm GW}$. 
The result (\ref{sp:dedt}) is consistent with the one obtained by the
standard post-Newtonian approach\cite{ref:KWW,ref:kidd} 
to the 2PN order in the limit $\mu/M \to 0$. 
The $\hat s$-dependent term of order $x^5$ is the 
one which is newly obtained by the black hole 
perturbation approach\cite{ref:TMSS}. 
\section{Black hole absorption}
\label{sec:bhabs}
When a particle moves around a Kerr black hole, it radiates
gravitational waves.  Some of those waves are absorbed by the black
hole.  We calculate such absorption of gravitational waves induced by a
particle of mass $\mu$ in circular orbit on the equatorial plane around
a Kerr black hole of Mass $M$.

The post-Newtonian approximation of the absorption of gravitational
waves into the black hole horizon was first calculated by Poisson and
Sasaki in the case when a test particle is in a circular orbit around a
Schwarzschild black hole\cite{ref:PoSa}.  In this case, the effect
of the black hole absorption is found to appear at $O(v^8)$ compared to
the flux emitted to infinity and it turns out to be negligible for the
orbital evolution of coalescing compact binaries in the near future
laser interferometer's band.  On the other hand, the black hole
absorption appears at $O(v^5)$ if a black hole is rotating. That
calculation was done by Tagoshi, Mano, and Takasugi\cite{ref:TMT}.

In order to calculate the post-Newtonian expansion of ingoing
gravitational waves into a Schwarzschild black hole, Poisson and
Sasaki\cite{ref:PoSa} used two type of representations of a solution of
the homogeneous Teukolsky equation.  One is expressed in terms of the
spherical Bessel functions which can be used at large radius, and the
other is expressed in terms of a hypergeometric function which can be
used near the horizon. Then two type of expressions are matched at some
region where both formulas can be applied.  They got formulas for a
solution of the Teukolsky equation which can be used to calculate
ingoing gravitational waves to $O(v^{13})$, although they gave formulas
for ingoing waves only to $O(v^8)$.

Here, we first review the method found by Mano, Suzuki and
Takasugi\cite{ref:MST}, since it is the only existing method by which
higher order post-Newtonian terms of the gravitational waves absorbed
into a rotating black hole can be calculated. We note that this method
is also the only existing method that can be used to calculate the
gravitational waves emitted to infinity to an arbitrarily high
post-Newtonian order. Then we calculate the energy flux absorbed into
the horizon to $O(v^{13})$, i.e., $O(v^8)$ beyond the lowest order
flux absorbed into the horizon, for circular orbits on the equatorial
plane of a Kerr black hole.

\subsection{Analytic solutions of the homogeneous Teukolsky equation}
Analytic series solutions of the homogeneous Teukolsky equation were
found by Mano, Suzuki and Takasugi\cite{ref:MST} and various properties
of the solution were discussed by Mano and Takasugi\cite{ref:MT}. 
Here, we follow the notation of Ref.~\citen{ref:MT} except that we focus 
on the case of spin weight $s=-2$.
In this method, the solution of the radial Teukolsky equation
(\ref{eq:radTeukolsky}) is represented by two kinds of expansion.  One
is given by a series of hypergeometric functions 
and the other by a series of Coulomb wave functions. The former is
convergent at horizon and the latter at infinity. Then the matching of
these two solutions are done exactly in the overlapping region of
convergence.

First we consider the solution expressed in terms of hypergeometric
functions. The solution which satisfies the ingoing wave boundary
condition at horizon is expressed as
\begin{eqnarray}
  \label{eq:hypinsol}
  R^{{\rm in}\,\nu}=
 &&e^{i\epsilon\kappa x}(-x)^{2-i\epsilon_+}(1-x)^{i\epsilon_-}
\nonumber\\
 &&\times \sum_{n=-\infty}^\infty a_n^\nu
 F(n+\nu+1-i\tau,-n-\nu-i\tau,3-2i\epsilon_+;x),
\end{eqnarray}
where $F(a,b,c,x)$ is the hypergeometric function and
\begin{eqnarray} 
&& x=-{\omega(r-r_+)\over\epsilon\kappa}\,,
\quad\kappa=\sqrt{1-q^2}\,,
\quad\tau={{\epsilon-m q}\over \kappa}\,,
\quad\epsilon_{\pm}={\epsilon\pm\tau\over2}\,.
\end{eqnarray}
The coefficients $a_n^{\nu}$ obey a three terms recurrence relation,
\begin{equation}
\alpha_n^\nu a_{n+1}^\nu+\beta^\nu_n a_n^\nu
              +\gamma^\nu_n a_{n-1}^\nu=0,
\label{eq:three}
\end{equation}
where
\begin{eqnarray}
\alpha_n^{\nu}&=&{i\epsilon \kappa (n+\nu-1+i\epsilon)
(n+\nu-1-i\epsilon)(n+\nu+1+i\tau)
\over{(n+\nu+1)(2n+2\nu+3)}}\,,
\cr
\beta_n^{\nu}&=&-\lambda-2+(n+\nu)(n+\nu+1)+\epsilon^2 
+\epsilon(\epsilon-m q)\cr
&&\hskip 3mm 
+{\epsilon (\epsilon-mq)(4+\epsilon^2) \over{(n+\nu)(n+\nu+1)}}\,, 
\cr
\gamma_n^{\nu}&=&-{i\epsilon \kappa
 (n+\nu+2+i\epsilon)(n+\nu+2-i\epsilon)(n+\nu-i\tau)
\over{(n+\nu)(2n+2\nu-1)}}\,.
\label{eq:alpbetgam}
\end{eqnarray}
The series converges if $\nu$ satisfies the equation,
\begin{equation}
R_n(\nu)L_{n-1}(\nu)=1, 
\label{eq:nueq}
\end{equation}
where $R_n(\nu)$ and $L_n(\nu)$ are the continued fractions defined by
\begin{eqnarray}
R_n(\nu)&\equiv&{a_n^\nu \over a_{n-1}^\nu}=
-{\gamma_n^\nu\over \beta_{n}^{\nu}+\alpha_n^\nu R_{n+1}(\nu)}\,,
\label{eq:cont1}
\cr
L_n(\nu)&\equiv&{a_n^\nu \over a_{n+1}^\nu}=
-{\alpha_n^\nu\over \beta_n^\nu+\gamma_n^\nu L_{n-1}(\nu)}\,.
\label{eq:cont2}
\end{eqnarray}
The range of convergence is $0\le (-x)<\infty$ for physical $x$.
We can prove that if $\nu$ is a solution to Eq.~(\ref{eq:nueq}),
then so is $-\nu-1$. Since we can set $n$ in Eq.~(\ref{eq:nueq}) to an
arbitrary integer, a convenient choice is to put $n=1$.
Further, for convenience, we may set $a_0^\nu=a_0^{-\nu-1}=1$.
It then follows that we have
\begin{equation}
  \label{eq:coeffsym}
  a_{-n}^{-\nu-1}=a_n^\nu\,.
\end{equation}
This implies $R^{{\rm in}\,\nu}=R^{{\rm in}\,-\nu-1}\equiv R^{\rm in}$.
Consequently, for $(-x)>0$, Eq.~(\ref{eq:hypinsol}) can be rewritten as
\begin{equation}
R^{\rm in}=e^{i\epsilon\kappa}(R_0^{\nu}+R_0^{-\nu-1}),
\label{eq:RinR0}
\end{equation}
where 
\begin{eqnarray}
{R}_{0}^{\nu} 
&=&e^{-i\epsilon\kappa \tilde{x}}(\tilde{x})^{\nu+i\epsilon_+}
(\tilde{x}-1)^{-s-i\epsilon_+}
\cr  &&
\times \sum_{n=-\infty}^{\infty}
\frac{\Gamma(3-2i\epsilon_+)\Gamma(2n+2\nu+1)}
{\Gamma(n+\nu+1-i\tau)\Gamma(n+\nu+3-i\epsilon)}a_{n}^{\nu}
\cr &&  \times 
\tilde{x}^n 
  F(-n-\nu-i\tau,-n-\nu+2-i\epsilon,-2n-2\nu;\frac 1{\tilde{x}}) ,
\label{eq:hyperex}
\end{eqnarray}
where 
\begin{equation}
\tilde{x}=1-x={\omega(r-r_-)\over {\epsilon\kappa}}\,.
\end{equation}
Since $\nu-(-\nu-1)=2\nu+1$ is not an integer in general, the solutions
$R_0^\nu$ and $R_0^{-\nu-1}$ form a pair of independent solutions. 

The other series solution which is convergent at infinity
is expressed in terms of the Coulomb wave functions;
\cite{ref:leaver}
\begin{equation} 
{R}_{C}^{\nu}= 
\tilde{z}
 \left(1-{\epsilon \kappa \over{\tilde{z}}}\right)^{2-i\epsilon_+}
\sum_{n=-\infty}^{\infty}(-i)^n
\frac{(\nu-1-i\epsilon)_n}{(\nu+3+i\epsilon)_n}
a_n^{\nu}F_{n+\nu}(\tilde{z}), 
\label{eq:coulombex} 
\end{equation} 
where $\tilde z=\omega(r-r_-)=\epsilon\kappa\tilde x$, 
$(a)_n=\Gamma(n+a)/\Gamma(a)$, and $F_{n+\nu}(z)$ is the
Coulomb wave function given by
\begin{eqnarray*} 
F_{n+\nu}(z)&=&e^{-iz}(2z)^{n+\nu} z 
{\Gamma(n+\nu+3+i\epsilon)\over{\Gamma(2n+2\nu+2)}} \cr
&&\times{\rm \Phi}(n+\nu+3+i\epsilon,2n+2\nu+2;2iz), 
\end{eqnarray*} 
where ${\rm \Phi}(a,b,z)$ is the regular confluent hypergeometric
function\cite{ref:handbook}.
A crucial observation made by Mano, Suzuki and Takasugi\cite{ref:MST} is
that the coefficients ${a_n^\nu}$ obey the same recurrence relation
as that for the hypergeometric type solution, Eq.~(\ref{eq:three}). 
The series (\ref{eq:coulombex}) converges in the range 
$\tilde z>\epsilon\kappa$ if $\nu$ is a solution of
Eq.~(\ref{eq:nueq}). 
The solution $R_C^\nu$ can be decomposed into a pair of
solutions, a purely incoming wave at infinity $R_{+}^{\nu}$
and a purely outgoing wave at infinity $R_{-}^{\nu}$.  
Explicitly, we have
\begin{equation}
R_{C}^{\nu}=R_{+}^{\nu}+R_{-}^{\nu},
\end{equation}
where
\begin{eqnarray}
R_{+}^{\nu}&&= 2^{\nu}e^{-\pi \epsilon}e^{i\pi(\nu+3)}
\frac{\Gamma(\nu+3+i\epsilon)}{\Gamma(\nu-1-i\epsilon)}
e^{-i\tilde{z}}\tilde{z}^{\nu+i\epsilon_+}
(\tilde{z}-\epsilon\kappa)^{2-i\epsilon_+}
\cr
&&\times \sum_{n=-\infty}^{\infty}i^n
a_n^{\nu}(2\tilde{z})^n{\rm \Psi}
(n+\nu+3+i\epsilon,2n+2\nu+2;2i\tilde{z}),
\\ 
R_{-}^{\nu}&&= 2^{\nu}e^{-\pi \epsilon}e^{-i\pi(\nu-1)}
e^{i\tilde{z}}\tilde{z}^{\nu+i\epsilon_+}
(\tilde{z}-\epsilon\kappa)^{2-i\epsilon_+}
\sum_{n=-\infty}^{\infty}i^n\cr
&&\times \frac{(\nu-1-i\epsilon)_n}{(\nu+3+i\epsilon)_n}
a_n^{\nu}(2\tilde{z})^n{\rm \Psi}
(n+\nu-1-i\epsilon,2n+2\nu+2;-2i\tilde{z}) ,
\label{eq:Rminus}
\end{eqnarray}
where ${\rm \Psi}(a,b,z)$ is the irregular confluent hypergeometric 
function\cite{ref:handbook}.
By definition, the upgoing solution $R^{\rm up}$ is given by 
\begin{equation}
R^{\rm up}=R_{-}^{\nu} \, . 
\end{equation}

We see that the above two kinds of solutions are convergent 
in the common region $1<\tilde x<\infty$.
Then comparing the asymptotic behaviors of $R^{\nu}_{0}$ and
$R^{\nu}_{C}$ for $\tilde x\to\infty$, we find that they have the same
characteristic exponent, $\sim\tilde{x}^\nu$, hence describe the
same solution up to the normalization factor.
Therefore by comparing each power of $\tilde x$ we have
\begin{equation}
R_{0}^\nu=K_\nu R^\nu_{C}, 
\label{eq:R0RC}
\end{equation}
where
\begin{eqnarray}
{K}_{\nu}&=&
\frac{(2\epsilon \kappa )^{-\nu-2-{\tilde r}}2^{2}i^{\tilde r} 
\Gamma(3-2i\epsilon_+)
\Gamma({\tilde r}+2\nu+1)\Gamma({\tilde r}+2\nu+2)}
{\Gamma(\nu+1-i\tau)\Gamma(\nu+3-i\epsilon)
\Gamma({\tilde r}+\nu+3+i\epsilon)} 
\cr &&
\times \frac{\Gamma(\nu+1+i\tau)\Gamma(\nu-1+i\epsilon)}
{\Gamma({\tilde r}+\nu+1+i\tau)\Gamma({\tilde r}+\nu-1+i\epsilon)}
\cr&&
\times \overline{\left ( \sum_{n={\tilde r}}^{\infty}
\frac{({\tilde r}+2\nu+1)_n}{(n-{\tilde r})!}
\frac{(\nu-1-i\epsilon)_n}{(\nu+3+i\epsilon)_n}a_n^{\nu}\right )}
\cr &&
\times \left(\sum_{n=-\infty}^{{\tilde r}}
\frac{(-1)^n}{({\tilde r}-n)!
({\tilde r}+2\nu+2)_n}\frac{(\nu-1-i\epsilon)_n}{(\nu+3+i\epsilon)_n}
a_n^{\nu}\right)^{-1}, 
\end{eqnarray}
where $\tilde r$ can be any integer and 
${K}_{\nu}$ is independent of the choice of  ${\tilde r}$.

The gravitational wave absorbed into the black hole is expressed by
Eq.~(\ref{eq:zhole}). Hence we need to know the amplitudes $B^{\rm inc}$
and $B^{\rm trans}$ of $R^{\rm in}$ and $C^{\rm trans}$ of $R^{\rm up}$,
defined in Eq.~(\ref{eq:Rin}).
The asymptotic ingoing amplitude at horizon, 
$B^{\rm trans}$, of $R^{\rm in}$ is readily obtained from
Eq.~(\ref{eq:hypinsol}) as 
\begin{equation} 
B^{\rm trans}=\left(\frac{\epsilon\kappa}{\omega}\right)^{2s} 
e^{i\epsilon_{+}\ln\kappa}\sum_{n=-\infty}^{\infty}a_{n}^{\nu}\,. 
\end{equation} 
Similarly, the asymptotic outgoing amplitude at infinity, 
$C^{\rm trans}$, of $R^{\rm up}$ is obtained from Eq.~(\ref{eq:Rminus})
as
\begin{eqnarray} 
C^{\rm trans} 
=&&2^{1+i\epsilon}e^{-\pi\epsilon/2} e^{-{\pi\over 2}i(\nu-1)} 
    \omega^3 e^{i\epsilon\ln\epsilon}
\nonumber\\
&&\quad\times\left[\sum_{n=-\infty}^{\infty}{(\nu-1-i\epsilon)_n
      \over (\nu+3+i\epsilon)_n}(-1)^na_n^\nu \right]. 
\end{eqnarray}
On the other hand, a bit of work is necessary to obtain the asymptotic
incoming amplitude at infinity, $B^{\rm inc}$, of $R^{\rm in}$. 
Setting the asymptotic behavior of $R_C^\nu$ at $z\to\infty$ as
\begin{equation}
  \label{RCasymp}
  R_C^\nu\to A_{+}^\nu z^{-1}e^{-i(z+\epsilon\ln z)}
+A_{-}^\nu z^3e^{i(z+\epsilon\ln z)},
\end{equation}
we find
\begin{eqnarray}
&&A_{+}^{\nu}=2^{-3-i\epsilon}
e^{i(\pi/2)(\nu+3)}e^{-\pi \epsilon/2}
\frac{\Gamma(\nu+3+i\epsilon)}{\Gamma(\nu-1-i\epsilon)}
\sum_{n=-\infty}^{\infty} 
a_n^{\nu}, 
\nonumber\\
&&A_{-}^{\nu}=2^{1+i\epsilon}e^{-i(\pi/2)(\nu-1)}e^{-\pi \epsilon/2}
\sum_{n=-\infty}^{\infty}(-1)^n
\frac{(\nu-1-i\epsilon)_n}{(\nu+3+i\epsilon)_n}a_n^{\nu}.
\end{eqnarray}
Because of Eq.~(\ref{eq:coeffsym}), $A_{\pm}^\nu$ and $A_{\pm}^{-\nu-1}$ 
are related to each other as
\begin{eqnarray}
  \label{eq:Apmsym}
&& A_{+}^{-\nu-1}=-ie^{-i\pi\nu}
{\sin\pi(\nu+i\epsilon)\over\sin\pi(\nu-i\epsilon)}A_{+}^\nu\,,
\nonumber\\
&& A_{-}^{-\nu-1}=ie^{i\pi\nu}A_{-}^\nu\,.
\end{eqnarray}
With the help of the above relations, we find from Eqs.~(\ref{eq:RinR0})
and (\ref{eq:R0RC}) the asymptotic amplitudes of $R^{\rm in}$ at
infinity as
\begin{eqnarray}
B^{\rm inc}
&=&\frac{ e^{i\epsilon\kappa}}{\omega}\left[{K}_{\nu}-
 ie^{-i\pi\nu} \frac{\sin \pi(\nu+i\epsilon)}
 {\sin \pi(\nu-i\epsilon)}
 {K}_{-\nu-1}\right]A_{+}^{\nu} e^{-i\epsilon\ln\epsilon}, 
\nonumber\\
B^{\rm ref}
&=&e^{i\epsilon\kappa}\omega^{3}\left[{K}_{\nu}
+ie^{i\pi\nu} {K}_{-\nu-1}\right]A_{-}^{\nu} 
e^{i\epsilon\ln\epsilon}.
\end{eqnarray}

So far our discussion has been on exact analytic series expressions for
the homogeneous Teukolsky functions. Now we consider their
post-Minkowski expansion by assuming $\epsilon\ll1$. Provided we set
$a_0^\nu=1$, we see from Eqs.~(\ref{eq:alpbetgam}) that $\alpha_n^\nu$,
$\gamma_n^\nu=O(\epsilon)$ and $\beta_n^\nu=O(1)$ unless the value of
$\nu$ is such that the denominator in the expression of $\alpha_n^\nu$
or $\gamma_n^\nu$ happens to vanish or $\beta_n^\nu$ happens to vanish
in the limit $\epsilon\to0$. Except for such an exceptional case, it is
easy to see from Eq.~(\ref{eq:three}) that the order of $a_n^\nu$ in
$\epsilon$ increases as $|n|$ increases. Thus the series solution
naturally gives the post-Minkowski expansion.

{}For the moment, let us assume that the above mentioned exceptional
case does not happen for $n=0$. Then we have $R_1(\nu)=O(\epsilon)$ and
$L_0(\nu)=O(1/\epsilon)$. This implies 
$\beta_0^\nu+\gamma_0^\nu L_{-1}(\nu)=O(\epsilon^2)$. Then assuming
$L_{-1}(\nu)=O(\epsilon)$,  we must have $\beta_0^\nu=O(\epsilon^2)$.
Using the expansion of $\lambda$ given by Eq.~(\ref{eq:pnlambda}), we
then find $\nu=\ell+O(\epsilon^2)$ or $\nu=-\ell-1+O(\epsilon^2)$.
Since we know that $-\nu-1$ is a solution if $\nu$ is so, we may
take the solution $\nu=\ell+O(\epsilon^2)$ without loss of generality.
Then the assumptions that $R_1(\nu)=1/L_0(\nu)=O(\epsilon)$ and
$L_{-1}(\nu)=O(\epsilon)$ are justified. 
Further it is easily seen that $R_n(\nu)=O(\epsilon)$ for all $n>0$.
On the other hand, for $n<0$, $\alpha_n^\nu=O(1)$ at $n=-\ell-1$ and 
$\beta_n^\nu=O(\epsilon^2)$ at $n=-2\ell-1$. Thus we have
$L_{-\ell-1}(\nu)=O(1)$ and $L_{-2\ell-1}(\nu)=O(1/\epsilon)$.
To summarize, we have
\begin{eqnarray}
 && R_n(\nu)={a_{n}^\nu\over a_{n-1}^\nu}=O(\epsilon)
\quad\hbox{\rm for all}~n>0,
\nonumber\\
 && L_{-\ell-1}(\nu)={a_{-\ell-1}^\nu\over a_{-\ell}^\nu}=O(1),
\quad 
L_{-2\ell-1}(\nu)={a_{-2\ell-1}^\nu\over a_{-2\ell}^\nu}=O(1/\epsilon),
\nonumber\\
 && L_{n}(\nu)={a_n^\nu\over a_{n+1}^\nu}=O(\epsilon)
\quad\hbox{\rm for all the other}~n<0.
\end{eqnarray}
With the above results, the post-Minkowski expansion of the homogeneous
Teukolsky functions can be obtained with arbitrary accuracy by solving
Eq.~(\ref{eq:nueq}) to a desired order and by summing 
up the terms to a sufficiently large $|n|$. For our present purpose, we
need $\nu$ which is accurate to $O(\epsilon^2)$. Solving
Eq.~(\ref{eq:nueq}) to this order, we find
\begin{eqnarray}
  \label{eq:nusol}
  \nu&=&\ell+{\epsilon^2\over2\ell+1}
\Biggl[-2-{4\over\ell(\ell+1)}
\nonumber\\
&&\qquad+{(\ell-1)^2(\ell+3)^2\over(2\ell+1)(2\ell+2)(2\ell+3)} 
-{(\ell-2)^2(\ell+2)^2\over(2\ell-1)2\ell(2\ell+1)}\Biggr].
\end{eqnarray}
Interestingly, $\nu$ is found to be independent of the azimuthal
eigenvalue $m$ to $O(\epsilon^2)$. 

The post-Newtonian expansion in the near zone is given by further
assuming $\epsilon\ll z\ll1$ in the series solution (\ref{eq:coulombex})
and expand it in powers $z$. For evaluation of the black hole
absorption, we need the post-Newtonian expansion of $R^{\rm up}$ which
is obtained from Eq.~(\ref{eq:Rminus}). The explicit post-Newtonian
formula for $R^{\rm up}$ and the asymptotic amplitudes 
$B^{\rm inc}$, $B^{\rm trans}$ and $C^{\rm trans}$ to $O(\epsilon^2)$
are given in Appendix \ref{ap:absfunc}.

\subsection{Absorption rate to $O(v^8)$}
In this subsection, we evaluate the energy absorption rate by a black
hole. The energy flux formula is given by Teukolsky and
Press\cite{ref:TP} as 
\begin{equation}
\left(dE_{\rm hole}\over dtd\Omega\right)=
\sum_{\ell m}\int d\omega {{_{2}S_{\ell m}^2}\over 2\pi} 
{{128 \omega k (k^2+4 {\tilde \epsilon}^2)(k^2+16 {\tilde \epsilon}^2)
(2 M r_+)^5}\over |C|^2}
|\tilde{Z}^{\rm H}_{\ell m\omega}|^2, 
\label{eq:dedthole}
\end{equation}
where ${\tilde \epsilon}=\kappa/(4r_+)$ and 
\begin{eqnarray}
|C|^2&=&\left((\lambda+2)^2+4 a \omega m -4 a^2 \omega^2\right)
  \left[\lambda^2+36 a \omega m - 36 a^2 \omega^2\right]
\nonumber\\
&&+(2\lambda+3)(96 a^2 \omega^2-48 a \omega m)+144 \omega^2(M^2-a^2).
\end{eqnarray}

The calculation of $\tilde{Z}^{\rm H}_{\ell m\omega}$ is parallel to the 
calculation of $\tilde{Z}^{\infty}_{\ell m\omega}$ except that 
$R^{\rm in}$ is replaced by $R^{\rm up}$.
The solution of the geodesic equations are given in 
section \ref{sec:kerrv8}. Using that solution, we have 
the amplitude of the Teukolsky function at the horizon 
in Eq.~(\ref{eq:zhole}) as 
\begin{eqnarray}
\tilde Z^{\rm H}_{\ell m\omega}&=&
{{2 \pi B^{\rm trans}\delta(\omega-m\Omega)}
\over {2i\omega B^{\rm inc}C^{\rm trans}}}
\Bigl[R^{{\rm up}}_{\ell m\omega}\{A_{n\,n\,0}+A_{{\bar m}\,n\,0}
+A_{{\bar m}\,{\bar m}\,0}\}
\nonumber\\
& &-{dR^{{\rm up}}_{\ell m\omega} \over dr}\{ A_{{\bar m}\,n\,1}
+A_{{\bar m}\,{\bar m}\,1}\}
+{d^2 R^{{\rm up}}_{\ell m\omega} \over dr^2}
A_{{\bar m}\,{\bar m}\,2}\Bigr]_{r=r_0,\theta=\pi/2},
\nonumber\\
&\equiv& 
\delta(\omega-m\Omega) Z^{\rm H}_{\ell m}\,.
\label{eq:zhole2}
\end{eqnarray}
{}From Eq.~(\ref{eq:dedthole}) and (\ref{eq:zhole2}), the time averaged 
energy absorption rate becomes
\begin{eqnarray} 
\left\langle{dE\over dt}\right\rangle_{\rm H}&=&
\sum_{\ell m}\left[
{{128 \omega k (k^2+4 {\tilde \epsilon}^2)(k^2+16 {\tilde \epsilon}^2)
(2 M r_+)^5}\over |C|^2}
|Z^{\rm H}_{\ell m}|^2\right]_{\omega=m\Omega} \cr
&\equiv&\sum_{\ell m}\left(dE\over dt\right)_{\ell,m}. 
\label{eq:dedthole2}
\end{eqnarray}
As in the case of the Teukolsky function at infinity, 
we can show that $\bar{Z}^{\rm H}_{\ell, -m, -\omega}$ 
$=(-1)^{\ell}Z^{\rm H}_{\ell,m,\omega}$. 
Then, from Eq.~(\ref{eq:dedthole2}), 
we have $(dE/dt)_{\ell,-m}=(dE/dt)_{\ell,m}$.

In order to express the post-Newtonian corrections to the black hole
absorption, 
we define $\eta_{\ell,m}^{\rm H}$ as 
\begin{equation}
\left( {dE \over dt} \right)_{\ell,m}
\equiv{1\over2}\left({dE\over dt}\right)_N v^5 \eta_{\ell,m}^{\rm H}\,,
\label{eq:etahole}
\end{equation}
where $(dE/dt)_N$ is the Newtonian quadrupole luminosity at infinity,
Eq.~(\ref{eq:Enewton}). 
In Appendix \ref{ap:absorption}, we show $\eta_{\ell m}$. 

The total absorption rate to $O(v^8)$ beyond the lowest order is given
by 
\begin{eqnarray}
\left\langle{dE\over dt}\right\rangle_{\rm H}&&=
\left(dE\over dt\right)_N\, v^5\, \Biggl[
-{3\over 4}\,{q}^{3}-{1\over 4}\,q
+\left (-q-{\frac {33}{16}}\,{q}^{3}\right ){v}^{2} \nonumber\\
&&
+\left ({7\over 2}\,{q}^{4}+2\,q{B_2}+{1\over 2}+6\,{q}^{3}{B_2
}+{\frac {85}{12}}\,{q}^{2}+3\,{q}^{4}\kappa+{1\over 2}\,\kappa
+{13\over 2}\,\kappa
\,{q}^{2}\right ){v}^{3} \nonumber\\
&&
+\left (-{\frac {
4651}{336}}\,{q}^{3}-{\frac {43}{7}}\,q-{\frac {17}{56}}\,{q}^{5}
\right ){v}^{4} \nonumber\\
&&
+\Bigl({\frac {569}{24}}\,{q}
^{2}+{\frac {371}{48}}\,{q}^{4}+18\,{q}^{3}{B_2}
     -{3\over 4}\,{q}^{3}{
B_1}+2\,\kappa+2+{\frac {33}{4}}\,{q}^{4}\kappa \nonumber\\
&&
+6\,q{B_2}+{\frac {163}{8}}\,\kappa\,{q}^{2}+q{B_1}\Bigr){v}^{5} 
\nonumber\\
&&
+\Bigl(-{
\frac {2718629}{44100}}\,q-4\,{B_2}+{\frac {428}{105}}\,\gamma\,q
+{2\over 3}\,{\pi }^{2}q+{\frac {428}{105}}\,q\ln 2-4\,q{C_2} 
\nonumber\\
&&
-12\,{q}^{3}{C_2}-36\,{q}^{4}{B_2}-56\,{q}^{2}{B_2}
+{\frac {428}{35}}\,{q}
^{3}\gamma+{\frac {428}{35}}\,{q}^{3}\ln 2+2\,{q}^{3}{\pi }^{2} 
\nonumber\\
&&
+{\frac {428}{105}}\,q\ln\kappa+{\frac {428}{105}}\,q{A_2}+{\frac 
{428}{35}}\,{q}^{3}\ln\kappa+6\,{\frac {{q}^{7}}{\kappa}}+{\frac {
428}{35}}\,{q}^{3}{A_2}-8\,q{{B_2}}^{2} \nonumber\\
&&
-24\,{q}^{3}{{B_2}}^{2}+{\frac {856}{105}}\,q\ln v
+{\frac {856}{35}}\,{q}^{3}\ln v-4\,{
\frac {{B_2}}{\kappa}}-32\,{\frac {{q}^{3}}{\kappa}}-31\,{\frac {q}
{\kappa}} \nonumber\\
&&
+57\,{\frac {{q}^{5}}{\kappa}}+{q}^{5}\kappa+{q}^{4}{B_1}-
{2\over 3}\,\kappa\,q-{7\over 6}\,\kappa\,{q}^{3}
-{4\over 3}\,{q}^{2}{B_1} \nonumber\\
&&
-48\,{\frac {{q}^{2}{B_2}}{\kappa}}
+28\,{\frac {{q}^{4}{B_2}}{\kappa}}-24\,{
\frac {{q}^{3}{C_2}}{\kappa}}-8\,{\frac {q{C_2}}{\kappa}}+24\,{
\frac {{q}^{6}{B_2}}{\kappa}} \nonumber\\
&&
-{\frac {2400247}{19600}}\,{q}^{3}+{
\frac {299}{16}}\,{q}^{5}\Bigr){v}^{6} \nonumber\\
&&
+\Bigl({\frac {225}{28}}\,{q}^{5}{B_3}-{\frac {41}{28}}\,\kappa\,
{q}^{6}+{\frac {86}{7}}+{\frac {8741}{56}}\,{q}^{2}+{\frac {3485}{42}}
\,{q}^{4}+{\frac {167}{112}}\,{q}^{6} \nonumber\\
&&
+{\frac {86}{7}}\,\kappa+{\frac {
45}{56}}\,q{B_3}-{\frac {9}{28}}\,{q}^{5}{B_1}+{\frac {899}{168}
}\,q{B_1}-{\frac {803}{224}}\,{q}^{3}{B_1}+{\frac {1665}{224}}\,
{q}^{3}{B_3} \nonumber\\
&&
+{\frac {2372}{21}}\,{q}^{3}{B_2}-{\frac {16}{7}}\,{
q}^{5}{B_2}+{\frac {719}{12}}\,{q}^{4}\kappa+{\frac {22201}{168}}\,
\kappa\,{q}^{2}+{\frac {796}{21}}\,q{B_2}\Bigr){v}^{7} 
\nonumber\\
&&
+\Bigl(-{\frac {20542807}{88200}}\,q-12\,{B_2}-2\,{B_1}+{\frac {
1061}{35}}\,\gamma\,q+{\frac {13}{6}}\,{\pi }^{2}q+{\frac {995}{21}}\,
q\ln 2\nonumber\\
&&
-12\,q{C_2}-36\,{q}^{3}{C_2}-{\frac {308}{3}}\,{q}^{4}{
B_2}-{\frac {1496}{9}}\,{q}^{2}{B_2}+{\frac {12197}{140}}\,{q}^{
3}\gamma+{\frac {3873}{28}}\,{q}^{3}\ln 2 \nonumber\\
&&
+{\frac {47}{8}}\,{q}^{3}{\pi }^{2}+{\frac {1391}{105}}\,q\ln\kappa
+{\frac {428}{35}}\,q{A_2}+{\frac {5029}{140}}\,{q}^{3}\ln\kappa
+{\frac {37}{6}}\,{\frac {{q}^{7}}{\kappa}}
+{\frac {1284}{35}}\,{q}^{3}{A_2} \nonumber\\
&&
-24\,q{{B_2}}^{2}-72\,{q}^{3}{{B_2}}^{2}
+{\frac {4574}{105}}\,q\ln v+{\frac {8613}{70}}\,{q}^{3}\ln v
-12\,{\frac {{B_2}}{\kappa}}
-{\frac {341}{4}}\,{\frac {{q}^{3}}{\kappa}} 
\nonumber\\
&&
-{\frac {637}{6}}\,{\frac {q}{\kappa}}+
{\frac {741}{4}}\,{\frac {{q}^{5}}{\kappa}}+{\frac {73}{6}}\,{q}^{4}{
B_1}-q{C_1}-{\frac {283}{18}}\,{q}^{2}{B_1} \nonumber\\
&&
+{3\over 2}\,{q}^{3}{{B_1}}^{2}+{\frac {107}{105}}\,q{A_1}
-{\frac {107}{140}}\,{q}^{3}
{A_1}-3\,{\frac {{q}^{6}{B_1}}{\kappa}}
+{13\over 2}\,{\frac {{q}^{4}{
B_1}}{\kappa}}-{3\over 2}\,{\frac {{q}^{2}{B_1}}{\kappa}} 
\nonumber\\
&&
-2\,{\frac {q{C_1}}{\kappa}}+{3\over 2}\,
{\frac {{q}^{3}{C_1}}{\kappa}}
-2\,q{{B_1}
}^{2}+{3\over 4}\,{q}^{3}{C_1}
-2\,{\frac {{B_1}}{\kappa}}-144\,{\frac {
{q}^{2}{B_2}}{\kappa}} \nonumber\\
&&
+84\,{\frac {{q}^{4}{B_2}}{\kappa}}-72\,{
\frac {{q}^{3}{C_2}}{\kappa}}-24\,{\frac {q{C_2}}{\kappa}}+72\,{
\frac {{q}^{6}{B_2}}{\kappa}}-{\frac {2945984497}{6350400}}\,{q}^{3}
\nonumber\\
&&
+{\frac {1385}{24}}\,{q}^{5}+{\frac {25}{252}}\,{q}^{7}\Bigr){v}^{8}
\Biggr]. 
\label{eq:dEdthole} 
\end{eqnarray}
where 
\begin{eqnarray}
A_n&=&{1\over 2}\left[\psi^{(0)}\left(3+{n i q\over \sqrt{1-q^2}}\right)
+\psi^{(0)}\left(3-{n i q\over \sqrt{1-q^2}}\right)
\right], \cr
B_n&=&{1\over 2i}\left[\psi^{(0)}\left(3+{n i q\over \sqrt{1-q^2}}\right)
-\psi^{(0)}\left(3-{n i q\over \sqrt{1-q^2}}\right)
\right], \cr
C_n&=&{1\over 2}\left[\psi^{(1)}\left(3+{n i q\over \sqrt{1-q^2}}\right)
+\psi^{(1)}\left(3-{n i q\over \sqrt{1-q^2}}\right)
\right], 
\end{eqnarray}
and $\psi^{(n)}(z)$ is the polygamma function. 
We see that the absorption effect starts at $O(v^5)$ beyond the
quadrapole formula in the case $q\neq 0$, while for $q=0$, the above
formula reduced to
\begin{equation}
\left({dE\over dt}\right)_{\rm H}=
\left(dE\over dt\right)_N \left(v^8+O(v^{10})\right), 
\end{equation}
as was found by Poisson and Sasaki\cite{ref:PoSa}. 
We note that the leading terms in $\langle dE/dt\rangle_{\rm H}$
are negative for $q>0$, i.e., the black hole loses the energy if the
particle is corotating. This is because of the superradiance for modes
with $k<0$. In Appendix \ref{ap:absorpinx}, we also show 
$\langle dE/dt\rangle_{\rm H}$ written in terms of
$x\equiv(M\Omega_\varphi)^{1/3}$. 

It is not manifest from Eq.~(\ref{eq:dEdthole}) that if
it has a finite limit for $|q|\to 1$. 
But by using the formulas, 
\begin{equation}
\lim_{q\to \pm 1}
\psi^{(0)}\left(3+{n i q\over \sqrt{1-q^2}}\right)
=\ln{n}-\ln\kappa+i{q\over |q|}{\pi\over 2}, 
\end{equation}
\begin{equation}
\lim_{q\to \pm 1}
\psi^{(k)}\left(3+{n i q\over \sqrt{1-q^2}}\right)=0, 
~~~(k\neq 0), 
\end{equation}
we obtain the limit of $\langle dE/dt\rangle_{\rm H}$ as 
\begin{eqnarray}
\lim_{q\to \pm 1}\left(dE\over dt\right)_{\rm H}
=&&\left(dE\over dt\right)_{\rm N} v^5 \Biggl[
-{q\over |q|}
-{\frac {49}{16}}\,{q\over |q|}\,{v}^{2}
+\left (4\,\pi +{\frac {133}{12}}\right ){v}^{3}
\nonumber \\
&&
-{\frac {6817}{336}}\,{q\over |q|}\,{v}^{4}
+\left ({\frac {535}{16}}+{\frac {97}{8}}\,\pi \right ){v}^{5}
\nonumber \\
&&
+\left ({\frac {3424}{105}}\,{q\over |q|}\,\ln (2)
+{\frac {1712}{105}}\,\gamma\,{q\over |q|}
+{\frac {3424}{105}}\,{q\over |q|}\,\ln (v)
\right.
\nonumber \\
&&
\left.
-{\frac {3647533}{22050}}\,{q\over |q|}
-{\frac {289}{6}}\,{q\over |q|}\,\pi 
-16/3\,{\pi }^{2}{q\over |q|}\right ){v}^{6}
\nonumber \\
&&
+\left ({\frac {84955}{336}}+{\frac {55873}{672}}\,\pi \right ){v}^{7}
\nonumber \\
&&
\left ({\frac {14077}{60}}\,{q\over |q|}\,\ln (2)+{\frac {16441}{140}}\,
\gamma\,{q\over |q|}+{\frac {34987}{210}}\,{q\over |q|}\,\ln (v)
-{\frac {193}{12}}\,{\pi }^{2}{q\over |q|}
\right.
\nonumber \\
&& 
\left.
-{\frac {4057965601}{6350400}}\,{q\over |q|}
-{\frac {1289}{9}}\,{q\over |q|}\,\pi \right ){v}^{8}\Biggr]. 
\end{eqnarray}

\appendix
\section{Spheroidal harmonics}
\label{ap:spheroid}

In this Appendix, we describe the expansion of the spheroidal 
harmonics $_{-2}S^{a\omega}_{\ell m}$ to $O((a\omega)^2)$. 

The spheroidal harmonics of spin weight $s=-2$ obey the equation,
\begin{eqnarray}
& &\Bigl[{1 \over \sin\theta}{d \over d\theta}
 \Bigl\{\sin\theta {d \over d\theta} \Bigr\}
-a^2\omega^2\sin^2\theta
 -{(m-2\cos\theta)^2 \over \sin^2\theta}
\nonumber\\
& &~~~~~~~~~~~~~~~
+4a\omega\cos\theta-2+2ma\omega+\lambda\Bigr] 
{}_{-2}S_{\ell m}^{a\omega}=0. \label{eq:spheroid2}
\end{eqnarray}
We expand $_{-2}S_{\ell m}^{a\omega}$ and $\lambda$ as
\begin{eqnarray} 
{}_{-2}S_{\ell m}^{a\omega}
&=&{}_{-2}P_{\ell m}+a\omega S_{\ell m}^{(1)}
+(a\omega)^2 S_{\ell m}^{(2)}
+O((a\omega)^3), \nonumber\\
\lambda&=&\lambda_0+a\omega\lambda_1
+a^2\omega^2 \lambda_2+O((a\omega)^3), \label{eq:slmexp2}
\end{eqnarray} 
where ${}_{-2}P_{\ell m}$ are the spherical harmonics of spin weight
$s=-2$. We set the normalizations of ${}_{-2}P_{\ell m}$ and
${}_{-2}S_{\ell m}^{a\omega}$ as
\begin{equation}
\int_0^{\pi} |_{-2}P_{\ell m}|^2 \sin\theta d\theta
=\int_0^{\pi} |_{-2}S_{\ell m}^{a\omega}|^2 \sin\theta d\theta=1.
\end{equation}

Inserting Eq.~(\ref{eq:slmexp2}) into Eq.~(\ref{eq:spheroid2}) and
collecting the terms of the same order to $(a\omega)^2$, we obtain
\begin{eqnarray}
&&\left[{\cal L}_0+\lambda_0\right]{}_{-2}P_{\ell m}=0\,,
\label{eq:slm0}\\
&&\left[{\cal L}_0+\lambda_0\right] S_{\ell m}^{(1)}
=-(4\cos\theta+2m+\lambda_1)\,_{-2}P_{\ell m} 
\label{eq:slm1}\\
&&\left[{\cal L}_0+\lambda_0\right] S^{(2)}_{\ell m}
=-(4\cos\theta+2m+\lambda_1)S_{\ell m}^{(1)}
-(\lambda_2-\sin^2\theta)\,_{-2}P_{\ell m}\,,
\label{eq:slm2}
\end{eqnarray}
where
\begin{equation}
{\cal L}_0\equiv{1 \over \sin\theta}{d \over d\theta}
\left(\sin\theta {d \over d\theta} \right)
-{(m-2\cos\theta)^2 \over \sin^2\theta}-2\,.
\end{equation}
The lowest order equation (\ref{eq:slm0}) says we have 
$\lambda_0=(\ell-1)(\ell+2)$. 

The first order correction to the
eigenvalue, $\lambda_1$, is obtained by multiplying Eq.~(\ref{eq:slm1}) by
$_{-2}P_{\ell m}$ from the left hand side and integrating it over
$\theta$. The result is 
\begin{equation}
  \label{eq:lambda1}
  \lambda_1=-2m{\ell(\ell+1)+4\over\ell(\ell+1)}\,.
\end{equation}
To obtain $S_{\ell m}^{(1)}$, we set 
\begin{equation}
S_{\ell m}^{(1)}=\sum_{\ell'} c_{\ell m}^{\ell'}~{}_{-2}P_{\ell'm}\,.
\label{eq:slm11}
\end{equation}
We insert this into Eq.~(\ref{eq:slm1}), multiply it by
${}_{-2}P_{\ell'm}$ and integrate it over $\theta$.
Then noting the normalization of the spheroidal harmonics, we have
$$
c_{\ell m}^{\ell'}=
\cases{\displaystyle
{4 \over (\ell'-1)(\ell'+2)-(\ell-1)(\ell+2)}\int d(\cos\theta)
\,{}_{-2}P_{\ell'm}\cos\theta\, {}_{-2}P_{\ell m}\,,
\quad&$\ell' \not=\ell$,
\cr
0,&$\ell'=\ell$.
\cr}
$$
Hence $c_{\ell m}^{\ell'}$ is non-zero only for $\ell'=\ell \pm 1$, and 
we obtain
\begin{eqnarray*}
c_{\ell m}^{\ell+1}&=&{2 \over (\ell+1)^2}\Bigl\lbrack 
{(\ell+3)(\ell-1)(\ell+m+1)(\ell-m+1) \over (2\ell+1)(2\ell+3)}
 \Bigr\rbrack^{1/2},\cr
c_{\ell m}^{\ell-1}&=&-{2 \over \ell^2}\Bigl\lbrack 
{(\ell+2)(\ell-2)(\ell+m)(\ell-m) \over (2\ell+1)(2\ell-1)} 
\Bigr\rbrack^{1/2}.
\end{eqnarray*} 

The next order equation can be solved similarly.
The second order correction to the eigenvalue, $\lambda_2$,
is obtained by multiplying Eq.~(\ref{eq:slm2}) by ${}_{-2}P_{\ell m}$
from the left hand side and integrating it over $\theta$. 
We find
\begin{eqnarray}
\lambda_2
&=&-4\int d(\cos\theta){}_{-2}P_{\ell m}\cos\theta\, S_{\ell m}^{(1)}
+\int d(\cos\theta){}_{-2}P_{\ell m}\sin^2\theta\,{}_{-2}P_{\ell m}\cr
&=&-2(\ell+1)(c_{\ell m}^{\ell+1})^2+2\ell(c_{\ell m}^{\ell-1})^2
+1-\int d(\cos\theta){}_{-2}P_{\ell m}\cos^2\theta\,{}_{-2}P_{\ell m}\,,
\end{eqnarray}
where the last integral becomes 
$$
\int d(\cos\theta){}_{-2}P_{\ell m}\cos^2\theta\,{}_{-2} P_{\ell m}=
{1 \over 3}+{2 \over 3}{(\ell+4)(\ell-3)(\ell^2+\ell-3m^2) \over 
\ell(\ell+1)(2\ell+3)(2\ell-1)}\,.
$$
As before, to obtain $S_{\ell m}^{(2)}$, we set
\begin{eqnarray}
S_{\ell m}^{(2)}&=&\sum_{\ell'} d_{\ell m}^{\ell'}~{}_{-2}P_{\ell'm}\,. 
\label{eq:slm22}
\end{eqnarray}
Inserting Eqs.~(\ref{eq:slm11}) and (\ref{eq:slm22}) into
Eq.~(\ref{eq:slm2}), multiplying it by ${}_{-2}P_{\ell' m}$ and
integrate it over $\theta$, we obtain
\begin{eqnarray}
d_{\ell m}^{\ell'}&=&{1\over {\lambda_0(\ell)-\lambda_0(\ell')}}
\left[-\left(2m+\lambda_1(\ell)\right)
\left(c^{\ell+1}_{\ell m}\delta_{\ell', \ell+1}
+c^{\ell-1}_{\ell m}\delta_{\ell', \ell-1}\right)
\right.\nonumber\\
& &
-4 c^{\ell+1}_{\ell m}\int d(\cos\theta) 
{}_{-2}P_{\ell' m}\cos\theta\,{}_{-2}P_{\ell+1 m}
-4 c^{\ell-1}_{\ell m}\int d(\cos\theta) 
{}_{-2}P_{\ell' m}\cos\theta\,{}_{-2}P_{\ell-1 m}
\nonumber\\
& &
\left.+\int d(\cos\theta) 
{}_{-2}P_{\ell' m}\sin^2\theta\,{}_{-2}P_{\ell m}\right],
\end{eqnarray}
for $\ell'\neq\ell$. 
The integrals in this equation are given by \cite{ref:press,ref:cam} 
\begin{eqnarray*}
\int d(\cos\theta){}_{-2}P_{\ell' m}\cos\theta\,{}_{-2}P_{\ell m}
&=&\sqrt{{2\ell+1}\over {2\ell'+1}}
<\ell,1,m,0|\ell',m><\ell,1,2,0|\ell',2>, \\
\int d(\cos\theta){}_{-2}P_{\ell' m}\sin^2\theta\,{}_{-2}P_{\ell m}
&=&{2\over 3}\delta_{\ell',\ell} \\
& &-{2\over 3}\sqrt{{2\ell+1}\over {2\ell'+1}}
<\ell,2,m,0|\ell',m><\ell,2,2,0|\ell',2>, \\
\end{eqnarray*} 
where $<j_1,j_2,m_1,m_2|J,M>$ is a Clebsch-Gordan coefficient. 
For $\ell=2$ and $3$, the non-vanishing $d^{\ell'}_{\ell m}$ 
$(\ell'\neq\ell)$ are given explicitly as 
\begin{eqnarray*}
d_{\ell m}^{\ell+1}&=&{m\over 324\sqrt{7}}(3-m)^{1/2}(3+m)^{1/2},\\
d_{\ell m}^{\ell+2}&=&{11\over 1764\sqrt{3}}(3-m)^{1/2}(3+m)^{1/2}
(4-m)^{1/2}(4+m)^{1/2},\\
\end{eqnarray*}
for $\ell=2$, and 
\begin{eqnarray*}
d_{\ell m}^{\ell+1}&=&{m\over 120\sqrt{21}}(4-m)^{1/2}(4+m)^{1/2},\\
d_{\ell m}^{\ell+2}&=&{1\over 180\sqrt{11}}(4-m)^{1/2}(4+m)^{1/2}
(5-m)^{1/2}(5+m)^{1/2},\\
d_{\ell m}^{\ell-1}&=&-{m\over 324\sqrt{7}}(3-m)^{1/2}(3+m)^{1/2},\\
\end{eqnarray*}
for $\ell=3$.
As for $d^{\ell}_{\ell m}$, it is determined by the normalization of
${}_{-2}S_{\ell m}^{a\omega}$, i.e.,
\begin{eqnarray*} 
1&=&\int d(\cos\theta) |{}_{-2}S_{\ell m}|^2\\ 
 &=&\int d(\cos\theta)
 \left\{({}_{-2}P_{\ell m})^2+2a\omega\sum_{\ell'}
c^{\ell'}_{\ell m}{}_{-2}P_{\ell' m}{}_{-2}P_{\ell m}
\right.\\
&+&
(a\omega)^2\sum_{\ell'\ell''}c^{\ell'}_{\ell m}c^{\ell''}_{\ell m}
{}_{-2}P_{\ell' m}{}_{-2}P_{\ell'' m} \\
&+&\left.  2(a\omega)^2\sum_{\ell'}
d^{\ell'}_{\ell m}{}_{-2}P_{\ell' m}{}_{-2}P_{\ell m}
+O\left((a\omega)^3\right)\right\} \\
&=&
1+(a\omega)^2\sum_{\ell'} \left(c^{\ell'}_{\ell m}\right)^2
+2(a\omega)^2d^\ell_{\ell m}+O((a\omega)^3).
\end{eqnarray*}
Then we have 
\begin{equation}
d^\ell_{\ell m}=-{1\over 2}\left\{
\left(c^{\ell+1}_{\ell m}\right)^2+\left(c^{\ell-1}_{\ell m}\right)^2
\right\}.
\end{equation}

\input ptapb.texin 
\section{4PN formulas for $R^{\rm in}_{\ell m}$}
\label{ap:Rin}

In this Appendix, we show the post-Newtonian expansion of $R^{\rm in}$
in the near zone, where $z=\omega r\ll1$, for a Kerr black hole which
are needed to evaluate gravitational waves at infinity to $O(v^8)$.
For convenience, we recover the indices $\ell m$ on 
$R^{\rm in}$ and give the formulas for $\omega R^{\rm in}_{\ell m}$.

\begin{eqnarray}
\omega R^{{\rm in}}_{2m}&=&
{{{z^4}}\over {30}} + {i\over {45}}\,{z^5} - {{11\,{z^6}}\over {1260}} - 
  {i\over {420}}\,{z^7} + {{23\,{z^8}}\over {45360}} +
  {i\over {11340}}\,{z^9} \nonumber\\
& &
- {{13\,{z^{10}}}\over {997920}} - {i\over {598752}}\,{z^{11}} + 
  {{59\,{z^{12}}}\over {311351040}} \nonumber\\
&+&\epsilon\left(
{{-{z^3}}\over {15}} - {i\over {60}}\,m\,q\,{z^3} 
- {i\over {60}}\,{z^4} + 
  {{m\,q\,{z^4}}\over {45}} - {{41\,{z^5}}\over {3780}} + 
  {{277\,i}\over {22680}}\,m\,q\,{z^5} 
\right.\nonumber\\
& &
- {{31\,i}\over {3780}}\,{z^6} - 
  {{7\,m\,q\,{z^6}}\over {1620}} 
+ {{17\,{z^7}}\over {5670}} - 
  {{61\,i}\over {54432}}\,m\,q\,{z^7} \nonumber\\
& &
\left.
+ {{41\,i}\over {54432}}\,{z^8} + 
  {{47\,m\,q\,{z^8}}\over {204120}} - {{1579\,{z^9}}\over {10692000}} + 
  {{703\,i}\over {17962560}}\,m\,q\,{z^9}
\right)\nonumber\\
&+&\epsilon^2\left(
{{{z^2}}\over {30}} + {i\over {40}}\,m\,q\,{z^2} + 
{{{q^2}\,{z^2}}\over {60}} - 
  {{{m^2}\,{q^2}\,{z^2}}\over {240}} - {i\over {60}}\,{z^3} - 
  {{m\,q\,{z^3}}\over {30}} + {i\over {90}}\,{q^2}\,{z^3} 
\right.\nonumber\\
& & 
- {i\over {120}}\,{m^2}\,{q^2}\,{z^3} 
+ {{7937\,{z^4}}\over {55125}} - 
  {{53\,i}\over {9072}}\,m\,q\,{z^4} 
- {{101\,{q^2}\,{z^4}}\over {35280}} + 
  {{4213\,{m^2}\,{q^2}\,{z^4}}\over {635040}} \nonumber\\
& &
+ {{4673\,i}\over {55125}}\,{z^5} - {{13\,m\,q\,{z^5}}\over {2835}} - 
  {{5\,i}\over {63504}}\,{q^2}\,{z^5} + 
  {{3503\,i}\over {1143072}}\,{m^2}\,{q^2}\,{z^5} 
- {{1665983\,{z^6}}\over {55566000}} \nonumber\\
& &
- {{1777\,i}\over {544320}}\,m\,q\,{z^6}
- {{{q^2}\,{z^6}}\over {5040}} - 
  {{643\,{m^2}\,{q^2}\,{z^6}}\over {653184}}
- {{107\,{z^4}\,\ln z}\over {6300}} \nonumber\\
& &\left.
- {{107\,i}\over {9450}}\,{z^5}\,\ln z + 
  {{1177\,{z^6}\,\ln z}\over {264600}}
\right)\nonumber\\
&+&\epsilon^3\left(
\left( {{-i}\over {180}}\,m\,q - {{{q^2}}\over {60}} + 
     {{{m^2}\,{q^2}}\over {240}} - {i\over {144}}\,m\,{q^3} + 
     {i\over {1440}}\,{m^3}\,{q^3} \right) \,z 
\right.\nonumber\\
& &
+ \left( {i\over {120}} + {{2\,m\,q}\over {135}} 
- {i\over {360}}\,{q^2} + 
     {{19\,i}\over {1440}}\,{m^2}\,{q^2} 
+ {{11\,m\,{q^3}}\over {1080}} - 
     {{{m^3}\,{q^3}}\over {540}} \right) \,{z^2} \nonumber\\
& &
+ {z^3}\,\left( -{{10933}\over {49000}} - 
     {{578569\,i}\over {7938000}}\,m\,q - 
     {{677\,{q^2}}\over {52920}} - 
     {{529\,{m^2}\,{q^2}}\over {63504}} 
\right.\nonumber\\
& &\left.\left.
   + {{317\,i}\over {63504}}\,m\,{q^3} - 
     {{167\,i}\over {84672}}\,{m^3}\,{q^3} + 
     {{107\,\ln z}\over {3150}} + 
     {{107\,i}\over {12600}}\,m\,q\,\ln z
      \right) 
\right)\nonumber\\
&+&\epsilon^4\left(
{{-i}\over {720}}\,m\,q + {{{m^2}\,{q^2}}\over {2880}} + 
  {i\over {288}}\,m\,{q^3} - {i\over {2880}}\,{m^3}\,{q^3} 
\right.
\nonumber\\
&&\left.+{{{q^4}}\over {480}} - {{{m^2}\,{q^4}}\over {720}} + 
  {{{m^4}\,{q^4}}\over {11520}}
\right).
\\
\omega R^{{\rm in}}_{3m}&=&
  {{{z^5}}\over {630}} + {i\over {1260}}\,{z^6} -
  {{{z^7}}\over {3780}} - 
  {i\over {16200}}\,{z^8} + {{29\,{z^9}}\over {2494800}} \nonumber\\
& &
+ {i\over {554400}}\,{z^{10}} - {{47\,{z^{11}}}\over {194594400}} \nonumber\\
&+&\epsilon\left(
  {{-{z^4}}\over {252}} - {i\over {1890}}\,m\,q\,{z^4} -
  {i\over {756}}\,{z^5} + 
  {{11\,m\,q\,{z^5}}\over {22680}} + 
  {{19\,i}\over {90720}}\,m\,q\,{z^6} 
\right.\nonumber\\
& &\left.
- {i\over {9450}}\,{z^7} - 
  {{m\,q\,{z^7}}\over {16200}} + 
  {{647\,{z^8}}\over {14968800}} - 
  {{247\,i}\over {17962560}}\,m\,q\,{z^8} 
\right)\nonumber\\
&+&\epsilon^2\left(
{{{z^3}}\over {315}} + {i\over {945}}\,m\,q\,{z^3} + 
  {{{q^2}\,{z^3}}\over {1260}} - {{{m^2}\,{q^2}\,{z^3}}\over {15120}} + 
  {i\over {2520}}\,{z^4} - {{17\,m\,q\,{z^4}}\over {15120}} 
\right.\nonumber\\
& &
+ {i\over {2160}}\,{q^2}\,{z^4} - 
  {{31\,i}\over {272160}}\,{m^2}\,{q^2}\,{z^4} + 
  {{81409\,{z^5}}\over {11113200}} \nonumber\\
& &\left.
- {{313\,i}\over {907200}}\,m\,q\,{z^5} - 
  {{41\,{q^2}\,{z^5}}\over {226800}} + 
  {{617\,{m^2}\,{q^2}\,{z^5}}\over {8164800}} - 
  {{13\,{z^5}\,\ln z}\over {26460}}
\right)\nonumber\\
&+&\epsilon^3\left(
  {{-{z^2}}\over {1260}} - {i\over {1680}}\,m\,q\,{z^2} - 
  {{{q^2}\,{z^2}}\over {840}} + {{{m^2}\,{q^2}\,{z^2}}\over {10080}} - 
  {i\over {5040}}\,m\,{q^3}\,{z^2}
\right).
\\
\omega R^{{\rm in}}_{4m}&=&
{{{z^6}}\over {11340}} + {i\over {28350}}\,{z^7} -
  {{13\,{z^8}}\over {1247400}} - {i\over {467775}}\,{z^9} +
  {{71\,{z^{10}}}\over {194594400}}\nonumber\\
&+&\epsilon\left(
{{-{z^5}}\over {3780}} - {i\over {45360}}\,m\,q\,{z^5} -
  {{11\,i}\over {136080}}\,{z^6} + {{m\,q\,{z^6}}\over {64800}} 
\right. \nonumber\\
& &\left.
+ {{131\,{z^7}}\over {18711000}} + {{697\,i}\over
{124740000}}\,m\,q\,{z^7}
\right)\nonumber\\
&+&\epsilon^2\left(
{{{z^4}}\over {3528}} + {i\over {18144}}\,m\,q\,{z^4} +
{{{q^2}\,{z^4}}\over {21168}} 
- {{{m^2}\,{q^2}\,{z^4}}\over{635040}}
\right).
\\
\omega R^{{\rm in}}_{5m}&=&
{{{z^7}}\over {207900}} + {i\over {623700}}\,{z^8} - 
  {{{z^9}}\over {2316600}}\nonumber\\
&+&\epsilon\left(
{{-{z^6}}\over {59400}} - {i\over {1039500}}\,m\,q\,{z^6}
\right).
\\
\omega R^{{\rm in}}_{6m}&=&
{{{z^8}}\over {4054050}}.
\end{eqnarray}

\section{The ingoing Regge-Wheeler functions to $O(\epsilon^3)$}
\label{ap:RWthird}

In this Appendix, we present a method to calculate the ingoing
Regge-Wheeler functions to $O(\epsilon^3)$ which are needed to calculate
the luminosity to $O(v^{11})$ beyond Newtonian in the Schwarzschild
case. In the Schwarzschild limit, $q\to0$, and solving
Eq.~(\ref{eq:PNexpand}) recursively is in principle straightforward.
Since the general homogeneous solution to the left-hand side of it
is given by a linear combination of the spherical Bessel functions
$j_\ell$ and $n_\ell$, one can immediately write the integral
expression for $\xi^{(n)}_\ell$. Noting that
 $j_\ell n_\ell'-n_\ell j_\ell'=1/z^2$, we have
\begin{equation}
 \xi_{\ell}^{(n)}
=n_\ell \int dz\,j_\ell W^{(n)}-j_\ell\int dz\,n_\ell W^{(n)},
\label{eq:SPNexpand}
\end{equation}
where the source term $W^{(n)}$ in the Schwarzschild case is given by
\begin{equation}
W^{(n)}=z^2 L^{(1)}[\xi^{(n-1)}]=e^{-iz}{d\over dz}\left[{1\over z^3}
   {d\over dz}\left(e^{iz}z^2\xi_\ell^{(n-1)}(z)\right)\right].
\end{equation}
We perform the above indefinite integral and set the appropriate
boundary condition by examining the asymptotic behavior at $z\to0$
order by order.

\subsection{General remarks}
\label{tamaremarks}
Let us first consider the boundary conditions of $\xi_\ell^{(n)}$.
In the Schwarzschild case, the original Regge-Wheeler ingoing wave
function $X^{\rm in}_{\ell}$ is related to $\xi^{\rm in}_{\ell}$ as
\begin{equation}
  \label{eq:Xinform}
 X^{\rm in}_{\ell}=ze^{-i\epsilon\ln(z-\epsilon)}
\xi^{\rm in}_{\ell},
\end{equation}
and it has the asymptotic behavior given by Eq.~(\ref{eq:xinasy}),
which in the present case reduces to
\begin{equation}
X^{\rm in}_{\ell}\rightarrow\cases{
A^{\rm ref}_{\ell} e^{i\omega r^*}+
A^{\rm inc}_{\ell} e^{-i\omega r^*}\,& 
for $z^* \rightarrow \infty$ \cr
A^{\rm trans}_{\ell} e^{-i\omega r^*}\, & for
$z^* \rightarrow -\infty$. }
\label{eq:RWxinasy}
\end{equation}
Thus, noting that $z^*=z+\epsilon\ln(z-\epsilon)$, the boundary
condition of $\xi^{\rm in}_{\ell}$ is that it is   
regular at $z^*\to-\infty$ ($z\to\epsilon$).
To implement this boundary condition to $\xi^{\rm in}_{\ell}$,
we need a different series expansion of it; a series in terms
of the variable $x:=r/2M=z/\epsilon$ around $x=1$.  
We write this expansion as 
\begin{equation}
 \xi^{\rm in}_{\ell}
=\xi_{\ell}^{\{0\}}(x)+\epsilon \xi_{\ell}^{\{1\}}(x)
          +\epsilon^2 \xi_{\ell}^{\{2\}}(x)+\cdots. 
\end{equation}
On the other hand, there are two independent solutions 
expanded by functions written in terms of $z$. 
We denote the solution whose zeroth order is given by 
$j_{\ell}$ ($n_{\ell}$) as $\xi_{j\ell}$ ($\xi_{n\ell}$). 
Then the general solution is given by 
$c_j \xi_{j\ell}+c_n \xi_{n\ell}$. 
As can be shown by using the result of Poisson and 
Sasaki\cite{ref:PoSa}, $\xi_{\ell}^{\{0\}}$ does not have terms matched
with $\xi_{n\ell}$.  A term matched with $\xi_{n\ell}$ first appears
from $\xi_{\ell}^{\{1\}}$. For $x\to\infty$ 
($\leftrightarrow \epsilon\ll z$), $\xi_{\ell}^{\{0\}}(x)$ behaves as
$x^{\ell}=z^{\ell} \epsilon^{-\ell}\sim\epsilon^{-\ell} \xi_{j\ell}$ 
On the other hand, $\epsilon \xi_{\ell}^{\{1\}}$ 
contains the term that behaves as 
$\epsilon x^{-\ell-1}=\epsilon^{\ell+2} z^{-\ell-1}
 \sim \epsilon^{\ell+2} \xi_{n\ell}$. 
Therefore, in the sense of the post-Minkowskian expansion, the inner
boundary condition affects the ingoing wave solution at and beyond
$O(\epsilon^{2\ell+2})$. However, in the post-Newtonian sense, 
since we evaluate $\xi_{\ell}$ in the near zone, i.e., for
$z=O(v)$, the contribution from $\xi_{n\ell}$ becomes $O(v^{4\ell+5})$
relative to $\xi_{j\ell}$. 
Since $\ell\geq2$, we find that the inner boundary condition 
affects the homogeneous solution at and beyond $O(v^{13})$ in the 
near zone.\footnote{In Ref.~\citen{ref:sasaki}, it was erroneously
argued that the outgoing gravitational waves are unaffected by the inner 
boundary condition until we reach $O(\epsilon^6)=O(v^{18})$. 
As shown here, this is true only in the post-Minkowskian sense.}

Since $j_\ell=O(z^\ell)$ as $z\rightarrow0$, we have
$X^{in}_\ell\rightarrow O(\epsilon^{\ell+1})e^{-iz^*}$,
or $A^{\rm trans}_\ell=O(\epsilon^{\ell+1})$.
On the other hand, from the asymptotic behavior of $j_\ell$ at
$z=\infty$, we find $A^{\rm inc}_\ell$ and $A^{\rm ref}_\ell$ are
of order unity. Then using the Wronskian argument,
we obtain
\begin{equation}
 |A^{\rm inc}_\ell|-|A^{\rm ref}_\ell|
={|A^{\rm trans}_\ell|^2\over|A^{\rm inc}_\ell|+|A^{\rm ref}_\ell|}
    =O(\epsilon^{2\ell+2}).
\end{equation}
Thus $|A^{\rm inc}_\ell|=|A^{\rm ref}_\ell|$ until we go to
$O(\epsilon^{2\ell+2})$ or more.
This fact implies that we can make 
$A^{\rm inc}_\ell$ and $A^{\rm ref}_\ell$ to be
complex conjugate to each other to $O(\epsilon^{2\ell+1})$. 
Hence the imaginary part of $X^{\rm in}_\ell$, which reflects the 
boundary condition at horizon, appears at 
$O(\epsilon^{2\ell+2})$ because the Regge-Wheeler equation is real. 
This is consistent with the argument given in the above paragraph. 
Provided we choose the phase of $X^{\rm in}_\ell$ in this way,
${\rm Im}\left(\xi^{(n)}_\ell\right)$ for a given $n\le 2\ell+1$
is completely determined in terms of 
${\rm Re}\left(\xi^{(r)}_\ell\right)$ for $r\le n-1$.

To see this explicitly, let us decompose $\xi^{(n)}_{\ell}$ into
the real and imaginary parts:
\begin{equation}
 \xi^{(n)}_\ell=f^{(n)}_\ell+ig^{(n)}_\ell\,.
\label{eq:tamafgdef}
\end{equation}
Inserting this expression into Eq.~(\ref{eq:Xinform}) 
and expanding the result
with respect to $\epsilon$ by assuming $z\gg\epsilon$, we find
\begin{eqnarray}
 X^{\rm in}_\ell
     &=&e^{-i\epsilon\ln(z-\epsilon)}z
         \left(j_\ell+\epsilon(f^{(1)}_\ell+ig^{(1)}_\ell)
             +\epsilon^2 (f^{(2)}_\ell+ig^{(2)}_\ell)
             +\epsilon^3 (f^{(3)}_\ell+ig^{(3)}_\ell)
      +\cdots\right)
\cr
    &=&z\Biggl(j_\ell+\epsilon f^{(1)}_\ell
        +\epsilon^2\Bigl(f^{(2)}_\ell+g^{(1)}_\ell\ln z
        -{1\over2}j_\ell(\ln z)^2\Bigr) 
\cr 
    && \quad   +\epsilon^3 \Bigl(f^{(3)}
           -{1\over 2}(\ln z)^2 f_{\ell}^{(1)}
           +{\ln z\over z} j_{\ell} +{\ln z} g_{\ell}^{(2)}
           -{1\over z}g_{\ell}^{(1)}\Bigr)+\cdots\Biggr)
\cr
     &&+iz\Biggl(\epsilon(g^{(1)}_\ell-j_\ell\ln z)
           +\epsilon^2\Bigl(g^{(2)}_\ell+{1\over z}j_\ell
                         -f^{(1)}_\ell\ln z\Bigr)
\cr
    &&\quad +\epsilon^3\Bigl(g^{(3)}_\ell-{1\over2}(\ln z)^2g^{(1)}_\ell 
     -\ln z\,f^{(2)}_\ell+{1\over z}f^{(1)}
\cr
&&\qquad\qquad
  +\left({1\over2z^2}+{1\over6}(\ln z)^3\right)j_\ell\Bigr)
+\cdots\Biggr).
\label{eq:tamaXinexpand}
\end{eqnarray}
Hence we must have
\begin{eqnarray}
&&g^{(1)}_\ell=j_\ell\ln z\,,\quad
g^{(2)}_\ell=-{1\over z}j_\ell+f^{(1)}_\ell\ln z\,,
\nonumber\\
&&g^{(3)}_\ell=\left({1\over3}(\ln z)^3-{1\over2z^2}\right)j_\ell
-{1\over z}f^{(1)}_\ell+\ln z\,f^{(2)}_\ell\,,\quad\cdots.
\label{eq:tamaImxi}
\end{eqnarray}
For completeness, we also give the relation between the
functions $f^{(n)}_\ell$ and
the conventional post-Newtonian expansion of $X^{\rm in}_\ell$:
\begin{eqnarray}
   &&X^{\rm in}_\ell=\sum_{n=0}^{\infty}\epsilon^n X^{(n)}_\ell\,;
\cr
   &&X^{(0)}_\ell=zf^{(0)}_\ell=zj_\ell\,,\quad
    X^{(1)}_\ell=zf^{(1)}_\ell\,,\quad
    X^{(2)}_\ell=z\left(f^{(2)}_\ell+{1\over2}j_\ell(\ln z)^2\right),
\cr
   &&X^{(3)}_\ell=z\left(f^{(3)}_\ell+{1\over2}f^{(1)}_\ell(\ln z)^2
                        -{\ln z\over z} j_{\ell}\right), 
   \quad\cdots.
\label{eq:tamaXxirel}
\end{eqnarray}

Now we turn to the asymptotic behavior at $z=\infty$.
Let the asymptotic form of $f^{(n)}_\ell$ be
\begin{equation}
 f^{(n)}_\ell\rightarrow P^{(n)}_\ell j_\ell+Q^{(n)}_\ell n_\ell
 \quad{\rm as}~z\rightarrow\infty\quad(n=1,2,3).
\label{eq:tamaPQdef}
\end{equation}
Then noting Eq.~(\ref{eq:tamaImxi}) and the equality
$e^{-i\epsilon\ln(z-\epsilon)}=e^{-iz^*}e^{iz}$,
the asymptotic form of $X^{\rm in}_\ell$ is expressed as
\begin{eqnarray}
 X^{\rm in}_\ell
   &\rightarrow&
   {1\over2}e^{-iz^*}\left(zh^{(2)}_\ell e^{iz}\right)
      \Biggl[1+\epsilon\left\{P^{(1)}_\ell
                       +i\left(Q^{(1)}_\ell+\ln z\right)\right\}
\cr
   &&\quad
    +\epsilon^2\left\{\left(P^{(2)}_\ell-Q^{(1)}_\ell\ln z\right)
               +i\left(Q^{(2)}_\ell+P^{(1)}_\ell\ln z\right)\right\}
\cr
   &&\quad
     +\epsilon^3\left\{\left(f_{\ell}^{(3)}-Q_{\ell}^{(2)}\ln z\right)
   +i\left(Q^{(3)}_\ell+P^{(2)}_\ell\ln z+{1\over3}(\ln z)^3\right)
     \right\} +\cdots\Biggr]
\cr
  &&+{1\over2}e^{iz^*}\left(zh^{(1)}_\ell e^{-iz}\right)
      e^{-2i\epsilon\ln(z-\epsilon)}
      \Biggl[1+\epsilon\left\{P^{(1)}_\ell
                       -i\left(Q^{(1)}_\ell-\ln z\right)\right\}
\cr
   &&\quad
    +\epsilon^2\left\{\left(P^{(2)}_\ell+Q^{(1)}_\ell\ln z\right)
                -i\left(Q^{(2)}_\ell-P^{(1)}_\ell\ln z\right)\right\}
\cr
   &&\quad
  +\epsilon^3\left\{\left(P_{\ell}^{(3)} +Q_{\ell}^{(2)}\ln z\right)
   -i\left(Q^{(3)}_\ell-P^{(2)}_\ell\ln z-{1\over3}(\ln z)^3\right)
 \right\} +\cdots\Biggr]. 
\label{eq:tamaAsymptXin}
\end{eqnarray}
Using the asymptotic behavior of $h^{(1)}_\ell$ and
$h^{(2)}_\ell$ given in Eq.~(\ref{eq:hankel}), 
the incident amplitude $A^{\rm inc}_\ell$ can be
readily extracted out as 
\begin{eqnarray}
   A^{\rm inc}_\ell&=&{1\over2}i^{\ell+1}e^{-i\epsilon\ln\epsilon}
      \Bigl[1+\epsilon\left\{P^{(1)}_\ell
                       +i\left(Q^{(1)}_\ell+\ln z\right)\right\}
\cr
   &&+\epsilon^2\left\{\left(P^{(2)}_\ell-Q^{(1)}_\ell\ln z\right)
            +i\left(Q^{(2)}_\ell+P^{(1)}_\ell\ln z\right)\right\}
\cr
   &&+\epsilon^3\left\{\left(P^{(3)}_\ell-Q^{(2)}_\ell\ln z\right)
   +i\left(Q^{(3)}_\ell+P^{(2)}_\ell\ln z+{1\over3}(\ln z)^3\right)
\right\} +\cdots\Bigr],
\label{eq:tamaAinlogform}
\end{eqnarray}
where note that the definition of $r^*$, Eq.~(\ref{eq:rst}), 
in the limit $q\to0$ is
\begin{equation}
 \omega r^*=\omega\left(r+2M\ln{r-2M\over2M}\right)
           =z^*-\epsilon\ln\epsilon,
\label{eq:tamazrstar}
\end{equation}
which gives rise to the phase $-i\epsilon\ln\epsilon$ of 
$A^{\rm inc}_\ell$. 

An important point to be noted in the above expression for
$A^{\rm inc}_\ell$ is that it contains $\ln z$-dependent terms.
Since $A^{\rm inc}_\ell$ should be constant,
$P^{(n)}_\ell$ and $Q^{(n)}_\ell$ should contain appropriate
$\ln z$-dependent terms which exactly cancel the $\ln z$-dependent
terms in the formula (\ref{eq:tamaAinlogform}). 

\subsection{Basic formalism for iteration}

Here we derive the formulas necessary to perform the iteration 
scheme.

\subsubsection{Definitions}
\label{tamadefinitions}

We introduce the following functions, 
\begin{eqnarray}
 B_{jj}&&:=\int_0^z z j_0 j_0 dz =-{1\over 2}C,
\nonumber \\
 B_{nj}&&:=\int_0^z z n_0 j_0 dz =-{1\over 2}S,
\nonumber \\
 B_{jn}&&:=\int_0^z z j_0 n_0 dz =-{1\over 2}S,
\nonumber \\
 B_{nn}&&:=\int_{z_*}^z z n_0 n_0 dz =-B_{jj} +\ln z,
\label{eq:Bjjetc}
\end{eqnarray}
where 
\begin{eqnarray}
S & =&\int_0^{2z} dy{\sin y\over y}  = 
-\int_x^{\infty} dy{\sin y\over y} +{\pi\over 2},
\cr
C & =&\int_0^{2z} dy{\cos y -1\over y} = 
-\int_{2z}^{\infty} dy{\cos y\over y} -\gamma-\ln 2z\,,
\end{eqnarray}
and the lower bound $z_*$ of the integral for the definition of $B_{nn}$
is adjusted so as to make $B_{nn}$ equal to the last expression of the
line.

As an extension of these integral sinusoidal functions, 
we further introduce the following functions:
\begin{eqnarray}
 B_{jJ}&&:=
        \int_{z_*}^z dz z j_0 D_0^{J}, 
\label{eq:tamadefBjJ} \\
 B_{nJ}&&:=
        \int_{z_*}^z dz z n_0 D_0^{J}, 
\label{eq:tamadefBnJ}
\end{eqnarray}
where $J$ stands for a sequence of $j$ and $n$, say, $J=jnnj$,
and we have also introduced an extension of the spherical 
Bessel functions by 
\begin{equation}
D^j_{\ell}:=j_{\ell},\quad D^n_{\ell}:=n_{\ell}\,, 
\end{equation}
and
\begin{eqnarray}
 D^{nJ}_{\ell}:=n_{\ell}B_{jJ}-j_{\ell}B_{nJ}\,,
\nonumber \\
 D^{jJ}_{\ell}:=j_{\ell}B_{jJ}+n_{\ell}B_{nJ}\,.
\label{eq:tamadefD}
\end{eqnarray}
We adopt the following rule to determine the lower bound of the
integrals in Eq.~(\ref{eq:tamadefBnJ}). Whenever we can put $z_*=0$,
we do so, which is always possible when the sequence $J$ ends with
$j$. On the other hand, in the case when $J$ ends with $n$, there may
appear in the integrand the square of $n_0$ which causes logarithmic
divergence if we set $z_*=0$. 
In such cases, we use the relation,
\begin{equation}
  \label{eq:n0j0rel}
  n_0^2={1\over z^2}-j_0^2\,,
\end{equation}
to replace $n_0^2$ with the right hand side and extract out the
logarithmically divergent term due to $1/z^2$. Then we set $z_*=0$ for
the $j_0^2$ term, while we set $z_*=1$ for the $1/z^2$ term so as to
make the resulting logarithmic term zero at $z=1$. 
For $J$ of two indices, this is how we have defined $B_{nn}$ in
Eq.~(\ref{eq:Bjjetc}). For $J$ of three indices, this applies to
$B_{njn}$. Specifically it is given by
\begin{equation}
  \label{eq:Bnjn}
  B_{njn}=B_{jjj}-B_{jj}\ln z+{1\over2}(\ln z)^2.
\end{equation}
For convenience, in what follows we call $B_{J}$ the generalized
integral sinusoidal functions and $D_\ell^{J}$ the generalized
spherical Bessel functions.

Note that all the $B_J$ whose $J$ end with $n$ can be expressed in
terms of those whose $J$ end with $j$. For example, for $J$ of three
indices, we have
\begin{eqnarray}
  \label{eq:BJn}
 && B_{jjn}=-B_{nnj}+\ln z B_{nj}\,,
\nonumber\\
 && B_{jnn}=2B_{jjj}+B_{nnj}-\ln z B_{jj}\,,
\nonumber\\
 && B_{nnn}=2B_{njj}-B_{jnj}-\ln z B_{nj}\,,
\end{eqnarray}
together with Eq.~(\ref{eq:Bnjn}). Using these relations, we can express 
all the $D_\ell^J$ whose $J$ end with $n$ in terms of those whose $J$
end with $j$.

Further we introduce the following indefinite integral operator,
\begin{equation}
F_{k,\ell}[X] := n_k\int dz j_{\ell} X
    -j_k\int dz n_{\ell} X,
\label{eq:Fkelldef}
\end{equation}
for a function $X$. Note that Eq.~(\ref{eq:SPNexpand}) is expressed in
terms of this operator as
\begin{equation}
  \label{eq:xiFellell}
  \xi^{(n)}_\ell=F_{\ell,\ell}[W^{(n)}].
\end{equation}
We also introduce the following operator,
\begin{eqnarray}
H_{k}^{J} [Y] & := & n_k\int dz D_{0}^{J}  Y
    -j_k\int dz D_{0}^{J}  \tilde Y, 
\end{eqnarray}
where $Y$ stands for a linear combination of the generalized Bessel
functions with the coefficients given by linear combinations of 
$z^m(\ln z)^n$ ($m\le1$, $n\ge0$) and $\tilde Y$ denotes the quantity
which is obtained by replacing $j$, $n$, $D^{jJ}$ and $D^{nJ}$ 
with $n$, $j$, $D^{nJ}$ and $D^{jJ}$, respectively, in the expression of
$Y$. By definition we see that  
\begin{equation}
 F_{\ell,0}[z^m(\ln z)^n j_{0}]=H_{\ell}^{j}[z^m(\ln z)^n j_{0}].
\end{equation}

\subsubsection{Basic formulas}

The spherical Bessel functions satisfy the recursion relation,
\begin{equation}
\zeta_{m-1} +\zeta_{m+1}={2m+1\over z}\zeta_m\,,
\label{eq:reczeta}
\end{equation}
where $\zeta_m=j_m$ or $n_m$. Note that 
\begin{equation}
n_m=(-1)^{m+1}j_{-m-1}\,,\quad j_m=(-1)^{m}n_{-m-1}\,.
\label{eq:njrel}
\end{equation}
The same recursion relation holds for the generalizes spherical
Bessel functions, 
\begin{equation}
D_{m-1}^{\zeta J} +D_{m+1}^{\zeta J} ={2m+1\over z}D_m^{\zeta J}\,,
\label{eq:recDellJ}
\end{equation}
where $\zeta=j$ or $n$. Further the relations the same as
Eqs.~(\ref{eq:njrel}) hold for $D_m^{\zeta J}$,
\begin{equation}
D_{m}^{nJ}=(-)^{m+1} D_{-m-1}^{jJ},\quad
D_{m}^{jJ}=(-)^{m} D_{-m-1}^{nJ}.
\end{equation}

The derivative recursion relation for the spherical Bessel functions is
\begin{equation}
{d\over dz}\zeta_{\ell}={1\over 2\ell +1}
   \left\{\ell\zeta_{\ell-1} -(\ell+1)\zeta_{\ell+1}\right\},
\label{eq:dzeta}
\end{equation}
and this extends to the generalized spherical Bessel functions as
\begin{eqnarray}
{d\over dz} D_{\ell}^{n\zeta J} & = & {1\over 2\ell +1}
   \left\{\ell D_{\ell-1}^{n\zeta J} 
   -(\ell+1) D_{\ell+1}^{n\zeta J}\right\}
    +{R_{\ell,0}\over z} D_{0}^{\zeta J}, 
\cr
{d\over dz} D_{\ell}^{j\zeta J} & = & {1\over 2\ell +1}
   \left\{\ell D_{\ell-1}^{j\zeta J} 
   -(\ell+1) D_{\ell+1}^{j\zeta J}\right\}
    -{R_{\ell,-1}\over z} D_{0}^{\zeta J}.
\label{eq:dDellJ}
 \end{eqnarray}

Useful integral formulas for the spherical Bessel functions are
\begin{eqnarray}
\int dz~ \zeta_m \zeta_n^{*}
 &=&{z^2 \over (m-n)(m+n+1)}
\left(\zeta_m \zeta_{n+1}^{*} +\zeta_{m-1} \zeta_{n}^{*}\right)
-{z\over m-n}\zeta_m \zeta_{n}^{*}, \cr
&&\hspace{4cm}\quad (m\ne n,\,-n-1), 
\cr
\int dz~ \zeta_{\ell} \zeta_{\ell}^{*}
 &=&{1 \over 2\ell +1}
\left\{\int dz~\zeta_{0} \zeta_{0}^{*} 
-z\left(\zeta_0 \zeta_{0}^{*} +2 \sum_{m=1}^{\ell-1} 
\zeta_{m} \zeta_{m}^{*}+\zeta_{\ell} \zeta_{\ell}^{*}\right)\right\}, 
\cr
\int dz~ z\zeta_m \zeta_n^{*}
 &=&\int dz~ z \zeta_{m-1} \zeta_{n-1}^{*}
-{z^2 \over m+n}
\left(\zeta_{m-1} \zeta_{n-1}^{*} +\zeta_{m} \zeta_{n}^{*}\right),
\cr
\int dz~ z \zeta_{\ell} \zeta_{\ell}^{*} &=&
\int dz~ z \zeta_{0} \zeta_{0}^{*} 
-{z^2\over 2} \left\{
\zeta_0 \zeta_{0}^{*} +\sum_{m=1}^{\ell-1} 
\left({1\over m}+{1\over m+1}\right)\zeta_{m} \zeta_{m}^{*}
+{1\over \ell}\zeta_{\ell} \zeta_{\ell}^{*}\right\}, 
\cr
\int dz~ z \zeta_{\ell} \zeta_{\ell-1}^{*} &=&
\int dz~ z \zeta_{0} \zeta_{-1}^{*} 
-z^2 \left\{
\zeta_0 \zeta_{-1}^{*} +\sum_{m=1}^{\ell-1} 
{4m\over 4m^2-1}\zeta_{m} \zeta_{m-1}^{*}
+{1\over 2\ell-1}\zeta_{\ell} \zeta_{\ell-1}^{*}\right\},
\cr
\int dz~ z \zeta_{\ell} \zeta_{\ell+1}^{*} &=&
\int dz~ z \zeta_{-1} \zeta_{0}^{*} 
-z^2 \left\{
\zeta_{-1} \zeta_{0}^{*} +\sum_{m=1}^{\ell} 
{4m\over 4m^2-1}\zeta_{m-1} \zeta_{m}^{*}
+{1\over 2\ell+1}\zeta_{\ell} \zeta_{\ell+1}^{*}\right\}, \cr&&
\label{eq:tamabasicf}
\end{eqnarray}
where $\zeta_m$ or $\zeta^*_m$ stands for $j_m$ or $n_m$.

The following polynomial of $1/z$ plays an important role in the
calculations:
\begin{eqnarray}
 R_{m,k}&& = z^2 (n_m j_k -j_m n_k) \cr
     = && -\sum_{r=0}^{[(m -k -1)/2]} (-1)^r
    {(m-k-1-r)!~ \Gamma\left(m+{1\over 2}-r \right)
    \over r!~(m -k -1 -2r)!~ \Gamma\left(k +{3\over 2}+r\right)}
    \left({2\over z}\right)^{m-k-1-2r},    
\label{eq:tamaR}
\end{eqnarray}
for $m>k$ and 
\begin{equation}
 R_{m,k}=-R_{k,m}.
\end{equation}
for $m<k$. 
By construction, a recursion formula similar to that 
satisfied by the spherical Bessel functions holds: 
\begin{equation}
R_{m,k-1}+R_{m,k+1}={2k+1\over z}R_{m,k}\,.
\end{equation}
Note that the indices $m$ and $k$ can be negative as well.
As examples, we write down the explicit forms for some special cases:
\begin{equation}
R_{k,k}=0, \quad R_{k,k+1}=1,\quad R_{k,k-1}=-1,
\quad R_{k,k-2}=-{2k-1\over z}. 
\end{equation}

\subsubsection{Source terms}
The source term can be rewritten as 
\begin{eqnarray}
W_{\ell}^{(n)}
=&&z^2 \left[{d^2 \over dz^2} +{2\over z}{d\over dz}
+\left(1-{\ell (\ell+1)\over z^2}\right)\right]{\xi^{(n-1)}_\ell\over z} 
\cr &&
+\left[{d\over dz}-{2 z}+{\ell (\ell+1)-4 \over z}\right]
\xi^{(n-1)}_\ell
+i\left[2z{d\over dz}+1\right]\xi^{(n-1)}_\ell\,.
\label{eq:Wnform}
\end{eqnarray}
The contribution to $\xi^{(n)}_\ell$ from the first term is 
given by $z^{-1}\xi^{(n-1)}_\ell$. So we focus on the second and third
terms. Note that the operators of the second and third terms have
opposite parities; the second term is odd while the third term is even
under the transformation $z\to-z$.
To perform the integration of these terms, we introduce the
concept of the standard form of source terms as follows. 

First consider the first iteration, $n=1$.
Since $\xi^{(0)}_\ell=j_\ell$ and since we only need to calculate the
real part of $\xi^{(1)}_\ell$, we only need to consider the second term.
Using the recursion relations (\ref{eq:reczeta}) and
(\ref{eq:dzeta}), it may be rewritten in the form,
\begin{equation}
  \alpha_0zj_\ell+\beta_0j_{\ell-1}+\beta_1j_{\ell+1}\,.
\label{eq:oddform1}
\end{equation}
Then using the integral formulas (\ref{eq:tamabasicf}), the integrals
$F_{\ell,\ell}[zj_\ell]$ and $F_{e\ll,\ell}[j_{\ell\pm1}]$ are readily
evaluated to give
\begin{eqnarray}
&&F_{\ell,\ell}\left[z j_{\ell}\right]
 =D_{\ell}^{nj} -{1\over 2} 
\left\{R_{\ell,0} j_{0}+\sum_{m=1}^{\ell -1}
 \left({1\over m}+{1\over m+1}\right) R_{\ell,m}j_{m}\right\},
\nonumber\\
&&F_{\ell,\ell}\left[j_{k}\right]
 =-{1\over (\ell -k)(\ell +k+1)}j_{k}\quad(k=\ell\pm1). 
\end{eqnarray}
The real part of $\xi^{(1)}_\ell$ is expressed in terms of these
functions, while the imaginary part is $(\ln z)j_\ell$ as given by
Eq.~(\ref{eq:tamaImxi}). The result is Eq.~(\ref{eq:xsi}) with $q=0$.

At the second iteration, $n=2$, we insert the real part of
$\xi^{(1)}_\ell$ to the second term in Eq.~(\ref{eq:Wnform}) and
the imaginary part $i(\ln z)j_\ell$ to the third term, to evaluate the
real part of $\xi^{(2)}_\ell$. Let us focus on the contribution of
the terms of the form $R_{\ell,m}j_m$ in $\xi^{(1)}_\ell$ for the
moment. Since $R_{\ell,m}$ are polynomials in $1/z$, we cannot apply
the integral formulas (\ref{eq:tamabasicf}) directly. So, by using the
recursion relation (\ref{eq:reczeta}) we get rid of the inverse powers
of $z$. Then we find the corresponding source term may be expressed in
the form, 
\begin{equation}
  z(\tilde\alpha_{-}j_{\ell-1}+\tilde\alpha_{+}j_{\ell+1})
 +\sum_{m\ne0}\tilde\beta_m j_{\ell+2m}\,.
\label{eq:evenform}
\end{equation}
Similarly, at the third iteration, $n=3$, the terms in
$\xi^{(2)}_\ell$ having the form $z^kj_n$ ($k\le0$) will give rise to
the source term which can be written in the form,
\begin{equation}
  \label{eq:oddform}
  z(\alpha_0j_\ell+\alpha_{-}j_{-\ell-2}+\alpha_{+}j_{-\ell})
  +\sum_{m\ne-\ell}\beta_mj_{\ell+2m-1}\,,
\end{equation}
which is a generalization of Eq.~(\ref{eq:oddform1}).
Because the operator of the second term in
Eq.~(\ref{eq:Wnform}) has the odd parity, the source terms
for $n=2$ and $3$ take different forms.
We call Eqs.~(\ref{eq:evenform}) and (\ref{eq:oddform}) the standard
forms. For convenience we call the former the even standard 
form and the latter the odd standard form.
Now turning to the terms with $D_m^{J}$ or $(\ln z)j_m$, since
they satisfy the same recursion relation as $j_m$ do, the same idea can
be extended to them in a natural sense. The standard form for them is
then defined by Eqs.~(\ref{eq:evenform}) and (\ref{eq:oddform}) with
$j_m$ replaced by $D_{m}^{J}$ or $(\ln z)j_m$. Note that
$D_\ell^{nj}$ at $n=2$ plays an analogous role of $j_\ell$ at
$n=1$. Hence the odd standard form of $D_m^{nj}$ appears at $n=2$. On
the other hand, since the second and third terms in
Eq.~(\ref{eq:Wnform}) have opposite parities, the parity of the standard
form of $(\ln z)j_m$ is equal to that of $j_m$.

To summarize, the source term at the second iteration consists of the
standard forms of
\begin{equation}
  j_m:~ \hbox{even},\quad (\ln z)j_m:~\hbox{even},\quad 
 D_m^{nj}:~\hbox{odd}.
\label{eq:2ndsource}
\end{equation}
The integration of these terms can be done by using the formulas 
given in subsection \ref{subsec:reductint} below. The resulting
$\xi^{(2)}_\ell$ are given by Eqs.~(\ref{eq:f2f3}) for $\ell=2$, $3$ and 
Eq.~(\ref{eq:f4}) for $\ell=4$. 
Then we find there appear new types of the source term at the third
iteration, which are
\begin{equation}
  zD_\ell^{nnj}\,,~~D_{\ell\pm1}^{nnj}\,,~~z(\ln z)^2j_\ell\,,
~~(\ln z)^2j_{\ell\pm1}\,,~~z(\ln z)D_{\ell\pm1}^{nj}\,,
\end{equation}
in addition to the opposite parity terms of Eq.~(\ref{eq:2ndsource}),
\begin{equation}
  j_m:~\hbox{odd},\quad (\ln z)j_m:~\hbox{odd},\quad 
 D_m^{nj}:~\hbox{even}.
\end{equation}

\subsection{Reduction of integrals}
\label{subsec:reductint}
In this subsection, we reduce the expressions
$F_{\ell,\ell}[\hbox{source terms}]$ to those written in terms of
$D_m^J$. For this purpose we need to evaluate integrals such as 
\begin{equation}
  F_{k,\ell}[z D_{\ell}^{nJ}]:= n_k\int dz\, z j_{\ell} D^{nJ}_{\ell}
    -j_k\int dz\, z n_{\ell} D^{nJ}_{\ell}.
\end{equation}
As an example let us show how this is evaluated. 
Using the basic integral formulas (\ref{eq:tamabasicf}),
we integrate it by part as
\begin{eqnarray}
 F_{k,\ell}[z D_{\ell}^{nJ}]
&&=n_k\int dz~ z j_{0} D^{nJ}_{0}-j_k\int dz~ z n_{0} D^{nJ}_{0}
\nonumber \\ &&
   +\Biggl[
    j_k\left(-{z^2\over 2}\right)\left\{
      n_0 j_0+\sum_{m=1}^{\ell -1}\left({1\over m}+{1\over m+1}\right)
      n_m j_m +{1\over \ell} n_{\ell} j_{\ell}\right\}
\nonumber \\ &&
   \quad -n_k\left(-{z^2\over 2}\right)\left\{
      j_0 j_0+\sum_{m=1}^{\ell -1}\left({1\over m}+{1\over m+1}\right)
      j_m j_m +{1\over \ell} j_{\ell} j_{\ell}\right\}\Biggr] B_{nJ}
\nonumber \\ &&
   -\Biggl[
    j_k\left(-{z^2\over 2}\right)\left\{
      n_0 n_0+\sum_{m=1}^{\ell -1}\left({1\over m}+{1\over m+1}\right)
      n_m n_m +{1\over \ell} n_{\ell} n_{\ell}\right\}
\nonumber \\ &&
   \quad -n_k\left(-{z^2\over 2}\right)\left\{
      j_0 n_0+\sum_{m=1}^{\ell -1}\left({1\over m}+{1\over m+1}\right)
      j_m n_m +{1\over \ell} j_{\ell} n_{\ell}\right\}\Biggr] B_{jJ}
\nonumber \\ &&
   -n_k\Biggl[
      \int dz \left(-{z^2\over 2}\right)\left\{
      j_0 n_0+\sum_{m=1}^{\ell -1}\left({1\over m}+{1\over m+1}\right)
      j_m n_m +{1\over \ell} j_{\ell} n_{\ell}\right\}
      z j_0 D^{J}_0
\nonumber \\ &&
   \quad -\int dz \left(-{z^2\over 2}\right)\left\{
      j_0 j_0+\sum_{m=1}^{\ell -1}\left({1\over m}+{1\over m+1}\right)
      j_m j_m +{1\over \ell} j_{\ell} j_{\ell}\right\}
      z n_0 D^{J}_0\Biggl]
\nonumber \\ &&
    +j_k\Biggl[
      \int dz \left(-{z^2\over 2}\right)\left\{
      n_0 n_0+\sum_{m=1}^{\ell -1}\left({1\over m}+{1\over m+1}\right)
      n_m n_m +{1\over \ell} n_{\ell} n_{\ell}\right\}
      z j_0 D^{J}_0
\nonumber \\ &&
     \quad -\int dz \left(-{z^2\over 2}\right)\left\{
      n_0 j_0+\sum_{m=1}^{\ell -1}\left({1\over m}+{1\over m+1}\right)
      n_m j_m +{1\over \ell} n_{\ell} j_{\ell}\right\}
      z n_0 D^{J}_0\Biggl]
\nonumber \\ 
     = &&D_k^{nnJ} 
      +{1\over 2} \left\{R_{0k} D^{nJ}_0 
      +\sum_{m=1}^{\ell -1}\left({1\over m}
                  +{1\over m+1}\right)R_{mk} D^{nJ}_m
      +{1\over \ell} R_{\ell k} D^{nJ}_{\ell}\right\}
\nonumber \\ &&
      +{1\over 2} n_k\int z dz\left(
       \sum_{m=1}^{\ell -1}\left({1\over m}
              +{1\over m+1}\right)R_{m0} j_m 
      +{1\over \ell} R_{\ell 0} j_{\ell} \right) D^{J}_{0}
\nonumber \\ &&
      -{1\over 2} j_k\int z dz\left(
       \sum_{m=1}^{\ell -1}\left({1\over m}
              +{1\over m+1}\right)R_{m0} n_m
      +{1\over \ell} R_{\ell 0} n_{\ell} \right) D^{J}_{0}
\nonumber\\
     = &&D_k^{nnJ} 
      +{1\over 2} \left\{R_{0k} D^{nJ}_0 
      +\sum_{m=1}^{\ell -1}\left({1\over m}
                  +{1\over m+1}\right)R_{mk} D^{nJ}_m
      +{1\over \ell} R_{\ell k} D^{nJ}_{\ell}\right\}
\nonumber \\ &&
      +{1\over 2} H^J_k\left[
       \sum_{m=1}^{\ell -1}\left({1\over m}
              +{1\over m+1}\right)R_{m0} j_m 
      +{1\over \ell} R_{\ell 0} j_{\ell} \right].
\end{eqnarray}
The reduction of the $H^J_k$ term in the last expression is done
similarly. In the following, we give formulas for each type of the
source terms.

\subsubsection{$j_m$-terms}
The source terms have the standard form (\ref{eq:evenform}) or
(\ref{eq:oddform}).
 
Using Eqs.~(\ref{eq:tamabasicf}), their integrals are evaluated as
\begin{equation}
F_{\ell,\ell}\left[\zeta_{m}\right]
 =-{1\over (\ell -m)(\ell +m+1)}\zeta_{m}\quad 
(m\ne \ell,~ -\ell-1),
\end{equation}
\begin{equation}
F_{\ell,\ell}\left[z \zeta_{\ell}\right]
 =D_{\ell}^{n\zeta} -{1\over 2} 
\left\{R_{\ell,0} \zeta_{0}+\sum_{m=1}^{\ell -1}
 \left({1\over m}+{1\over m+1}\right) R_{\ell,m} \zeta_{m}\right\},
\end{equation}
\begin{equation}
F_{\ell,\ell}\left[z j_{\ell-1}\right]
 =-D_{\ell}^{nn} -
\left\{-R_{\ell,0} n_{0}+\sum_{m=1}^{\ell -1}
 {4m\over 4m^2-1} R_{\ell,m} j_{m-1}\right\},
\end{equation}
\begin{equation}
F_{\ell,\ell}\left[z n_{\ell-1}\right]
 =D_{\ell}^{nj} -
\left\{R_{\ell,0} j_{0}+\sum_{m=1}^{\ell -1}
 {4m\over 4m^2-1} R_{\ell,m} n_{m-1}\right\},
\end{equation}
\begin{equation}
F_{\ell,\ell}\left[z \zeta_{\ell+1}\right]
 =-D_{\ell}^{j\zeta} -
\left\{R_{\ell,-1} \zeta_{0}+\sum_{m=1}^{\ell}
 {4m\over 4m^2-1} R_{\ell,m-1} \zeta_{m}\right\}, 
\end{equation}
where $\zeta$ represents $j$ or $n$. 
Note also that rather general formulas,
\begin{eqnarray}
F_{k,\ell}[\zeta_m] & = &{1\over (\ell -m)(\ell +m +1)}\left\{
R_{k,\ell}\zeta_{m+1}+R_{k,\ell-1} \zeta_{m}\right\}
\cr
&&\qquad-{1\over \ell -m}{R_{k,\ell}\over z}\zeta_{m}
\quad(m\ne \ell,~-\ell-1),
\\
F_{k,\ell}[z\zeta_{\ell}] & = & D_k ^{n\zeta} -{1\over 2}
\left\{R_{k,0}\zeta_{0}+\sum_{m=1}^{\ell-1}
\left({1\over m}+{1\over m+1}\right) R_{k,m}\zeta_{m}+
{1\over \ell}R_{k,\ell} \zeta_{\ell}\right\}, 
\end{eqnarray}
hold.  

\subsubsection{$(\ln z)j_m$-terms}
The source terms are in the form (\ref{eq:oddform1}) or
(\ref{eq:evenform}) with $j_m$ replaced by $(\ln z)j_m$.

First we give the general formula for $F_{\ell,\ell}[(\ln z)j_m]$.
Using the first formula in Eqs.~(\ref{eq:tamabasicf}), we obtain
\begin{eqnarray}
F_{\ell,\ell}[(\ln z) j_{m}]
 &=&{1 \over (\ell -m)(\ell +m+1)}
  \left(-\ln z+
    {1 \over (\ell +m+1)}\right) j_{m}
\cr
&&+{1\over \ell-m}F_{\ell,\ell}\left[
    -{2z \over \ell +m+1}j_{m+1}+j_{m}\right],
\end{eqnarray}
for $m\neq\ell,~-\ell-1$.

Next we consider the remaining term of the odd parity.
With the aid of the second formula in Eqs.~(\ref{eq:tamabasicf}), 
after the integration by part, we have 
\begin{eqnarray}
F_{\ell,\ell}[z (\ln z)&& j_{\ell}]
 = F_{\ell,0}[z (\ln z) j_{0}]
  -{\ln z\over 2}\left\{R_{\ell,0} j_{0}+\sum_{m=1}^{\ell-1}
 \left({1\over m}+{1\over m+1}\right) R_{\ell,m} j_{m}\right\}
 \cr
 && +\left(\sum_{m=1}^{\ell}{1\over m}\right)D_{\ell}^{nj}
 - {1\over 4}\left\{2\left(\sum_{m=1}^{\ell}{1\over m}\right)
    -1\right\} R_{\ell,0} j_{0}
\cr &&
 - {1\over 4}\sum_{m=1}^{\ell-1} \left({1\over m}
    +{1\over m+1}\right)\left\{2\left
   (\sum_{k=m+1}^{\ell}{1\over k}\right)
    +{1\over m(m+1)}\right\} R_{\ell,m} j_{m}.
\end{eqnarray}

To evaluate the first term, we use the following trick.
Note that 
\begin{eqnarray}
B_{jnJ} & = & \int_{0}^{z} dz~ z j_0\left(
   n_0 B_{jJ} -j_0 B_{nJ}\right)
\cr
  & = & \int_{0}^{z} dz~ z n_0\left(
   j_0 B_{jJ} +n_0 B_{nJ}\right)
    -\int_{0}^{z} {dz\over z}B_{nJ}
\cr
  & = & B_{njJ}-(\ln z) B_{nJ} +\int_{0}^{z} dz (\ln z) 
   {dB_{nJ}\over dz}, 
\end{eqnarray}
where we assumed $J$ do not end with $n$, i.e., $J\ne n, jn,\cdots$. 
In the same way, 
\begin{eqnarray}
B_{jjJ} 
  & = & -B_{nnJ}+(\ln z) B_{jJ} -\int_{0}^{z} dz (\ln z )
   {d B_{jJ}\over dz}. 
\end{eqnarray}
Thus we obtain 
\begin{eqnarray}
\int_{0}^{z} dz (\ln z) 
   {d B_{nJ}\over dz}
& = & B_{jnJ} -B_{njJ}+(\ln z) B_{nJ},
\cr
\int_{0}^{z} dz (\ln z) 
   {d B_{jJ}\over dz}
& = & -B_{jjJ} -B_{nnJ}+(\ln z) B_{jJ}.
\label{eq:tamalogtrick1}
\end{eqnarray}
Using Eqs.~(\ref{eq:tamalogtrick1}), we find 
\begin{equation}
F_{\ell,0}[z (\ln z) j_{0}]  
 =-D_{\ell}^{njj}-D_{\ell}^{jnj}+(\ln z) D_{\ell}^{nj}. 
\label{eq:Fell0log}
\end{equation}

As for the remaining terms of the even parity, we have 
\begin{eqnarray}
  F_{\ell,\ell}[z(\ln z) j_{\ell+1}]
  &=& -D_{\ell}^{nnj}+D_{\ell}^{jjj}-\ln z D_{\ell}^{jj}\cr
  && -R_{\ell,-1}(\ln z) j_0 -\ln z \sum_{m=1}^{\ell} 
   {4m\over 4m^2-1}R_{\ell,m-1} j_m \cr
  && +F_{\ell,-1}[z j_0]+\sum_{m=1}^{\ell} 
     {4m\over 4m^2-1} F_{\ell,m-1}[z j_m]
     +{1\over 2\ell+1}F_{\ell,\ell}[z j_{\ell+1}], 
\cr
  F_{\ell,\ell}[z(\ln z) j_{\ell-1}]
  &=& D_{\ell}^{jjj}-D_{\ell}^{nnj}-(\ln z)D_{\ell}^{jj}
             +{1\over2}(\ln z)^2j_\ell \cr
  && -R_{\ell,-1}(\ln z) j_0 +\ln z \sum_{m=1}^{\ell-1} 
   {4m\over 4m^2-1}R_{\ell,m} j_{m-1} \cr
  && -F_{\ell,0}[z n_0]+\sum_{m=1}^{\ell-1} 
     {4m\over 4m^2-1} F_{\ell,m}[z j_{m-1}]
     +{1\over 2\ell+1}F_{\ell,\ell}[z j_{\ell-1}].
\end{eqnarray}

\subsubsection{$(\ln z)^2j_m$-terms}
The source terms we have to evaluate are $z(\ln z)^2j_\ell$ and 
$(\ln z)^2j_{\ell\pm1}$. 

Using the first formula in Eqs.~(\ref{eq:tamabasicf}),
we obtain
\begin{eqnarray}
F_{\ell,\ell}[(\ln z)^2 j_{m}]
 &=&{1\over (\ell-m)(\ell+m+1)}\biggl\{
  \left[-(\ln z)^2 
 +{2 \ln z \over \ell+m+1} -{2\over (\ell+m+1)^2}\right] j_m
\cr &&
 \quad +2F_{\ell,\ell}
  \left[(\ln z)\left(-2z j_{m+1}+(\ell+m+1) j_m
  \right)+{2 z j_{m+1}\over \ell +m+1} \right]
   \biggr\},
\end{eqnarray}
for $m\neq\ell,\,-\ell-1$.

Also, with the aid of the second formula in Eqs.~(\ref{eq:tamabasicf}),
we find
\begin{eqnarray}
F_{\ell,\ell}[z (\ln z)^2 j_{\ell}]
 =&& F_{\ell,0}[z (\ln z)^2 j_{0}]
  -{(\ln z)^2\over 2}\left\{R_{\ell,0} j_{0}+\sum_{m=1}^{\ell-1}
 \left({1\over m}+{1\over m+1}\right) R_{\ell,m} j_{m}\right\}
 \cr
 && +F_{\ell,0}\left[z (\ln z) j_{0}\right]
  +\sum_{m=1}^{\ell-1}\left({1\over m}+{1\over m+1}\right) 
   F_{\ell,m}\left[z (\ln z) j_{m}\right]
\cr
&&\quad 
  +{1\over \ell}F_{\ell,\ell}\left[z (\ln z) j_{\ell}\right].
\label{eq:tamalog2}
\end{eqnarray}

The evaluation of the first term in the above equation is done as
follows. Using a technique similar to the one used to derive
Eqs.~(\ref{eq:tamalogtrick1}), we obtain 
\begin{eqnarray}
\int_{0}^{z} dz \left(\ln z\right)^2 
   {d B_{nJ}\over dz}
& = & 2\left[-B_{nnnJ} -B_{jjnJ}+B_{njjJ} -B_{jnjJ} \right]
\cr
&&\quad 
 +2 \left(\ln z\right) \left[B_{jnJ} -B_{njJ}\right]
 +\left(\ln z\right)^2 B_{nJ},
\cr
\int_{0}^{z} dz \left(\ln z\right)^2 
   {d B_{jJ}\over dz}
& = & 2\left[-B_{jnnJ} + B_{njnJ}+ B_{jjjJ} +B_{nnjJ}\right]
\cr
&&\quad 
 -2 \left(\ln z\right) \left[B_{nnJ} +B_{jjJ}\right]
 +\left(\ln z\right)^2 B_{jJ}.
\label{eq:tamalogtrick2}
\end{eqnarray}
Using these, we can rewrite the first term as 
\begin{eqnarray}
F_{\ell,0}[z (\ln z)^2 j_{0}]  
 &=&2\left(-D_{\ell}^{nnnj}
  +D_{\ell}^{jjnj}+D_{\ell}^{njjj}+D_{\ell}^{jnjj}\right)
\cr
&&\quad 
  -2(\ln z) \left(D_{\ell}^{jnj}+D_{\ell}^{njj}\right)
  +(\ln z)^2 D_{\ell}^{nj}.  
\end{eqnarray}

We need one more formula to evaluate (\ref{eq:tamalog2}):  
\begin{eqnarray} 
 F_{\ell,m}[z (\ln z) j_{m}]
  =&& F_{\ell,0}[z (\ln z) j_{0}]
   -{1\over 4m^2} R_{\ell,m} j_{m}
\cr
&&  -{\ln z\over 2}\left\{R_{\ell,0} j_{0}+\sum_{k=1}^{m-1}
 \left({1\over k}+{1\over k+1}\right) R_{\ell,k} j_{k}
   +{1\over m} R_{\ell,m} j_{m}\right\}
\cr
 && +\left(\sum_{k=1}^{m}{1\over k}\right)D_{\ell}^{nj}
 - {1\over 4}\left\{2\left(\sum_{k=1}^{m}{1\over k}\right)
    -1\right\} R_{\ell,0} j_{0}
\cr &&
 - {1\over 4}\sum_{k=1}^{m-1} \left({1\over k}
     +{1\over k+1}\right)\left\{2\left
   (\sum_{p=k+1}^{m}{1\over p}\right)
    +{1\over k(k+1)}\right\}R_{\ell,k} j_{k}.
\end{eqnarray}

\subsubsection{$D_k^{nJ}$-terms}
The source terms are the even and odd standard forms of $D_m^{nj}$,
and $zD_\ell^{nnj}$ and $D_{\ell\pm1}^{nnj}$.

Necessary formulas are 
\begin{eqnarray}
F_{\ell,\ell}[ z D_{\ell +1}^{nJ}]
& = & -D_{\ell}^{jnJ}-R_{\ell, -1} D_{0}^{nJ}
-\sum_{m=1}^{\ell} {4m\over 4m^2-1} 
 R_{\ell, m-1} D_{m}^{nJ}
\cr
&&+H_{\ell}^{J}\left[z\left\{\sum_{m=1}^{\ell} 
{4m\over 4m^2-1} R_{m,0} j_{m-1} 
+{R_{\ell +1,0}\over 2\ell +1} j_{\ell}
\right\}\right],
\cr
F_{\ell,\ell}[ z D_{\ell -1}^{nJ}]
& = & D_{\ell}^{njJ}-R_{\ell, 0} D_{0}^{jJ}
-\sum_{m=1}^{\ell-1} 
{4m\over 4m^2-1} R_{\ell, m} D_{m-1}^{nJ} +D_{\ell}^{nJ}
\cr
&&+H_{\ell}^{J}\left[z \left\{\sum_{m=2}^{\ell-1} 
{4m\over 4m^2-1} R_{m-1,0} j_{m} 
+{R_{\ell-1,0}\over 2\ell-1} j_{\ell}
\right\}\right],
\cr
F_{\ell,\ell}[D_{m}^{nJ}]
& = & -{1\over (\ell-m)(\ell+m+1)} D_{m}^{nJ}
\cr
&&\quad 
-H_{\ell}^{J}\left[{z \over (\ell -m)(\ell+m+1)} \left( 
 R_{m+1,0} j_{\ell} +R_{m,0} j_{\ell-1}\right) 
 -{R_{m,0}\over \ell-m} j_{\ell}
\right].
\cr
F_{\ell,\ell}[z D_{\ell}^{nJ}] & = &
D_{\ell}^{nnJ}+{1\over 2}\left\{R_{0,\ell}D_{0}^{nJ}
     +\sum_{m=1}^{\ell -1}
\left({1\over m}+{1\over m+1}\right) R_{m,k} D_{m}^{nJ}\right\}
 \cr &&
+H_{\ell}^{J}\left[{z\over 2}\left(\sum_{m=1}^{\ell -1}
\left({1\over m}+{1\over m+1}\right) R_{m,0} j_{m}
+{1\over \ell}R_{\ell,0} j_{\ell}\right)\right],
\cr
F_{\ell,\ell}[D_{\ell+1}^{nJ}] & = &
{1\over 2(\ell +1)} D_{\ell +1}^{nJ}
+H_{\ell}^{J}\left[
-{z \over 2(\ell+1)} \left(R_{0,\ell +2} j_{\ell} 
+R_{0,\ell +1} j_{\ell -1}\right)
+R_{0,\ell+1} j_{\ell}\right],
\cr
F_{\ell,\ell}[D_{\ell-1}^{nJ}] & = &
-{1\over 2\ell} D_{\ell -1}^{nJ}
+H_{\ell}^{J}\left[
{z \over 2\ell} \left(R_{0,\ell} j_{\ell} 
+ R_{0,\ell -1} j_{\ell -1}\right)
-R_{0,\ell-1} j_{\ell}\right],
\end{eqnarray}

In order to evaluate the terms involving the integral
operator $H_{\ell}^{J}$, we recast its argument into the 
form,
\begin{eqnarray}
&& z(\alpha_{-2} j_{-2} +\alpha_{-1} j_{-1} +
   \alpha_{0} j_{0}+ \alpha_{1} j_{1}) 
   +\sum_{n\neq0,-1}\beta_n j_n.
\label{eq:tamanstform}
\end{eqnarray}
Then all the necessary terms can be easily evaluated as 
\begin{eqnarray}
 H_{\ell}^j [j_m]&=&{1\over m(m+1)}
 \left(-R_{\ell,m+1} j_{0} +R_{\ell,m} n_{0} \right)
 +{R_{\ell,m}\over m z} j_{0} \quad (m\ne 0,\,-1), 
\cr
 H_{\ell}^j [z j_{1}] &=&
 -D_{\ell}^{nn} +R_{\ell,0} n_{0}- R_{\ell,1} j_{0},
\cr
 H_{\ell}^j [z j_{0}] &=&
   D_{\ell}^{nj},
\cr
 H_{\ell}^j [z j_{-1}] &=&
 -D_{\ell}^{jj},
\cr
 H_{\ell}^j [z j_{-2}]&=&
   -D_{\ell}^{jn}-j_{0} R_{\ell,-2}+j_{-1} R_{\ell,-1}, 
\label{eq:Hjell}
\end{eqnarray}
and 
\begin{eqnarray}
H_{\ell}^{nj}[z j_1]&=&D_{\ell}^{njj} 
- {R_{\ell,0}\over z} D_{0}^{nj}, \cr
H_{\ell}^{nj}[z j_{0}]&=&
D_{\ell}^{nnj}, \cr 
H_{\ell}^{nj}[z j_{-1}]&=&
- D_{\ell}^{jnj}, \cr 
H_{\ell}^{nj}[z j_{-2}]&=&
\hbox{(unnecessary for our present calculation)},\cr  
H_{k}^{nj}[j_{m}]&=&
{1\over m(m+1)}\left(R_{k,m} D_{1}^{nj} + R_{k,m-1} D_{0}^{nj}\right)
\cr
&&\quad 
-{1\over m z} R_{k,m} D_{0}^{nj} +{1\over m(m+1)} H_{k}^{j}[z j_m].  
\end{eqnarray}
The last term $H_k^j[z j_m]$ can be reduced recursively to 
those given in Eq.~(\ref{eq:Hjell}).

\subsubsection{$z(\ln z) D_{m}^{nj}$-terms}
What we need to evaluate are $F_{\ell,\ell}[z(\ln z)D_{\ell\pm1}^{nj}]$.

First we consider 
$F_{\ell,\ell}\left[ z (\ln z) D_{\ell +1}^{nj}\right]$. 
It is evaluated as
\begin{eqnarray}
F_{\ell,\ell}\left[ z (\ln z) D_{\ell +1}^{nj}\right]
     & = & F_{\ell,-1}\left[ z (\ln z) D_{0}^{nj}\right]
        -(\ln z) \left\{R_{\ell,-1} D_{0}^{nj}
         +\sum_{m=1}^{\ell}
         {4m \over 4m^2 -1}
         R_{\ell,m-1} D_{m}^{nj} \right\}
\cr
   && +F_{\ell,-1}\left[ z D_{0}^{nj}\right]
      +\sum_{m=1}^{\ell} {4m \over 4m^2 -1}
      F_{\ell,m-1}\left[ z D_{m}^{nj}\right]
      +{1\over 2\ell +1} 
      F_{\ell,\ell}\left[ z D_{\ell +1}^{nj}\right]
\cr
  && +H_{\ell}^{j}\left[z (\ln z)\left\{
      \sum_{m=1}^{\ell} {4m \over 4m^2 -1}
      R_{m,0} j_{m-1} +{1\over 2\ell +1} 
      R_{\ell+1,0}j_{\ell}\right\}\right],
\cr&&
\label{eq:FlogDp1}
\end{eqnarray}
where the first term is further expressed in terms of $D_\ell^J$ as
\begin{equation}
  \label{eq:FlogD}
 F_{\ell,-1}\left[ z (\ln z) D_{0}^{nj}\right]
  = -D_{\ell}^{nnnj}+D_{\ell}^{jjnj}-(\ln z)D_{\ell}^{jnj}\,.
\end{equation}

The other unknown terms in Eq.~(\ref{eq:FlogDp1}) are also evaluated as
\begin{eqnarray}
 F_{\ell,m-1}\left[z D_{m}^{nj}\right]
   & = & -D_{\ell}^{jnj}-\left\{
   R_{\ell,-1}D_{0}^{nj}+\sum_{k=1}^{m-1}
   {4k \over 4k^2 -1}
   R_{\ell,k-1}D_{k}^{nj}
   +{R_{\ell,m-1}\over 2m-1} D_{m}^{nj}\right\}
\cr
   && +H_{\ell}^{j}\left[z \left\{
      \sum_{k=1}^{m-1} {4k \over 4k^2 -1}
      R_{k,0} j_{k-1} +{R_{m,0}\over 2m -1} 
       j_{m-1}\right\}\right]\quad( m\ge 1),
\cr
 F_{\ell,-1}\left[z D_{0}^{nj}\right]
   & = & -D_{\ell}^{jnj}, 
\cr
 H_{\ell}^{j}\left[z(\ln z) j_0 \right]
&=&F_{\ell,0}[z(\ln z)j_0]
=-D_\ell^{njj}-D_\ell^{jnj}+(\ln z)D_\ell^{nj}\,,
\cr
 H_{\ell}^{j}\left[(\ln z) j_m \right]
  & = & (\ln z)\left\{{1\over m(m+1)} \left(
     -R_{\ell,m+1} j_{0} +R_{\ell,m} n_{0}\right)
    +{R_{\ell,m}\over mz} j_{0}\right\}
\cr
  &&+H_{\ell}^{j}\left[{2 z\over m(m+1)} j_{m+1}\right]
    +{2m+1\over m^2 (m+1)^2} 
     \left(R_{\ell,m+1} j_{0}-R_{\ell,m} n_{0}\right)
\cr
&&     -{R_{\ell,m}\over m^2 z} j_{0}\,.
\end{eqnarray}

In the same way, we can evaluate 
$F_{\ell,\ell}\left[ z (\ln z) D_{\ell -1}^{nj}\right]$
as 
\begin{eqnarray}
F_{\ell,\ell}\left[ z (\ln z) D_{\ell -1}^{nj}\right]
     & = & F_{\ell,0}\left[ z (\ln z) D_{0}^{jj}\right]
        -(\ln z) \left\{R_{\ell,0} D_{-1}^{nj}
         +\sum_{m=1}^{\ell-1} {4m\over 4m^2 -1}
         R_{\ell,m} D_{m-1}^{nj} \right\}
\cr
   && +F_{\ell,0}\left[ z D_{-1}^{nj}\right]
      +\sum_{m=1}^{\ell-1} {4m\over 4m^2 -1}
      F_{\ell,m}\left[ z D_{m-1}^{nj}\right]
      +{1\over 2\ell -1} 
      F_{\ell,\ell}\left[ z D_{\ell -1}^{nj}\right]
\cr
  && +H_{\ell}^{j}\left[z (\ln z)\left(
      j_{0} +\sum_{m=2}^{\ell-1} {4m\over 4m^2 -1}
      R_{m-1,0} j_{m} +{1\over 2\ell -1} 
      R_{\ell-1,0}j_{\ell}\right)\right], \cr&&
\end{eqnarray}
where the unknown terms in the right hand side are evaluated as 
%
%
\begin{eqnarray}
 F_{\ell,0}\left[ z (\ln z) D_{0}^{jj}\right]
 & = & -D_{\ell}^{njjj}-D_{\ell}^{jnjj}+(\ln z)
      D_{\ell}^{njj} \,,
\cr
 F_{\ell,m}\left[z D_{m-1}^{nj}\right]
   & = & D_{\ell}^{njj}-\left\{
   R_{\ell,0}D_{-1}^{nj}+\sum_{k=1}^{m-1}
   {4k\over 4k^2 -1} R_{\ell,k}D_{k-1}^{nj}
   +{R_{\ell,m}\over 2m-1} D_{m-1}^{nj}\right\}
\cr
   && +H_{\ell}^{j}\left[z \left(j_{0}
      +\sum_{k=1}^{m-1} {4k\over 4k^2 -1}
      R_{k-1,0} j_{k} +{R_{m-1,0}\over 2m -1} 
       j_{m}\right)\right]\quad(m\ge 1),
\cr
F_{\ell,0}\left[z D_{-1}^{nj}\right]
   & = & D_{\ell}^{njj}\,.
\end{eqnarray}

\subsection{The asymptotic behavior}
\label{tamaasympf}
Using the results of the preceding subsection, we obtain the analytic
expression for $\xi^{(3)}_\ell$. The real part of it is given in
Eq.~(\ref{eq:thirdf2}) for $\ell=2$ and in Eq.~(\ref{eq:thirdf3}) for
$\ell=3$, while the imaginary part is determined by
Eq.~(\ref{eq:tamaImxi}). To obtain $A^{\rm inc}_\ell$ to
$O(\epsilon^3)$, we then see that all what we need to evaluate are the
asymptotic behaviors of $D_\ell^{nj}$, $D_\ell^{nnj}$, $D_\ell^{nnnj}$
and $F_{\ell,0}[z(\ln z)j_0]$. Although the last of these can be
expressed in terms of $D_\ell^J$ as given by Eq.~(\ref{eq:Fell0log}), we
find it is easier to evaluate the integral directly as it is. 

Here in order to evaluate the asymptotic behaviors of these functions,
we first give necessary basic formulas.
Then we evaluate the asymptotic behavior of all the necessary 
$B_J$ and $F_{\ell,0}[z(\ln z)j_0]$. In this subsection $x=2z$.

\subsubsection{Basic asymptotic formulas}
First, we give the most basic formulas:
\begin{eqnarray}
&& S=\int_0^x dy {\sin y\over y} \rightarrow {\pi\over2},
\cr
&& C=\int_0^x dy {\cos y-1\over y} \rightarrow 
  -\ln x -\gamma,
\cr
&& \int_0^x dy {\sin y\over y} \ln y \rightarrow -{\pi\over2}\gamma,
\cr
&& \int_0^x dy {\cos y-1\over y} \ln y \rightarrow
  -{1\over 2}(\ln x)^2-{\pi^2\over 24}+{\gamma^2\over 2},
\cr
&& \int_0^x dy {\sin y\over y} (\ln y)^2 \rightarrow 
     {\pi\over2}\gamma^2+{\pi^3\over 24},
\cr
&& \int_0^x dy {\cos y-1\over y} (\ln y)^2 \rightarrow
  -{1\over 3}(\ln x)^3-{\gamma^3\over 3}
   +{\gamma\pi^2\over 12}-{2\over 3}\zeta(3),
\label{eq:tamacat1}
\end{eqnarray}
where $\zeta(z)$ is the Riemann zeta function and $\zeta(3)=1.202\cdots$.

There appear several expressions to be estimated 
that diverge if we evaluate them term by term. But they give finite
results when combined together. They are
\begin{eqnarray}
&& \int_0^{x} {dy\over y+1}\left\{\ln (y+1)-\ln y \right\} 
    \rightarrow {\pi^2 \over 6},
\cr
&& \int_0^{x} {dy\over y+1}\left\{\left(\ln (y+1)\right)^2-
   \left(\ln y\right)^2 \right\} 
    \rightarrow 0,
\cr
&& \int_0^{x} {dy\over y+1}\left\{\ln (y+2)-\ln (y+1)\right\} 
    \rightarrow {\pi^2\over 12},
\cr
&& \int_0^{x} {dy\over y+1}
\left\{(\ln (y+2))^2-(\ln (y+1))^2\right\}
    \rightarrow 2\phi(3,1/2)+{\pi^2\over 6}\ln 2
     -{1\over 3}(\ln 2)^3.
\cr
&& \int_0^{x} {dy\over y+2}\ln(y+1)
                   \left\{\ln (y+2)-\ln (y+1)\right\}
\cr
&& \rightarrow -\phi(3,1/2)-{\pi^2\over 12}\ln 2+{1\over 6}(\ln 2)^3
                +{3\over 2}\zeta(3), 
\label{eq:tamacat2}
\end{eqnarray}
where $\phi(a,b)$ represents the modified zeta function. We mention that 
the $\phi(3,1/2)$ terms are found to cancel out in the final
expression for $A^{\rm inc}_\ell$.

There are several formulas which require multiple integrations.  
They are evaluated as
\begin{equation}
 \int_0^{x} {dy\over y}\left[
     \int_y^{\infty} {du\over u}(\cos (u-y) -\cos u )\right]
    \rightarrow {\pi^2 \over 6},
\label{eq:tama57}
\end{equation}
\begin{equation}
 \int_0^{x} {dy\over y}\ln y\left[
     \int_y^{\infty} {du\over u}(\cos (u-y) -\cos u )\right]
    \rightarrow -{\pi^2\gamma\over 6},
\label{eq:tama58}
\end{equation}
\begin{eqnarray}
 \int_0^{x} {dy\over y}\ln y&&\left[
     \int_y^{\infty}{du\over u}(\cos (u+y) -\cos u )\right]
\cr
&& \rightarrow \phi(3,1/2)+{\pi^2\over12}\ln 2-{1\over 6}(\ln 2)^3 
          +{\pi^2\gamma\over 12},
\label{eq:tama59}
\end{eqnarray}
\begin{equation}
 \int_0^{x} {dy\over y}\ln y\left[
     \int_y^{\infty} {du\over u}(\sin (u-y) -\sin u )\right]
    \rightarrow {\pi^3\over 12},
\label{eq:tama60}
\end{equation}
\begin{equation}
 \int_0^{x} {dy\over y}\ln y\left[
     \int_y^{\infty} {du\over u}(\sin (u+y) -\sin u )\right]
     \rightarrow -{\pi^3\over 24},
\label{eq:tama61}
\end{equation}
\begin{equation}
 \int_0^{x} {dy\over y}\left[
     \int_y^{\infty} {du\over u}(\sin (u-y) -\sin u )\right]
    \rightarrow 0.
\label{eq:tama62}
\end{equation}
These are obtained by changing the variable from $u$ to 
$\displaystyle u'={u\over y}-1$, performing the 
$dy$-integration first and using the formulas (\ref{eq:tamacat2}).

Next we give several formulas that contain $S$ and $C$. 
Let us recall the definition of $S$ and $C$: 
\begin{equation}
 S=\int_0^x {dy \over y}\sin y
  =-\int_x^{\infty} {dy \over y}\sin y +{\pi\over 2}, 
\end{equation}
\begin{equation}
 C=\int_0^x {dy \over y}(\cos y-1)
  =-\int_x^{\infty} {dy \over y}\cos y -\ln x -\gamma.
\end{equation}
By using the integration by part, we have 
\begin{eqnarray}
 \int_{0}^{x} dy \ln & y & \left({\sin y\over y} S 
  +{\cos y -1\over y} C\right) 
  \cr &&= {\pi\over 2} \int_{0}^{x} dy {\sin y\over y} \ln y
   -\int_{0}^{x} dy {\cos y -1\over y}
   \left( \gamma \ln y +(\ln y)^2\right)
\cr &&
   - \int_{0}^{x} dy {\ln y\over y} 
 \left[\int_{y}^{\infty} {du\over u} (\cos (u-y) -\cos u)\right],
\label{eq:tama65}
\end{eqnarray} 
whose asymptotic behavior is determined by Eqs.~(\ref{eq:tamacat1}) and 
(\ref{eq:tama58}).  
Similarly we have
\begin{eqnarray}
 \int_{0}^{x} dy \ln & y & \left({\sin y\over y} S 
  -{\cos y -1\over y} C\right) 
  \cr &&= {\pi\over 2} \int_{0}^{x} dy {\sin y\over y} \ln y
   +\int_{0}^{x} dy {\cos y -1\over y}
   \left( \gamma \ln y +(\ln y)^2\right)
\cr &&
   + \int_{0}^{x} dy {\ln y\over y} 
 \left[\int_{y}^{\infty} {du\over u} (\cos (u+y) -\cos u)\right],
\label{eq:tama66}
\end{eqnarray} 
and 
\begin{eqnarray}
 \int_{0}^{x} dy \ln & y & \left({\sin y\over y} C 
  \pm {\cos y -1\over y} S \right) 
  \cr &&= - \int_{0}^{x} dy {\sin y\over y} \ln y(\ln y+\gamma)
   \pm {\pi\over 2}\int_{0}^{x} dy {\cos y -1\over y}\ln y
\cr &&
   \mp \int_{0}^{x} dy {\ln y\over y} 
 \left[\int_{y}^{\infty} {du\over u} (\sin (u\pm y) -\sin u)\right],
\label{eq:tama67}
\end{eqnarray}    
whose asymptotic behaviors are determined by Eqs.~(\ref{eq:tamacat1}),
(\ref{eq:tama59}), (\ref{eq:tama60}) and (\ref{eq:tama61}). 
Then we obtain
\begin{eqnarray}
 \int_0^{x} {dy\over y}\left(S^2+C^2\right)
    \rightarrow &&{\pi^2\over 4}(\ln x +\gamma)
        +{1\over 3}(\ln x+\gamma)^3 
     -{4\over 3}\zeta(3), 
\label{eq:tama68}
\end{eqnarray}
\begin{eqnarray}
 \int_0^{x} {dy\over y}\left(S^2-C^2\right)
    \rightarrow && {\pi^2\over 4}\ln x -{1\over 3}(\ln x+\gamma)^3
     +{\pi^2\gamma\over 4}+{4\over 3}\zeta(3)
\cr &&
      -2\left(\phi(3,1/2)+{\pi^2\over12}\ln 2
              -{1\over 6}(\ln 2)^3\right).
\label{eq:tama69}
\end{eqnarray}
These two formulas are obtained by reducing them to the forms to
which Eqs.~(\ref{eq:tama65}) and (\ref{eq:tama66}) can be applied,
respectively.

Finally, we present two complicated formulas. 
The first is 
\begin{eqnarray}
I_1:= && \int_0^{x} {dy\over y}
\left(-2\sin y SC+(\cos y -1)(S^2-C^2)\right)
\cr    \rightarrow && 2\left(\phi(3,1/2)
      +{\pi^2\over12}\ln 2-{1\over 6}(\ln 2)^3\right)
\nonumber \\ &&
  -{7\over 3}\zeta(3) -{1\over 4}\pi^2(\ln x+\gamma)
    +{1\over 3}(\ln x +\gamma)^3   
\label{eq:tama70}
\end{eqnarray}
where we have used the equalities,
\begin{equation}
SC=\int_0^{\infty} {\sin(t+2) x\over t+2}\ln(t+1) dt 
    +{\pi\over 2}C-(\ln x +\gamma)S+{\pi\over 2}(\ln x +\gamma),
\end{equation}
\begin{equation}
S^2-C^2=-2\int_0^{\infty} {\cos(t+2) x\over t+2}\ln(t+1) dt
    +{\pi}S-{\pi^2\over 4}+2(\ln x +\gamma)C+(\ln x +\gamma)^2, 
\end{equation}
and applied the formulas (\ref{eq:tamacat1}), (\ref{eq:tama65}) and the 
last one of Eqs.~(\ref{eq:tamacat2}). 
The second is 
\begin{eqnarray}
I_2 :=&&\int_0^{x} {dy\over y}
\left(2 (\cos y -1)SC+ \sin y (S^2-C^2)\right)
\cr
    \rightarrow &&{\pi \over 2}\left((\ln x +\gamma)^2
     +{\pi^2\over 4}\right),    
\label{eq:tama71}
\end{eqnarray}
which is obtained in the same way by applying the formulas 
(\ref{eq:tamacat1}), (\ref{eq:tama62}) and (\ref{eq:tama67}). 

\subsubsection{The asymptotic behavior of $B_J$}
\label{tamaBasymp}

As we have mentioned, what we have to evaluate are the asymptotic
behaviors of $D_\ell^{nj}$, $D_\ell^{nnj}$, $D_\ell^{nnnj}$ and 
$F_{\ell,0}[z(\ln z)j_0]$. Hence, recalling the definition of
$D_\ell^J$, we need to evaluate $B_{nj}, B_{jj}, B_{nnj}, B_{jnj}, 
B_{nnnj}$ and $B_{jnnj}$ in addition to $F_{\ell,0}[z(\ln z)j_0]$.

The formulas for $B_J$ with two indices are given by 
the first two equations of (\ref{eq:tamacat1}):
\begin{equation}
  B_{nj}=-{1\over2}S\to-{\pi\over4}\,,\quad
 B_{jj}=-{1\over2}C\to{1\over2}(\ln x+\gamma)\,.
\end{equation}
As for $F_{\ell,0}[z(\ln z)j_0]$, its asymptotic behavior is directly
evaluated as
\begin{eqnarray}
   F_{\ell,0}[z(\ln z)j_0]
&=&n_\ell\int_0^z dz z(\ln z)j_0^2
-j_\ell\int_0^z dz z(\ln z)j_0n_0
\nonumber\\
&\to&
{1\over4}\left((\ln z)^2+{\pi^2\over12}-(\gamma+\ln2)^2\right)n_\ell
-{\pi\over4}(\gamma+\ln2)j_\ell\,.
\end{eqnarray}

The formulas for $B_J$ with three indices are given by 
\begin{eqnarray}
B_{jnj}
& = & {1\over 4}\int_{0}^{x} dy \left({\sin y\over y}C 
     -{\cos y -1\over y}S\right) 
\cr
& \rightarrow & {\pi\over 8}(\ln x+\gamma), 
\end{eqnarray}
where we have used Eqs.~(\ref{eq:tamacat1}) and (\ref{eq:tama62}), and
\begin{eqnarray}
B_{nnj} 
     & = & -{1\over 4}\int_{0}^{x} dy \left({\cos y +1\over y}C 
    +{\sin y\over y}S \right) 
\cr
     & \rightarrow &-{1\over 8}\left({5\over 12}\pi^2
        - (\ln x +\gamma)^2\right), 
\end{eqnarray}
where we have used Eqs.~(\ref{eq:tamacat1}). 

For $B_J$ with four indices, we have 
\begin{eqnarray}
B_{jnnj} & = & -{1\over 2}\int_{0}^{x} dy
    {\sin y\over y} B_{jnj} 
   +{1\over 2}\int_{0}^{x} dy {\cos y -1\over y} B_{nnj}
\cr
   & = &
  -{1\over 2} \left[S B_{jnj} -C B_{nnj}\right]
  +{1\over 8} \int_{0}^{x} dy \left(
    {\sin y\over y} C -{\cos y -1\over y} S\right)S
\cr
&&\qquad
  +{1\over 8} \int_{0}^{x} dy \left(
    {\sin y\over y} S +{\cos y +1\over y} C\right)C
\cr
   & = & 
  -{1\over 2} \left[S B_{jnj} -C B_{nnj}\right]
  -{1\over 8} I_1
  +{1\over 4} \int_{0}^{x} dy {C^2 \over y} 
\cr
& \rightarrow & {1\over 24}\left[
 {5\over 8}\pi^2 (\ln x+\gamma)
 -{1\over 2}(\ln x+\gamma)^3-\zeta(3)\right], 
\end{eqnarray}
where we have used Eqs.~(\ref{eq:tama68}), (\ref{eq:tama69}) and
(\ref{eq:tama70}), and
\begin{eqnarray}
B_{nnnj} & = & {1\over 2}\int_{0}^{x} dy
    {\sin y\over y} B_{nnj} 
   +{1\over 2}\int_{0}^{x} dy {\cos y +1\over y} B_{jnj}
\cr
   & = &
  \int_{0}^{x} dy B_{jnj} +
   {1\over 2} \left[C B_{jnj} +S B_{nnj}\right]
\cr && \quad
  -{1\over 8} \int_{0}^{x} dy \left(
    {\sin y\over y} C -{\cos y -1\over y} S\right)C
  +{1\over 8} \int_{0}^{x} dy \left(
    {\sin y\over y} S +{\cos y +1\over y} C\right)S
\cr
   & = & \ln x B_{jnj}
  -{1\over 4} \int_{0}^{x} dy \ln y\left(
    {\sin y\over y} C -{\cos y -1\over y} S\right)
\cr
&&\qquad  
  +{1\over 2} \left[C B_{jnj} +S B_{nnj}\right]
  +{1\over 8} I_2
  +{1\over 4} \int_{0}^{x} dy {SC \over y} 
\cr
& \rightarrow & {\pi\over 32}\left[
 -{1\over 4}\pi^2 + (\ln x+\gamma)^2\right], 
\end{eqnarray}
where we have used Eqs.~(\ref{eq:tama67}) and (\ref{eq:tama71}). 
We should note that Eq.~(\ref{eq:tama67}) has been also used 
in evaluating the term $\displaystyle 
{1\over 4} \int_{0}^{x} dy {SC \over y}$.

With these results, the asymptotic incoming amplitudes 
$A^{\rm inc}_\ell$ to the required order are obtained,
which are given in Eqs.~(\ref{eq:Aincto3}) in the text.
\section{5.5PN formulas for $R^{\rm in}_{\ell}$ and $(dE/dt)_{\ell m}$
  in the Schwarzschild case}
\label{ap:v11}

In this Appendix, we show the post-Newtonian expansion of $R^{\rm in}$ 
which are necessary to evaluate the 5.5PN gravitational wave luminosity 
in the case of a Schwarzschild black hole. Then we show each
$(\ell,m)$-mode contribution to the total luminosity from a particle in
circular orbit. 

For convenience, we give the formulas for $c_0\omega R^{\rm in}_\ell$,
where we recover the index $\ell$ on $R^{\rm in}$
and $c_0=(\ell-1)\ell(\ell+1)(\ell+2)-6i\epsilon$.
\begin{eqnarray} 
c_0\omega R^{\rm in}_{2}=
&& 
\Biggl({{4\,{z^4}}\over 5} + {{8\,i}\over {15}}\,{z^5} 
  - {{22\,{z^6}}\over {105}} - 
    {{2\,i}\over {35}}\,{z^7} + {{23\,{z^8}}\over {1890}} + 
    {{2\,i}\over {945}}\,{z^9} - {{13\,{z^{10}}}\over {41580}} 
\cr && \quad - 
    {i\over {24948}}\,{z^{11}} + {{59\,{z^{12}}}\over {12972960}} + 
    {i\over {2162160}}\,{z^{13}} - {{83\,{z^{14}}}\over {1945944000}} - 
    {i\over {277992000}}\,{z^{15}}\Biggr) 
\cr &&+ 
  \Biggl( {{-8\,{z^3}}\over 5} - {{3\,i}\over 5}\,{z^4} - 
     {{8\,{z^5}}\over {63}} - {{13\,i}\over {90}}\,{z^6} + 
     {{109\,{z^7}}\over {1890}} + {{341\,i}\over {22680}}\,{z^8} 
\cr && \quad - 
     {{9403\,{z^9}}\over {3118500}} 
         - {{293\,i}\over {594000}}\,{z^{10}} + 
     {{38963\,{z^{11}}}\over {567567000}} + 
     {{75529\,i}\over {9081072000}}\,{z^{12}} \Biggr) \,\epsilon  
\cr && + 
  \Biggl( {{4\,{z^2}}\over 5} + {{123317\,{z^4}}\over {36750}} + 
     {{231479\,i}\over {110250}}\,{z^5} 
          - {{889954\,{z^6}}\over {1157625}} - 
     {{454499\,i}\over {2315250}}\,{z^7} 
\cr && \quad+ 
     {{215321483\,{z^8}}\over {5501034000}} + 
     {{35106811\,i}\over {5501034000}}\,{z^9} - 
     {{214\,{z^4}\,\ln z}\over {525}} 
- {{428\,i}\over {1575}}\,{z^5}\,\ln z 
\cr && \quad 
+ {{1177\,{z^6}\,\ln z}\over {11025}} + 
     {{107\,i}\over {3675}}\,{z^7}\,\ln z - 
     {{2461\,{z^8}\,\ln z}\over {396900}} - 
     {{107\,i}\over {99225}}\,{z^9}\,\ln z \Biggr) \,{{\epsilon }^2} 
\cr && + 
  \Biggl( {{-66823\,{z^3}}\over {12250}} 
         - {{99851\,i}\over {55125}}\,{z^4} - 
     {{504569\,{z^5}}\over {694575}} 
          - {{2488639\,i}\over {3969000}}\,{z^6} + 
     {{428\,{z^3}\,\ln z}\over {525}} 
\cr && \quad + 
     {{107\,i}\over {350}}\,{z^4}\,\ln z + 
     {{428\,{z^5}\,\ln z}\over {6615}} 
   + {{1391\,i}\over {18900}}\,{z^6}\,\ln z \Biggr) \,{{\epsilon }^3} 
\cr && \quad + 
  \left( {{471487\,{z^2}}\over {220500}} 
            - {{263\,i}\over {1260}}\,{z^3} - 
     {{214\,{z^2}\,\ln z}\over {525}} \right) \,{{\epsilon }^4} ,
\\
c_0\omega R^{\rm in}_{3}=
&&
\Biggl(
 {{4\,{z^5}}\over {21}} + {{2\,i}\over {21}}\,{z^6} 
                  - {{2\,{z^7}}\over {63}} - 
    {i\over {135}}\,{z^8} + {{29\,{z^9}}\over {20790}} + 
    {i\over {4620}}\,{z^{10}} - {{47\,{z^{11}}}\over {1621620}} 
\cr && \quad\quad - 
    {i\over {294840}}\,{z^{12}} + {{23\,{z^{13}}}\over {64864800}} + 
    {i\over {29937600}}\,{z^{14}}\Biggr) 
\cr && + 
  \Biggl( {{-10\,{z^4}}\over {21}} - {{53\,i}\over {315}}\,{z^5} + 
     {{{z^6}}\over {210}} - {i\over {90}}\,{z^7} + 
     {{751\,{z^8}}\over {155925}} + {{1483\,i}\over {1247400}}\,{z^9} 
\cr &&\quad- 
     {{23\,{z^{10}}}\over {102960}} 
                - {{367\,i}\over {10810800}}\,{z^{11}}
      \Biggr) \,\epsilon  
\cr && + \Biggl( {{8\,{z^3}}\over {21}} + 
     {i\over {14}}\,{z^4} + {{40337\,{z^5}}\over {46305}} + 
     {{79099\,i}\over {185220}}\,{z^6} 
              - {{12562147\,{z^7}}\over {91683900}} - 
     {{4840537\,i}\over {157172400}}\,{z^8}
\cr &&\quad
     -{{26\,{z^5}\,\ln z}\over {441}} - 
     {{13\,i}\over {441}}\,{z^6}\,\ln z + 
     {{13\,{z^7}\,\ln z}\over {1323}} + 
     {{13\,i}\over {5670}}\,{z^8}\,\ln z \Biggr) \,{{\epsilon }^2} 
\cr && + 
  \left( {{-2\,{z^2}}\over {21}} - {{182981\,{z^4}}\over {92610}} - 
     {{3753697\,i}\over {5556600}}\,{z^5} \right.
\cr&&\quad
\left.+ {{65\,{z^4}\,\ln z}\over {441}} + 
     {{689\,i}\over {13230}}\,{z^5}\,\ln z \right) \,{{\epsilon }^3}, 
\\
c_0\omega R^{\rm in}_{4}=
&&
\Biggl({{2\,{z^6}}\over {63}} + {{4\,i}\over {315}}\,{z^7} - 
    {{13\,{z^8}}\over {3465}} - {{8\,i}\over {10395}}\,{z^9} 
\cr &&\quad+ 
    {{71\,{z^{10}}}\over {540540}} + {i\over {54054}}\,{z^{11}} - 
    {{37\,{z^{12}}}\over {16216200}} 
               - {i\over {4054050}}\,{z^{13}}\Biggr) 
\cr &&+ 
  \Biggl( {{-2\,{z^5}}\over {21}} - {{4\,i}\over {135}}\,{z^6} + 
     {{142\,{z^7}}\over {51975}} - {{31\,i}\over {51975}}\,{z^8} + 
     {{929\,{z^9}}\over {2702700}} 
          + {{8\,i}\over {96525}}\,{z^{10}} \Biggr) \,
   \epsilon  
\cr &&
   + \Biggl( {{5\,{z^4}}\over {49}} + {{97\,i}\over {4410}}\,{z^5} + 
     {{958223891\,{z^6}}\over {6051137400}} + 
     {{239560304\,i}\over {3781960875}}\,{z^7} 
\cr &&\quad
 - {{1571\,{z^6}\,\ln z}\over {218295}} - 
     {{3142\,i}\over {1091475}}\,{z^7}\,\ln z \Biggr) \,{{\epsilon }^2} 
\cr && + 
  \Biggl( {{-20\,{z^3}}\over {441}} - {i\over {196}}\,{z^4} \Biggr) \,
   {{\epsilon }^3} ,
\\
c_0\omega R^{in}_{5}=
&&\Biggl({{2\,{z^7}}\over {495}} + {{2\,i}\over {1485}}\,{z^8} - 
    {{7\,{z^9}}\over {19305}} - {i\over {15015}}\,{z^{10}} + 
    {{17\,{z^{11}}}\over {1621620}} + {i\over {737100}}\,{z^{12}}\Biggr) 
\cr &&+ 
  \Biggl( {{-7\,{z^6}}\over {495}} - {{67\,i}\over {17325}}\,{z^7} + 
     {{1831\,{z^8}}\over {4054050}} 
           - {{43\,i}\over {4054050}}\,{z^9} \Biggr) 
    \,\epsilon  
\cr &&
   + \Biggl( {{28\,{z^5}}\over {1485}} + 
     {{59\,i}\over {14850}}\,{z^6} \Biggr) \,{{\epsilon }^2} ,
\\
c_0\omega R^{in}_{6}=
&&\Biggl({{8\,{z^8}}\over {19305}} + {{16\,i}\over {135135}}\,{z^9} - 
    {{4\,{z^{10}}}\over {135135}} 
           - {{2\,i}\over {405405}}\,{z^{11}}\Biggr) 
\cr &&+ 
  \Biggl( {{-32\,{z^7}}\over {19305}} 
           - {{163\,i}\over {405405}}\,{z^8} \Biggr)
    \,\epsilon  ,
\\
c_0\omega R^{\rm in}_{7}=
&& 
{{8\,{z^9}}\over {225225}} + {{2\,i}\over {225225}}\,{z^{10}}\,.
\end{eqnarray}

Next we show the contribution of each $(\ell,m)$-mode to the
gravitational wave luminosity to $O(v^{11})$ in the case of a circular
orbit around a Schwarzschild black hole.
We set 
\begin{equation}
\left\langle{dE\over dt}\right\rangle
={1\over2}\left({dE\over dt}\right)_{\rm N}
\sum_{\ell,m} \eta_{\ell,m}\,,
\end{equation}
where $(dE/dt)_{\rm N}$ is the Newtonian quadrupole luminosity,
Eq.~(\ref{eq:Enewton}).
In the present case we have $\eta_{\ell,m}=\eta_{\ell,-m}$. 
Hence we show only the modes $m>0$.
\begin{eqnarray}
\eta_{2,2}&=&1 - {{107\,{v^2}}\over {21}} + 4\,\pi \,{v^3} + 
  {{4784\,{v^4}}\over {1323}} - {{428\,\pi \,{v^5}}\over {21}} 
\nonumber\\
&+& {v^6}\,\left( {{99210071}\over {1091475}} - 
     {{1712\,\gamma }\over {105}} + {{16\,{{\pi }^2}}\over 3} - 
     {{3424\,\ln 2}\over {105}} - {{1712\,\ln v}\over {105}}
      \right)  
\nonumber\\
&+& {{19136\,\pi \,{v^7}}\over {1323}} 
\nonumber\\
&+& {v^8}\,\left( -{{27956920577}\over {81265275}} + 
     {{183184\,\gamma }\over {2205}} - 
     {{1712\,{{\pi }^2}}\over {63}} + 
     {{366368\,\ln 2}\over {2205}} + 
     {{183184\,\ln v}\over {2205}} \right)  
\nonumber\\
&+& {v^9}\,\left( {{396840284\,\pi }\over {1091475}} - 
     {{6848\,\gamma \,\pi }\over {105}} - 
     {{13696\,\pi \,\ln 2}\over {105}} - 
     {{6848\,\pi \,\ln v}\over {105}} \right)  
\nonumber\\
&+& {v^{10}}\,\left( {{187037845924}\over {6257426175}} - 
     {{8190208\,\gamma }\over {138915}} + 
     {{76544\,{{\pi }^2}}\over {3969}} 
\right.
\nonumber\\ & &\left. \hspace{1cm}
-    {{16380416\,\ln 2}\over {138915}} - 
     {{8190208\,\ln v}\over {138915}} \right)  
\nonumber\\
&+& {v^{11}}\,\left( {{-111827682308\,\pi }\over {81265275}} + 
     {{732736\,\gamma \,\pi }\over {2205}} 
\right.
\nonumber\\ & &\left. \hspace{1cm}
+ 
     {{1465472\,\pi \,\ln 2}\over {2205}} + 
     {{732736\,\pi \,\ln v}\over {2205}} \right), 
\\
\eta_{2,1}&=&
{{{v^2}}\over {36}} - {{17\,{v^4}}\over {504}} + 
  {{\pi \,{v^5}}\over {18}} - {{2215\,{v^6}}\over {254016}} - 
  {{17\,\pi \,{v^7}}\over {252}} 
\nonumber\\
&+& {v^8}\,\left( {{15707221}\over {26195400}} - 
     {{107\,\gamma }\over {945}} + {{{{\pi }^2}}\over {27}} - 
     {{107\,\ln 2}\over {945}} - {{107\,\ln v}\over {945}}
      \right)  
\nonumber\\
&-& {{2215\,\pi \,{v^9}}\over {127008}} 
\nonumber\\
&+& {v^{10}}\,
   \left( -{{6435768121}\over {57210753600}} + 
     {{1819\,\gamma }\over {13230}} - 
     {{17\,{{\pi }^2}}\over {378}} + 
     {{1819\,\ln 2}\over {13230}} + 
     {{1819\,\ln v}\over {13230}} \right)  
\nonumber\\
&+& {v^{11}}\,\left( {{15707221\,\pi }\over {13097700}} - 
     {{214\,\gamma \,\pi }\over {945}} - 
     {{214\,\pi \,\ln 2}\over {945}} - 
     {{214\,\pi \,\ln v}\over {945}} \right) , 
\\
\eta_{3,3}&=&{{1215\,{v^2}}\over {896}} - {{1215\,{v^4}}\over {112}} + 
  {{3645\,\pi \,{v^5}}\over {448}} + 
  {{243729\,{v^6}}\over {9856}} - 
  {{3645\,\pi \,{v^7}}\over {56}} 
\nonumber\\
&+& {v^8}\,\left( {{25037019729}\over {125565440}} - 
     {{47385\,\gamma }\over {1568}} + 
     {{3645\,{{\pi }^2}}\over {224}} - 
     {{47385\,\ln 2}\over {1568}} 
\right.
\nonumber\\
& &\left.
-    {{47385\,\ln 3}\over {1568}} - 
     {{47385\,\ln v}\over {1568}} \right)  
\nonumber\\
&+& {{731187\,\pi \,{v^9}}\over {4928}} 
\nonumber\\
&+& {v^{10}}\,\left( -{{2074855555161}\over {1381219840}} + 
     {{47385\,\gamma }\over {196}} - 
     {{3645\,{{\pi }^2}}\over {28}} + 
     {{47385\,\ln 2}\over {196}}
\right.
\nonumber\\
& &\left.
    +{{47385\,\ln 3}\over {196}} + 
     {{47385\,\ln v}\over {196}} \right)  
\nonumber\\
&+& {v^{11}}\,\left( {{75111059187\,\pi }\over {62782720}} - 
     {{142155\,\gamma \,\pi }\over {784}} - 
     {{142155\,\pi \,\ln 2}\over {784}} 
\right.
\nonumber\\
& &\left.
-    {{142155\,\pi \,\ln 3}\over {784}} - 
     {{142155\,\pi \,\ln v}\over {784}} \right) ,
\\
\eta_{3,2}&=&
{{5\,{v^4}}\over {63}} - {{193\,{v^6}}\over {567}} + 
  {{20\,\pi \,{v^7}}\over {63}} + 
  {{86111\,{v^8}}\over {280665}} - 
  {{772\,\pi \,{v^9}}\over {567}} 
\nonumber\\
&+& 
  {v^{10}}\,\left( {{960188809}\over {178783605}} - 
     {{1040\,\gamma }\over {1323}} + 
     {{80\,{{\pi }^2}}\over {189}} - 
     {{2080\,\ln 2}\over {1323}} - 
     {{1040\,\ln v}\over {1323}} \right) 
\nonumber\\
&+& 
  {{344444\,\pi \,{v^{11}}}\over {280665}} ,
\\
\eta_{3,1}&=&
{{{v^2}}\over {8064}} - {{{v^4}}\over {1512}} + 
  {{\pi \,{v^5}}\over {4032}} + {{437\,{v^6}}\over {266112}} - 
  {{\pi \,{v^7}}\over {756}} 
\nonumber\\
&+& {v^8}\,\left( -{{1137077}\over {50854003200}} - 
     {{13\,\gamma }\over {42336}} + {{{{\pi }^2}}\over {6048}} - 
     {{13\,\ln 2}\over {42336}} - {{13\,\ln v}\over {42336}}
      \right)  
\nonumber\\
&+& {{437\,\pi \,{v^9}}\over {133056}} 
\nonumber\\
&+& {v^{10}}\,
   \left( -{{38943317051}\over {5034546316800}} + 
     {{13\,\gamma }\over {7938}} - {{{{\pi }^2}}\over {1134}} + 
     {{13\,\ln 2}\over {7938}} + {{13\,\ln v}\over {7938}}
      \right)  
\nonumber\\
&+& {v^{11}}\,
   \left( {{-1137077\,\pi }\over {25427001600}} - 
     {{13\,\gamma \,\pi }\over {21168}} - 
     {{13\,\pi \,\ln 2}\over {21168}} - 
     {{13\,\pi \,\ln v}\over {21168}} \right) ,
\\
\eta_{4,4}&=&
{{1280\,{v^4}}\over {567}} - {{151808\,{v^6}}\over {6237}} + 
  {{10240\,\pi \,{v^7}}\over {567}} + 
  {{560069632\,{v^8}}\over {6243237}} - 
  {{1214464\,\pi \,{v^9}}\over {6237}} 
\nonumber\\
&+& {v^{10}}\,\left( {{36825600631808}\over {88497884475}} - 
     {{25739264\,\gamma }\over {392931}} + 
     {{81920\,{{\pi }^2}}\over {1701}} - 
     {{25739264\,\ln 2}\over {130977}} 
\right.
\nonumber\\
& &
\left.
 - {{25739264\,\ln v}\over {392931}} \right) 
+ {{4480557056\,\pi \,{v^{11}}}\over {6243237}} ,
\\
\eta_{4,3}&=&
{{729\,{v^6}}\over {4480}} - {{28431\,{v^8}}\over {24640}} + 
  {{2187\,\pi \,{v^9}}\over {2240}} + 
  {{620077923\,{v^{10}}}\over {246646400}} - 
  {{85293\,\pi \,{v^{11}}}\over {12320}} ,
\\
\eta_{4,2}&=&
{{5\,{v^4}}\over {3969}} - {{437\,{v^6}}\over {43659}} + 
  {{20\,\pi \,{v^7}}\over {3969}} + 
  {{7199152\,{v^8}}\over {218513295}} - 
  {{1748\,\pi \,{v^9}}\over {43659}} 
\nonumber\\
&+& {v^{10}}\,\left( {{9729776708}\over {619485191325}} - 
     {{25136\,\gamma }\over {2750517}} + 
     {{80\,{{\pi }^2}}\over {11907}} - 
     {{50272\,\ln 2}\over {2750517}} - 
     {{25136\,\ln v}\over {2750517}} \right) 
\nonumber\\
&+& {{28796608\,\pi \,{v^{11}}}\over {218513295}} ,
\\
\eta_{4,1}&=&
{{{v^6}}\over {282240}} - {{101\,{v^8}}\over {4656960}} + 
  {{\pi \,{v^9}}\over {141120}} + 
  {{7478267\,{v^{10}}}\over {139848508800}} - 
  {{101\,\pi \,{v^{11}}}\over {2328480}} ,
\\
\eta_{5,5}&=&
{{9765625\,{v^6}}\over {2433024}} - 
  {{2568359375\,{v^8}}\over {47443968}} + 
  {{48828125\,\pi \,{v^9}}\over {1216512}} 
\nonumber\\
& &+ 
  {{7060478515625\,{v^{10}}}\over {25904406528}} - 
  {{12841796875\,\pi \,{v^{11}}}\over {23721984}} , 
\\
\eta_{5,4}&=&
{{4096\,{v^8}}\over {13365}} - 
  {{18231296\,{v^{10}}}\over {6081075}} + 
  {{32768\,\pi \,{v^{11}}}\over {13365}} ,
\\
\eta_{5,3}&=&
{{2187\,{v^6}}\over {450560}} - 
  {{150903\,{v^8}}\over {2928640}} + 
  {{6561\,\pi \,{v^9}}\over {225280}} + 
  {{600654447\,{v^{10}}}\over {2665062400}} - 
  {{452709\,\pi \,{v^{11}}}\over {1464320}} ,
\\
\eta_{5,2}&=&
{{4\,{v^8}}\over {40095}} - {{15644\,{v^{10}}}\over {18243225}} + 
  {{16\,\pi \,{v^{11}}}\over {40095}} ,
\\
\eta_{5,1}&=&
{{{v^6}}\over {127733760}} - {{179\,{v^8}}\over {2490808320}} + 
  {{\pi \,{v^9}}\over {63866880}} 
\cr & &
+ 
  {{290803\,{v^{10}}}\over {971415244800}} - 
  {{179\,\pi \,{v^{11}}}\over {1245404160}} ,
\\
\eta_{6,6}&=&
{{26244\,{v^8}}\over {3575}} - 
  {{2965572\,{v^{10}}}\over {25025}} + 
  {{314928\,\pi \,{v^{11}}}\over {3575}} ,
\\
\eta_{6,5}&=&
{{244140625\,{v^{10}}}\over {435891456}}, 
\\
\eta_{6,4}&=&
{{131072\,{v^8}}\over {9555975}} - 
  {{4063232\,{v^{10}}}\over {22297275}} + 
  {{1048576\,\pi \,{v^{11}}}\over {9555975}}, 
\\
\eta_{6,3}&=&
{{59049\,{v^{10}}}\over {98658560}} ,
\\
\eta_{6,2}&=&
{{4\,{v^8}}\over {5733585}} - {{4\,{v^{10}}}\over {495495}} + 
  {{16\,\pi \,{v^{11}}}\over {5733585}} ,
\\
\eta_{6,1}&=&
{{{v^{10}}}\over {7192209024}} ,
\\
\eta_{7,7}&=&
{{96889010407\,{v^{10}}}\over {7116595200}} ,
\\
\eta_{7,5}&=&
{{6103515625\,{v^{10}}}\over {181330845696}} ,
\\
\eta_{7,3}&=&
{{1594323\,{v^{10}}}\over {205209804800}} ,
\\
\eta_{7,1}&=&
{{{v^{10}}}\over {5983917907968}} ,
\end{eqnarray}

\section{4PN luminosity in terms of the orbital frequency}
\label{ap:v8}

Here we present the $(\ell,m)$-mode contributions to the gravitational
wave luminosity for a circular orbit on the equatorial
plane around a Kerr black hole.  In stead of $v\equiv(M/r_0)^{1/2}$,
the formulas are expressed in terms of the parameter 
$x\equiv(M\Omega_\varphi)^{1/3}$, where $\Omega_\varphi$ is the orbital
angular frequency, which is more relevant in the actual analysis of
observed gravitational wave signals. We express the partial mode
contributions as 
\begin{equation}
\left\langle{dE\over dt}\right\rangle
\equiv{16\over 5}\left({\mu\over M}\right)^2 x^{10}
\sum_{\ell,m} \eta_{\ell,m}\,.
\end{equation}
Since $\eta_{\ell,m}=\eta_{\ell,-m}$, we show only the modes $m>0$
below. 
\begin{eqnarray}
\eta_{2,2}&=&
1 - {{107\,{x^2}}\over {21}} + 
      \left( 4\,\pi  - {{8\,q}\over 3} \right) \,{x^3} + 
      \left( {{4784}\over {1323}} + 2\,{q^2} \right) \,{x^4} 
\nonumber\\
&+&   \left( {{-428\,\pi }\over {21}} + {{52\,q}\over {27}} \right) \,
       {x^5} 
\nonumber\\
&+& \left( {{99210071}\over {1091475}} - 
         {{1712\,\gamma }\over {105}} + {{16\,{{\pi }^2}}\over 3} - 
         {{32\,\pi \,q}\over 3} - {{1817\,{q^2}}\over {567}} 
\right.
\nonumber\\
&-&\left.
  {{3424\,\ln 2}\over {105}} - {{1712\,\ln x}\over {105}}
          \right)\,{x^6}
+ \left( {{19136\,\pi }\over {1323}} 
+  {{364856\,q}\over {11907}} + 8\,\pi \,{q^2} - {{8\,{q^3}}\over 3}
          \right) \,{x^7} 
\nonumber\\
&+& \left( -{{27956920577}\over {81265275}} + 
         {{183184\,\gamma }\over {2205}} 
           - {{1712\,{{\pi }^2}}\over {63}} + 
         {{208\,\pi \,q}\over {27}} 
\right.
\nonumber\\
&+&\left.
         {{105022\,{q^2}}\over {9261}} + {q^4} + 
         {{366368\,\ln 2}\over {2205}} + 
         {{183184\,\ln x}\over {2205}} \right)\,{x^8}\,.
\\
\eta_{2,1}&=&
{{{x^2}}\over {36}} - {{q\,{x^3}}\over {12}} + 
      \left( -{{17}\over {504}} + {{{q^2}}\over {16}} \right) \,
       {x^4} 
\nonumber\\
&+& \left( {{\pi }\over {18}} + 
         {{215\,q}\over {9072}} \right) \,{x^5} 
\nonumber\\
&+& \left( -{{2215}\over {254016}} - {{\pi \,q}\over 6} + 
         {{313\,{q^2}}\over {1512}} \right) \,{x^6} 
\nonumber\\
&+& \left( {{-17\,\pi }\over {252}} - {{18127\,q}\over {190512}} + 
         {{\pi \,{q^2}}\over 8} - {{7\,{q^3}}\over {24}} \right) \,
       {x^7} 
\nonumber\\
&+& \left( {{15707221}\over {26195400}} - {{107\,\gamma }\over {945}} + 
         {{{{\pi }^2}}\over {27}} + {{215\,\pi \,q}\over {4536}} 
\right.
\nonumber\\
&+& \left.
         {{44299\,{q^2}}\over {95256}} + {{{q^4}}\over {16}} - 
         {{107\,\ln 2}\over {945}} - {{107\,\ln x}\over {945}}
          \right)\, x^8\,.
\\
\eta_{3,3}&=&
     {{1215\,{x^2}}\over {896}} - 
      {{1215\,{x^4}}\over {112}} + 
      \left( {{3645\,\pi }\over {448}} 
        - {{1215\,q}\over {224}} \right) \,
       {x^5} 
\nonumber\\
&+& \left( {{243729}\over {9856}} + 
         {{3645\,{q^2}}\over {896}} \right) \,{x^6} + 
      \left( {{-3645\,\pi }\over {56}} 
         + {{41229\,q}\over {1792}} \right) \,
       {x^7} 
\nonumber\\
&+& \left( {{25037019729}\over {125565440}} - 
         {{47385\,\gamma }\over {1568}} 
             + {{3645\,{{\pi }^2}}\over {224}} - 
         {{3645\,\pi \,q}\over {112}} 
\right.
\nonumber\\
&-&\left.
         {{236925\,{q^2}}\over {14336}} - 
         {{47385\,\ln 2}\over {1568}} - {{47385\,\ln 3}\over {1568}} - 
         {{47385\,\ln x}\over {1568}} \right)\,{x^8}\,.
\\
\eta_{3,2}&=&{{5\,{x^4}}\over {63}} - 
      {{40\,q\,{x^5}}\over {189}} + 
      \left( -{{193}\over {567}} + {{80\,{q^2}}\over {567}} \right) \,
       {x^6} 
\nonumber\\
&+& \left( {{20\,\pi }\over {63}} + 
         {{982\,q}\over {1701}} \right) \,{x^7} + 
      \left( {{86111}\over {280665}} - {{160\,\pi \,q}\over {189}} + 
         {{80\,{q^2}}\over {189}} \right) \,{x^8}\,.
\\
\eta_{3,1}&=&{{{x^2}}\over {8064}} - {{{x^4}}\over {1512}} + 
      \left( {{\pi }\over {4032}} - {{25\,q}\over {18144}} \right) \,
       {x^5} 
\nonumber\\
&+& \left( {{437}\over {266112}} + 
         {{17\,{q^2}}\over {24192}} \right) \,{x^6} + 
      \left( {{-\pi }\over {756}} + {{2257\,q}\over {435456}} \right) \,
       {x^7} 
\nonumber\\
&+& \left( -{{1137077}\over {50854003200}} - 
         {{13\,\gamma }\over {42336}} + {{{{\pi }^2}}\over {6048}} 
\right.
\nonumber\\
&-&\left.
         {{25\,\pi \,q}\over {9072}} + {{12863\,{q^2}}\over {3483648}} - 
         {{13\,\ln 2}\over {42336}} - {{13\,\ln x}\over {42336}}
          \right)\,{x^8}\,.
\\
\eta_{4,4}&=& {{1280\,{x^4}}\over {567}} - 
      {{151808\,{x^6}}\over {6237}} + 
      \left({{10240\,\pi}\over {567}}-{{20480\,q}\over{1701}}\right) \,
       {x^7} 
\nonumber\\
&+& \left( {{560069632}\over {6243237}} + 
         {{5120\,{q^2}}\over {567}} \right) \,{x^8}\,.
\\
\eta_{4,3}&=&{{729\,{x^6}}\over {4480}} - 
      {{729\,q\,{x^7}}\over {1792}} + 
      \left( -{{28431}\over {24640}} 
         + {{3645\,{q^2}}\over {14336}} \right) \,
       {x^8}\,.
\\
\eta_{4,2}&=&{{5\,{x^4}}\over {3969}} - 
      {{437\,{x^6}}\over {43659}} + 
      \left( {{20\,\pi }\over {3969}} 
       - {{170\,q}\over {11907}} \right) \,
       {x^7} 
\nonumber\\
&+& \left( {{7199152}\over {218513295}} + 
         {{200\,{q^2}}\over {27783}} \right) \,{x^8}\,.
\\
\eta_{4,1}&=& {{{x^6}}\over {282240}} - 
      {{q\,{x^7}}\over {112896}} + 
      \left( -{{101}\over {4656960}} 
      + {{5\,{q^2}}\over {903168}} \right) \,
       {x^8}\,.
\\
\eta_{5,5}&=& {{9765625\,{{{\it x}}^6}}\over {2433024}} - 
      {{2568359375\,{{{\it x}}^8}}\over {47443968}}\,.
\\
\eta_{5,4}&=&{{4096\,{x^8}}\over {13365}}\,.
\\
\eta_{5,3}&=& {{2187\,{{{\it x}}^6}}\over {450560}} - 
      {{150903\,{{{\it x}}^8}}\over {2928640}}\,.
\\
\eta_{5,2}&=&{{4\,{x^8}}\over {40095}}\,.
\\
\eta_{5,1}&=&{{{x^6}}\over {127733760}} - 
      {{179\,{x^8}}\over {2490808320}}\,.
\\
\eta_{6,6}&=&
{{26244\,{x^8}}\over {3575}}\,.
\\
\eta_{6,4}&=&{{131072\,{x^8}}\over {9555975}}\,.
\\
\eta_{6,2}&=&{{4\,{x^8}}\over {5733585}}\,.
\end{eqnarray}

\section{4PN luminosity for a slightly 
eccentric orbit in the Schwarzschild case}
\label{ap:v8ecc}

We define the partial mode contribution $\eta_{\ell,m,n}$ as 
\begin{equation}
\left\langle{dE \over dt}\right\rangle
={1 \over 2}\left({dE\over dt}\right)_{\rm N} \sum_{\ell,m,n} 
\eta_{\ell,m,n},
\end{equation}
where $(dE/dt)_{\rm N}$ is the Newtonian quadrupole luminosity,
Eq.~(\ref{eq:Enewton}), and
$\eta_{\ell,-m,-n}=\eta_{\ell,m,n}$ because of the symmetry in 
the Teukolsky equation. 
Up to $O(v^8)$, $\eta_{\ell,1,-1}$ do not appear irrespective of 
$\ell$, and other modes are as follows. 

\noindent
For $\ell=2$, 
\begin{eqnarray}
\eta_{2,2,0}&&=
1 - {{107\,{v^2}}\over {21}} + 4\,\pi \,{v^3} + 
  {{4784\,{v^4}}\over {1323}} - 
  {{428\,\pi \,{v^5}}\over {21}} + 
  {{99210071\,{v^6}}\over {1091475}} - 
  {{1712\,{\gamma}\,{v^6}}\over {105}} 
 \nonumber \\
&& +{{16\,{{\pi }^2}\,{v^6}}\over 3}+ 
  {{19136\,\pi \,{v^7}}\over {1323}} - 
  {{27956920577\,{v^8}}\over {81265275}} + 
  {{183184\,{\gamma}\,{v^8}}\over {2205}} - 
  {{1712\,{{\pi }^2}\,{v^8}}\over {63}} \nonumber \\
&& - 
  {{3424\,{v^6}\,\ln (2)}\over {105}} + 
  {{366368\,{v^8}\,\ln (2)}\over {2205}} - 
  {{1712\,{v^6}\,\ln (v)}\over {105}} + 
  {{183184\,{v^8}\,\ln (v)}\over {2205}} \nonumber \\
&& + 
  {{{e}}^2}\,\biggl( -10 + {{932\,{v^2}}\over {21}} - 
     46\,\pi \,{v^3} - {{14270\,{v^4}}\over {147}} + 
     {{4748\,\pi \,{v^5}}\over {21}} - 
     {{516582901\,{v^6}}\over {363825}}  \nonumber \\
&&+{{22256\,{\gamma}\,{v^6}}\over {105}}
-{{208\,{{\pi }^2}\,{v^6}}\over 3} - 
     {{189502\,\pi \,{v^7}}\over {441}} + 
     {{405725734982\,{v^8}}\over {127702575}} - 
     {{352672\,{\gamma}\,{v^8}}\over {315}} \nonumber \\
&& +{{3296\,{{\pi }^2}\,{v^8}}\over 9} 
+{{44512\,{v^6}\,\ln (2)}\over {105}} - 
     {{705344\,{v^8}\,\ln (2)}\over {315}} 
\nonumber\\
&&+ {{22256\,{v^6}\,\ln (v)}\over {105}} - 
     {{352672\,{v^8}\,\ln (v)}\over {315}} \biggr), \\
\eta_{2,1,0}&&=
{{{v^2}}\over {36}} - {{17\,{v^4}}\over {504}} + 
  {{\pi \,{v^5}}\over {18}} - {{2215\,{v^6}}\over {254016}} - 
  {{17\,\pi \,{v^7}}\over {252}} + 
  {{15707221\,{v^8}}\over {26195400}} \nonumber \\ 
&&-{{107\,{\gamma}\,{v^8}}\over {945}} 
+{{{{\pi }^2}\,{v^8}}\over {27}}
-{{107\,{v^8}\,\ln (2)}\over {945}} 
-{{107\,{v^8}\,\ln (v)}\over {945}} \nonumber \\ 
&& +{{{e}}^2}\,\biggl( {{-2\,{v^2}}\over 9} + 
     {{93\,{v^4}}\over {112}} - 
     {{19\,\pi \,{v^5}}\over {36}} + 
     {{60667\,{v^6}}\over {84672}} + 
     {{169\,\pi \,{v^7}}\over {84}} - 
     {{1877507981\,{v^8}}\over {419126400}} \nonumber \\
&&+{{1177\,{\gamma}\,{v^8}}\over {945}} - 
     {{11\,{{\pi }^2}\,{v^8}}\over {27}} + 
     {{1177\,{v^8}\,\ln (2)}\over {945}} + 
     {{1177\,{v^8}\,\ln (v)}\over {945}} \biggr), \\
\eta_{2,2,1}&&=e^2\biggl(
{{729}\over {64}} - {{3645\,{v^2}}\over {64}} + 
  {{2187\,\pi \,{v^3}}\over {32}} + 
  {{24057\,{v^4}}\over {256}} - 
  {{6561\,\pi \,{v^5}}\over {16}} + 
  {{9067629321\,{v^6}}\over {3449600}} \nonumber \\ 
&&- 
  {{234009\,{\gamma}\,{v^6}}\over {560}} + 
  {{2187\,{{\pi }^2}\,{v^6}}\over {16}} + 
  {{102789\,\pi \,{v^7}}\over {128}} - 
  {{545827954239\,{v^8}}\over {44844800}} \nonumber \\ 
&&+ 
  {{234009\,{\gamma}\,{v^8}}\over {80}} - 
  {{15309\,{{\pi }^2}\,{v^8}}\over {16}} - 
  {{234009\,{v^6}\,\ln (2)}\over {560}} + 
  {{234009\,{v^8}\,\ln (2)}\over {80}} \nonumber \\
&&- 
  {{234009\,{v^6}\,\ln (3)}\over {560}} + 
  {{234009\,{v^8}\,\ln (3)}\over {80}} \nonumber \\
&&- 
  {{234009\,{v^6}\,\ln (v)}\over {560}} + 
  {{234009\,{v^8}\,\ln (v)}\over {80}} \biggr) ,\\
\eta_{2,2,-1}&&=e^2\biggl( 
{9\over {64}} + {{1041\,{v^2}}\over {448}} + 
  {{9\,\pi \,{v^3}}\over {32}} + 
  {{2224681\,{v^4}}\over {112896}} + 
  {{615\,\pi \,{v^5}}\over {112}} + 
  {{11918100443\,{v^6}}\over {93139200}}  \nonumber \\
&&- 
  {{321\,{\gamma}\,{v^6}}\over {560}}+ 
  {{3\,{{\pi }^2}\,{v^6}}\over {16}} + 
  {{3083119\,\pi \,{v^7}}\over {56448}} + 
  {{181180743580847\,{v^8}}\over {228843014400}} - 
  {{50611\,{\gamma}\,{v^8}}\over {3920}}\nonumber \\
&&  + 
  {{473\,{{\pi }^2}\,{v^8}}\over {112}} - 
  {{321\,{v^6}\,\ln (2)}\over {560}} - 
  {{50611\,{v^8}\,\ln (2)}\over {3920}}\nonumber \\
&& - 
  {{321\,{v^6}\,\ln (v)}\over {560}} - 
  {{50611\,{v^8}\,\ln (v)}\over {3920}} \biggr) ,\\
\eta_{2,1,1}&&=e^2\biggl(
{{4\,{v^2}}\over 9} - {{172\,{v^4}}\over {63}} + 
  {{16\,\pi \,{v^5}}\over 9} + {{10300\,{v^6}}\over {1323}} - 
  {{856\,\pi \,{v^7}}\over {63}} + 
  {{145520063\,{v^8}}\over {3274425}} \nonumber \\
&&- {{6848\,{\gamma}\,{v^8}}\over {945}} + 
  {{64\,{{\pi }^2}\,{v^8}}\over {27}} - 
  {{13696\,{v^8}\,\ln (2)}\over {945}} - 
  {{6848\,{v^8}\,\ln (v)}\over {945}} \biggr) ,\\
\eta_{2,0,1}&&=e^2\biggl(
{1\over {96}} - {{145\,{v^2}}\over {672}} + 
  {{\pi \,{v^3}}\over {48}} + 
  {{282521\,{v^4}}\over {169344}} - 
  {{83\,\pi \,{v^5}}\over {168}} - 
  {{776764633\,{v^6}}\over {139708800}} - 
  {{107\,{\gamma}\,{v^6}}\over {2520}} \nonumber \\
&&+{{{{\pi }^2}\,{v^6}}\over {72}} + 
  {{384203\,\pi \,{v^7}}\over {84672}} + 
  {{6362456713\,{v^8}}\over {3467318400}} + 
  {{20009\,{\gamma}\,{v^8}}\over {17640}} - 
  {{187\,{{\pi }^2}\,{v^8}}\over {504}} \nonumber \\
&&-{{107\,{v^6}\,\ln (2)}\over {2520}}  + 
  {{20009\,{v^8}\,\ln (2)}\over {17640}} - 
  {{107\,{v^6}\,\ln (v)}\over {2520}} + 
  {{20009\,{v^8}\,\ln (v)}\over {17640}} \biggr). 
\end{eqnarray}
For $\ell=3$, 
\begin{eqnarray}
\eta_{3,3,0}&&=
{{1215\,{v^2}}\over {896}} - {{1215\,{v^4}}\over {112}} + 
  {{3645\,\pi \,{v^5}}\over {448}} + 
  {{243729\,{v^6}}\over {9856}} - 
  {{3645\,\pi \,{v^7}}\over {56}} + 
  {{25037019729\,{v^8}}\over {125565440}} \nonumber \\
&&-{{47385\,{\gamma}\,{v^8}}\over {1568}} + 
  {{3645\,{{\pi }^2}\,{v^8}}\over {224}} - 
  {{47385\,{v^8}\,\ln (2)}\over {1568}} - 
  {{47385\,{v^8}\,\ln (3)}\over {1568}} - 
  {{47385\,{v^8}\,\ln (v)}\over {1568}} \nonumber \\
&&+{{{e}}^2}\,\biggl( {{-10935\,{v^2}}\over {448}} + 
     {{37665\,{v^4}}\over {256}} - 
     {{142155\,\pi \,{v^5}}\over {896}} - 
     {{4428189\,{v^6}}\over {9856}} + 
     {{455625\,\pi \,{v^7}}\over {448}} \nonumber \\ 
&&-{{81628707987\,{v^8}}\over {17937920}} + 
     {{142155\,{\gamma}\,{v^8}}\over {224}} - 
     {{10935\,{{\pi }^2}\,{v^8}}\over {32}} \nonumber \\
&&+{{142155\,{v^8}\,\ln (2)}\over {224}} + 
     {{142155\,{v^8}\,\ln (3)}\over {224}} + 
     {{142155\,{v^8}\,\ln (v)}\over {224}} \biggr), \\
\eta_{3,2,0}&&=
{{5\,{v^4}}\over {63}} - {{193\,{v^6}}\over {567}} + 
  {{20\,\pi \,{v^7}}\over {63}} + 
  {{86111\,{v^8}}\over {280665}} \nonumber \\ 
&& +{{{e}}^2}\,\biggl( {{-65\,{v^4}}\over {63}} + 
     {{947\,{v^6}}\over {189}} - 
     {{290\,\pi \,{v^7}}\over {63}} - 
     {{442003\,{v^8}}\over {40095}} \biggr), \\
\eta_{3,1,0}&&=
{{{v^2}}\over {8064}} - {{{v^4}}\over {1512}} + 
  {{\pi \,{v^5}}\over {4032}} + {{437\,{v^6}}\over {266112}} - 
  {{\pi \,{v^7}}\over {756}} - 
  {{1137077\,{v^8}}\over {50854003200}} \nonumber \\ 
&& -{{13\,{\gamma}\,{v^8}}\over {42336}} + 
  {{{{\pi }^2}\,{v^8}}\over {6048}} - 
  {{13\,{v^8}\,\ln (2)}\over {42336}} - 
  {{13\,{v^8}\,\ln (v)}\over {42336}} \nonumber \\
&& + {{{e}}^2}\,\biggl( {{-{v^2}}\over {4032}} + 
     {{65\,{v^4}}\over {16128}} - 
     {{\pi \,{v^5}}\over {1152}} - 
     {{11063\,{v^6}}\over {798336}} + 
     {{5\,\pi \,{v^7}}\over {448}} + 
     {{614545391\,{v^8}}\over {30512401920}} \nonumber \\
&&+{{65\,{\gamma}\,{v^8}}\over {42336}} - 
     {{5\,{{\pi }^2}\,{v^8}}\over {6048}} + 
     {{65\,{v^8}\,\ln (2)}\over {42336}} + 
     {{65\,{v^8}\,\ln (v)}\over {42336}} \biggr),\\
\eta_{3,3,1}&&
=e^2\biggl(
{{640\,{v^2}}\over {21}} - {{46720\,{v^4}}\over {189}} + 
  {{5120\,\pi \,{v^5}}\over {21}} + 
  {{2135648\,{v^6}}\over {2673}} - 
  {{408320\,\pi \,{v^7}}\over {189}} \nonumber \\
&& + {{485145507664\,{v^8}}\over {59594535}} - 
  {{532480\,{\gamma}\,{v^8}}\over {441}} + 
  {{40960\,{{\pi }^2}\,{v^8}}\over {63}} \nonumber \\
&& -{{532480\,{v^8}\,\ln (2)}\over {441}} - 
  {{532480\,{v^8}\,\ln (4)}\over {441}} - 
  {{532480\,{v^8}\,\ln (v)}\over {441}} \biggr),\\
\eta_{3,3,-1}&&
=e^2 \biggl(
{{15\,{v^2}}\over {14}} + {{1055\,{v^4}}\over {126}} + 
  {{30\,\pi \,{v^5}}\over 7} + 
  {{1530967\,{v^6}}\over {37422}} + 
  {{2515\,\pi \,{v^7}}\over {63}} \nonumber \\
&& +{{5409381833\,{v^8}}\over {19864845}} - 
  {{520\,{\gamma}\,{v^8}}\over {49}} + 
  {{40\,{{\pi }^2}\,{v^8}}\over 7} - 
  {{1040\,{v^8}\,\ln (2)}\over {49}} - 
  {{520\,{v^8}\,\ln (v)}\over {49}}\biggr),\\
\eta_{3,2,1}&&
=e^2\biggl(
{{3645\,{v^4}}\over {1792}} - {{13851\,{v^6}}\over {896}} + 
  {{10935\,\pi \,{v^7}}\over {896}} + 
  {{2808837\,{v^8}}\over {49280}} \biggr),\\
\eta_{3,2,-1}&&
=e^2\biggl(
{{5\,{v^4}}\over {1792}} + {{1609\,{v^6}}\over {24192}} + 
  {{5\,\pi \,{v^7}}\over {896}} + 
  {{4153669\,{v^8}}\over {5132160}} \biggr),\\
\eta_{3,1,1}&&
=e^2\biggl(
{{{v^2}}\over {126}} - {{23\,{v^4}}\over {126}} + 
  {{2\,\pi \,{v^5}}\over {63}} + {{6437\,{v^6}}\over {4158}} - 
  {{7\,\pi \,{v^7}}\over 9} - 
  {{1688449328\,{v^8}}\over {297972675}} \nonumber \\
&& -{{104\,{\gamma}\,{v^8}}\over {1323}} + 
  {{8\,{{\pi }^2}\,{v^8}}\over {189}} - 
  {{208\,{v^8}\,\ln (2)}\over {1323}} - 
  {{104\,{v^8}\,\ln (v)}\over {1323}} \biggr),\\
\eta_{3,0,1}&&
= e^2\biggl(
{{{v^4}}\over {2688}} - {{55\,{v^6}}\over {6048}} + 
  {{\pi \,{v^7}}\over {1344}} + 
  {{203417\,{v^8}}\over {2395008}} \biggr). 
\end{eqnarray}
For $\ell=4$, 
\begin{eqnarray}
\eta_{4,4,0}&&=
{{1280\,{v^4}}\over {567}} - {{151808\,{v^6}}\over {6237}} + 
  {{10240\,\pi \,{v^7}}\over {567}} + 
  {{560069632\,{v^8}}\over {6243237}} \nonumber \\
&&+{{{e}}^2}\,\biggl( {{-37120\,{v^4}}\over {567}} + 
     {{3284224\,{v^6}}\over {6237}} - 
     {{312320\,\pi \,{v^7}}\over {567}} - 
     {{60497401856\,{v^8}}\over {31216185}} \biggr),\\
\eta_{4,3,0}&&=
{{729\,{v^6}}\over {4480}} - {{28431\,{v^8}}\over {24640}} + 
  {{{e}}^2}\,\biggl( {{-2187\,{v^6}}\over {640}} + 
     {{2062341\,{v^8}}\over {98560}} \biggr), \\
\eta_{4,2,0}&&=
{{5\,{v^4}}\over {3969}} - {{437\,{v^6}}\over {43659}} + 
  {{20\,\pi \,{v^7}}\over {3969}} + 
  {{7199152\,{v^8}}\over {218513295} }\nonumber \\
&&+ {{{e}}^2}\,\biggl( {{-25\,{v^4}}\over {3969}} + 
     {{2743\,{v^6}}\over {43659}} - 
     {{130\,\pi \,{v^7}}\over {3969}} - 
     {{27410\,{v^8}}\over {77077}} \biggr),\\
\eta_{4,1,0}&&=
{{{v^6}}\over {282240}} - {{101\,{v^8}}\over {4656960}} + 
  {{{e}}^2}\,\biggl( {{-{v^6}}\over {56448}} + 
     {{4951\,{v^8}}\over {18627840}} \biggr), \\
\eta_{4,4,1}&&=e^2 \biggl(
{{48828125\,{v^4}}\over {580608}} - 
  {{11767578125\,{v^6}}\over {12773376}} + 
  {{244140625\,\pi \,{v^7}}\over {290304}} \cr
&&\hskip 1cm + 
  {{3640732421875\,{v^8}}\over {874266624}}\biggr),\\
\eta_{4,4,-1}&&=e^2 \biggl(
{{32805\,{v^4}}\over {7168}} + 
  {{321489\,{v^6}}\over {22528}} + 
  {{98415\,\pi \,{v^7}}\over {3584}} + 
  {{251783118603\,{v^8}}\over {6314147840}}\biggr),\\
\eta_{4,3,1}&&=e^2 \biggl(
{{2048\,{v^6}}\over {315}} - {{216064\,{v^8}}\over {3465}}\biggr),\\
\eta_{4,3,-1}&&=e^2 \biggl(
{{2\,{v^6}}\over {35}} + {{24\,{v^8}}\over {35}}\biggr),\\
\eta_{4,2,1}&&=e^2 \biggl(
{{295245\,{v^8}}\over {495616}}\biggr),\\
\eta_{4,2,-1}&&=e^2 \biggl(
{{5\,{v^4}}\over {254016}} + {{5501\,{v^6}}\over {11176704}} + 
  {{5\,\pi \,{v^7}}\over {127008}} + 
  {{5568858419\,{v^8}}\over {895030456320}}\biggr),\\
\eta_{4,1,1}&&=e^2 \biggl(
{{2\,{v^6}}\over {2205}} - {{28\,{v^8}}\over {1485}} \biggr),\\
\eta_{4,0,1}&&=e^2 \biggl(
{{{v^4}}\over {225792}} - {{113\,{v^6}}\over {709632}} + 
  {{\pi \,{v^7}}\over {112896}} + 
  {{1469592037\,{v^8}}\over {596686970880}} \biggr).
\end{eqnarray}
For $\ell=5$, 
\begin{eqnarray}
\eta_{5,5,0}&&=
{{9765625\,{v^6}}\over {2433024}} - 
  {{2568359375\,{v^8}}\over {47443968}} \cr
&&\hskip 2cm + 
  {{{e}}^2}\,\biggl( {{-419921875\,{v^6}}\over 
       {2433024}} + {{336376953125\,{v^8}}\over {189775872}} \biggr),\\
\eta_{5,4,0}&&=
{{4096\,{v^8}}\over {13365}} - 
  {{131072\,{{{e}}^2}\,{v^8}}\over {13365}},\\
\eta_{5,3,0}&&=
{{2187\,{v^6}}\over {450560}} - 
  {{150903\,{v^8}}\over {2928640}} + 
  {{{e}}^2}\,\biggl( {{-2187\,{v^6}}\over {40960}} + 
     {{5736501\,{v^8}}\over {11714560}} \biggr),\\
\eta_{5,2,0}&&=
{{4\,{v^8}}\over {40095}} - 
  {{32\,{{{e}}^2}\,{v^8}}\over {40095}},\\
\eta_{5,1,0}&&=
{{{v^6}}\over {127733760}} - 
  {{179\,{v^8}}\over {2490808320}} + 
  {{{e}}^2}\,\biggl( {{{v^6}}\over {25546752}} - 
     {{43\,{v^8}}\over {9963233280}} \biggr), \\
\eta_{5,5,1}&&=e^2\biggl(
{{19683\,{v^6}}\over {88}} - {{17498187\,{v^8}}\over {5720}}\biggr),\\
\eta_{5,5,-1}&&=e^2\biggl(
{{512\,{v^6}}\over {33}} - {{54272\,{v^8}}\over {6435}}\biggr),\\
\eta_{5,4,1}&&=e^2\biggl(
{{48828125\,{v^8}}\over {2737152}}\biggr),\\
\eta_{5,4,-1}&&=e^2\biggl(
{{19683\,{v^8}}\over {56320}}\biggr),\\
\eta_{5,3,1}&&=e^2\biggl(
{{512\,{v^6}}\over {13365}} - {{16384\,{v^8}}\over {173745}}\biggr),\\
\eta_{5,3,-1}&&=e^2\biggl(
{{11\,{v^6}}\over {9720}} + {{13\,{v^8}}\over {1080}}\biggr), \\
\eta_{5,2,-1}&&=e^2\biggl(
{{{v^8}}\over {2566080}}\biggr),\\
\eta_{5,1,1}&&=e^2\biggl(
{{{v^6}}\over {13860}} - {{1091\,{v^8}}\over {540540}}\biggr),\\
\eta_{5,0,1}&&=e^2\biggl(
{{{v^8}}\over {9580032}} \biggr).
\end{eqnarray}
Note that up to $O(v^8)$, $\eta_{5,2,1}$ does not appear. 

\noindent
For $\ell=6$, 
\begin{eqnarray}
\eta_{6,6,0}&&=
{{26244\,{v^8}}\over {3575}} - 
  {{314928\,{{{e}}^2}\,{v^8}}\over {715}},\hskip 7cm \\
\eta_{6,4,0}&&=
{{131072\,{v^8}}\over {9555975}} - 
  {{524288\,{{{e}}^2}\,{v^8}}\over {1911195}},\\
\eta_{6,2,0}&&=
{{4\,{v^8}}\over {5733585}} + 
  {{16\,{{{e}}^2}\,{v^8}}\over {5733585}},\\
\eta_{6,6,1}&&=e^2
{{678223072849\,{v^8}}\over {1186099200}},\\
\eta_{6,6,-1}&&=e^2
{{244140625\,{v^8}}\over {5271552}},\\
\eta_{6,4,1}&&=e^2
{{244140625\,{v^8}}\over {782825472}},\\
\eta_{6,4,-1}&&=e^2
{{964467\,{v^8}}\over {80537600}},\\
\eta_{6,2,1}&&=e^2
{{6561\,{v^8}}\over {32215040}},\\
\eta_{6,2,-1}&&=e^2
{{5\,{v^8}}\over {4696952832}}, \\
\eta_{6,0,1}&&=e^2
{{{v^8}}\over 2739889152}. 
\end{eqnarray}
Note that up to $O(v^8)$, $\eta_{6,5,k}$, $\eta_{6,3,k}$ and 
$\eta_{6,1,k}$ do not appear. 

Now, we define $\eta_{\ell}^{(0)}$, $\eta_{\ell}^{(2)}$, 
and $\eta_{z\ell}^{(2)}$ as 
\begin{eqnarray}
&&\left\langle{dE \over dt}\right\rangle
=\left({dE\over dt}\right)_{\rm N}
\sum_{\ell} \biggl(\eta_{\ell}^{(0)}+\eta_{\ell}^{(2)}e^2\biggr), \\
&&\left\langle{dJ_z \over dt}\right\rangle
={1\over \Omega}\left({dE\over dt}\right)_{\rm N}
\sum_{\ell} \biggl(\eta_{\ell}^{(0)}+\eta_{z \ell}^{(2)}e^2\biggr). 
\end{eqnarray}
Then, 
\begin{eqnarray}
\eta_2^{(0)} &&=1 - {{1277\,{v^2}}\over {252}} + 4\,\pi \,{v^3} + 
  {{37915\,{v^4}}\over {10584}} - 
  {{2561\,\pi \,{v^5}}\over {126}} + 
  {{76187\,\pi \,{v^7}}\over {5292}} \nonumber \\
&&+ 
  {v^6}\,\biggl( {{2116278473}\over {23284800}} - 
     {{1712\,{\gamma}}\over {105}} + 
     {{16\,{{\pi }^2}}\over 3} - 
     {{3424\,\ln (2)}\over {105}} - 
     {{1712\,\ln (v)}\over {105}} \biggr) \nonumber \\
&&+ 
  {v^8}\,\biggl( -{{2455920939443}\over {7151344200}} + 
     {{548803\,{\gamma}}\over {6615}} - 
     {{5129\,{{\pi }^2}}\over {189}} \cr
&& \hskip 1cm + 
     {{219671\,\ln (2)}\over {1323}} + 
     {{548803\,\ln (v)}\over {6615}} \biggr),\\
\eta_2^{(2)}&& =
   {{37}\over {24}} - 
     {{2581\,{v^2}}\over {252}} + 
     {{1087\,\pi \,{v^3}}\over {48}} + 
     {{346561\,{v^4}}\over {21168}} - 
     {{29857\,\pi \,{v^5}}\over {168}} + 
     {{35639309\,\pi \,{v^7}}\over {84672}} \nonumber \\
&&+ 
     {v^6}\,\biggl( {{93579660049}\over {69854400}} - 
        {{65056\,{\gamma}}\over {315}} + 
        {{608\,{{\pi }^2}}\over 9} + 
        {{1712\,\ln (2)}\over {315}} \cr
&& \hskip 1cm - 
        {{234009\,\ln (3)}\over {560}} -
        {{65056\,\ln (v)}\over {315}} \biggr)\nonumber \\
&&   + 
     {v^8}\,\biggl( -{{3395552510663}\over {416078208}} + 
        {{4730363\,{\gamma}}\over {2646}} - 
        {{221045\,{{\pi }^2}}\over {378}} \nonumber \\
&& \hskip 1cm  +
        {{2914573\,\ln (2)}\over {4410}} + 
        {{234009\,\ln (3)}\over {80}} + 
        {{4730363\,\ln (v)}\over {2646}} \biggr),\\
\eta_3^{(0)}&& =
{{1367\,{v^2}}\over {1008}} - {{32567\,{v^4}}\over {3024}} + 
  {{16403\,\pi \,{v^5}}\over {2016}} + 
  {{152122\,{v^6}}\over {6237}} - 
  {{13991\,\pi \,{v^7}}\over {216}} \nonumber \\
&& +
  {v^8}\,\biggl( {{5712521850527}\over {28605376800}} - 
     {{79963\,{\gamma}}\over {2646}} + 
     {{6151\,{{\pi }^2}}\over {378}} -
     {{79963\,\ln (2)}\over {2646}} \nonumber \\ 
&&\hskip 3cm - 
     {{47385\,\ln (3)}\over {1568}} - 
     {{79963\,\ln (v)}\over {2646}} \biggr) \\
\eta_3^{(2)}&&=
   {{1801\,{v^2}}\over {252}} - 
     {{78509\,{v^4}}\over {864}} + 
     {{40083\,\pi \,{v^5}}\over {448}} + 
     {{8163047\,{v^6}}\over {21384}} - 
     {{1894997\,\pi \,{v^7}}\over {1728}} \nonumber \\
&& + 
     {v^8}\,\biggl( {{446664927141403}\over {114421507200}} - 
        {{771979\,{\gamma}}\over {1323}} + 
        {{59383\,{{\pi }^2}}\over {189}} - 
        {{87347\,\ln (2)}\over {147}}  \nonumber \\
&&\hskip 2cm +
        {{142155\,\ln (3)}\over {224}} - 
        {{532480\,\ln (4)}\over {441}} - 
        {{771979\,\ln (v)}\over {1323}} \biggr),\\
\eta_4^{(0)}&&=
{{8965\,{v^4}}\over {3969}} - 
  {{84479081\,{v^6}}\over {3492720}} + 
  {{23900\,\pi \,{v^7}}\over {1323}} + 
  {{51619996697\,{v^8}}\over {582702120}},\\
\eta_4^{(2)}&& =
   {{2946739\,{v^4}}\over {127008}} - 
     {{58555205\,{v^6}}\over {155232}} + 
     {{107560723\,\pi \,{v^7}}\over {338688}} + 
     {{5187619686371\,{v^8}}\over {2330808480}},\\
\eta_5^{(0)}&&=
{{1002569\,{v^6}}\over {249480}} - 
  {{3145396841\,{v^8}}\over {58378320}},\\ 
\eta_5^{(2)}&&= {{2491525\,{v^6}}\over {37422}} - 
     {{150181214159\,{v^8}}\over {116756640}} ,\\
\eta_6^{(0)}&&=
{{210843872\,{v^8}}\over {28667925}},\\
\eta_6^{(2)}&&=  {142651028551 ~v^8 \over 802701900 },
\end{eqnarray}
and
\begin{eqnarray}
\eta_{z2}^{(2)}&&=
 -{5\over 8} + 
     {{137\,{v^2}}\over {24}} + {{49\,\pi \,{v^3}}\over 8} - 
     {{235675\,{v^4}}\over {14112}} - 
     {{20437\,\pi \,{v^5}}\over {504}} + 
     {{883609\,\pi \,{v^7}}\over {14112}} \nonumber \\
&&+ 
     {v^6}\,\biggl( {{303218627}\over {470400}} - 
        {{19367\,{\gamma}}\over {210}} + 
        {{181\,{{\pi }^2}}\over 6} + 
        {{20009\,\ln (2)}\over {210}} - 
        {{78003\,\ln (3)}\over {280}} -
        {{19367\,\ln (v)}\over {210}} \biggr) \nonumber \\
&&+ 
     {v^8}\,\biggl( -{{1553872210987}\over {488980800}} + 
        {{3197053\,{\gamma}}\over {4410}} - 
        {{29879\,{{\pi }^2}}\over {126}} \nonumber \\ 
&& \hskip 2cm -
        {{2649641\,\ln (2)}\over {13230}} + 
        {{234009\,\ln (3)}\over {140}} + 
        {{3197053\,\ln (v)}\over {4410}} \biggr), \\
\eta_{z3}^{(2)}&&=
   {{67\,{v^2}}\over {32}} - 
     {{66497\,{v^4}}\over {2016}} + 
     {{43193\,\pi \,{v^5}}\over {1008}} + 
     {{2711543\,{v^6}}\over {18144}} - 
     {{2203487\,\pi \,{v^7}}\over {4032}} \nonumber \\ 
&&+ 
     {v^8}\,\biggl( {{321229428757}\over {133358400}} - 
        {{586079\,{\gamma}}\over {1764}} + 
        {{45083\,{{\pi }^2}}\over {252}} - 
        {{1842685\,\ln (2)}\over {5292}}  \nonumber \\
&&\hskip 2cm + 
        {{1848015\,\ln (3)}\over {3136}} - 
        {{133120\,\ln (4)}\over {147}} - 
        {{586079\,\ln (v)}\over {1764}} \biggr),\\
\eta_{z4}^{(2)}&&=
 {{478195\,{v^4}}\over {42336}} - 
     {{64132457\,{v^6}}\over {317520}} + 
     {{5239355\,\pi \,{v^7}}\over {28224}} + 
     {{670696042069\,{v^8}}\over {537878880}}, \\
\eta_{z5}^{(2)}&&=
 {{1778041\,{v^6}}\over {45360}} - 
     {{300193429\,{v^8}}\over {374220}},\\
\eta_{z6}^{(2)}&&= 
  {{113949013\,{{{e}}^2}\,{v^8}}\over {980100}}. 
\end{eqnarray} 
\section{Asymptotic amplitudes and $R^{\rm up}$}
\label{ap:absfunc}

In this Appendix, we show the asymptotic amplitudes $B^{\rm inc}$,
$B^{\rm trans}$ and $C^{\rm trans}$ and the post-Newtonian expansion of
$R^{\rm up}$ which are used in section \ref{sec:bhabs} to evaluate
the black hole absorption rate to $O(v^{13})$ relative to the quadrupole
energy flux at infinity in section.

\noindent
(a) $\ell=2$

\begin{eqnarray}
B^{\rm inc}&&=
{1\over \omega}{1\over \kappa^4 \epsilon^4}
e^{{1\over 2}\,\pi i \,\left(\nu+3\right)} e^{i \epsilon \kappa} 
e^{-i\epsilon\ln\epsilon}\nonumber\\
&&
\Biggl\{ {\frac {15}{4}}
+\Biggl(-{15\over 2}\,{i}\gamma-{\frac {25}{12
}}\,mq-{\frac {15}{8}}\,\pi -{\frac {15}{4}}\,{i}{\psi^{(0)}}(3+{\frac 
{{i}mq}{\kappa}})-{\frac {15}{4}}\,{i}\ln (2)+{\frac {
125}{16}}\,{i}\Biggr){\epsilon} \nonumber\\
&&
+\Biggl({\frac {1089}{56}}\,\gamma+{\frac {725}{2352}}\,{\kappa}^{2}
+{\frac {1089}{112}}\,\ln (2)-{15\over 2}\,{\gamma}^{2}
+{\frac {12625}{21168}}\,{m}^{2}{q}^{2}
+{\frac {35}{32}}\,{\pi }^{2} \nonumber\\
&&
-{\frac {535}{144}}\,{i}mq
+{\frac {15}{4}}\,{i}\gamma\,\pi -{\frac {125}{32}}\,{i}\pi 
+{\frac {25}{6}}\,{i}\gamma\,mq+{\frac {25}{24}}\,m\pi \,q
-{15\over 2}\,\gamma\,\ln (2) \nonumber\\
&&
-{15\over 2}\,\gamma\,{\psi^{(0)}}(3+{\frac {{i}mq}{\kappa}})
-{\frac {15}{4}}\,\ln (2){\psi^{(0)}}(3+{\frac {{i}mq}{\kappa}})
+{\frac {107}{56}}\,\ln ({\epsilon})+{\frac {107}{56}}\,
\ln (\kappa) \nonumber\\
&&
-{\frac {20573}{960}}-{\frac {15}{8}}\,\left (\ln (2)\right )^{2}
+{\frac {1089}{112}}\,{\psi^{(0)}}(3+{\frac {{i}mq}{\kappa}})
-{\frac {15}{8}}\,
  \left ({\psi^{(0)}}(3+{\frac {{i}mq}{\kappa}})\right )^{2} \nonumber\\
&&
-{\frac {15}{8}}\,\psi^{(1)}(3+{\frac {{i}mq}{\kappa}})
+{\frac {25}{12}}\,{i}mq\ln (2) 
+{\frac {25}{12}}\,{i}mq{\psi^{(0)}}(3+{\frac {{i}mq}{\kappa}}) \nonumber\\
&&
+{\frac {15}{8}}\,{i}\pi \,\ln (2)
+{\frac {15}{8}}\,{i}\pi \,{\psi^{(0)}}(3+{\frac {{i}mq}{\kappa}})
-{\frac {15}{4}}\,\psi^{(1)}(3+{\frac {{i}mq}{\kappa}}){\kappa}^{-1}
\Biggr){{\epsilon}}^{2} \Biggr\}.
\\
B^{\rm trans}&&=\left(\omega\over \epsilon\kappa\right)^{4}
e^{i\epsilon_{+}\ln\kappa}
\Biggl\{
1+\left ({5\over 6}\,{i}\kappa-{\frac {5}{18}}\,mq\right )\epsilon 
\nonumber\\
&&
+\left ({\frac {325}{7938}}\,{m}^{2}{q}^{2}+{\frac {5}{18}}-{\frac {
15}{49}}\,{\kappa}^{2}-{\frac {85}{378}}\,{i}\kappa\,mq\right )
{\epsilon}^{2} \Biggr\}.
\\
C^{\rm trans}&&=\omega^3 2^{i\epsilon} e^{-{\pi\over 2}i(\nu-1)}
e^{i\epsilon\ln\epsilon}
\Biggl\{
2+\left (-{1\over 3}\,{i}\kappa-\pi +{1\over 9}\,mq\right )
\epsilon \nonumber\\
&&
+\Biggl({1\over 4}\,{\pi }^{2}-{\frac {1}{189}}\,{i}\kappa\,mq
+{\frac {2}{49}}\,{q}^{2}-{1\over 3}\,{i}mq
-{\frac {11}{3969}}\,{m}^{2}{q}^{2} \nonumber\\
&&
-{1\over 18}\,m\pi \,q+{1\over 6}\,{i}\kappa\,\pi -{\frac {67}{441}}
\Biggr){\epsilon}^{2} \Biggr\}.
\\
R^{\rm up}&&=
-3\,{\frac {{i}}{z}}-3+{3\over 2}\,{i}z+{1\over 2}\,{z}^{2}
-{1\over 8}\,{i}{z}^{3}
+{\frac {31}{40}}\,{z}^{4}+{\frac {43}{80}}\,{i}{z}^{5}
-{\frac {117}{560}}\,{z}^{6}-{\frac {769}{13440}}\,{i}{z}^{7}
\nonumber\\
&&
+\epsilon\,\Biggl(
\Bigl({-{3\over 2}\,{i}+mq}\Bigr){1\over {z}^{2}} 
+\left({-3\,\gamma+\kappa+3\,{i}\pi -{1\over 6}\,{i}mq}\right){1\over
{z}} 
-{9\over 4}\,{i}+3\,{i}\gamma
\nonumber\\
&&
-{i}\kappa+3\,\pi +{1\over 3}\,mq
+\Bigl (-{5\over 2}+{3\over 2}\,\gamma-{1\over 2}\,\kappa-{3\over
2}\,{i}\pi 
-{1\over 4}\,{i}mq\Bigr )z 
\nonumber\\
&&
+\Bigl ({\frac {85}{48}}\,{i}-{1\over 2}\,{i}\gamma
+{1\over 6}\,{i}\kappa-{1\over 2}\,\pi -{\frac {7}{72}}\,mq\Bigr
){z}^{2} 
\nonumber\\
&&
+\Bigl (-{\frac {13}{60}}-{1\over 8}\,\gamma+{1\over 24}\,\kappa+{1\over 
8}\,{i}\pi 
-{\frac {269}{720}}\,{i}mq\Bigr ){z}^{3}
\nonumber\\
&&
+\Bigl (-{\frac {1559}{2400}}\,{i}+{1\over 40}\,{i}\gamma
+{\frac {31}{120}}\,{i}\kappa-{3\over 8}\,\pi +{\frac {49}{180}}\,mq
+{4\over 5}\,{i}\ln (2)+{4\over 5}\,{i}\ln (z)\Bigr ){z}^{4}\Biggr) \nonumber\\
&&
+{\epsilon}^{2}\Biggl(
\Bigl(-{3\over 4}\,{i}-{\frac {3}{28}}\,{i}{\kappa}^{2}+mq
+{\frac {11}{56}}\,{i}{m}^{2}{q}^{2}\Bigr){z}^{-3} 
+\Bigl(-1-{3\over 2}\,\gamma+{1\over 2}\,\kappa+{1\over
14}\,{\kappa}^{2}
\nonumber\\
&&
+{3\over 2}\,{i}\pi 
-{\frac {7}{12}}\,{i}mq-{i}\gamma\,mq+{1\over 3}\,{i}\kappa\,mq 
-m\pi \,q-{\frac {31}{504}}\,{m}^{2}{q}^{2}\Bigr){z}^{-2} 
\nonumber\\
&&
+\Bigl({\frac {183}{28}}\,{i}-{\frac {107}{70}}\,{i}\gamma 
+{3\over 2}\,{i}{\gamma}^{2}-{i}\gamma\,\kappa
-{\frac {4}{49}}\,{i}{\kappa}^{2}+3\,\gamma\,\pi 
-\kappa\,\pi -{7\over 4}\,{i}{\pi }^{2}
-{1\over 4}\,mq
\nonumber\\
&&
-{1\over 6}\,\gamma\,mq
+{\frac {19}{252}}\,\kappa\,mq+{1\over 6}\,{i}m\pi \,q 
+{\frac {781}{21168}}\,{i}{m}^{2}{q}^{2}-{\frac {107}{70}}\,{i}\ln (2)
\nonumber\\
&&
-{\frac {107}{70}}\,{i}\ln (z)\Bigr){z}^{-1}
+{\frac {1791}{280}}-{\frac {529}{140}}\,\gamma
+{3\over 2}\,{\gamma}^{2}+{3\over 4}\,\kappa-\gamma\,\kappa
-{\frac {11}{392}}\,{\kappa}^{2}+{9\over 4}\,{i}\pi 
\nonumber\\ &&
-3\,{i}\gamma\,\pi
+{i}\kappa\,\pi -{7\over 4}\,{\pi }^{2}
+{\frac {13}{24}}\,{i}mq-{1\over 3}\,{i}\gamma\,mq
+{\frac {23}{252}}\,{i}\kappa\,mq-{1\over 3}\,m\pi \,q 
\nonumber\\ &&
-{\frac {563}{21168}}\,{m}^{2}{q}^{2}
-{\frac {107}{70}}\,\ln (2)
-{\frac {107}{70}}\,\ln (z)
+\Bigl (-{\frac {13751}{3360}}\,{i} 
\nonumber\\ &&
+{\frac {457}{140}}\,{i}\gamma-{3\over 4}\,{i}{\gamma}^{2}
-{5\over 6}\,{i}\kappa+{1\over 2}\,{i}\gamma\,\kappa 
+{\frac {17}{4704}}\,{i}{\kappa}^{2}+{5\over 2}\,\pi 
\nonumber\\ &&
-{3\over 2}\,\gamma\,\pi +{1\over 2}\,\kappa\,\pi 
+{\frac {7}{8}}\,{i}{\pi }^{2}+{2\over 3}\,mq 
-{1\over 4}\,\gamma\,mq
\nonumber\\ &&
+{\frac {37}{504}}\,\kappa\,mq
+{1\over 4}\,{i}m\pi \,q+{\frac {751}{28224}}\,{i}{m}^{2}{q}^{2} 
+{\frac {107}{140}}\,{i}\ln (2)+{\frac {107}{140}}\,{i}\ln (z)\Bigr )z 
\Biggr) 
\nonumber\\ &&
+{\epsilon}^{3}\Biggl(\Bigl(-{3\over 8}\,{i}
-{\frac {9}{56}}\,{i}{\kappa}^{2}+{3\over 4}\,mq
+{\frac {15}{112}}\,{\kappa}^{2}mq
+{\frac {33}{112}}\,{i}{m}^{2}{q}^{2} \nonumber\\
&&
-{\frac {3}{112}}\,{m}^{3}{q}^{3}\Bigr){z}^{-4}
+\Bigl(-1-{3\over 4}\,\gamma+{1\over 4}\,\kappa-{1\over 7}\,{\kappa}^{2}
-{\frac {3}{28}}\,\gamma\,{\kappa}^{2}+{1\over 28}\,{\kappa}^{3} \nonumber\\
&&
+{3\over 4}\,{i}\pi +{\frac {3}{28}}\,{i}{\kappa}^{2}\pi 
-{\frac {157}{168}}\,{i}mq-{i}\gamma\,mq+{1\over 3}\,{i}\kappa\,mq
-{\frac {5}{84}}\,{i}{\kappa}^{2}mq-m\pi \,q \nonumber\\
&&
+{\frac {41}{504}}\,{m}^{2}{q}^{2}+{\frac
{11}{56}}\,\gamma\,{m}^{2}{q}^{2}
-{\frac {11}{168}}\,\kappa\,{m}^{2}{q}^{2}
-{\frac {11}{56}}\,{i}{m}^{2}\pi \,{q}^{2} \nonumber\\
&&
-{\frac {23}{1008}}\,{i}{m}^{3}{q}^{3}\Bigr){z}^{-3}
+\Bigl(-{1\over 3}\,{i}\kappa\,m\pi \,q
+{\frac {31}{504}}\,{i}\gamma\,{m}^{2}{q}^{2} \nonumber\\
&&
-{\frac {1}{72}}\,{i}\kappa\,{m}^{2}{q}^{2}+{i}\gamma\,m\pi \,q
-{1\over 2}\,{\gamma}^{2}mq+{\frac {7}{12}}\,m{\pi }^{2}q
+{\frac {31}{504}}\,{m}^{2}\pi \,{q}^{2}+{\frac {107}{210}}\,mq\ln (2)
\nonumber\\
&&
+{\frac {107}{210}}\,mq\ln (z)+{\frac {2335}{42336}}\,{i}{m}^{2}{q}^{2}
-{1\over 2}\,{i}\gamma\,\kappa+{1\over 3}\,\gamma\,\kappa\,mq
+{\frac {7}{12}}\,{i}m\pi \,q
-{\frac {25}{168}}\,{i}\kappa\,{q}^{2} \nonumber\\
&&
-{1\over 14}\,{i}\gamma\,{\kappa}^{2}-{\frac
{239}{14112}}\,{\kappa}^{2}mq+\pi 
+{\frac {1269}{280}}\,{i}+{3\over 2}\,\gamma\,\pi -{1\over
2}\,\kappa\,\pi 
-{1\over 14}\,{\kappa}^{2}\pi +{\frac {25}{504}}\,m{q}^{3} \nonumber\\
&&
+{\frac {69}{196}}\,{i}{\kappa}^{2}+{\frac {33}{140}}\,{i}\gamma
-{\frac {31}{168}}\,{i}\kappa-{\frac {107}{140}}\,{i}\ln (2)
-{\frac {107}{140}}\,{i}\ln (z)+{3\over 4}\,{i}{\gamma}^{2} \nonumber\\
&&
-{\frac {7}{8}}\,{i}{\pi }^{2}-{1\over 8}\,{i}{\kappa}^{3}
-{\frac {31}{420}}\,\gamma\,mq+{\frac {103}{504}}\,\kappa\,mq 
-{\frac {421}{210}}\,mq-{\frac
{239}{127008}}\,{m}^{3}{q}^{3}\Bigr){z}^{-2}
\Biggr) \nonumber\\
&&
+{\epsilon}^{4}\Bigl(\bigl(-{3\over 16}\,{i}-{\frac
{9}{56}}\,{i}{\kappa}^{2}
-{\frac {1}{112}}\,{i}{\kappa}^{4}+{1\over 2}\,mq
+{\frac {15}{56}}\,{\kappa}^{2}mq+{\frac {33}{112}}\,{i}{m}^{2}{q}^{2}
\nonumber\\
&&
+{\frac {11}{252}}\,{i}{\kappa}^{2}{m}^{2}{q}^{2}
-{\frac {3}{56}}\,{m}^{3}{q}^{3}
-{\frac {23}{8064}}\,{i}{m}^{4}{q}^{4}\Bigr ){z}^{-5}\Bigr) 
\end{eqnarray}.

\noindent
(b) $\ell=3$

\begin{eqnarray}
B^{\rm inc}&&=
{1\over \omega}{1\over \kappa^5 \epsilon^5}
e^{{1\over 2}\,\pi i \,\left(\nu+3\right)} e^{i \epsilon \kappa} 
e^{-i\epsilon\ln\epsilon}
\Biggl\{
{\frac {945}{2}}\,{\frac {\kappa}{3\,\kappa+{i}mq}} \nonumber\\
&&
-{\frac {63}{8}}\,{\frac {\kappa}{(3\,\kappa+{i}mq)^2}} 
\Biggl(90\,\kappa\,\pi +30\,{i}mq\pi 
+180\,{i}\kappa\,\ln (2)-60\,mq\ln (2)\nonumber\\
&&
+45\,\kappa\,mq 
+15\,{i}{m}^{2}{q}^{2}+360\,{i}\kappa\,\gamma
-120\,\gamma\,mq-591\,{i}\kappa
+197\,mq \nonumber\\
&&
+180\,{i}{\psi^{(0)}}({\frac {3\,\kappa+{i}mq}{\kappa}})\kappa 
-60\,{\psi^{(0)}}({\frac {3\,\kappa+{i}mq}{\kappa}})mq
-60\,{i}\Biggr)\epsilon\Biggl\}.
\\
B^{\rm trans}&&=
\left(\omega\over \epsilon\kappa\right)^{4}
e^{i\epsilon_{+}\ln\kappa}
\left (1+\left (-{\frac {11}{72}}\,mq+{2\over 3}\, i 
\kappa\right )\epsilon\right ).
\\
C^{\rm trans}&&=
{\omega}^{3}2^{i\epsilon} e^{-{\pi\over 2}i(\nu-1)}
e^{i\epsilon\ln\epsilon}
\left (2+\left (-\pi +{1\over 36}\,mq-{2\over 3}\,{i}\kappa
\right )\epsilon\right ).
\\
R^{\rm up}&&=
-45\,{\frac {{i}}{{z}^{2}}}-30\,{z}^{-1}+{15\over 2}\,{i}
+{5\over 8}\,{i}{z}^{2} \nonumber\\
&&
+\epsilon\,\Biggl(\Bigl({\frac {45}{4}}\,mq-45\,{i}\kappa
+90\,{i}\Bigl (-{1\over 2}+{1\over 2}\,\kappa\Bigr )\Bigr){z}^{-3} 
\nonumber\\
&&
+\Bigl(45\,{i}\pi +30+{15\over 2}\,\kappa-45\,\gamma
-{\frac {15}{8}}\,{i}mq\Bigr){z}^{-2}+\Bigl(30\,\pi -45\,{i} \nonumber\\
&&
+{\frac {35}{24}}\,mq+30\,{i}\gamma-5\,{i}\kappa\Bigr){z}^{-1}\Biggr) 
\nonumber\\
&&
+{\epsilon}^{2}\Biggl({\frac {135}{8}}\,\kappa\,mq-{
\frac {75}{2}}\,{i}{\kappa}^{2} \nonumber\\
&&
+135\,{i}\kappa\,\Bigl (-{1\over 2}+{1\over 2}\,\kappa\Bigr )
-{\frac {135}{4}}\,\Bigl (-{1\over 2}+{1\over 2}\,\kappa\Bigr )mq
+{\frac {15}{8}}\,{i}{m}^{2}{q}^{2} \nonumber\\
&&
-45\,{i}\Bigl (-\Bigl (-{1\over 2}+{1\over 2}\,\kappa\Bigr )^{2}
+2\,\Bigl (-1+\kappa\Bigr )\Bigl (
-{1\over 2}+{1\over 2}\,\kappa\Bigr )\Bigr )\Biggr){z}^{-4}.
\end{eqnarray}

\noindent
(c) $\ell=4$

\begin{eqnarray}
B^{\rm inc}&&=
{1\over \omega}{1\over \kappa^6 \epsilon^6}
e^{{1\over 2}\,\pi i \,\left(\nu+3\right)} e^{i \epsilon \kappa} 
e^{-i\epsilon\ln\epsilon}
\Bigl\{
79380\,{\frac {{\kappa}^{2}}{\left (3\,\kappa+{i}mq\right )
\left (4\,\kappa+{i}mq\right )}}
\Bigr\}.
\\
B^{\rm trans}&&=
\left(\omega\over \epsilon\kappa\right)^{4}
e^{i\epsilon_{+}\ln\kappa}
\left (1+O(\epsilon) \right ).
\\
C^{\rm trans}&&=
{\omega}^{3}2^{i\epsilon} e^{-{\pi\over 2}i(\nu-1)}
e^{i\epsilon\ln\epsilon}
\left (2+O(\epsilon)\right ).
\\
R^{\rm up}&&=-{630 i\over z^3}\,.
\end{eqnarray}
\section{Energy absorption by a Kerr black hole}
\label{ap:absorption}
Here we give the $(\ell,m)$-components of the energy absorption rate to
$O(v^8)$ beyond the lowest order for the Kerr black hole, that is
$O(v^{13})$ relative to the quadrupole luminosity at infinity.
\begin{eqnarray}
\eta^{H}_{2,2}&&=
-{1\over 4}\,q-{3\over 4}\,{q}^{3}+{v}^{2}\left (-{3\over 4}\,q
-{9\over 4}\,{q}^{3}\right )
\nonumber\\
&&
+\left({\frac {27}{4}}\,{q}^{2}+2\,q{B_2}+{\frac {15}{4}}\,{q}^{4}
+6\,{q}^{3}{B_2}+{13\over 2}\,\kappa\,{q}^{2}+3\,{q}^{4}\kappa
+{1\over 2}\,\kappa+{1\over 2}\right ){v}^{3}\nonumber\\
&&
+\left (-{\frac {199}{42}}\,q-{\frac {593}{42}}\,{q}^{3}
+{2\over 7}\,{q}^{5}\right ){v}^{4} \nonumber\\
&&
+\Bigl({\frac {721}{36}}\,{q}^{2}
+6\,q{B_2}+{\frac {127}{12}}\,{q}^{4}
+18\,{q}^{3}{B_2}+{\frac {39}{2}}
\,\kappa\,{q}^{2}+9\,{q}^{4}\kappa \nonumber\\
&&
+{3\over 2}\,\kappa+{3\over 2}\Bigr){v}^{5} \nonumber\\
&&
+\Bigl(-{\frac {607076}{11025}}\,q-4\,{B_2}+{\frac {428}{105}}\,
\gamma\,q+{2\over 3}\,{\pi }^{2}q+{\frac {428}{105}}\,q\ln 2 \nonumber\\
&&
-4\,q{C_2}-12\,{q}^{3}{C_2}
-36\,{q}^{4}{B_2}-56\,{q}^{2}{B_2}
+{\frac {428}{35}}\,{q}^{3}\gamma
+{\frac {428}{35}}\,{q}^{3}\ln 2\nonumber\\
&&
+2\,{q}^{3}{\pi }^{2}
+{\frac {428}{105}}\,q\ln\kappa+{\frac {428}{105}}\,q{A_2}
+{\frac {428}{35}}\,{q}^{3}\ln\kappa+6\,{\frac {{q}^{7}}{\kappa}}
+{\frac {428}{35}}\,{q}^{3}{A_2}\nonumber\\
&&
-8\,q{{B_2}}^{2}-24\,{q}^{3}{{B_2}}^{2}
+{\frac {856}{105}}\,q\ln v+{\frac {856}{35}}\,{q}^{3}
\ln v-4\,{\frac {{B_2}}{\kappa}}-32\,{\frac {{q}^{3}}{\kappa}}
\nonumber\\
&&
-31\,{\frac {q}{\kappa}}+57\,{\frac {{q}^{5}}{\kappa}}-48\,{\frac {{q}^{2
}{B_2}}{\kappa}}+28\,{\frac {{q}^{4}{B_2}}{\kappa}}-24\,{\frac {
{q}^{3}{C_2}}{\kappa}}-8\,{\frac {q{C_2}}{\kappa}}
\nonumber\\
&&
+24\,{\frac {{
q}^{6}{B_2}}{\kappa}}-{\frac {11883052}{99225}}\,{q}^{3}+{\frac {
548}{27}}\,{q}^{5}\Bigr){v}^{6}
\nonumber\\
&&
+\Bigl({\frac {16747}{126}}\,{q}^{2}
+{\frac {15893}{189}}\,{q}^{4}-{\frac {155}{126}}\,{q}^{6}+123\,\kappa
\,{q}^{2}+{\frac {1142}{21}}\,{q}^{4}\kappa-{\frac {8}{7}}\,\kappa\,{q
}^{6}
\nonumber\\
&&
+{\frac {199}{21}}\,\kappa+{\frac {199}{21}}+{\frac {796}{21}}\,q
{B_2}+{\frac {2372}{21}}\,{q}^{3}{B_2}-{\frac {16}{7}}\,{q}^{5}{
B_2}\Bigr){v}^{7}
\nonumber\\
&&
+\Bigl(-{\frac {761349}{3920}}\,q-12\,{B_2}+
{\frac {3076}{105}}\,\gamma\,q+2\,{\pi }^{2}q+{\frac {4868}{105}}\,q
\ln 2-12\,q{C_2}
\nonumber\\
&&
-36\,{q}^{3}{C_2}-{\frac {308}{3}}\,{q}^{4}{B_2}
-{\frac {1496}{9}}\,{q}^{2}{B_2}+{\frac {3076}{35}}\,{q}^{3}
\gamma+{\frac {4868}{35}}\,{q}^{3}\ln 2+6\,{q}^{3}{\pi }^{2}
\nonumber\\
&&
+{\frac {428}{35}}\,q\ln\kappa+{\frac {428}{35}}\,q{A_2}
+{\frac {1284}{35}}\,{q}^{3}\ln\kappa
+{\frac {46}{3}}\,{\frac {{q}^{7}}{\kappa}}+{
\frac {1284}{35}}\,{q}^{3}{A_2}-24\,q{{B_2}}^{2}
\nonumber\\
&&
-72\,{q}^{3}{{B_2}}^{2}
+{\frac {872}{21}}\,q\ln v+{\frac {872}{7}}\,{q}^{3}\ln 
(v)-12\,{\frac {{B_2}}{\kappa}}-{\frac {272}{3}}\,{\frac {{q}^{3}}{
\kappa}}-{\frac {833}{9}}\,{\frac {q}{\kappa}}
\nonumber\\
&&
+{\frac {1511}{9}}\,{
\frac {{q}^{5}}{\kappa}}-144\,{\frac {{q}^{2}{B_2}}{\kappa}}+84\,{
\frac {{q}^{4}{B_2}}{\kappa}}-72\,{\frac {{q}^{3}{C_2}}{\kappa}}
-24\,{\frac {q{C_2}}{\kappa}}
\nonumber\\
&&
+72\,{\frac {{q}^{6}{B_2}}{\kappa}}
-{\frac {140529967}{317520}}\,{q}^{3}+{\frac {46465}{756}}\,{q}^{5}-{
\frac {191}{588}}\,{q}^{7}\Bigr){v}^{8}. 
\\
&&\nonumber\\
\eta^H_{2,1}&&=
{v}^{2}\left ({3\over 16}\,{q}^{3}-{1\over 4}\,q\right )
+\left (-{1\over 4}\,{q}^{4}+{1\over 3}\,{
q}^{2}\right ){v}^{3} \nonumber\\
&&
+\left ({1\over 12}\,{q}^{5}+{\frac {8}{9}}\,{q}^{3}-{4\over 3}
\,q\right ){v}^{4}\nonumber\\
&&
+\left (-{\frac {137}{48}}\,{q}^{4}-{3\over 4}\,{q}^{3}{
B_1}+{\frac {265}{72}}\,{q}^{2}+q{B_1}
-{3\over 4}\,{q}^{4}\kappa+{1\over 2}\,
\kappa+{\frac {7}{8}}\,\kappa\,{q}^{2}+{1\over 2}\right ){v}^{5} 
\nonumber\\
&&
+\left (-{\frac {382}{63}}\,q
+{\frac {865}{756}}\,{q}^{3}+{\frac {2321}{1008}}\,
{q}^{5}-{2\over 3}\,\kappa\,q-{7\over 6}\,\kappa\,{q}^{3}
+{q}^{5}\kappa+{q}^{4}{
B_1}-{4\over 3}\,{q}^{2}{B_1}\right ){v}^{6} \nonumber\\
&&
+\Bigl({8\over 3}+{\frac {4643}{252}}
\,{q}^{2}-{\frac {32}{9}}\,{q}^{3}{B_1}
-{\frac {1405}{3024}}\,{q}^{6}
-{\frac {65}{18}}\,{q}^{4}\kappa-{\frac {123269}{9072}}\,{q}^{4} 
\nonumber\\
&&
+{\frac {44}{9}}\,\kappa\,{q}^{2}-{1\over 3}\,{q}^{5}{B_1}
-{1\over 3}\,\kappa\,{q}^{6}
+{16\over 3}\,q{B_1}+{8\over 3}\,\kappa\Bigr){v}^{7} \nonumber\\
&&
+\Bigl(-{\frac {6292409
}{176400}}\,q-2\,{B_1}+{\frac {107}{105}}\,\gamma\,q
+{1\over 6}\,{\pi }^{2}q
+{\frac {107}{105}}\,q\ln 2-{\frac {107}{140}}\,{q}^{3}\gamma 
\nonumber\\
&&
-{\frac {107}{140}}\,{q}^{3}\ln 2
-{1\over 8}\,{q}^{3}{\pi }^{2}+{\frac {107}{
105}}\,q\ln\kappa-{\frac {107}{140}}\,{q}^{3}
 \ln\kappa-{\frac {55}{6}}\,{\frac {{q}^{7}}{\kappa}}
+{\frac {214}{105}}\,q\ln v 
\nonumber\\
&&
-{\frac {107}{70}}\,{q}^{3}\ln v+{\frac {65}{12}}\,{\frac {{q}^{3}}{
\kappa}}-{\frac {245}{18}}\,{\frac {q}{\kappa}}+{\frac {625}{36}}\,{
\frac {{q}^{5}}{\kappa}}+{\frac {73}{6}}\,{q}^{4}{B_1}-q{C_1} 
\nonumber\\
&&
-{\frac {283}{18}}\,{q}^{2}{B_1}
+{3\over 2}\,{q}^{3}{{B_1}}^{2}+{\frac {
107}{105}}\,q{A_1}-{\frac {107}{140}}\,{q}^{3}{A_1}-3\,{\frac {{
q}^{6}{B_1}}{\kappa}}
+{13\over 2}\,{\frac {{q}^{4}{B_1}}{\kappa}} \nonumber\\
&&
-{3\over 2}\,{\frac {{q}^{2}{B_1}}{\kappa}}
-2\,{\frac {q{C_1}}{\kappa}}+{3\over 2}\,
{\frac {{q}^{3}{C_1}}{\kappa}}
-2\,q{{B_1}}^{2}+{3\over 4}\,{q}^{3}{
C_1}-2\,{\frac {{B_1}}{\kappa}} \nonumber\\
&&
+{\frac {381643}{6350400}}\,{q}^{3}+{
\frac {6439}{378}}\,{q}^{5}-{\frac {11}{168}}\,{q}^{7}\Bigr){v}^{8}. 
\\
&&\nonumber\\
\eta^H_{3,3}&&=
\left (-{\frac {75}{112}}\,{q}^{5}-{\frac {555}{896}}\,{q}^{3}-{\frac 
{15}{224}}\,q\right ){v}^{4}+\left (-{\frac {45}{112}}\,q-{\frac {1665
}{448}}\,{q}^{3}-{\frac {225}{56}}\,{q}^{5}\right ){v}^{6} \nonumber\\
&&
+\Bigl({\frac {225}{28}}\,{q}^{5}{B_3}+{\frac {375}{112}}\,{q}^{6}
+{\frac {15}{112}}+{\frac {1905}{448}}\,\kappa\,{q}^{2}
+{\frac {2055}{224}}\,{q}^{4}\kappa
+{\frac {15}{112}}\,\kappa \nonumber\\
&&
+{\frac {1665}{224}}\,{q}^{3}{B_3}
+{\frac {45}{56}}\,q{B_3}+{\frac {10995}{896}}\,{q}^{4}
+{\frac {2055}{448}}\,{q}^{2}\Bigr){v}^{7} \nonumber\\
&&
+\left (-{\frac {17697}{896}
}\,{q}^{3}-{\frac {18315}{896}}\,{q}^{5}+{\frac {125}{112}}\,{q}^{7}-{
\frac {481}{224}}\,q\right ){v}^{8}.
\\
&&\nonumber\\
\eta^H_{3,2}&&=
\left ({\frac {25}{189}}\,{q}^{5}-{\frac {5}{63}}\,q-{\frac {110}{567}
}\,{q}^{3}\right ){v}^{6} \nonumber\\
&&
+\left ({\frac {5}{42}}\,{q}^{2}+{\frac {55}{
189}}\,{q}^{4}-{\frac {25}{126}}\,{q}^{6}\right ){v}^{7} \nonumber\\
&&
+\left ({\frac {25}{336}}\,{q}^{7}-{\frac {40}{63}}\,q
-{\frac {14485}{9072}}\,{
q}^{3}+{\frac {205}{216}}\,{q}^{5}\right ){v}^{8} 
\\
&&\nonumber\\
\eta^H_{3,1}&&=
\left (-{\frac {1}{224}}\,q+{\frac {59}{8064}}\,{q}^{3}-{\frac {1}{336
}}\,{q}^{5}\right ){v}^{4} \nonumber\\
&&
+\left ({\frac {767}{12096}}\,{q}^{3}-{
\frac {13}{504}}\,{q}^{5}-{\frac {13}{336}}\,q\right ){v}^{6} \nonumber\\
&&
+\Bigl({
\frac {1}{84}}\,{q}^{5}{B_1}+{\frac {1}{56}}\,q{B_1}+{\frac {31}
{4032}}\,\kappa\,{q}^{2}-{\frac {55}{2016}}\,{q}^{4}\kappa+{\frac {1}{
84}}\,\kappa\,{q}^{6}+{\frac {109}{3024}}\,{q}^{6} \nonumber\\
&&
-{\frac {6395}{72576
}}\,{q}^{4}-{\frac {59}{2016}}\,{q}^{3}{B_1}+{\frac {1}{112}}+{
\frac {185}{4032}}\,{q}^{2}+{\frac {1}{112}}\,\kappa\Bigr){v}^{7} 
\nonumber\\
&&
+\left (-{\frac {1}{336}}\,{q}^{7}-{\frac {431}{2016}}\,q+{\frac {25105
}{72576}}\,{q}^{3}-{\frac {3271}{24192}}\,{q}^{5}\right ){v}^{8}
\\
&&\nonumber\\
\eta^H_{4,4}&&=
-{\frac {5}{2268}}\,{v}^{8}q\left (9+7\,{q}^{2}\right )\left (3\,{q}^{
2}+1\right )\left (15\,{q}^{2}+1\right ).
\\
&&\nonumber\\
\eta^H_{4,2}&&=
-{\frac {5}{63504}}\,{v}^{8}q\left (5\,{q}^{2}-9\right )\left (3\,{q}^
{2}-4\right )\left (3\,{q}^{2}+1\right ).
\end{eqnarray}

\section{$(dE/dt)_{\rm H}$ in terms of the orbital frequency}
\label{ap:absorpinx}
In this Appendix, we describe the absorption 
rate $\langle dE/dt\rangle_{\rm H}$ by a Kerr black hole in terms of 
$x\equiv(M\Omega_\varphi)^{1/3}$. 
Using the relation,
\begin{equation}
v=x\bigl(1+{1\over 3} q x^3+{2\over 9} q^2 x^6 +O(x^9)\bigr), 
\end{equation}
we have 
\begin{eqnarray}
\left( {dE\over dt}\right)_{\rm H}&&=
{32\over 5}\left(\mu\over M\right)^2 x^{10}\, x^5\, \Biggl[
-{1\over 4}\,q-{3\over 4}\,{q}^{3}
+\bigl (-q-{\frac {33}{16}}\,{q}^{3}\bigr ){x}^{2} \nonumber\\
&&
+\bigl (2\,q{B_2}+{1\over 2}+{13\over 2}\,
\kappa\,{q}^{2}+{\frac {35}{6}}\,{q}^{2}-{1\over 4}\,{q}^{4}
+{1\over 2}\,\kappa \nonumber\\
&&
+3\,{q}^{4}\kappa+6\,{q}^{3}{B_2}\bigr ){x}^{3} \nonumber\\
&&
+\bigl (-{\frac {43}{7}}\,q-{\frac {17}{56}}\,{q}^{5}-{\frac {
4651}{336}}\,{q}^{3}\bigr ){x}^{4}\nonumber\\
&&
+\bigl ({\frac {433}{24}}\,{q}^{2}-{\frac {95}{24}}\,{q}^{4}+2
-{3\over 4}\,{q}
^{3}{B_1}+2\,\kappa+{\frac {33}{4}}\,{q}^{4}\kappa+6\,q{B_2} \nonumber\\
&&
+18\,{q}^{3}{B_2}+{\frac {163}{8}}\,\kappa\,{q}^{2}
+q{B_1}\bigr){x}^{5} \nonumber\\
&&
+\bigl (-{\frac {2586329}{44100}}\,q-4\,{B_2}
-{\frac {1640747}{19600}}\,{q}^{3}+19\,{q}^{5}\kappa 
+{\frac {428}{105}}\,\gamma\,q 
+{2\over 3}\,{\pi }^{2}q \nonumber\\
&&
+{\frac {428}{105}}\,q\ln (2)-4\,q{C_2}
-12\,{q}^{3}{C_2}-44\,{q}^{2}{B_2}+{\frac {428}{35}}\,{q}^{3}
\gamma+{\frac {428}{35}}\,{q}^{3}\ln (2) \nonumber\\
&&
+2\,{q}^{3}{\pi }^{2} 
+{\frac {428}{105}}\,q\ln (\kappa)+{\frac {428}{105}}\,q{A_2}
+{\frac {428}{35}}\,{q}^{3}\ln (\kappa)
+6\,{\frac {{q}^{7}}{\kappa}} \nonumber\\
&&
+{\frac {428}{35}}\,{q}^{3}{A_2}-8\,q{{B_2}}^{2}
-24\,{q}^{3}{{B_2}}^{2}-4\,{\frac {{B_2}}{\kappa}}
-32\,{\frac {{q}^{3}}{\kappa}}-31\,{\frac {q}{\kappa}} \nonumber\\
&&
+57\,{\frac {{q}^{5}}{\kappa}}+{q}^{4}{B_1}
-{4\over 3}\,{q}^{2}{B_1}+{7\over 3}\,\kappa\,q
+{\frac {227}{6}}\,\kappa\,{q}^{3}
+{\frac {455}{16}}\,{q}^{5}\nonumber\\
&&
-48\,{\frac {{q}^{2}{B_2}}{\kappa}}
+28\,{\frac {{q}^{4}{B_2}}{\kappa}}
-24\,{\frac {{q}^{3}{C_2}}{\kappa}}
-8\,{\frac {q{C_2}}{\kappa}} \nonumber\\
&&
+24\,{\frac {{q}^{6}{B_2}}{\kappa}}
+{\frac {856}{105}}\,q\ln (x)
+{\frac {856}{35}}\,{q}^{3}\ln (x)\bigr ){x}^{6} \nonumber\\
&&
+\bigl ({\frac {19687}{168}}\,{q}^{2}
-{\frac {145}{336}}\,{q}^{6}-{\frac {4729}{1008}}\,{q}^{4}
+{\frac {899}{168}}\,q{B_1}
-{\frac {41}{28}}\,\kappa\,{q}^{6} \nonumber\\
&&
+{\frac {45}{56}}\,q{B_3}
-{\frac {803}{224}}\,{q}^{3}{B_1}
+{\frac {1665}{224}}\,{q}^{3}{B_3}
+{\frac {86}{7}}+{\frac {719}{12}}\,{q}^{4}\kappa \nonumber\\
&&
+{\frac {796}{21}}\,q{B_2}+{\frac {86}{7}}\,\kappa
-{\frac {16}{7}}\,{q}^{5}{B_2}+{\frac {225}{28}}\,{q}^{5}{B_3} \nonumber\\
&&
-{\frac {9}{28}}\,{q}^{5}{B_1}+{\frac {22201}{168}}\,\kappa\,{q}^{2}
+{\frac {2372}{21}}\,{q}^{3}{B_2}\bigr ){x}^{7} \nonumber\\
&&
\bigl (-{\frac {19366807}{88200}}\,q-12\,{B_2}-2\,{B_1}
-{\frac {2062220497}{6350400}}\,{q}^{3}-q{C_1}+55\,{q}^{5}\kappa \nonumber\\
&&
+{\frac {1061}{35}}\,\gamma\,q+{\frac {13}{6}}\,{\pi }^{2}q
+{\frac {995}{21}}\,q\ln (2)-12\,q{C_2}-36\,{q}^{3}{C_2} \nonumber\\
&&
+{\frac {52}{3}}\,{q}^{4}{B_2}-{\frac {1136}{9}}\,{q}^{2}{B_2}
+{\frac {12197}{140}}\,{q}^{3}\gamma
+{\frac {3873}{28}}\,{q}^{3}\ln (2) \nonumber\\
&&
+{\frac {47}{8}}\,{q}^{3}{\pi }^{2}
+{\frac {1391}{105}}\,q\ln (\kappa)+{\frac {428}{35}}\,q{A_2}
+{\frac {5029}{140}}\,{q}^{3}\ln (\kappa) \nonumber\\
&&
+{\frac {37}{6}}\,{\frac {{q}^{7}}{\kappa}}
+{\frac {1284}{35}}\,{q}^{3}{A_2}-24\,q{{B_2}}^{2}
-72\,{q}^{3}{{B_2}}^{2}-12\,{\frac {{B_2}}{\kappa}} \nonumber\\
&&
-{\frac {341}{4}}\,{\frac {{q}^{3}}{\kappa}}
-{\frac {637}{6}}\,{\frac {q}{\kappa}}
+{\frac {741}{4}}\,{\frac {{q}^{5}}{\kappa}} \nonumber\\
&&
+{\frac {43}{6}}\,{q}^{4}{B_1}
-{\frac {163}{18}}\,{q}^{2}{B_1}
+{3\over 2}\,{q}^{3}{{B_1}}^{2}+{\frac {107}{105}}\,q{A_1} \nonumber\\
&&
-{\frac {107}{140}}\,{q}^{3}{A_1}-2\,q{{B_1}}^{2}
+{3\over 4}\,{q}^{3}{C_1}-2\,{\frac {{B_1}}{\kappa}}
+{\frac {40}{3}}\,\kappa\,q+{\frac {815}{6}}\,\kappa\,{q}^{3} \nonumber\\
&&
+{\frac {1265}{18}}\,{q}^{5}+{\frac {25}{252}}\,{q}^{7}
-144\,{\frac {{q}^{2}{B_2}}{\kappa}}
+84\,{\frac {{q}^{4}{B_2}}{\kappa}}
-72\,{\frac {{q}^{3}{C_2}}{\kappa}}
-24\,{\frac {q{C_2}}{\kappa}} \nonumber\\
&&
+72\,{\frac {{q}^{6}{B_2}}{\kappa}}
-3\,{\frac {{q}^{6}{B_1}}{\kappa}}
+{13\over 2}\,{\frac {{q}^{4}{B_1}}{\kappa}} 
-{3\over 2}\,{\frac {{q}^{2}{B_1}}{\kappa}}
-2\,{\frac {q{C_1}}{\kappa}}
+{3\over 2}\,{\frac {{q}^{3}{C_1}}{\kappa}}\nonumber\\
&&
+{\frac {4574}{105}}\,q\ln (x)
+{\frac {8613}{70}}\,{q}^{3}\ln (x)\bigr ){x}^{8}\,.
\end{eqnarray}
 

\begin{thebibliography}{99}
\bibitem{ref:NP} E.T. Newman and R. Penrose, J. Math. Phys. 
{\bf 7}, 863(1966).
\bibitem{ref:BaPr} J.M. Bardeen and W.H. Press, J. Math. Phys. 
{\bf 14}, 7(1973). 
\bibitem{ref:Teu} S. A. Teukolsky, Astrophys. J. {\bf 185}, 635(1973).
\bibitem{ref:breuer} R. A. Breuer, {\it Gravitational Perturbation 
Theory and Synchrotron Radiation}, Lecture Notes in Physics 
{\bf 44}, 1 (1975), Springer Verlag.
\bibitem{ref:chandra} S. Chandrasekhar, {\it The Mathematical Theory of 
Black Holes}, (1983), Oxford University Press. 
\bibitem{ref:NOK} T. Nakamura, K. Oohara, and Y. Kojima, Prog. Theor. 
Phys. Suppl. {\bf 90}, (1987), Part I and II. 
\bibitem{ref:GMP} D.V. Gal'tsov, A.A. Matiukhin and V.I. Petukhov, 
Phys. Lett. {\bf 77A} 387 (1980). 
\bibitem{ref:poisson} E. Poisson, Phys. Rev. {\bf D47}, 1497(1993).
\bibitem{ref:CFPS} C. Cutler, L.S. Finn, E. Poisson and G.J. Sussman, 
Phys. Rev. D{\bf 47}, 1511 (1993).
\bibitem{ref:TN} H.Tagoshi and T. Nakamura, Phys. Rev. {\bf D49}, 
4016 (1994). 
\bibitem{ref:sasaki} M. Sasaki, Prog. Theor. Phys. {\bf 92}, 17(1994). 
\bibitem{ref:TS} H.Tagoshi and M.Sasaki, Prog. Theor. Phys. {\bf 92}, 
745 (1994).
\bibitem{ref:TTS}
T. Tanaka, H. Tagoshi, and M. Sasaki, Prog. Theor. Phys. {\bf 96},
1087 (1996). 
\bibitem{ref:poisson2} E. Poisson, Phys. Rev. {\bf D48}, Brief Reports, 
1860 (1993). 
\bibitem{ref:SSTT} M. Shibata, M. Sasaki, H. Tagoshi and T. Tanaka, 
Phys. Rev. {\bf D51}, 1646 (1995).
\bibitem{ref:TSTS}
H. Tagoshi, M. Shibata, T. Tanaka, and M. Sasaki, 
Phy. Rev. {\bf D54}, 1439 (1996).
\bibitem{ref:tagoshi} H. Tagoshi, Prog. Theor. Phys. {\bf 93}, 
307 (1995).
\bibitem{ref:TMSS} T. Tanaka, Y. Mino, M. Sasaki and M. Shibata, 
Phys. Rev. {\bf D54}, 3762 (1996). 
\bibitem{ref:papa} A. Papapetrou, Proc. Roy. Soc. Lond. {\bf A209}, 
243 (1951).
\bibitem{ref:dixon} W.G. Dixon, in {\it Isolated Gravitating Systems in 
General relativity}, pp156-219, ed. J. Ehlers, North-Holland (1979). 
\bibitem{ref:PoSa} E. Poisson and M. Sasaki, Phys. Rev.{\bf D51}, 
5753 (1995).
\bibitem{ref:galtsov} D.V. Gal'tsov, J. Phys. A. {\bf 15}, 3737 (1982). 
\bibitem{ref:MST}
S. Mano, H. Suzuki and E. Takasugi, 
Prog. Theor. Phys. {\bf 95}, 1079 (1996); Prog. Theor. Phys. 
{\bf 96}, 549 (1996); S. Mano and E. Takasugi, 
Prog. Theor. Phys. {\bf 97}, 213 (1997). 
\bibitem{ref:TMT} H.Tagoshi, S.Mano, and E.Takasugi, submitted to 
Prog. Theor. Phys. \,.
\bibitem{ref:OTS} A. Ohashi, H. Tagoshi, and M. Sasaki, 
Prog. Theor. Phys. {\bf 96}, 713 (1996). 
\bibitem{ref:PT} W.H. Press and S.A. Teukolsky,
Astrophys. J. {\bf 185}, 649(1973). 
\bibitem{ref:fackerell} E.D. Fackerell and R.G. Crossman, 
J. Math. Phys. {\bf 9}, 1849(1977).
\bibitem{ref:SN}
M. Sasaki and T. Nakamura, Prog. Theor. Phys. {\bf 67}, 1788(1982). 
\bibitem{ref:chandratrans} S. Chandrasekhar, Proc. R. Soc. London A 
{\bf 343}, 289 (1975). 
\bibitem{ref:spin} J. N. Goldberg, A.J. MacFarlane, E. T. Newman, 
F. Rohrlich, and E. C. G. Sudarshan, J. Math. Phys. {\bf 8}, 2155
(1967).
\bibitem{ref:RW} T. Regge and J.A. Wheeler, Phys. Rev. {\bf 108}, 
1063 (1957).
\bibitem{ref:Chan} S. Chandrasekhar, Proc. R. Soc. London {\bf A343}, 
289 (1975). 
\bibitem{ref:WagonerW} R.V. Wagoner and C.M. Will, Astrophys. J. 
{\bf 210}, 764(1976). 
\bibitem{ref:wisem} A.G. Wiseman, Phys. Rev. {\bf D48}, 4757 (1993).
\bibitem{ref:BDIWW}  L. Blanchet, T. Damour, B.R. Iyer, C.M. Will 
and A.G. Wiseman, Phys. Rev. Lett. {\bf 74}, L3515 (1995). 
\bibitem{ref:BDI} L. Blanchet, T. Damour and B.R. Iyer, 
Phys. Rev. {\bf D51}, 5360 (1995). 
\bibitem{ref:blanchet2} 
L. Blanchet, Phys. Rev. {\bf D54}, 1417 (1996)
\bibitem{ref:WW} C.M. Will and A.G. Wiseman, Phys. Rev. {\bf D54}, 
4813 (1996). 
\bibitem{ref:three}
C.~Cutler et al., Phys.~Rev.~Lett. {\bf 70} (1993), 2984. 
\bibitem{ref:poisson3} 
E.~Poisson, Phys.~Rev. {\bf D52} (1995), 5719.
\bibitem{ref:KWW} L.E. Kidder, C.M. Will and A.G. Wiseman, 
Phys. Rev. {\bf D47}, R4183 (1993). 
\bibitem{ref:Ryan} F.D. Ryan, Phys. Rev. {\bf D52}, 5707 (1995). 
\bibitem{ref:AKOP} T. Apostolatos, D. Kennefick, A. Ori 
and E. Poisson, Phys. Rev. {\bf D47}, 5376 (1993).
\bibitem{ref:peters} P. C. Peters, Phys. Rev. {\bf 136}, 1224 (1963).
\bibitem{ref:PM} P.C. Peters and J. Mathews, Phys. Rev. {\bf 131}, 435
(1963). 
\bibitem{ref:BS} L. Blanchet and G. Sch\"{a}fer, 
Mon. Not. R. astr. Soc. {\bf 239}, 845 (1989). 
\bibitem{ref:BS2} L. Blanchet and G Sch\"{a}fer, 
Class. Quantum Grav. {\bf 10}, 2699 (1993). 
\bibitem{ref:shibata} M. Shibata, Phys. Rev. {\bf D50} 6297 (1994). 
\bibitem{ref:membran} K.S. Thorne, R.M. Price and D. MacDonald, 
{\it Black Holes: The Membrance Paradigm} (Yale University Press,
New Haven, 1986). 
\bibitem{ref:Ryan2} F.D. Ryan, Phys. Rev. {\bf D53}, 3064 (1996).
\bibitem{ref:KenOri} D. Kennefick and A. Ori, Phys. Rev. {\bf D53},
4319 (1996).
\bibitem{ref:wald} R. Wald, Phys. Rev. {\bf D6}, 406(1972).
\bibitem{ref:MSTspin} Y. Mino, M. Shibata and T. Tanaka, 
Phys. Rev. {\bf D52}, 622 (1996).
\bibitem{ref:kidd} L. E. Kidder, Phys. Rev. {\bf D52} 821(1995).
\bibitem{ref:MT}
S.~Mano and E.~Takasugi, Prog.~Theor.~Phys. {\bf 97}, 213 (1997). 
\bibitem{ref:leaver} E.W.~Leaver, J. Math. Phys. {\bf 27}, 1238 (1986). 
\bibitem{ref:handbook}
M.~Abramowitz and I.A.~Stegun eds., 
{\it Handbook of Mathematical Functions}, Dover, New York (1972), 
Chapter 13. 
\bibitem{ref:TP} S.A.~Teukolsky and W.H.~Press, 
Astrophys. J. {\bf 193}, 443 (1974).
\bibitem{ref:press} W.H.~Press and S.A.~Teukolsky, 
Astrophys. J. {\bf 185}, 649(1973). 
\bibitem{ref:cam} W.B.~Campbell and T.~Morgan, Physica {\bf 53}, 
264(1971). 
\end{thebibliography}
\end{document}